\documentclass{article}
\usepackage{arxiv}

\usepackage[utf8]{inputenc} 
\usepackage[T1]{fontenc}    
\usepackage{hyperref}       
\usepackage{url}            
\usepackage{booktabs}       
\usepackage{amsfonts}       
\usepackage{nicefrac}       
\usepackage{microtype}      
\usepackage{amsmath}
\usepackage{cleveref}       
\usepackage{lipsum}         
\usepackage{graphicx}
\usepackage[numbers]{natbib}
\usepackage{doi}
\usepackage{dcolumn}
\usepackage{bm}
\usepackage{mathptmx}
\usepackage{etoolbox}
\usepackage{placeins}
\usepackage{xcolor}

\title{Machine Learning Visualization Tool for Exploring Parameterized Hydrodynamics}

\date{\today}

\newif\ifuniqueAffiliation
\uniqueAffiliationtrue

\ifuniqueAffiliation 
\author{ 
  C. F. Jekel\thanks{jekel1@llnl.gov} \\
  \And
  D. M. Sterbentz \\
  \And
  T. M. Stitt \\
  \And
  P. Mocz \\
  \And
  R. N. Rieben \\
  \And
  D. A. White \\
  \And
  J. L. Belof \\
  \And
  Lawrence Livermore National Laboratory, PO Box 808, Livermore, CA,94551, USA\thanks{This manuscript has been authored by Lawrence Livermore National Security, LLC under Contract  No. DE-AC52-07NA2 7344 with the US. Department of Energy. The United States Government retains, and the publisher, by accepting the article for publication, acknowledges that the United States Government retains a non-exclusive, paid-up, irrevocable, world-wide license to publish or reproduce the published form of this manuscript, or allow others to do so, for United States Government purposes. LLNL-JRNL-865692} \\
}
\else
\fi
\renewcommand{\headeright}{}
\renewcommand{\undertitle}{}
\renewcommand{\shorttitle}{Machine Learning Visualization Tool for Exploring Parameterized Hydrodynamics}

\hypersetup{
pdftitle={Machine Learning Visualization Tool for Exploring Parameterized Hydrodynamics},
pdfsubject={physics.flu-dyn, physics.comp-ph, cs.LG},
pdfauthor={C. F. Jekel, D. M. Sterbentz, T. M. Stitt, P. Mocz, R. N. Rieben, D. A. White, J. L. Belof},
pdfkeywords={Machine learning visualization, Ensemble of hydrodynamics, Ensemble visualization},
}

\begin{document}
\maketitle

\begin{abstract}
	We are interested in the computational study of shock hydrodynamics, i.e. problems
  involving compressible solids, liquids, and gases that undergo large deformation. These problems
  are dynamic and nonlinear and can exhibit complex instabilities. 
  Due to advances in high performance computing it is possible to parameterize a hydrodynamic
  problem and perform a computational study yielding $\mathcal{O}\left({\rm TB}\right)$ of simulation state 
  data. We present an interactive machine learning tool that can be used to compress, browse, 
  and interpolate these large simulation datasets. This tool allows computational scientists and 
  researchers to quickly visualize ``what-if'' situations, 
  perform sensitivity analyses, and optimize complex hydrodynamic experiments.
\end{abstract}

\section{Introduction}

For hydrodynamics, it can be very difficult to understand the sensitivity of physical instabilities to small perturbations in initial conditions. It is possible for a human to understand the impact of one or two inputs on the temporal evolution of an instability. However, as the number of system parameters grow, so does the complexity of the system. Consider the Rayleigh-Taylor instability (RTI) and Richtmyer-Meshkov instabilities (RMI) which can have several inputs that influence the transient state. In these cases, ensembles of simulations are required to understand the sensitivity of these instabilities with respect to their initial states. Often these simulation results are computationally expensive, and it is difficult for researchers to look at every simulation result. The Cinema project \cite{7013022} set out to aid researchers in understanding ensemble calculations by providing tools to quickly look through simulation results. The tools provided an intuitive interface for researchers to explore pictures of simulation results. While this tool is quite useful, the results are limited to only the performed calculations. Our work builds upon this concept of allowing researchers to quickly explore ensemble calculations, with the main advantage being the ability to quickly visualize results that were not previously calculated. This interpolation is accomplished with a machine learning (ML) model that allows a user to view the temporal evolution of instabilities by seamlessly changing initial conditions.

The RTI and RMI are closely related. A RTI occurs at the interface of two fluids mixing with different densities. A RMI occurs when a shock wave amplifies perturbations at a material interface, causing large jet-like growths \cite{PhysRevE.99.053102SHORT,PhysRevLett.104.135504,doi:10.1063/1.4971669,buttler_2012}. The use and understanding of the transient behavior of these instabilities is important in many applications. For example, experimentally measuring RMI formations is useful for calibrating high strain rate material models \cite{PhysRevLett.107.264502,doi:10.1063/1.3686572}. Additionally, in inertial confinement fusion (ICF) experiments, where lasers are used to heat and compress a fuel capsule to the point of starting a self-sustaining fusion reaction \cite{Zylstra2022}. Unfortunately, RMI have been known to form within ICF capsules. The RMI often destroys the fuel target before fusion ignition is achieved \cite{DESJARDINS2019100705}. Increasing our ability to design and control RMI would have profound impacts in fusion research.

Our ability to perform simulations with the onset of exascale computing\cite{10.1007/978-3-031-32041-5_10} may exceed our ability to comprehend those results. Several tools, for instance Merlin\cite{peterson2022enabling}, currently exist to orchestrate millions of simulations with high performance computing (HPC). Typically,  a researcher processes simulation results one at a time. The tool Cinema \cite{7013022} can be used to quickly swap between images of the ensemble of simulation results, but is limited to only those results. Our work proposed to use ML to interpolate and quickly infer from datasets of several simulations, while still allowing researchers to quickly build their own intuition for these complex systems.

There happen to be several applications of ML work applied specifically to the flow around airfoils. These simulations are typically 2D and are utilized to understand the flow properties (e.g. lift, drag) of the airfoil. \citet{WU2020104393} and \citet{WU2022470} trained General Adversarial Networks (GAN) to airfoil flow fields. The airfoil shape was represented as a vector of 14 values, which the generator model then learned the mapping from the parameterized geometry to the CFD solution. For more detail on GANs we refer the reader to the review by \citet{Chakraborty_2024}. \citet{LI2023108398} used the same methodology to learn force fields of a hypersonic vehicle from parameterized flight conditions. \citet{WANG2023105738} then goes on to demonstrate how transfer learning can be used on subsequent datasets in a related domain. \citet{hariansyah2022deep} used Deep Convolutional Generative Adversarial Networks (DCGAN) model to produce fake airfoils in an optimization. \citet{nandal2023synergistic} trained a GAN model to pressure fields around parameterized airfoils. \citet{doi:10.1080/19942060.2022.2030802} predicted hydrodynamic solutions around a submarine using a time series of previous fluid flow states. A prediction at an arbitrary time requires iteratively feeding past ML predictions into the model. \citet{10.1063/5.0063904} used neural networks to predict the permeability from 2D or 3D point clouds of the boundary surface of porous media. These studies all had very specific applications in mind, and our work will build upon these methods to show how to approach a general hydrodynamics problem.

It is also important to mention that ML has not been limited to just simulation results, as there has been recent work applying to experiments. \citet{10.1063/5.0140624} trained a neural network to predict how a flame would develop within a scramjet from pictures of the previous states of the flame. \citet{10.1063/1.5124133} trained a system of neural networks to learn the mapping of pressure sensors to images of the shockwave structure. The structure of the network also follows the DCGAN model. \citet{10.1121/10.0016896} used GANs for the reconstruction of acoustic fields. The work shows that the potentially machine learning can be used to enhance bandwidth limited acoustic sensors by being able to recover some lost sound energy at high frequencies.

A number of the previous literature utilized the Deep Convolutional Generative Adversarial Networks (DCGAN)\cite{DBLP:journals/corr/RadfordMC15} architecture on physics based datasets. \citet{akkari2020deep} trained DCGAN models to CFD solutions of flow fields around a parameterized square obstacle. Instead of using the parameterized position of the obstacle, the DCGAN model learned an unsupervised latent representations of the obstacle placement.  \citet{CHENG2020113000} used a DCGAN to predict fluid flow solutions. \citet{10.1063/5.0082562} trained a DCGAN model to fluid flow around a cylinder and a low pressure turbine stator, with the goal focusing on the model being able to quickly generate several synthetic fluid flows.

Our work will also utilize the DCGAN model architecture as it is fast in inference, easy to train, and simple to construct. It also works well for predicting multiple 2D physical fields. The focus of this work is largely limited to 2D simulations to demonstrate how the real-time tool can be used to intuitively explore the complex transient space of these instabilities. \citet{GraneroBelinchon_2024} used GANs to investigate 1D stochastic fields for multiscale physics of turbulence, and expressed a desire to extend their methods to 2D. 3D ML methods are currently a bit more expensive in inference time. For 3D applications, we refer the reader to the work of \citet{https://doi.org/10.1111/cgf.13620}, who used an autoencoder to compress 3D Eulerian solutions of dropping liquid and buoyant smoke. Then a LSTM network was used to learn concurrent time sequences in the compressed latent space. The two models working together allowed for the generation of quick 3D fluid predictions. 

The generic multi-material hydrodynamic problem of interest consists of materials (in the solid, liquid, or gas state) that undergo large deformations. These problems are driven by high velocity impacts or rapid energy deposition e.g. chemical explosives or incident laser pulses. These problems are dynamic and nonlinear and exhibit complex behavior such as RTI and RMI. A key aspect of this work is studying the control, e.g. enhancement or suppression, of these hydrodynamic instabilities, while most of the previous literature applied ML to steady-state fluid dynamics.

The outline is as follows:  Section \ref{sec:Hydrodynamic Simulations} reviews the hydrodynamic simulation methodology, Section \ref{sec:hydroexamples} reviews specific applications (high-velocity impacts, RTI mixing, linear shaped charges), and Section \ref{sec:machinelearning} details the machine learning approach. The key results are discussed in Section \ref{sec:results}, and ``screenshots'' of the real-time interactive visualization tool  are shown in Section \ref{sec:visualization}.  Finally, we will discuss some of the limitations of the approach.

\section{Hydrodynamic simulations}\label{sec:Hydrodynamic simulations}
\begin{figure*}
  \includegraphics[width=1.0\textwidth]{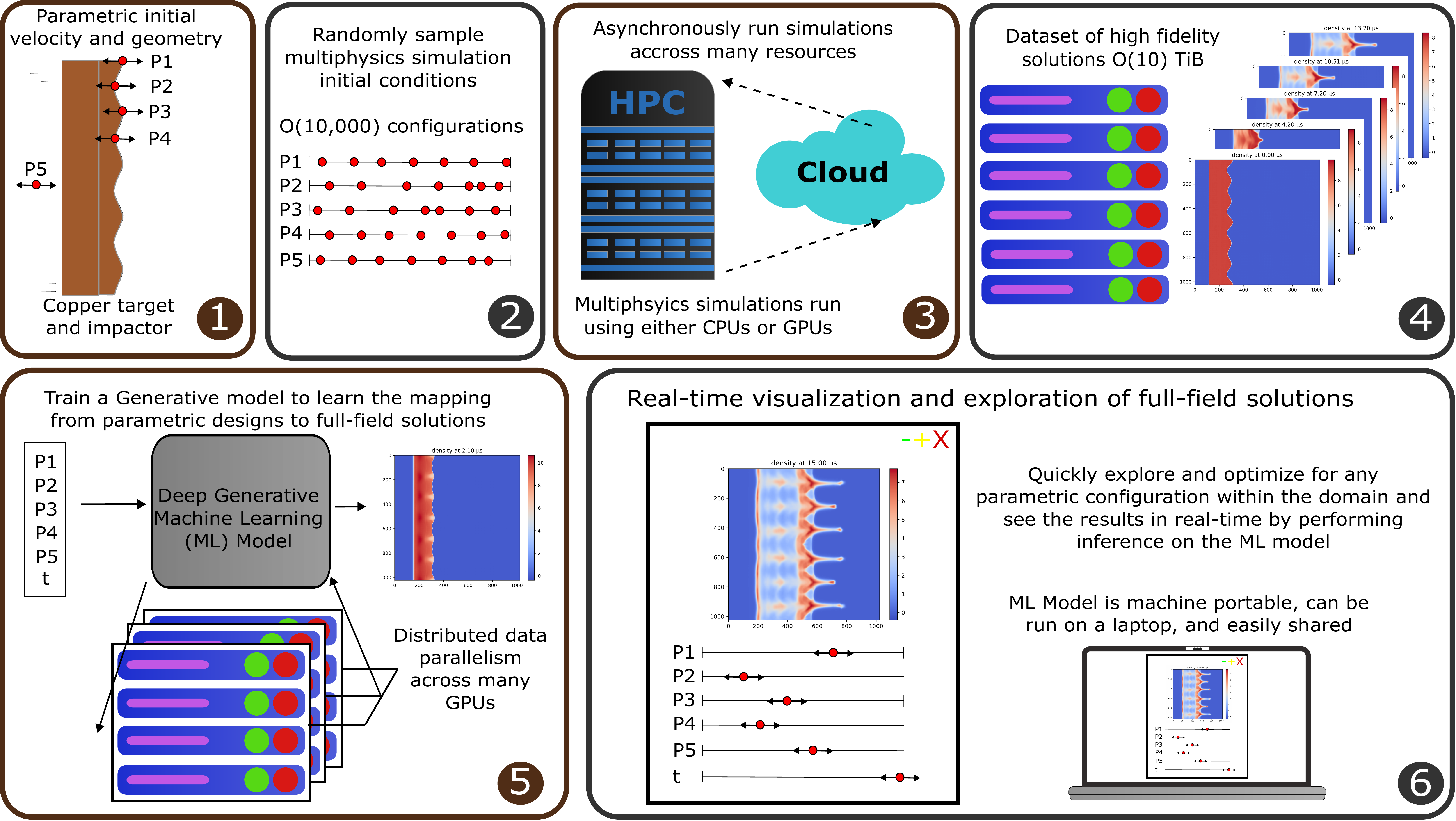}
  \caption{\label{fig:overal_method}This is an overview of the proposed method. 1) A hydrodynamic simulation is parameterized to study the effect of these parameters on the resulting instabilities. 2) These parameters are randomly sampled. 3) Simulations are performed on HPC using an asynchronous queue. 4) This yields a large dataset of full-field hydrodynamic solutions. 5) A generative machine learning model is then trained to learn the temporal mapping from the parameters to the full-field solutions. 6) This machine learning model can be used for real-time visualization, dissemination of results, optimization, and more.}
\end{figure*}

A general overview of the methodology used to create the machine learning
model is shown in Figure~\ref{fig:overal_method}. The key steps are: (1) parameterize initial conditions of a hydrodynamic problem, (2) define sample points in parameter space for which to perform high-fidelity hydrodynamic simulations, (3) asynchronously run the independent hydrodynamic simulations on large computational resources, (4) create a dataset of the full-field solutions, which is typically $O\left(TB\right)$ of data, (5) train a generative deep convolutional neural network to ``learn'' the hydrodynamic response. The final step (6) is to use the ML model to quickly predict time dependent full-field results for arbitrary parameters, including parameters \emph{not included} the training space.

The hydrodynamic model consists of the Euler Equations 
(conservation of mass, momentum, and energy) combined with equation of state formula and 
strength models such as Steinberg-Guinan\cite{doi:10.1142/S0217979208046396}. The initial condition consists of the geometry of
the materials and the initial density, velocity, and internal energy. Boundary conditions such as wall,
periodic, or outflow are enforced. Source terms such as energy deposition may be applied.
The Euler Equations are solved using an Arbitrary Lagrangian-Eulerian (ALE) Finite Element Method (FEM). The Euler Equations are integrated in time
using explicit conditionally stable Runge-Kutta with adaptive time stepping. At fixed intervals, e.g.
1 $\mu$s, the fields are projected from the high order ALE-FEM mesh onto a uniform Cartesian mesh, i.e. an image, and
are recorded to disk. A single hydrodynamics simulation thus results in a sequence of images of $l$ physical fields (e.g. density, velocity, energy) 
at fixed snapshots of time. The images are of a constant $n \times m$ size, and
we have $k$ temporal instances of these images. Each value in the image arrays is a single precision float.

A hydrodynamic study consists of a hydrodynamic problem with an initial condition that is parameterized 
via $d$ real-valued parameters. Clearly these parameters must have bounds and it is assumed these
parameters can be normalized to the interval $\left[ 0,1 \right]$. The overall goal is to construct an accurate
and compact model of the form 
\begin{equation}\label{eq:mlmodel}
f_{ML} : R_\square^{d+1} \rightarrow R^{l \times n \times m}
\end{equation}
where $R$ is the set of real numbers in a specified dimension.
The input space consists of the $d$ initial condition parameters plus time (where time has also been normalized to $\left[ 0,1 \right]$.
The output consists of images
of hydrodynamic fields e.g. density, velocity, energy. The model function $f_{Ml}$ has millions of free
parameters, the optimal value of these parameters are ``learned'' from simulation data without
knowledge of the Euler Equations, material properties, etc.

The machine learning tool that we have developed has three major advantages that have the potential to revolutionize the way analysis and postprocessing of large sets of hydrodynamic simulations.
One key point is that the size of this model (in bytes) is many orders of magnitude smaller than the raw image data, this enables
efficient archiving and sharing of large ensembles of hydrodynamic studies with colleagues. A second key point is that
the model can be evaluated at some new point in parameter space many orders of magnitude faster
than performing a new hydrodynamic simulation, enabling a researcher to quickly visualize a hydrodynamics
simulation that was not actually performed. A third point is that since by construction the model
 $f_{Ml}$ is continuous, the derivative can be computed (exactly, using the automatic differentiation
capability of ML libraries) enabling sensitivity analysis and inverse design optimization.
The architecture of the model function $f_{Ml}$, and the
training process, is described in more detail in Section \ref{sec:mchinelearning}.

The hydrodynamic simulations were performed using an Arbitrary Lagrangian-Eulerian (ALE) Finite Element Method (FEM) simulation code called MARBL. Generally speaking, Lagrangian formulations allow
for precise tracking of material interfaces, but the computational mesh can become
excessively distorted. On the other hand, Eulerian methods do not attempt to maintain sharp
interfaces between material interfaces, and materials are allowed to mix. In the ALE
method the simulation begins as pure Lagrangian but at later time allows mesh relaxation and remap 
as the material deformation
becomes large, the flow becomes turbulent, or material mixing occurs. A particular feature of MARBL 
is higher order elements
\cite{marbl2011},\cite{marbl2012},\cite{marbl2018}, this allows for greater accuracy
for a given mesh, and results in high efficiency on Graphical Processing Units (GPU) due to high ratio
of flops per mesh element. Other considerations include the need for high order artificial viscosity 
\cite{marbl2009}
and of course the important ALE remap step \cite{marbl2015}. 
The Livermore Equation of State Library (LEOS) \cite{LEOS} is used for the equation of state for
all materials, and Steinberg-Cochran-Guinan strength model \cite{Steinberg80} is used for
solid materials.
The MARBL simulation code is not in the public domain, but a more limited
high-order Lagrangian-FEM code restricted to ideal gases and single materials, named Laghos \cite{laghos}, is publicly available.

\section{Parameterized hydrodynamic examples}\label{sec:hydroexamples}

Four parameterized hydrodynamic problems are presented here. An ensemble of simulations is performed to understand the influence of the parameters on the complex time-dependent instabilities.  Latin Hypercube Sampling \cite{Viana16} (LHS) was used to generate samples from the bounded parameters. The workflow utility Merlin\cite{peterson2022enabling} was used to manage the ensemble of simulations. The physics simulations were either run in a single large allocation, or multiple smaller allocations. Merlin executes the simulations asynchronously as soon as either resources became available, or a simulation completed. The studies were performed on the LLNL Lassen\footnote{Lassen is a HPC with 795 number of nodes. Each node contains 2 power 9 ppc64 CPU sockets and 4 NVidia 16 GB V100 GPUs. \protect\url{https://hpc.llnl.gov/hardware/compute-platforms/lassen}} HPC.

An overview of the datasets generated is shown in Table~\ref{tab:tabdata}. The full-field solutions were stored as float 32 data using hdf5~\cite{folk2011overview}. All datasets were on order of a hundred billion pixels, with the Rayleigh-Taylor dataset having the fewest number of pixels at 120 billion. Lossless compression with gzip was used to reduce storage requirements of the data. The Rayleigh-Taylor dataset is 328~GB on disk. 

\begin{table}
  \center
  \caption{\label{tab:tabdata}An overview of the datasets created.}
  \begin{tabular}{lccccc}
    Name & \# ~of sims. & Time Steps & Fields & Pixels & \#~of pixels (millions) \\ 
    \hline
    PCHIP impact & 2,985 & 51 & 6 & 1024 x 1024 & 957,779 \\
    Double sine wave & 1,626 & 51 & 3 & 1024 x 1024 & 260,862 \\
    Linear shaped-charge & 2,299 & 41 & 7 & 512 x 1664 & 566,153 \\
    Rayleigh-Taylor & 2,000 & 51 & 6 & 768 x 256 & 120,324 \\
  \end{tabular}
\end{table}




\subsection{High velocity impact study}\label{sec:imapct}

\begin{figure*}[h]
  \center
  \includegraphics[scale=0.5]{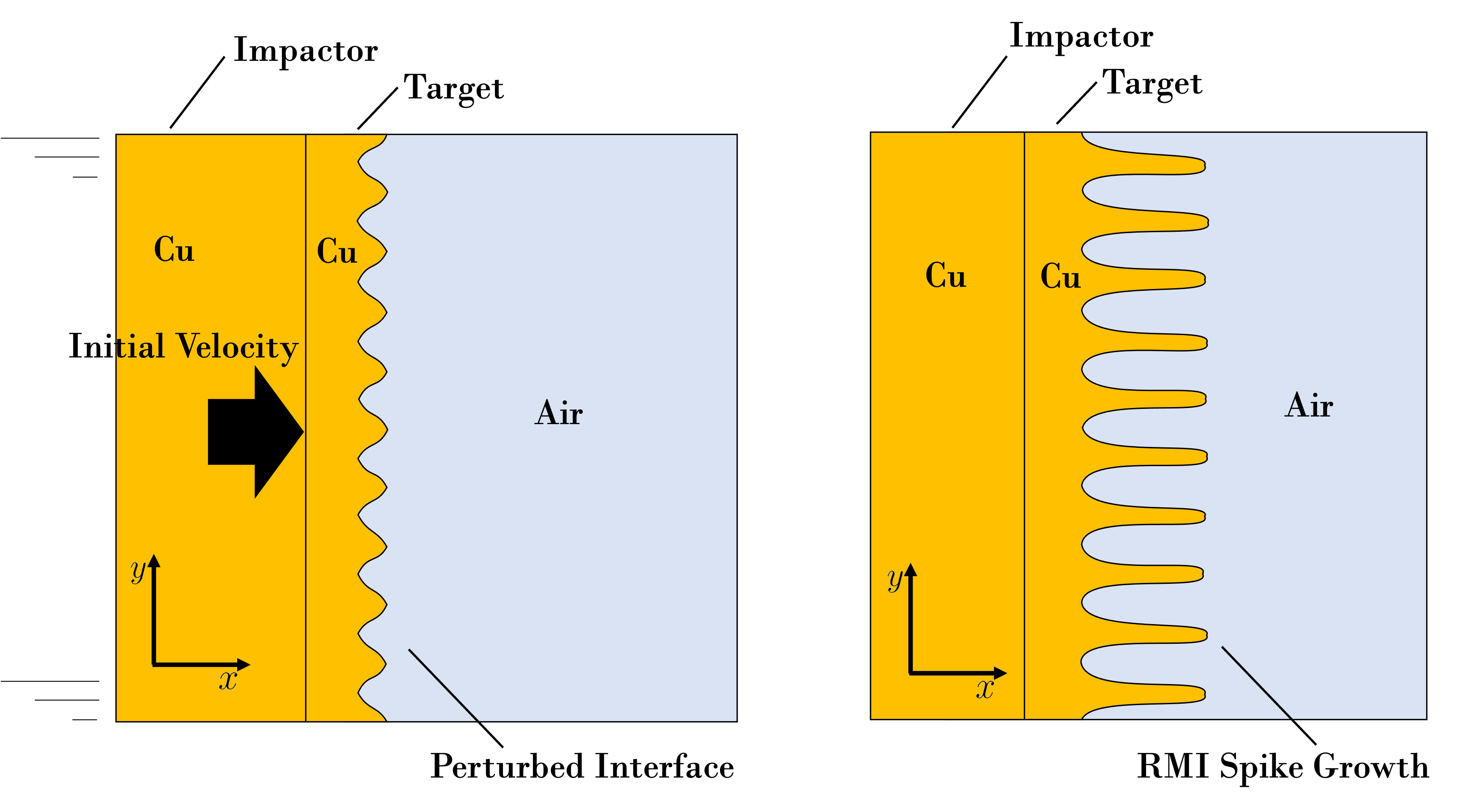}
  \caption{\label{fig:rmi1}Illustration of the high velocity impact. The left figure
is at $t=0$, the right figure is at some later time i.e. $t=10 \mu s$.}
\end{figure*}

The high velocity impact studies consist of an initially stationary copper target with a 
 perturbation machined
into the right-hand side (the copper-air interface), and a copper impactor with velocity of $2 km/s$.
As the shock wave reaches the interface perturbations, vorticity deposition occurs along the interface due to misalignments between pressure and density gradients at the perturbations. This creates the RMI that generally results in the jetting of the copper target material.
The impactor is $1\times 9$~cm and the target is nominally $0.5\times 9$~cm. These dimensions and velocities are chosen to be compatible with the
two-stage gas gun at LLNL's High Explosive Applications Facility (HEAF)\cite{armstrong2022use,nguyen2022modulation}. Whereas in actual HEAF
experiments the impactor/target are circular with 9cm diameter, the simulations are 2D. 
The simulations begin at time $t=0$ with the impactor and target in contact with a discontinuous velocity. An example figure of one of these impacts 
is shown in Figure~\ref{fig:rmi1}.

Two different parameterizations of these high velocity impacts were studied to investigate the formation of Richtmyer-Meshkov
Instabilities (RMI) that occur at the target-air interface, as small perturbations evolve
into a sharp jet. The first study is the PCHIP impact which looks at how changes in the perturbation in the target-air interface influence the RMI. The second study is the double sine wave study, which looks at how a sine wave can be placed on the impactor side of the target to influence RMI, when there is a fixed sine wave along the target-air interface. 

\subsubsection{PCHIP impact}

\begin{figure*}
  \center
 \includegraphics[scale=0.5]{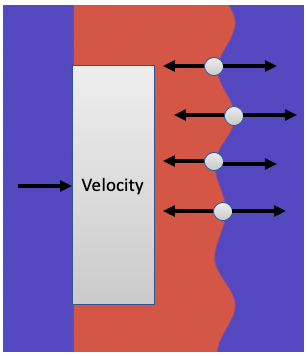}
 \caption{\label{fig:pchip_rhs}An illustrative depiction of the parameterized PCHIP impact.}
\end{figure*}

An example of the experimental setup of the PCHIP impact is shown in Figure \ref{fig:pchip_rhs}. The target-air interface is parameterized with four parameters
defining a Piecewise Cubic Hermite Interpolating
Polynomial (PCHIP) \cite{10.1063/5.0100100}. The PCHIP parameter range was $[-0.25, 0.25]$~cm.

The hydrodynamic solutions were computed using a nominal mesh of $144\times 144$ quadratic ($Q_2 Q_1$) elements, the mesh was morphed to have conformal interfaces between air and copper.
During the simulation the fields density, velocity $x$, velocity $y$, energy, pressure, and the material indicator are projected onto a $1024 \times 1024$ Cartesian image and exported at 51 uniform timesteps from 0 to 15~$\mu$s.



\subsubsection{Double sine wave}


The other high velocity impact involved sinusoidal waves on both sides of the impactor. The impactor side of the target was parameterized with the sinusoidal wave

\begin{equation}
  B \cos \bigg ( \frac{2\pi Q x - s\pi}{9.0} \bigg)
\end{equation}

to seed initial RMI growths. The free side of the target (copper-air interface) utilized a fixed wave of

\begin{equation}
 0.5 + 0.1 \cos \bigg ( \frac{2\pi 10 x}{9.0} \bigg)
\end{equation}

which also seed RMI growth. The purpose of this parameterization was to study how parameterized RMI growths interact with a known RMI seed (on the free side), the notion is that an optimized sinusoidal perturbation can initiate vorticity that ``cancels out'' the primary RMI. The following bounds were placed on the three parameters of the impactor side wave: $B$ from $[0.1, 0.25]$, $Q$ from $[5.0, 25.]$, and $s$ from $[0.0, 3.14]$.

\begin{figure*}
  \center
  \includegraphics[scale=0.3]{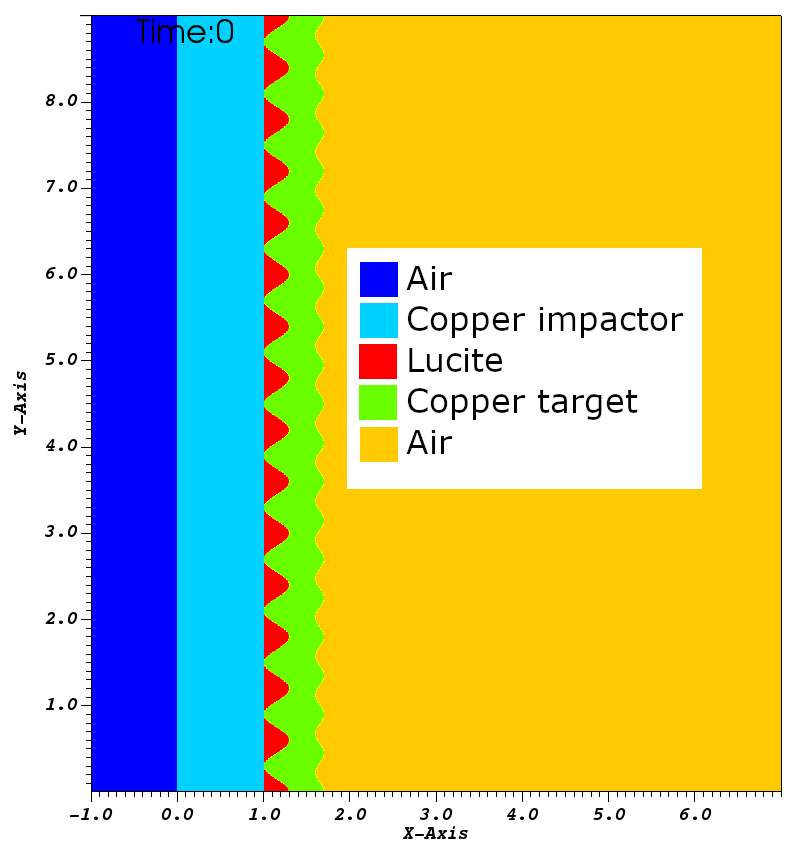}
  \caption{\label{fig:dswave}This shows the geometry and materials of the double sine wave impact.}
\end{figure*}

The copper impactor was $1\times 9$~cm and traveling at 2~km/s. Lucite was used to fill in the material between the target and the impactor, creating a flush interface for an initial impact. The simulations were ran out for 7~$\mu$s after the initial impact. An overview of the simulation setup is seen in Figure~\ref{fig:dswave}.

The solutions were computed using quadratic ($Q_2 Q_1$) elements on a nominal grid of $144\times 144$. The mesh was morphed to have conformal interfaces. The simulation results were saved on a $1024 \times 1024$ uniform grid for the following fields: density, velocity $x$, velocity $y$. These fields were exported along 51 ideally uniform timesteps from 0 to 7~$\mu$s.


\subsection{Linear shaped charge study}\label{sec:shapedcharge}

Linear shaped charges have a variety of industrial use cases such as structural demolition \cite{Duan17},
geo-engineering \cite{Sher17}, \cite{Wu21}, and aerospace \cite{Novotny07}. They utilize an explosive which propels a liner (typically copper) into a high velocity jet that will penetrate or cut into materials. Like the previous high velocity impactor study, the shaped charge jet is initiated by RMI. Design exploration of shaped charge jet formation can be non-intuitive and require many thousands of hydrodynamic simulations to explore the parameter space \cite{10.1063/5.0156373}. As a case study, this work proposes a parameterized linear shaped charge design involving liner shape and detonator locations.

\begin{figure*}[h]
  \includegraphics[scale=1.0]{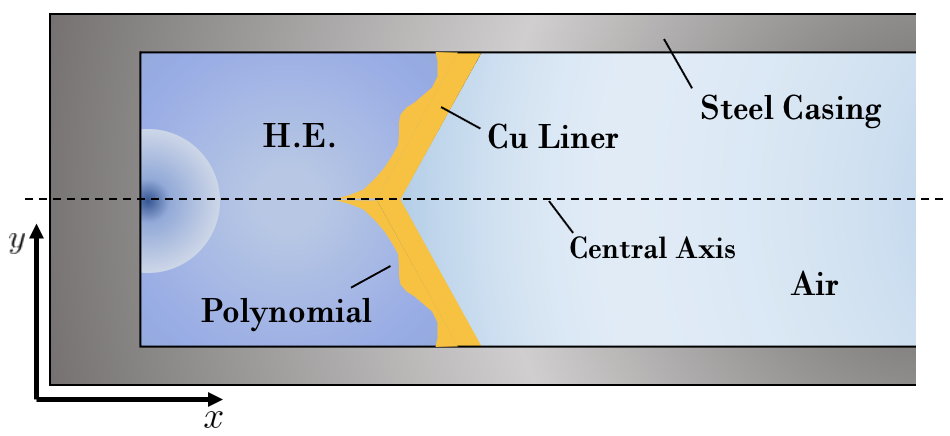}
  \caption{\label{fig:shapedcharge}The planar shaped charge in this study consists of a steel case,
explosive, and a copper liner. One face of the copper liner is fixed, the other face is parameterized
with a polynomial; the polynomial coefficients can be optimized to enhance performance.}
\end{figure*}

An overview of the parameterized linear shaped charge is shown in Figure~\ref{fig:shapedcharge}. The shape
of the copper liner facing the explosive is parameterized with four spline parameters ranging from $[0.05, 0.3]$~cm, while jet side of the liner is fixed at a $60^\circ$angle. One additional parameter controls the location of two detonation points, which are kept symmetric about the center of the liner. The detonator location was parameterized along the steel case, where 0 represents a placement along the center of the liner, and 1 represents a placement against the liner.

The linear shaped charge hydrodynamic simulations used a mesh of 11,000 quadratic ($Q_2 Q_1$) elements. The full field solutions were saved on a $512 \times 1664$ uniform grid. The results contain the following fields: density of the liner, density, velocity $x$, velocity $y$, energy, pressure, and the volume fraction of the liner. The simulations were run from initial detonation to t=20~$\mu$s, and results were saved every 0.5~$\mu$s. The high explosive was modeled using the Cochran-Tarver ignition and growth reactive flow model.

\subsection{Rayleigh-Taylor study}\label{sec:RT}


The final study is a single-mode Rayleigh-Taylor instability (RTI). The setup for the initial RTI was based off of the example in Athena$++$\cite{Stone2020} and the work of \citet{doi:10.1137/S1064827502402120}. The problem has a $x$ domain of $[-1/6, 1/6]$~cm and a $y$ domain of $[-0.5, 0.5]$~cm. A heavier ideal gas is placed on-top of a lighter ideal gas. An initial velocity was applied in $y$ direction to seed the instability growth as
\begin{equation}
  v_y = v_{\text{init}}(u_0*(1+\cos(6\pi x))(1+\cos(2\pi y))/4) 
\end{equation}
with $u_0=0.01$~cm/s. There is a constant gravitational acceleration of 1.0~cm/s$^2$.


The simulations were parameterized for three physical parameters: density ratio of the two gases, the heat capacity ratio of $\gamma$ for both gases, and the initial velocity of $v_\text{init}$. The parameters were randomly sampled in the following ranges respectively: $[1.1, 6.7]$, $[1.1, 1.6]$, and $[0.6, 10.0]$.

It is well known that this RTI problem becomes turbulent with decreasing feature size as time progresses.
Thus, the simulation based on the Euler equations is never fully resolved. Our solutions were computed on a $192\times 64$ grid using cubic ($Q_3 Q_2$) elements.

The full field solutions were saved on a $768 \times 256$ uniform grid. The results contain the following fields: density, velocity $x$, velocity $y$, energy, pressure, and the materials. The simulations were run to 10~s, and results were saved every 0.2~s (with 51 total snapshots per simulation).



\section{Machine learning}\label{sec:machinelearning}

The purpose of the proposed Machine Learning (ML) model is to learn the mapping Eq \ref{eq:mlmodel} which maps the parameterized simulation to 2D image arrays of physics fields. The ML architecture originates from a generative model based on deep, but sparse, convolutional layers. The typical application of these deep convolutional networks is in unsupervised learning of pictures e.g. human faces\cite{DBLP:journals/corr/RadfordMC15}. In 
these applications, the faces have no labels and no
known \emph{a-prior} parameterization. In our work, neural network layers are added to the generative model to connect the simulation parameters to the layers that have the smallest image representation. This allows us to take just about any generative model architecture and perform regression to learn the parameterized simulation solutions.

The DCGAN \cite{DBLP:journals/corr/RadfordMC15} generator architecture is perhaps one of the simplest generative models to learn the mapping of parameters to full-field solutions. The models require only a couple milliseconds for inference which is approximately a million times faster than a full hydrodynamic solution. The model uses transposed convolution layers (sometimes also called inverse convolutional, or deconvolutional \cite{odena2016deconvolution}). The first transposed convolutional layer creates an initial kernel representation (e.g. $4\times4$ pixels) of the entire field. Then, each subsequent transposed convolutional layer doubles the previous layer's full field representation (e.g.$4\times4 \to 8\times8$). These layers can be stacked until the output is the correct size of the final images.

Our use of the DCGAN model differs in a couple key characteristics from \cite{jekel2022using}. We discovered that we could achieve superior accuracy by using the image channels (normally used for e.g. red, green, blue) to learn multiple physical fields (e.g. density, velocity, pressure) rather than using an entirely separate generator models for each physical field. The model takes advantage of the many correlations between fields. Each physical field has different units that may be orders of magnitude different, which would cause issues with the transposed convolution layers. To address the imbalance of units within the physical fields, we propose the model learn the following linear mapping as the final layer,
\begin{equation}
  \alpha \bm{X} + \beta
\end{equation}
where $\alpha$ and $\beta$ are vectors that are each the size of the number of $l$ physical fields. This allows to output the physical fields of different orders of magnitude into the correct units, which would be important when applying physics informed constraints\cite{HANSEN2024133952}.

\begin{figure}[!htb]
    \centering
        \includegraphics[width=1.0\textwidth]{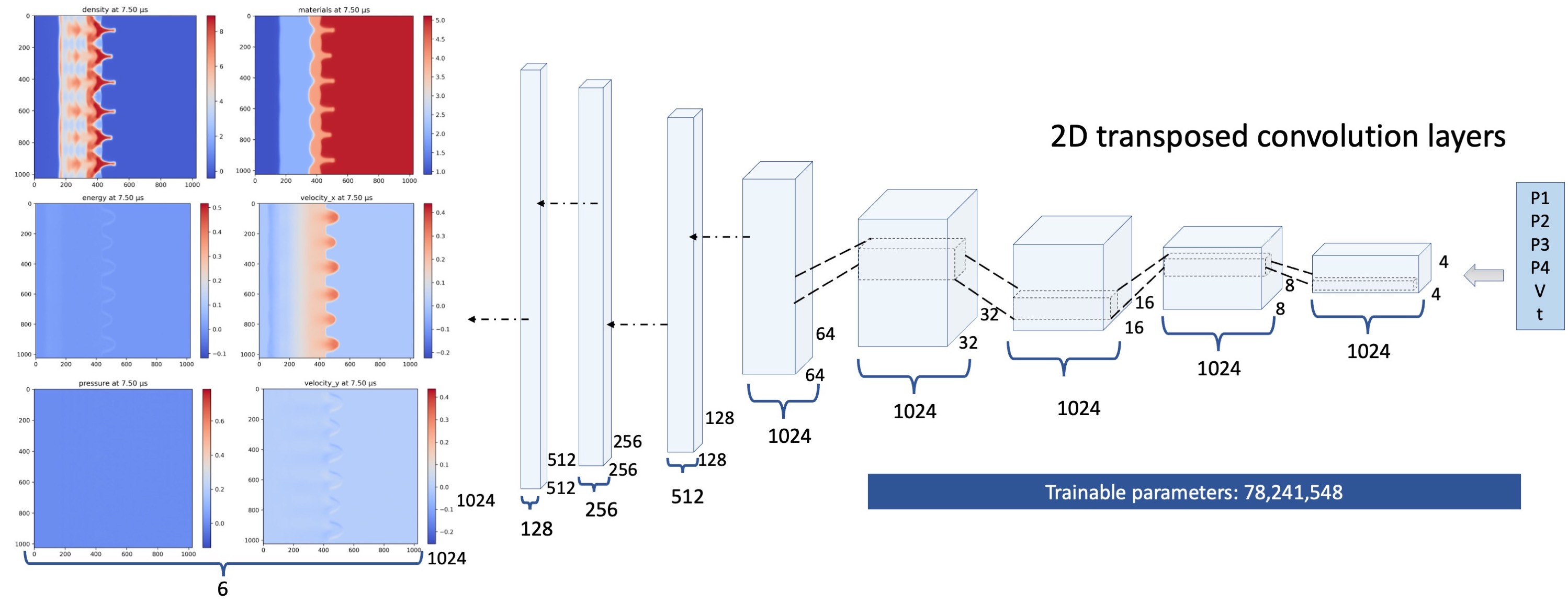}
    \caption{
        The convolution neural network architecture of the generator model starts with a $4\times 4$ kernel (right). Each subsequent layer doubles pixels until the final $1024\times 1024$ fields are created (left). This is actual ML output for the PCHIP impact, showing fields of density, energy, pressure, materials, velocity~$x$, and velocity~$y$.
        }
    \label{fig:cnnarch}
\end{figure}

A visual overview of the model architecture for the PCHIP impact is shown in Figure~\ref{fig:cnnarch}, and ML models for the other datasets was similar. The initial kernel was $4\times 4$, and the number of channels ranged from 1024 down to the final 6 fields. Each transposed convolution layer is followed by batch norm \cite{ioffe2015batch} and ReLU activations \cite{nair2010rectified}. The layer by layer\footnote{The PyTorch layers, shapes, and learnable parameters are outputs of a packaged called torchinfo available online at \protect\url{https://github.com/TylerYep/torchinfo} } breakdown with learnable parameters for each ML model is shown in Appendix~\ref{app:modelarch}. All models generally have around 80~million learnable parameters.

The number of features in the model were limited to between 128 through 1024 (in the dimension where RGB is typically used in images). Reducing to 64 through 512 reduces model quality while also reducing the total number of learnable parameters. Increasing these features to 256 through 2048 greatly increases the total number of learnable parameters while making a negligible impact on accuracy, however this remains an area of active investigation. It is still not well understood how the sizes in this dimension affect the accuracy of the model, as well as which datasets would benefit from the additional features. 

The models were implemented using PyTorch \cite{paszke2019pytorch}. Training was performed using Adam\cite{kingma2014adam} to minimize the mean absolute error. All models were trained using distributed data-parallel training and a zero redundancy optimizer \cite{rajbhandari2020zero}. There is a copy of the model on each GPU which processes some unique $n$ mini-batch fraction of the data. The mini-batch size was chosen as the largest value that maximizes the available GPU memory. The effective batch size for $k$ GPUs is given by $nk$. An epoch represents one complete training iteration through the entire dataset. The learning rate $\eta$ was selected to be $1e-5$. The learning rate scaling is inspired by \citet{goyal2017accurate} and follows 
\begin{equation}
  \eta_\text{effective} = \eta n k
\end{equation}  
which should result in similar models when trained on different resources. The models were trained on NVIDIA 16GB V100 GPUs on the Lassen HPC using 12 hour allocations.

\FloatBarrier

\section{Results}\label{sec:results}

Machine learning (ML) models were trained on the four datasets. The training details including batch size, number of GPUs, learning rates, and effective batch size are reported in Table~\ref{tab:tabdata}. A single image array from the Rayleigh-Taylor dataset was roughly a third of the size of the images from the other datasets, and thus could support approximately three times the batch size on a single GPU compared to the others.

\begin{table}
  \center
  \caption{\label{tab:tabdata}An overview of the datasets created.}
  \begin{tabular}{lccccc}
    Name & Batch size & \# of GPUs & $\eta$ & $\eta_\text{effective}$ & effective batch size \\ 
    \hline
    PCHIP impact & 14 & 160 & 1e-5 & 2.24e-2 & 2240 \\
    Double sine wave & 14 & 40 & 1e-5 & 5.6e-3 & 560 \\
    Linear shaped-charge & 14 & 60 & 1e-5  & 8.4e-3 & 840 \\
    Rayleigh-Taylor & 48 & 32 & 1e-5 & 1.536e-2 & 1536 \\
  \end{tabular}
\end{table}

\begin{figure}[!htb]
  \centering
    \includegraphics[width=0.32\textwidth]{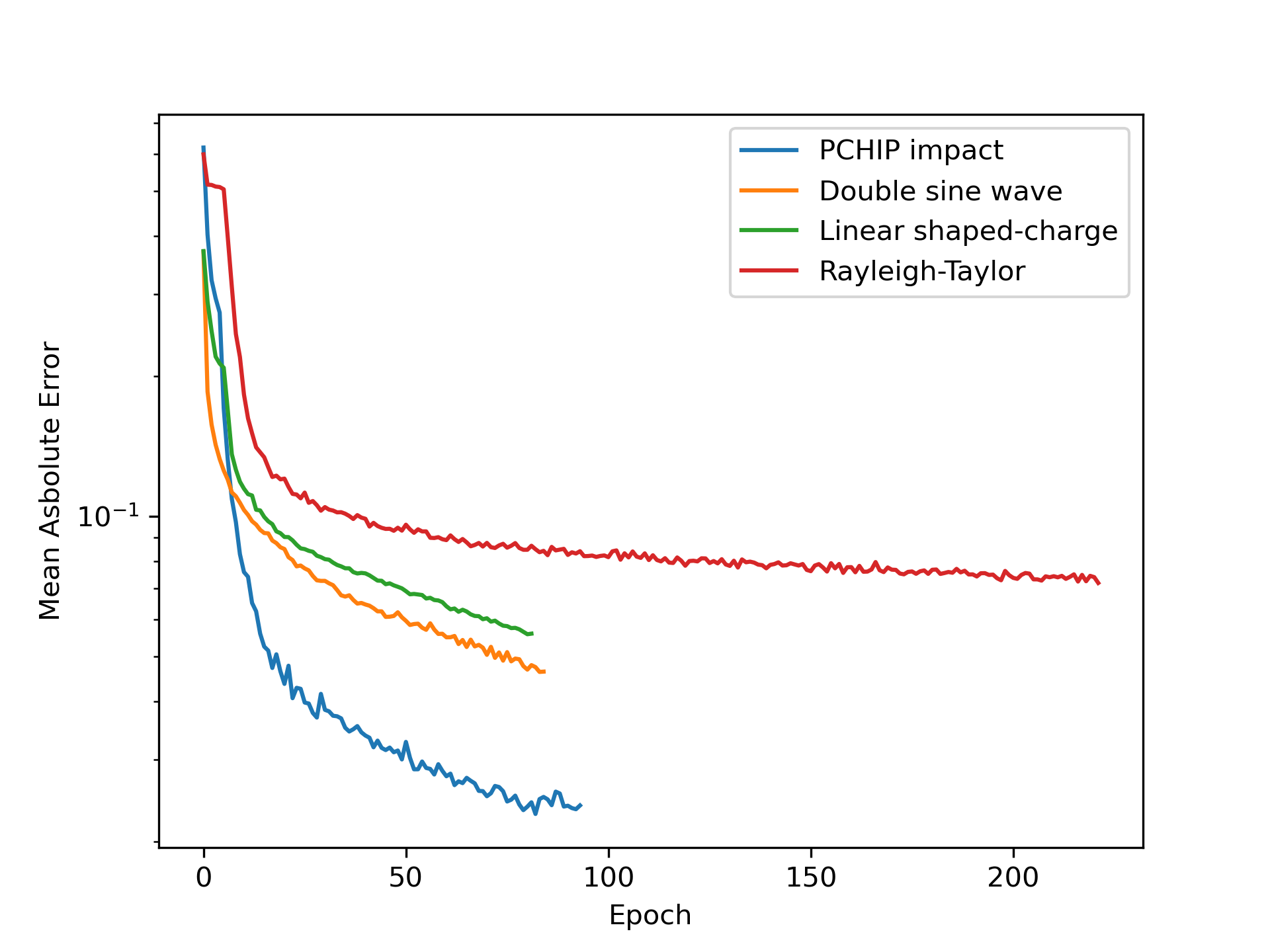}
    \includegraphics[width=0.32\textwidth]{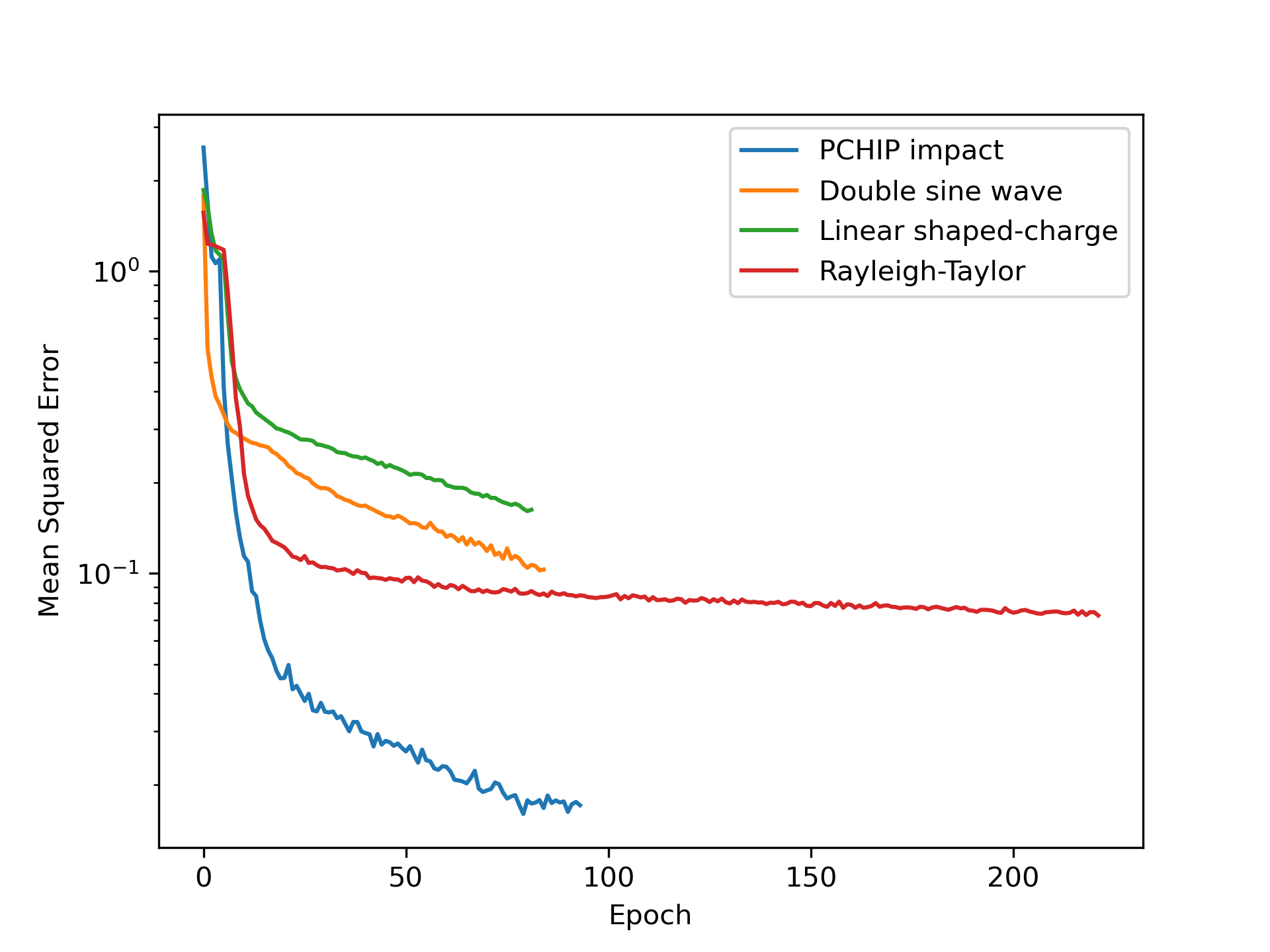}
    \includegraphics[width=0.32\textwidth]{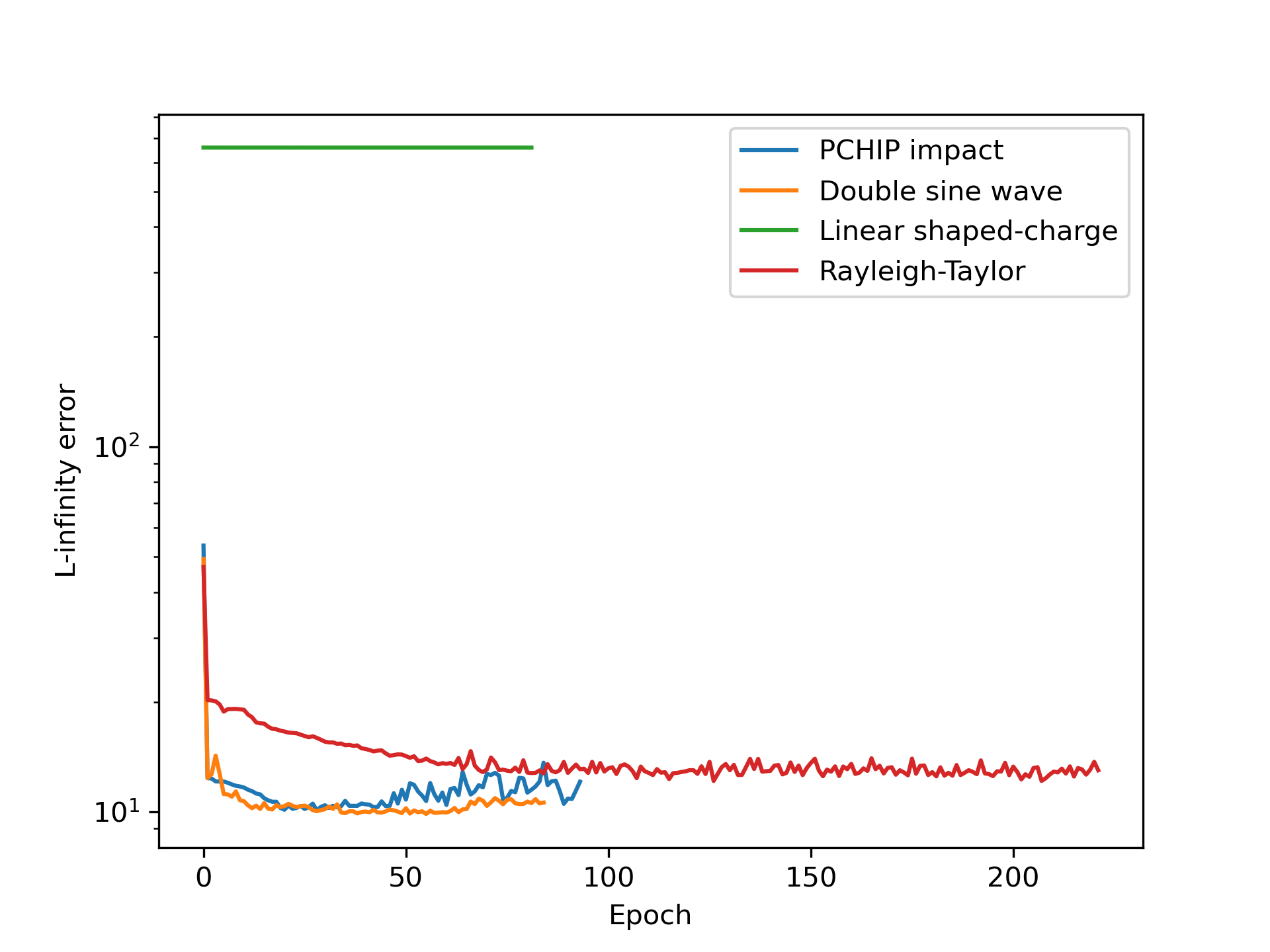}
  \caption{
      The mean absolute error, mean squared error, and L-infinity training errors for each dataset.
      }
  \label{fig:mltraining}
\end{figure}

The training errors for each epoch are shown in Figure~\ref{fig:mltraining}. All models were trained on mean absolute error, and there is strong correlation with mean absolute error, mean squared error, and L-infinity error. Diminishing training returns with respect to additional epochs is observed with all models. This is most pronounced in the Rayleigh-Taylor case which shows negligible training improvements for the last 100 epochs.

One necessary aspect that enables the ML models to be machine portable is having a reasonable file size. Table~\ref{tab:size} shows the file sizes of the ML model weights compared to the dataset size. The ML models were generally around 0.9 gigibytes, while the dataset were on the order of a tebibyte\footnote{One tebibyte is 1.099511627776 terabytes.}. In this aspect, the ML model can be thought of as a lossy compression of the data. In these cases, the achieved compression factor was around 1,000 times. The dataset sizes discussed here is the size on disk of the hdf5 files which utilize gzip lossless compression.

\begin{table}
  \center
  \caption{\label{tab:size}Dataset vs machine learning model size in gigibytes.}
  \begin{tabular}{lcc}
    Name & Dataset size (GiB) & ML model size (GiB) \\ 
    \hline
    PCHIP impact & 2098 & 0.900 \\
    Double sine wave & 729 & 0.896 \\
    Linear shaped-charge & 1151 & 0.709 \\
    Rayleigh-Taylor & 328 & 0.899 \\
  \end{tabular}
\end{table}

\begin{figure}[!htb]
  \begin{center}
  \begin{tabular}{*{5}{c}}
  Ml & \includegraphics[width=0.175\textwidth]{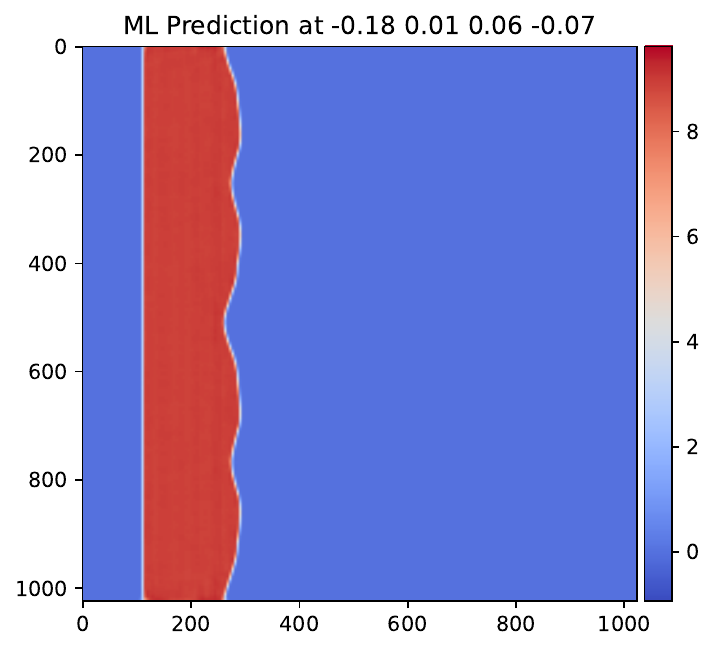} & \includegraphics[width=0.175\textwidth]{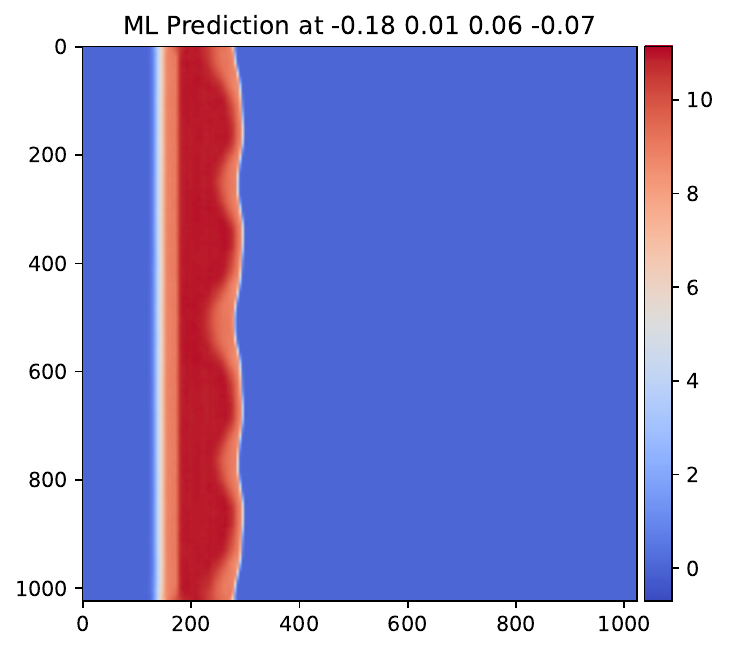}  & \includegraphics[width=0.175\textwidth]{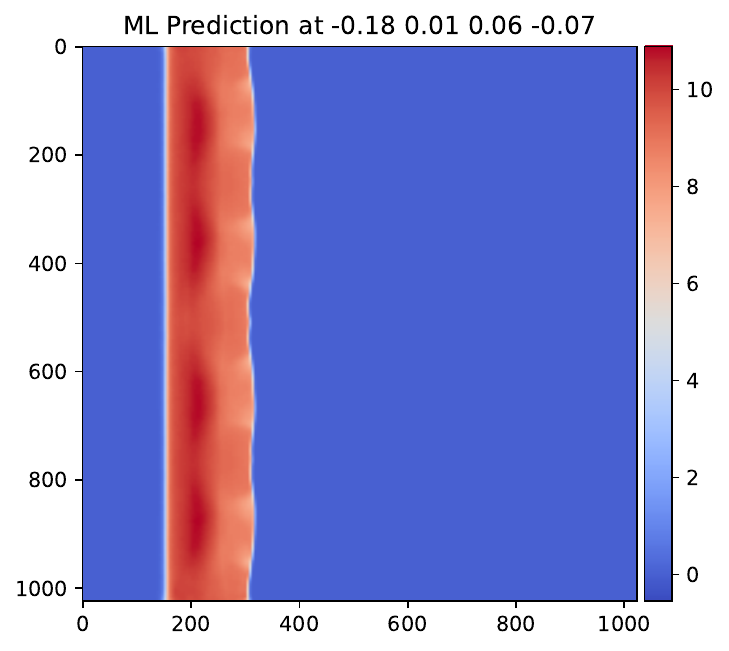}  & \includegraphics[width=0.175\textwidth]{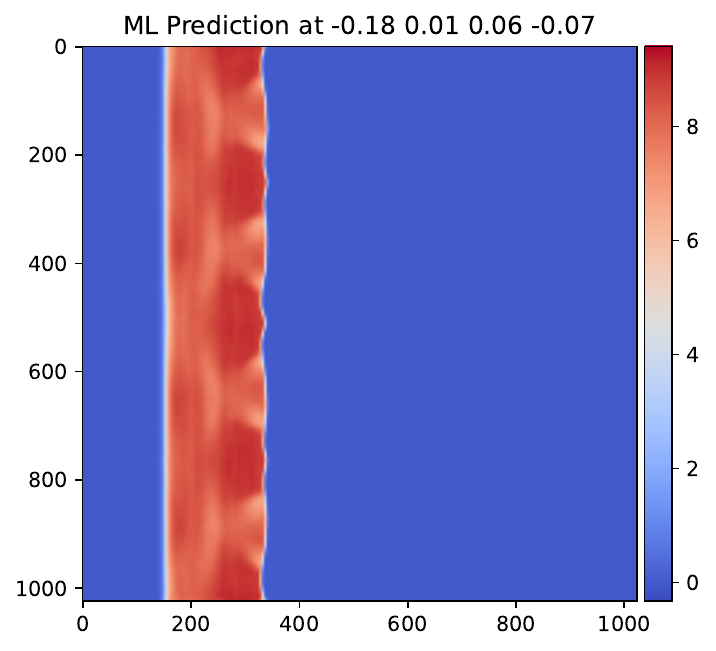}  \\
  Sim & \includegraphics[width=0.175\textwidth]{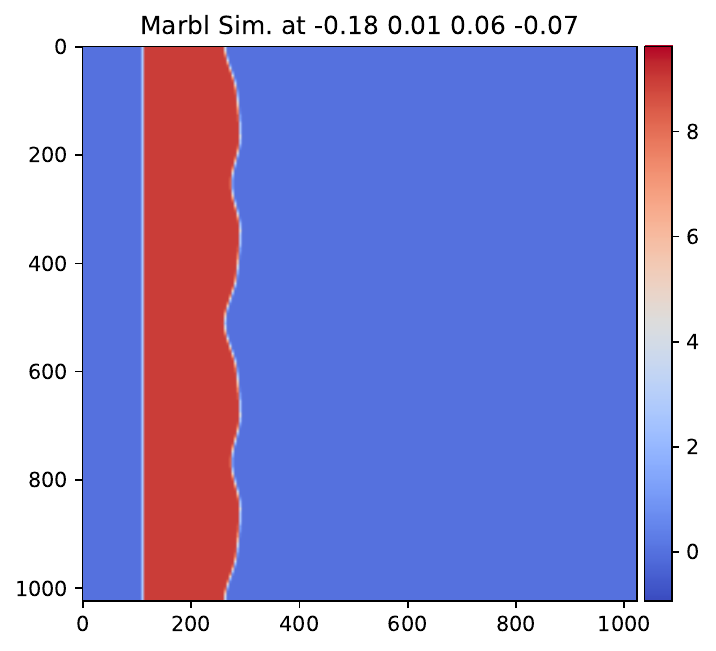} & \includegraphics[width=0.175\textwidth]{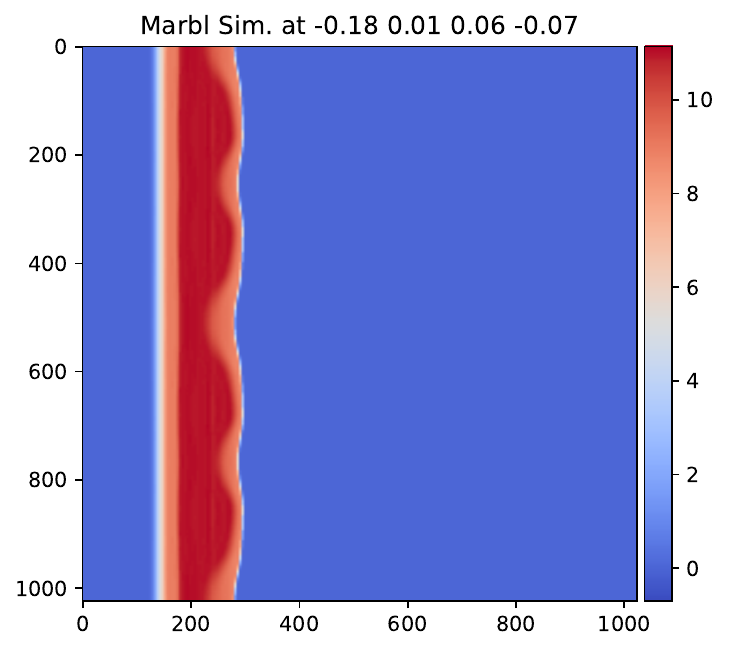}  & \includegraphics[width=0.175\textwidth]{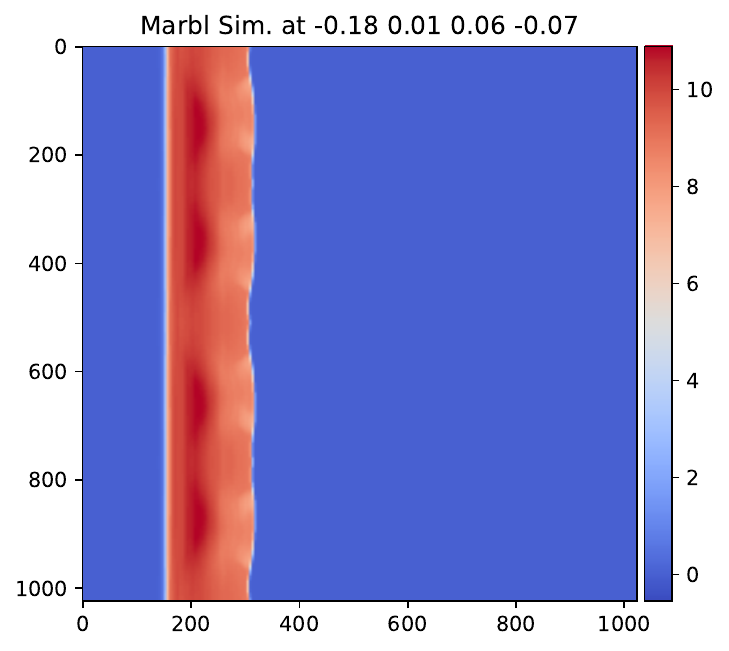}  & \includegraphics[width=0.175\textwidth]{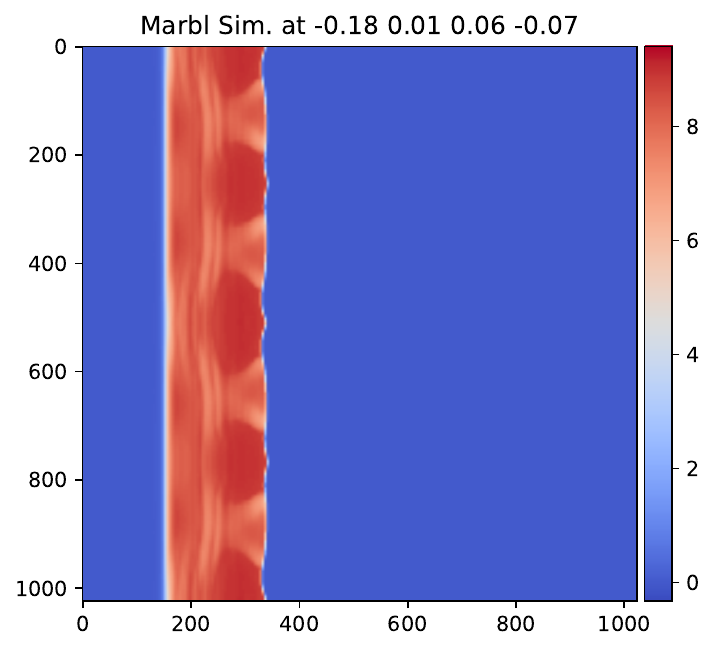}  \\
  \hline
  Ml  & \includegraphics[width=0.175\textwidth]{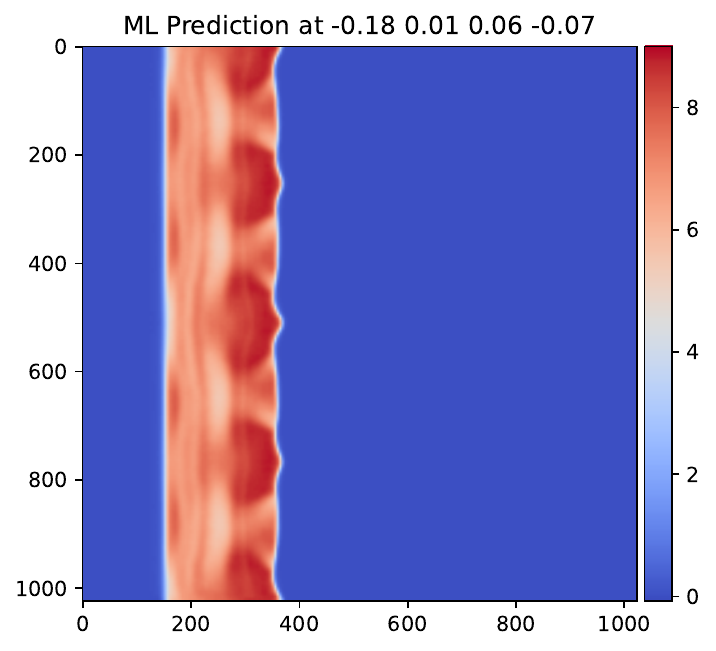} & \includegraphics[width=0.175\textwidth]{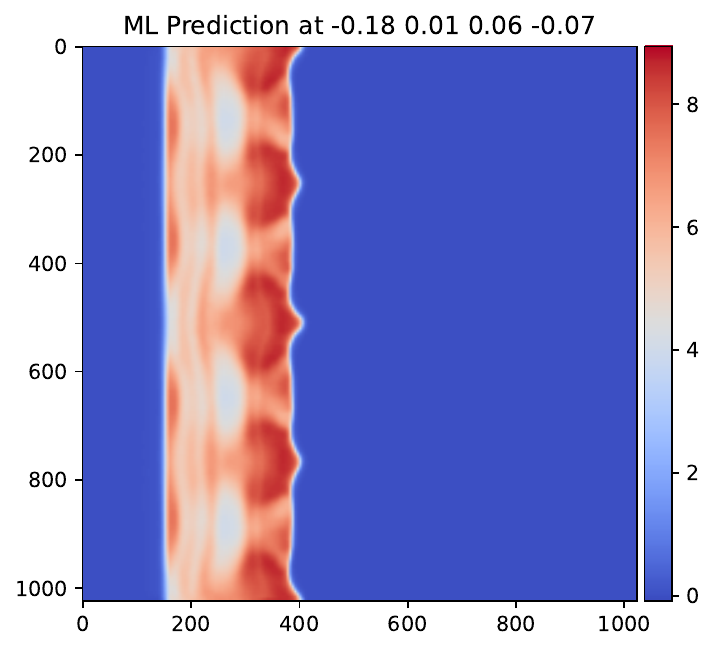} & \includegraphics[width=0.175\textwidth]{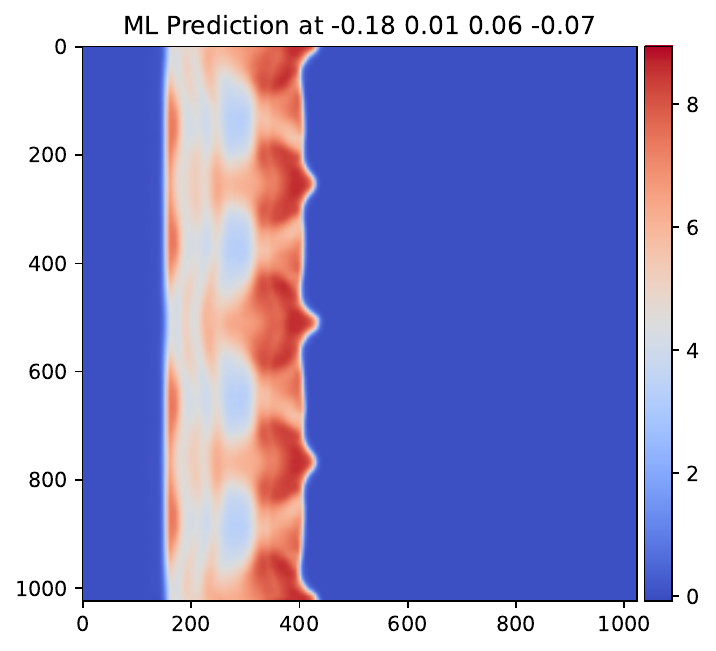} & \includegraphics[width=0.175\textwidth]{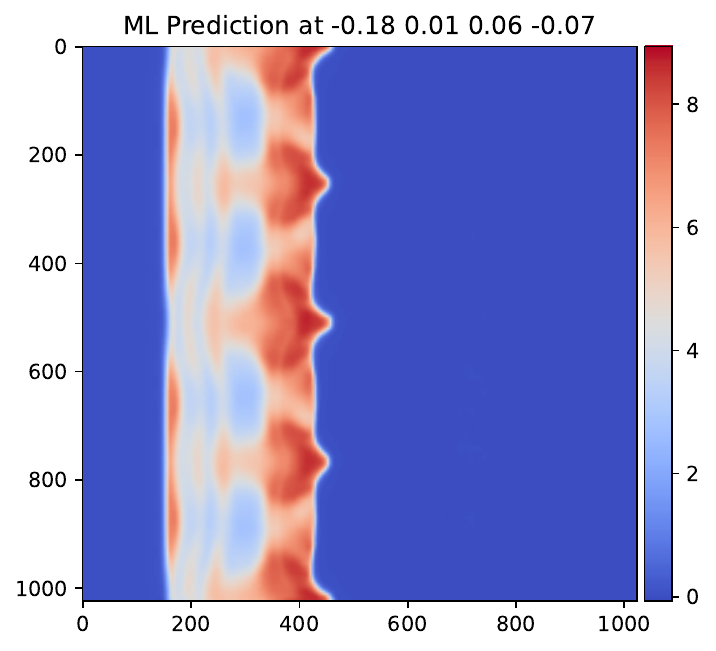} \\
  Sim & \includegraphics[width=0.175\textwidth]{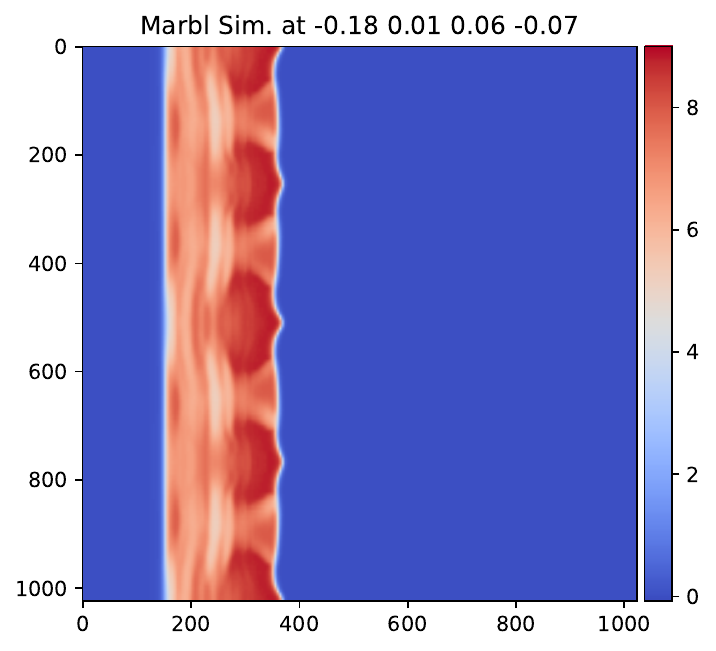} & \includegraphics[width=0.175\textwidth]{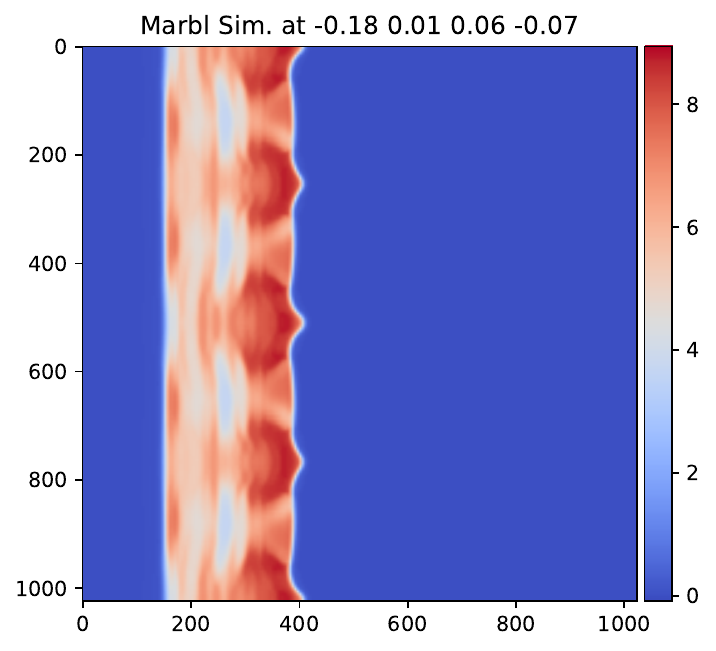} & \includegraphics[width=0.175\textwidth]{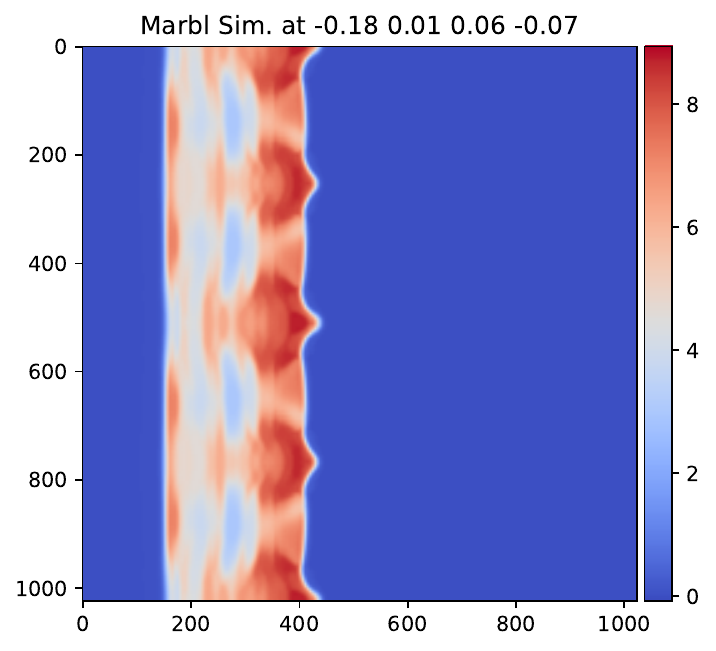} & \includegraphics[width=0.175\textwidth]{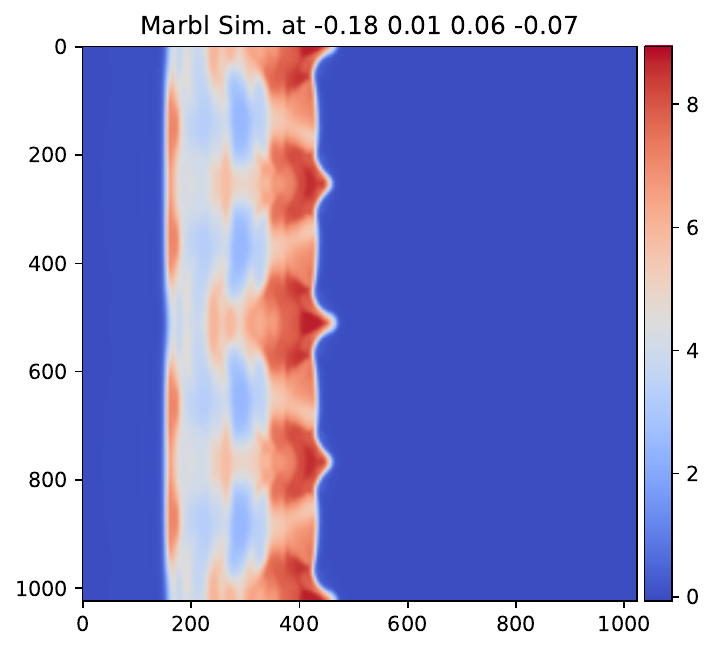}   \\
  \hline
  Ml & \includegraphics[width=0.175\textwidth]{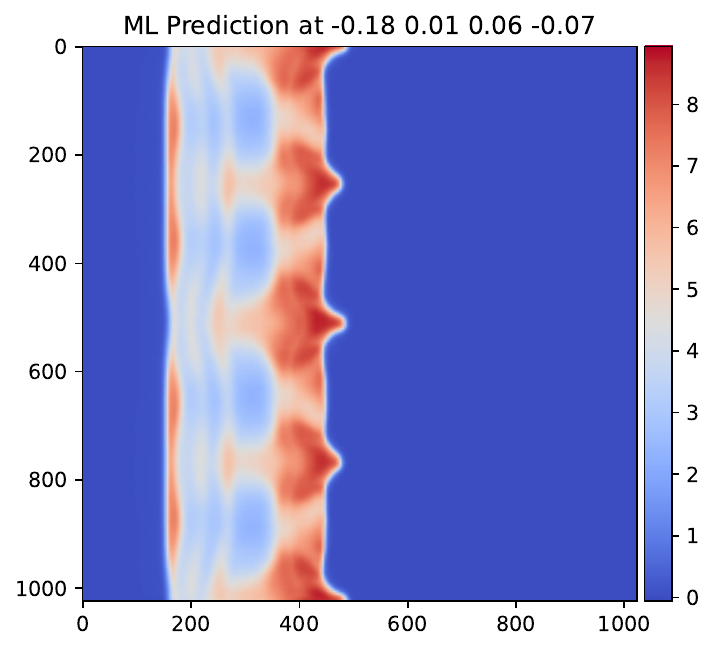} & \includegraphics[width=0.175\textwidth]{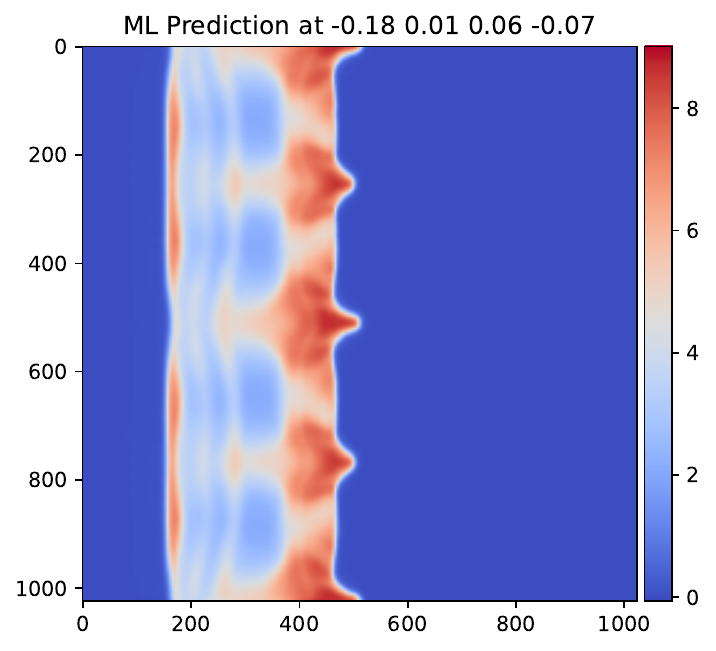} & \includegraphics[width=0.175\textwidth]{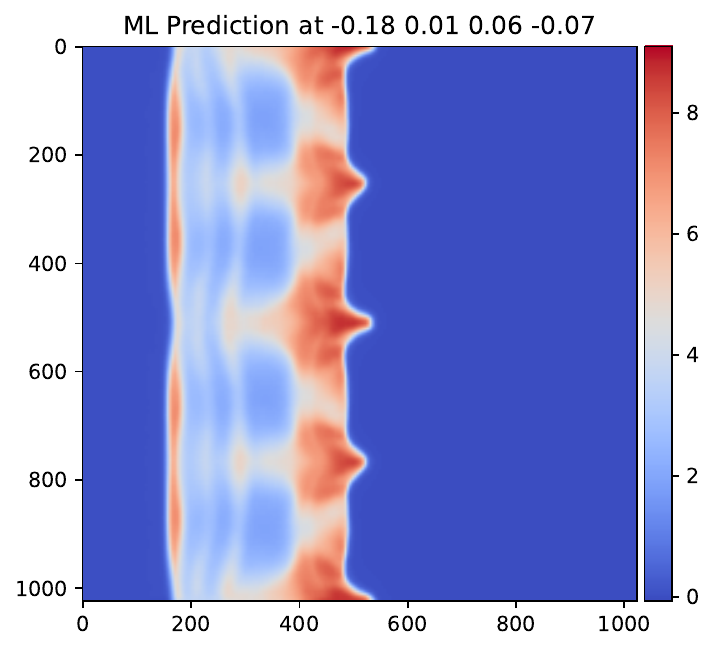} & \includegraphics[width=0.175\textwidth]{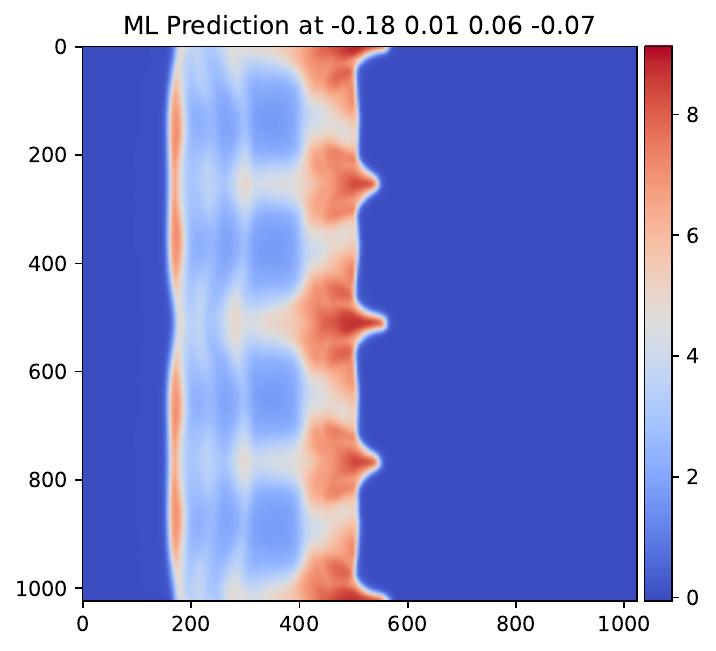} \\
  Sim & \includegraphics[width=0.175\textwidth]{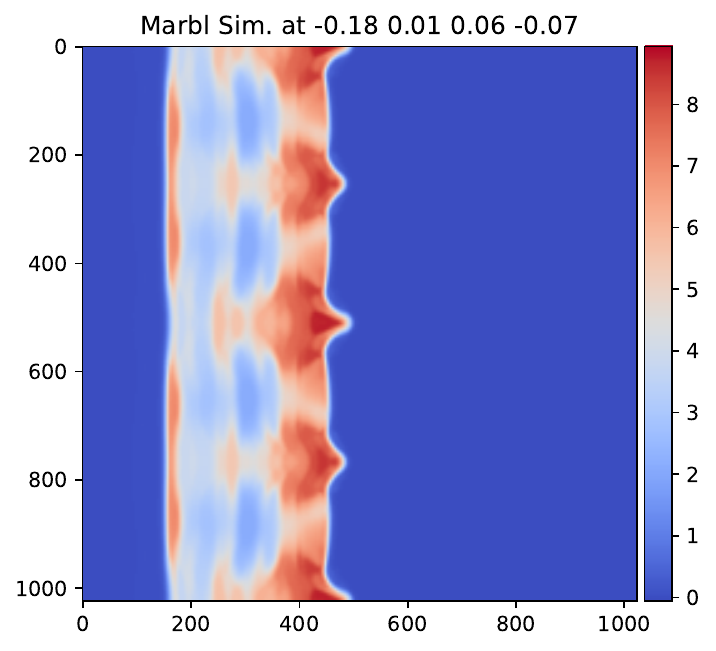} & \includegraphics[width=0.175\textwidth]{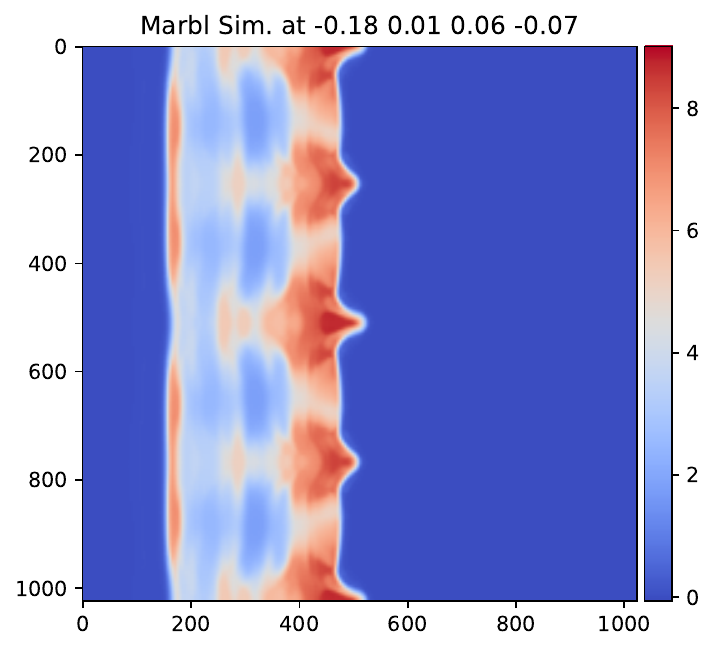} & \includegraphics[width=0.175\textwidth]{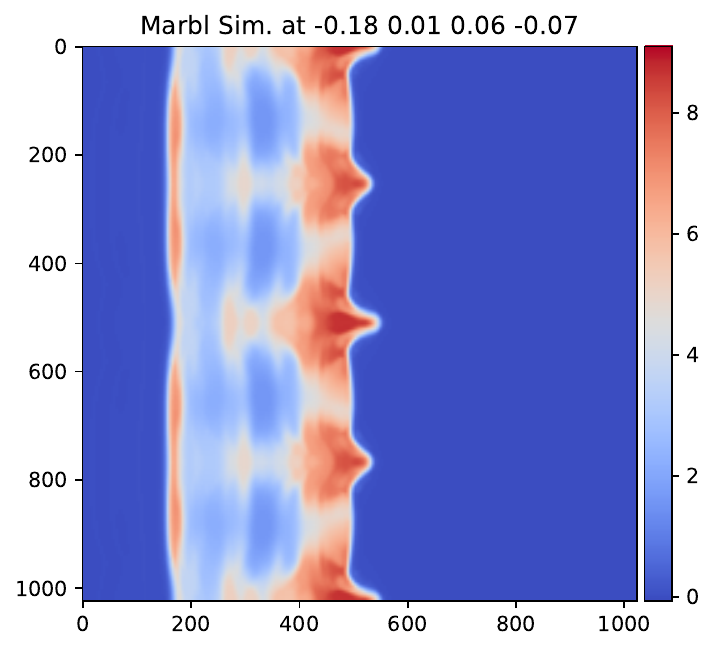} & \includegraphics[width=0.175\textwidth]{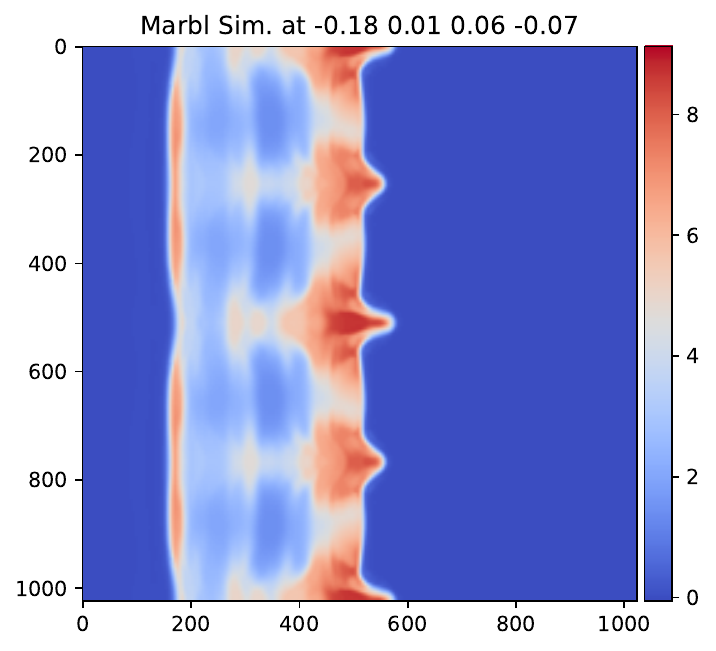}   \\
  \hline
  Ml & \includegraphics[width=0.175\textwidth]{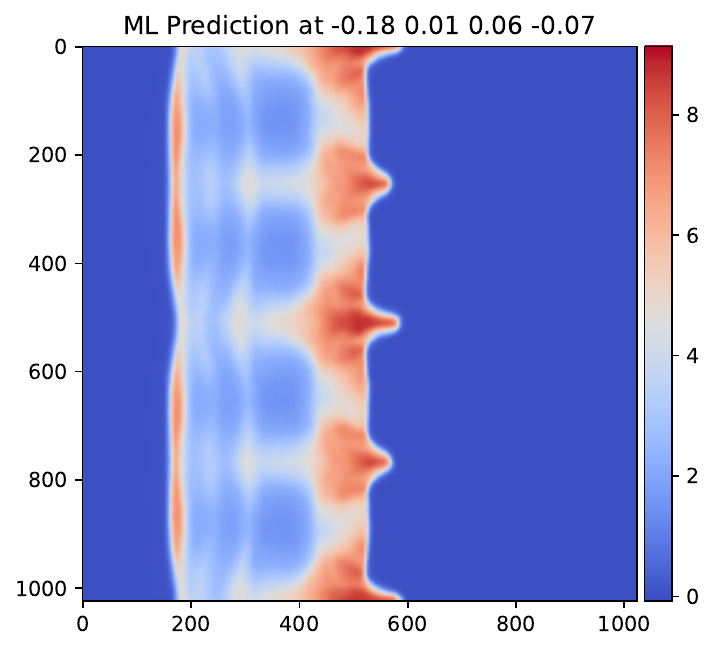} & \includegraphics[width=0.175\textwidth]{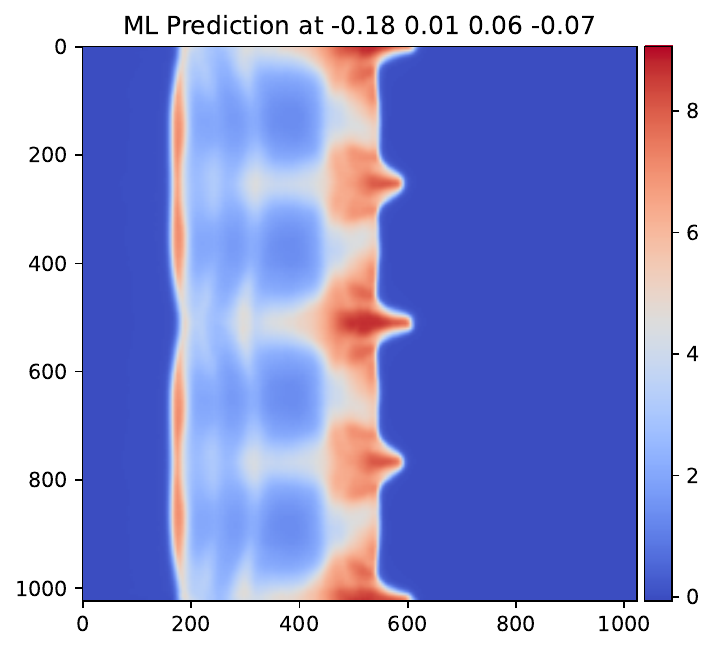} & \includegraphics[width=0.175\textwidth]{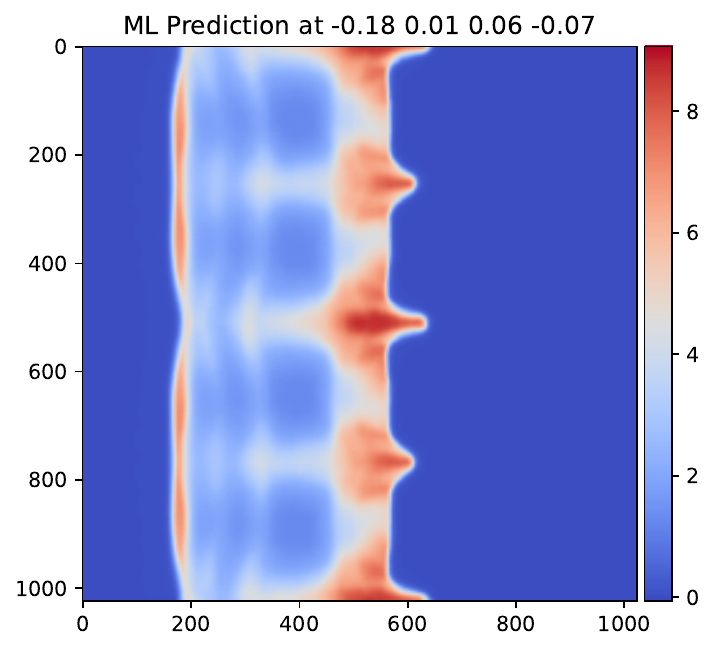} & \includegraphics[width=0.175\textwidth]{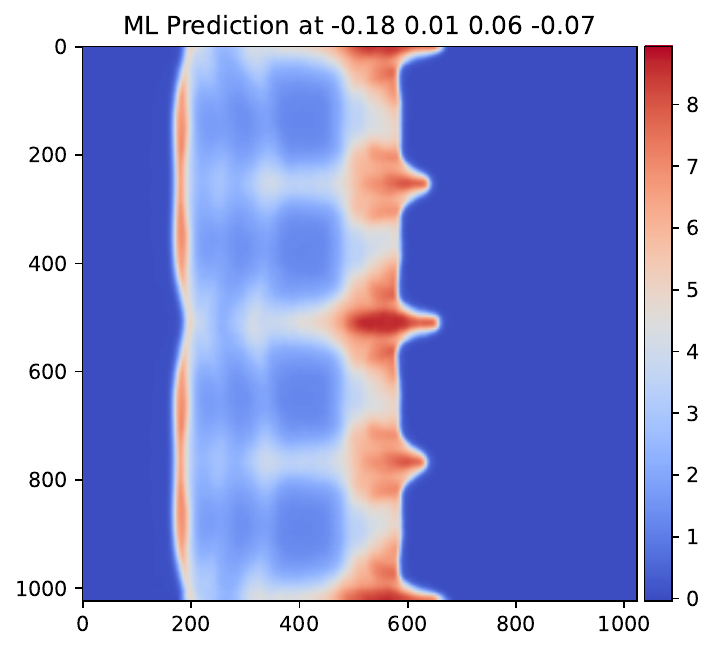} \\
  Sim & \includegraphics[width=0.175\textwidth]{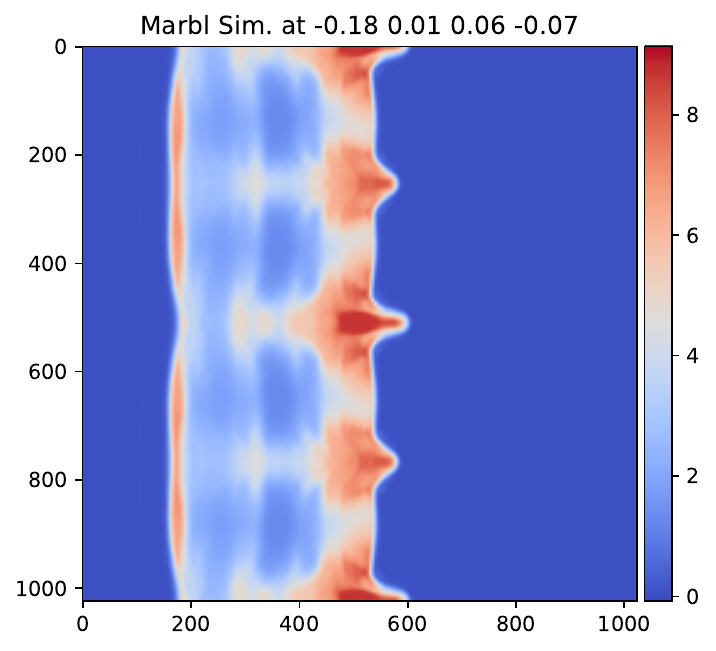} & \includegraphics[width=0.175\textwidth]{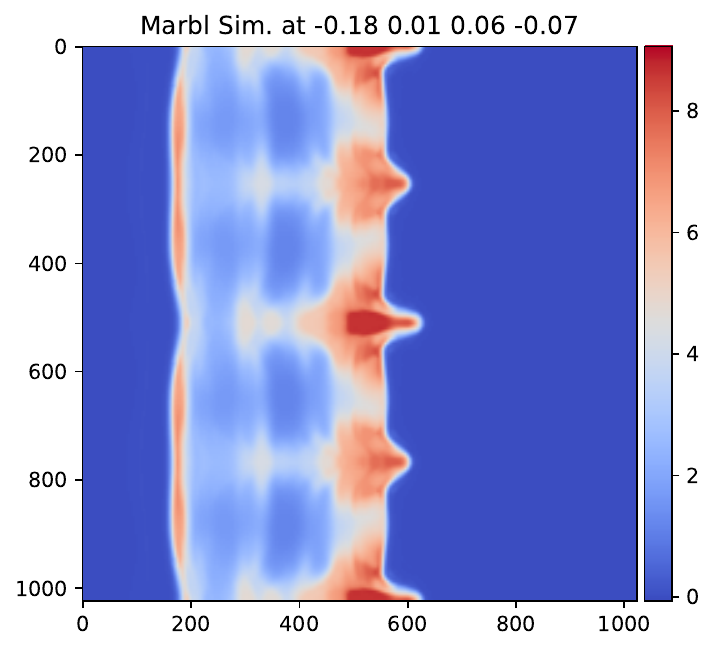} & \includegraphics[width=0.175\textwidth]{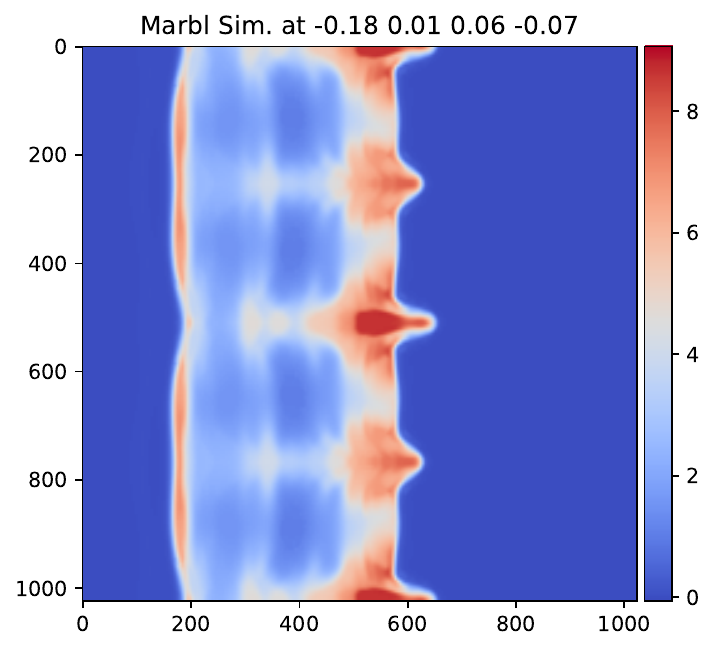} & \includegraphics[width=0.175\textwidth]{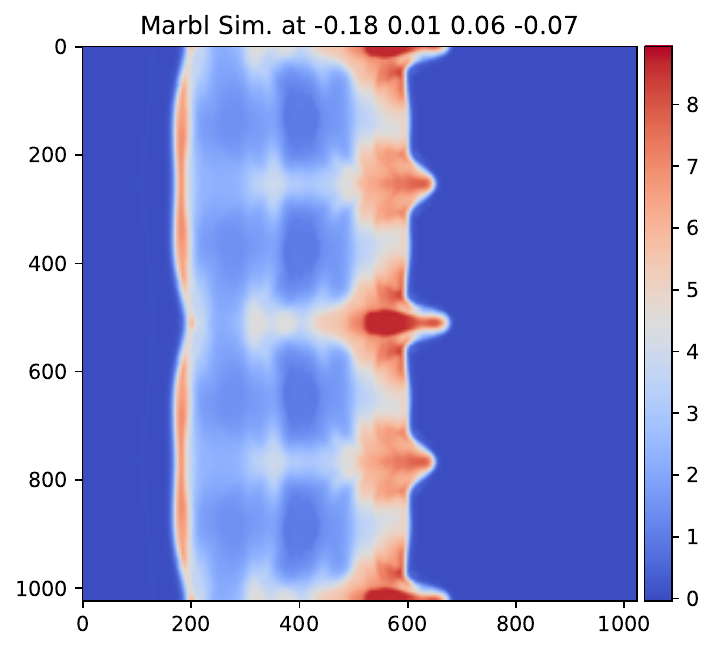}   \\
  \hline
  \end{tabular}
  \caption{
      Figure of density predictions and  truth for the PCHIP impact. Rows show machine learning model (ML) next to simulation (Sim) results.
      }\label{fig:pchip_density}
  \end{center}
\end{figure}

\begin{figure}[!htb]
  \begin{center}
  \begin{tabular}{*{5}{c}}
  Ml & \includegraphics[width=0.175\textwidth]{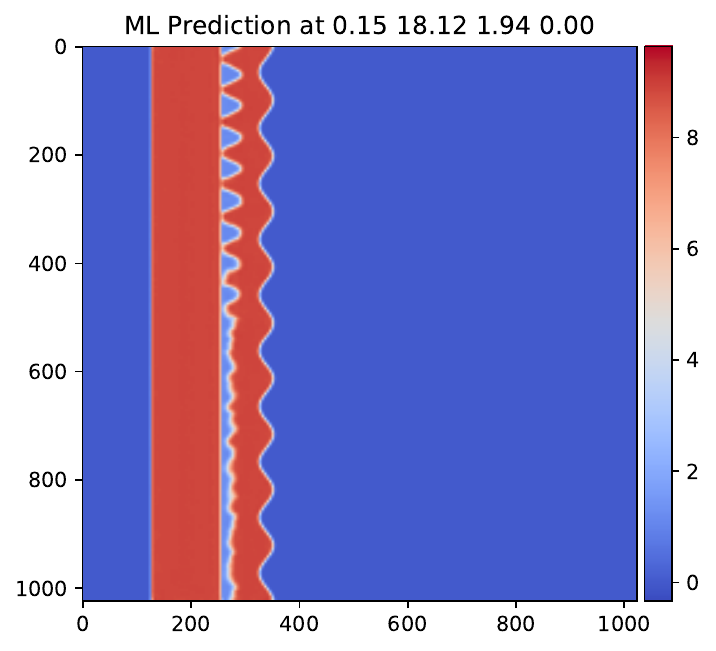} & \includegraphics[width=0.175\textwidth]{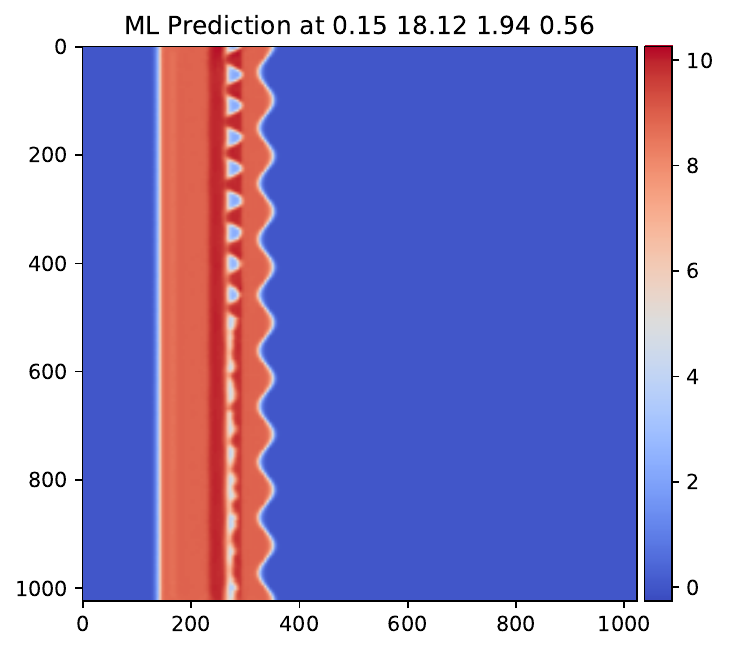}  & \includegraphics[width=0.175\textwidth]{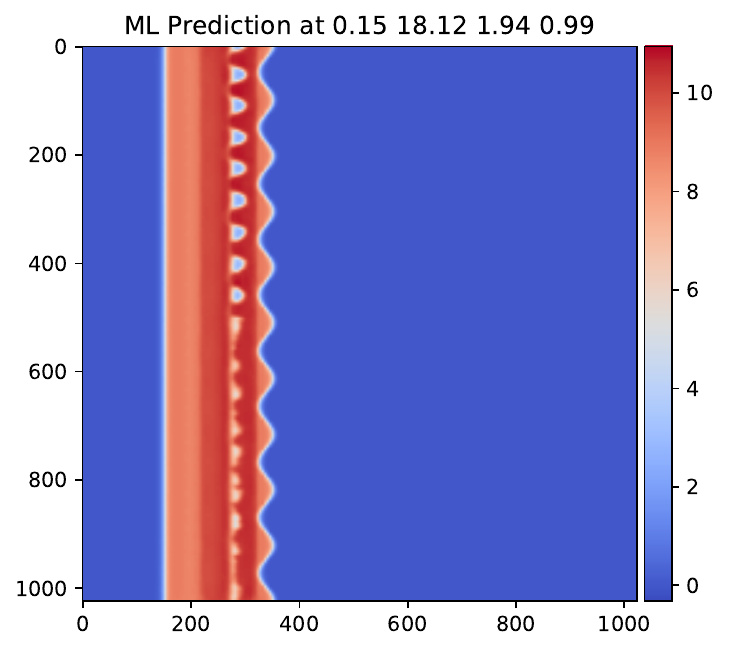}  & \includegraphics[width=0.175\textwidth]{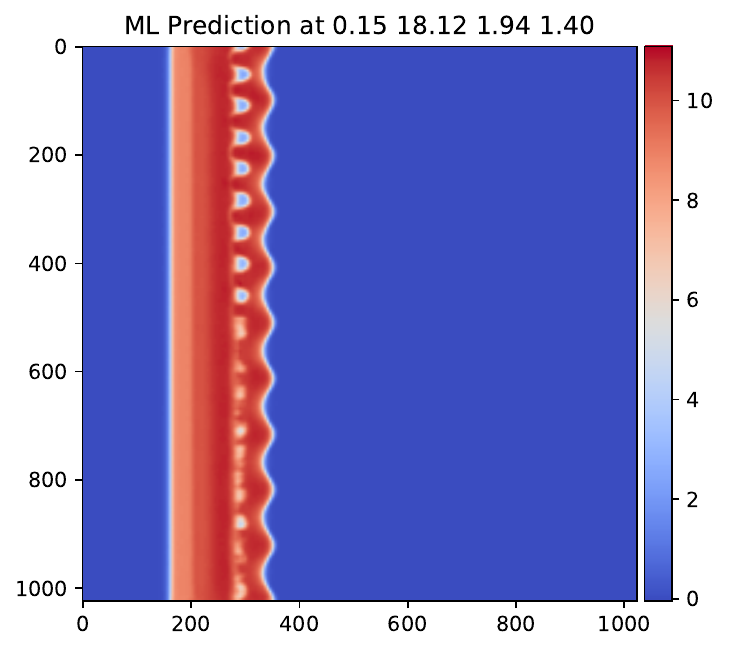}  \\
  Sim & \includegraphics[width=0.175\textwidth]{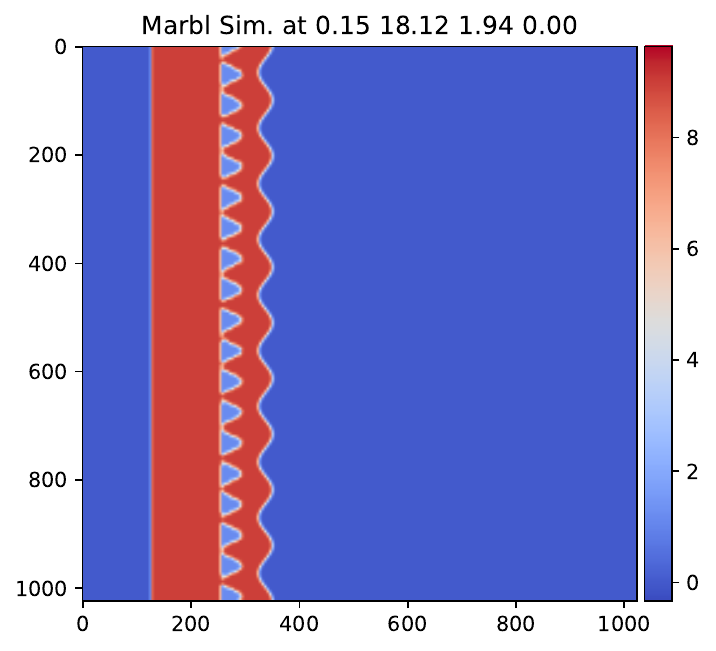} & \includegraphics[width=0.175\textwidth]{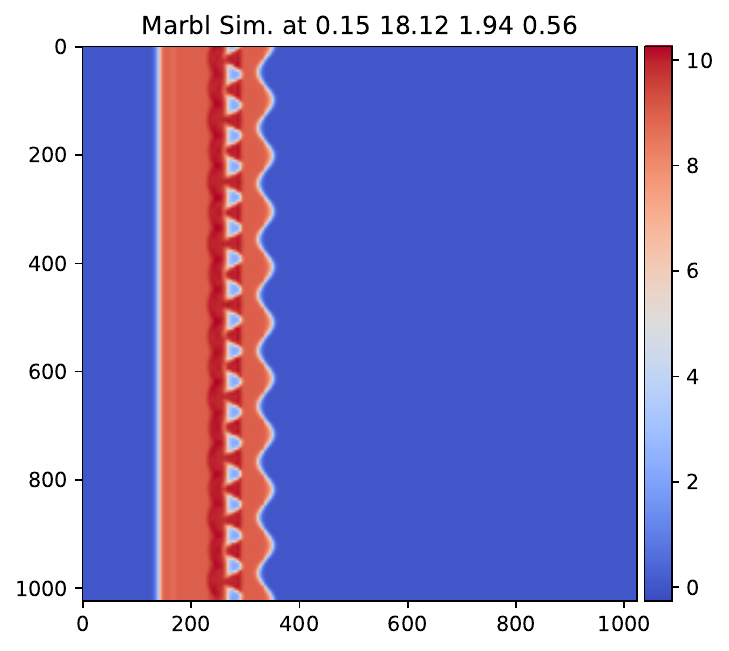}  & \includegraphics[width=0.175\textwidth]{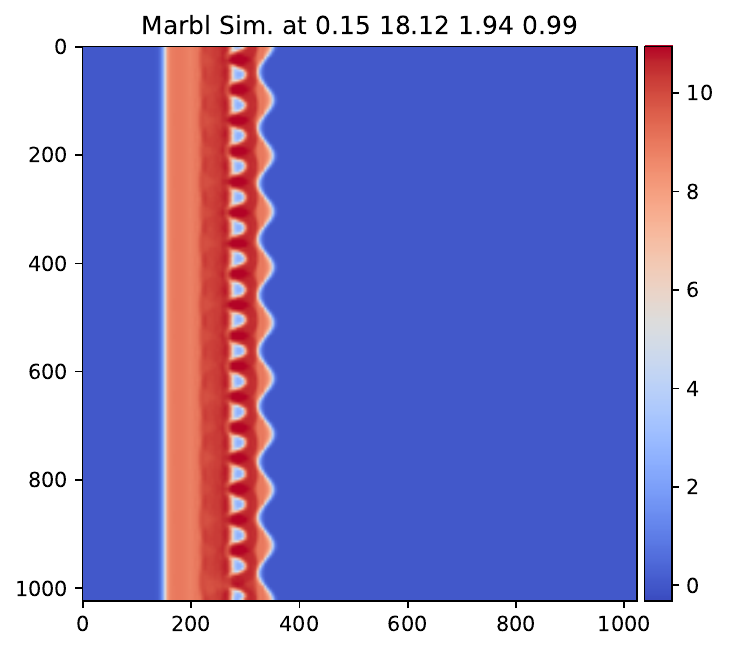}  & \includegraphics[width=0.175\textwidth]{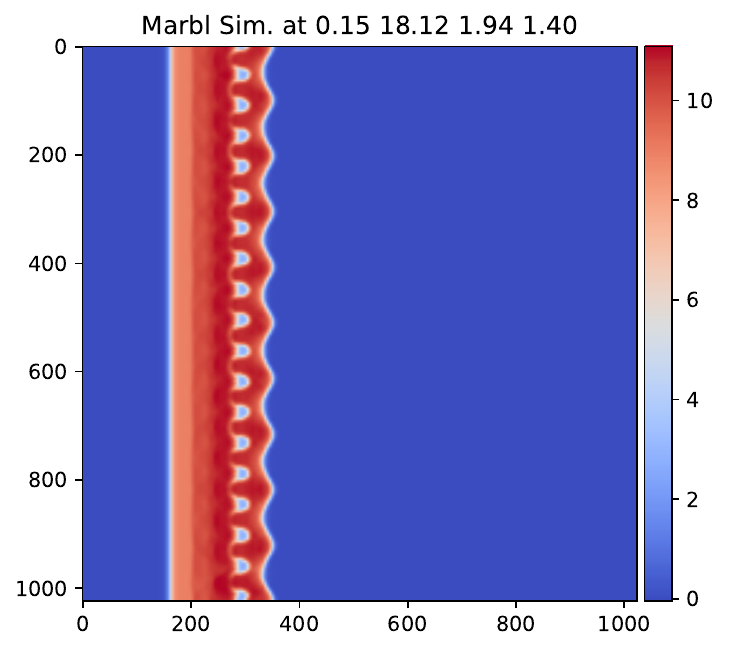}  \\
  \hline
  Ml & \includegraphics[width=0.175\textwidth]{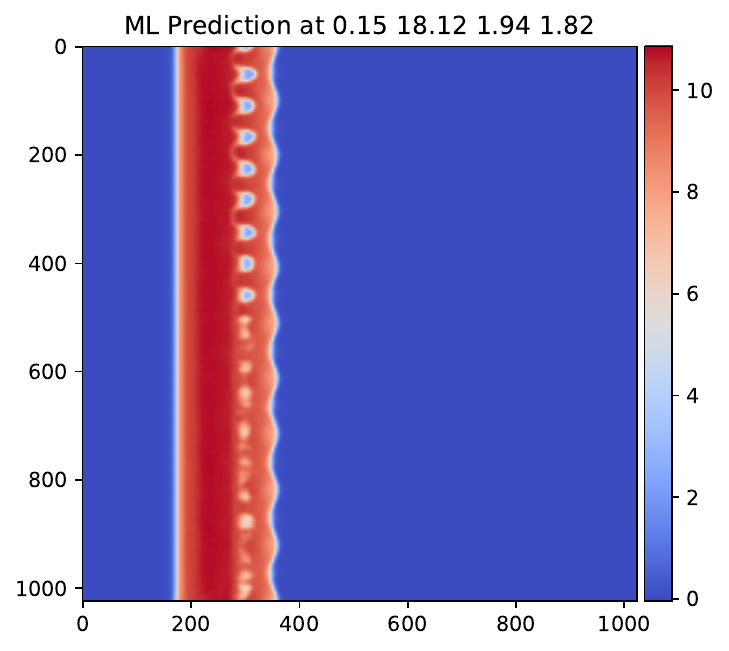} & \includegraphics[width=0.175\textwidth]{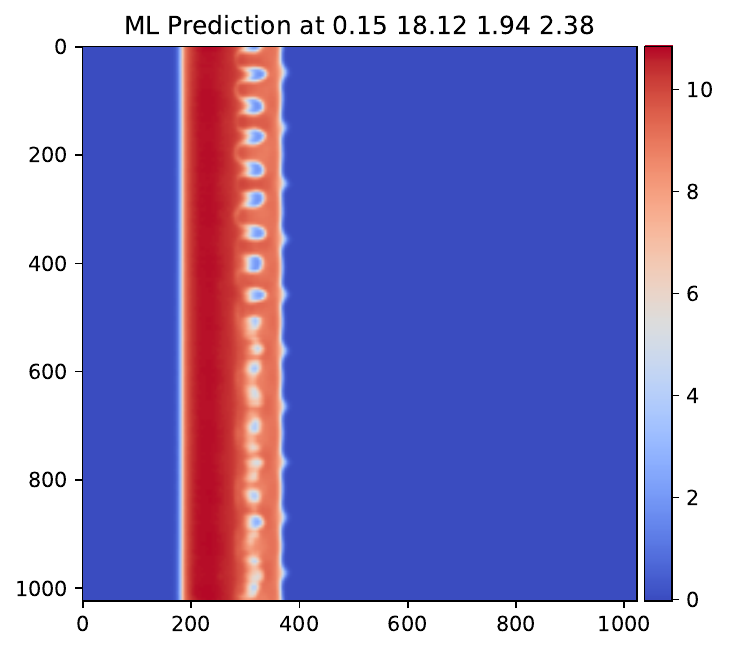} & \includegraphics[width=0.175\textwidth]{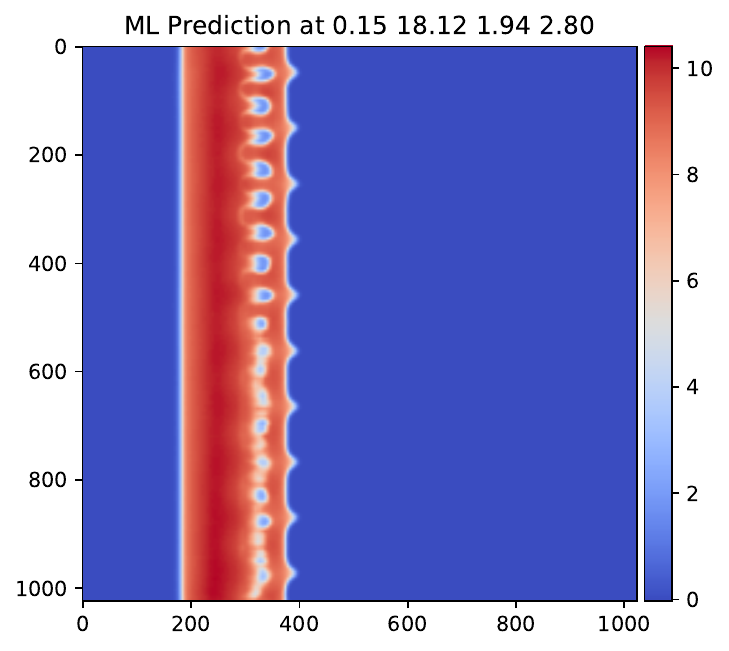} & \includegraphics[width=0.175\textwidth]{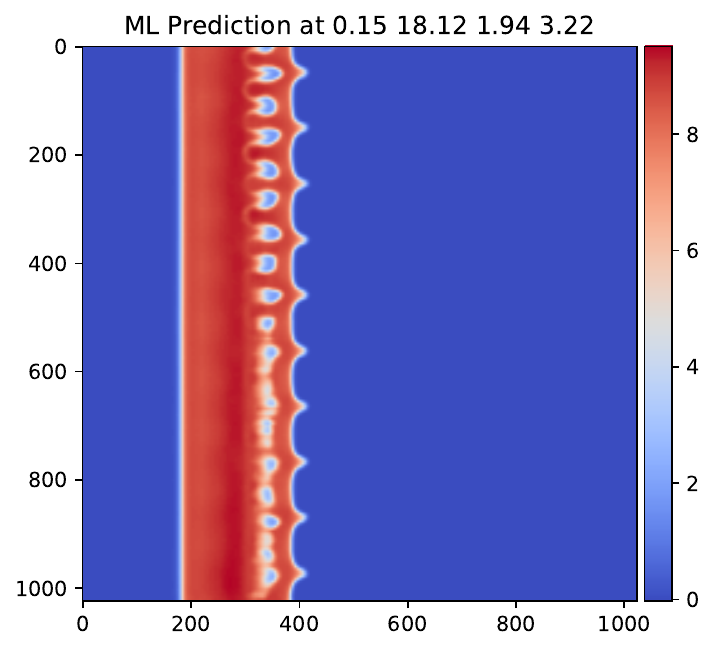} \\
  Sim & \includegraphics[width=0.175\textwidth]{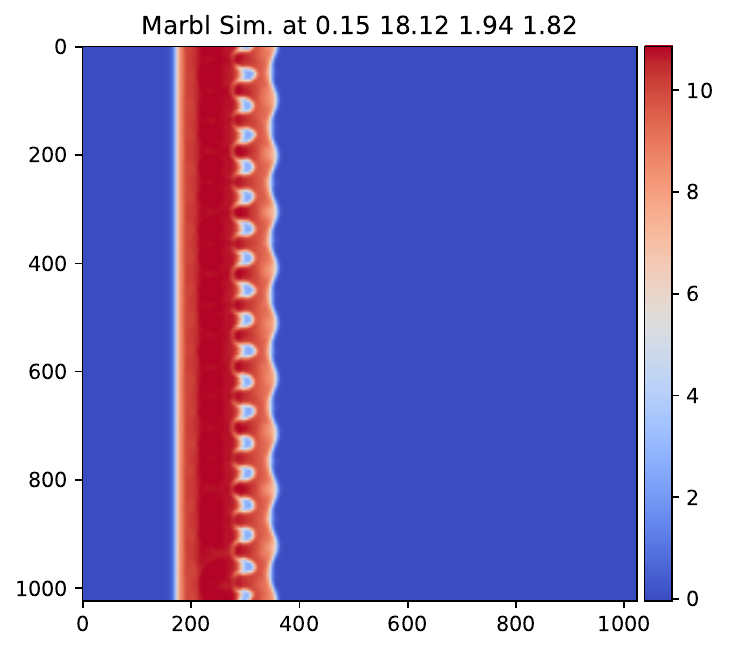} & \includegraphics[width=0.175\textwidth]{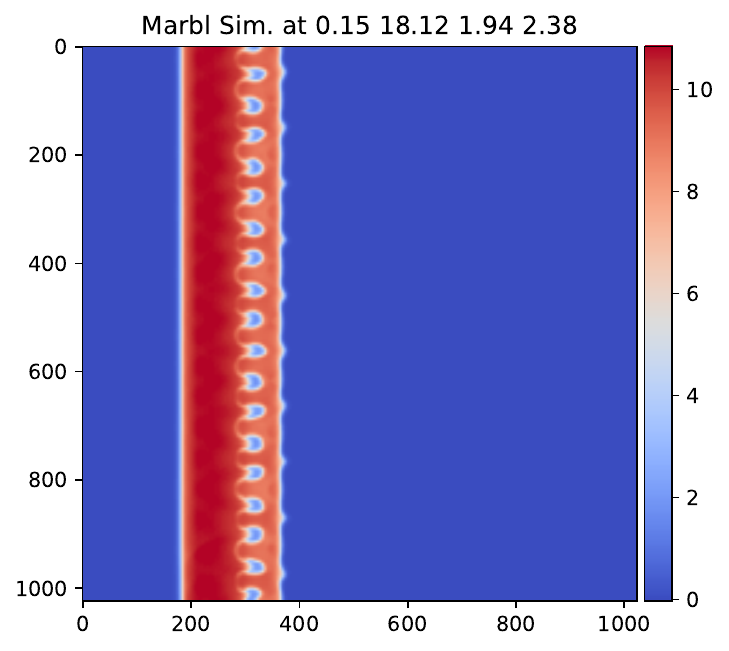} & \includegraphics[width=0.175\textwidth]{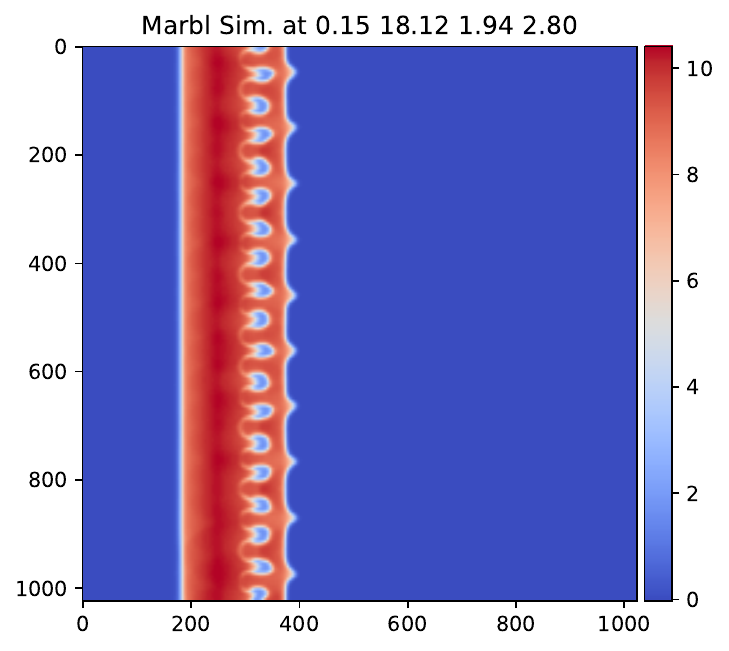} & \includegraphics[width=0.175\textwidth]{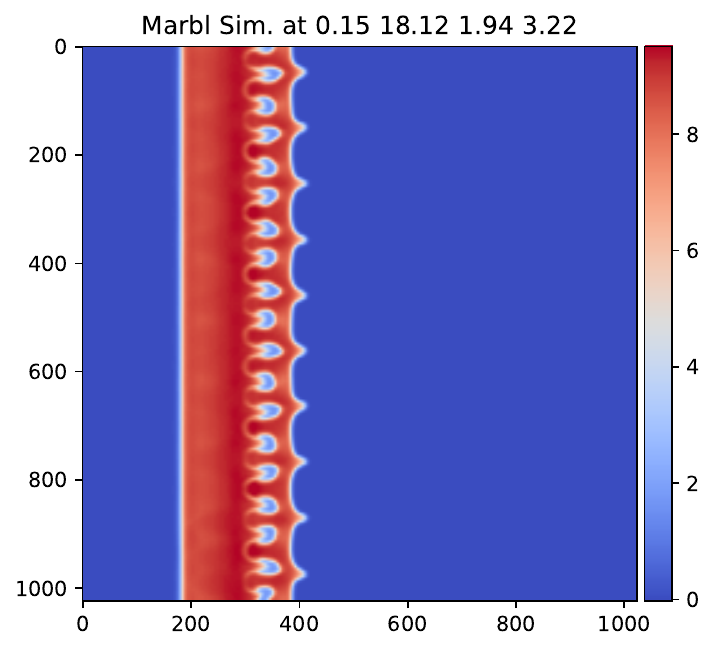}   \\
  \hline
  Ml & \includegraphics[width=0.175\textwidth]{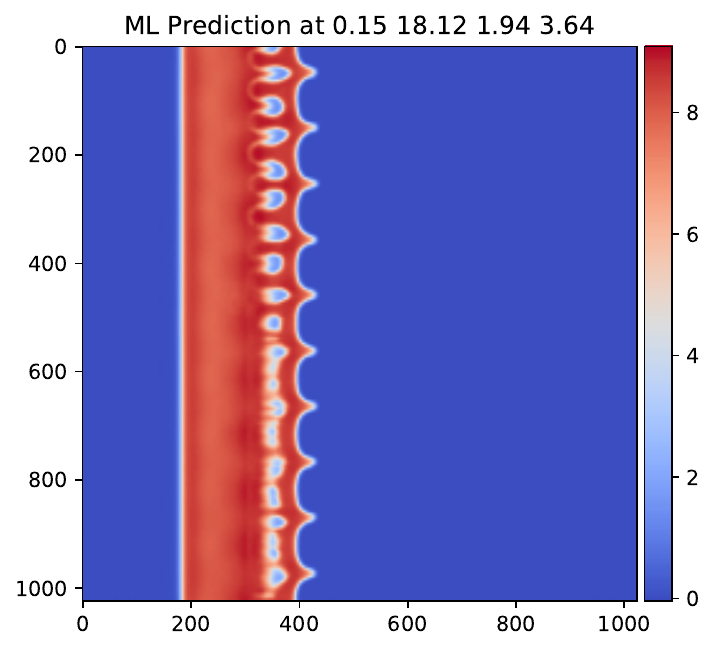} & \includegraphics[width=0.175\textwidth]{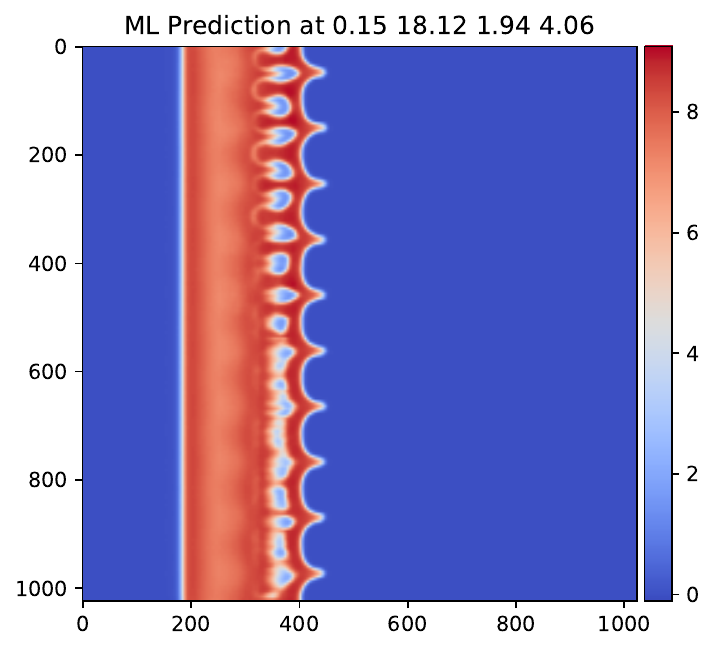} & \includegraphics[width=0.175\textwidth]{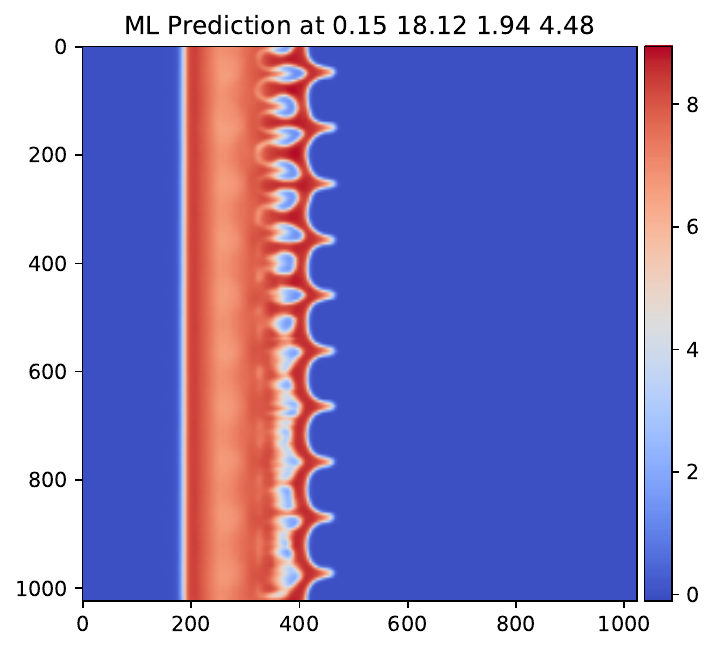} & \includegraphics[width=0.175\textwidth]{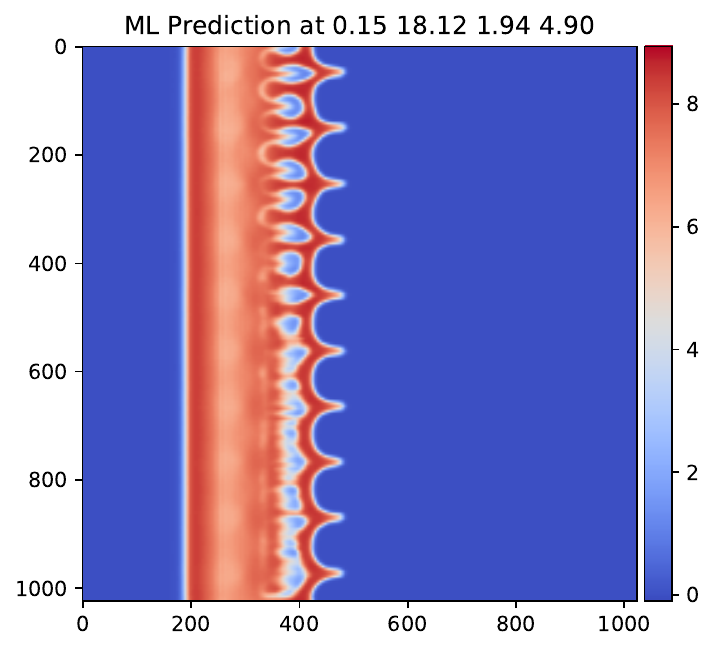} \\
  Sim & \includegraphics[width=0.175\textwidth]{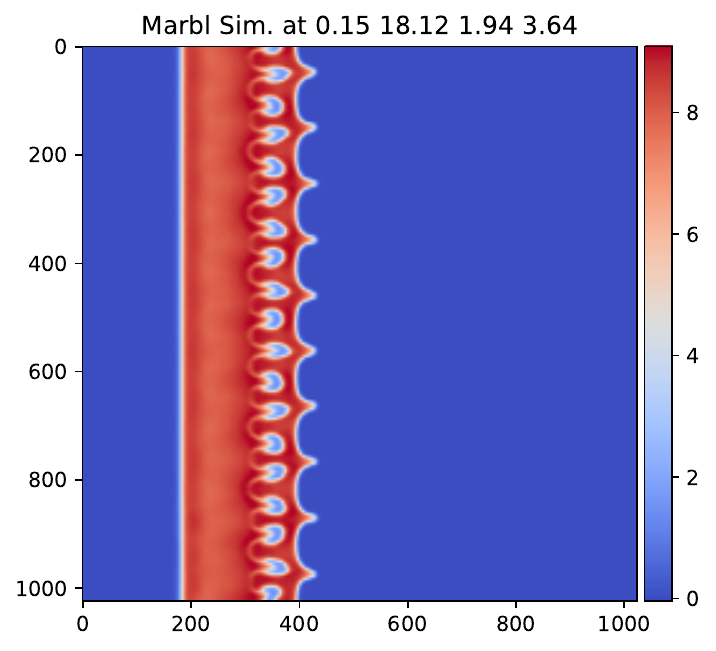} & \includegraphics[width=0.175\textwidth]{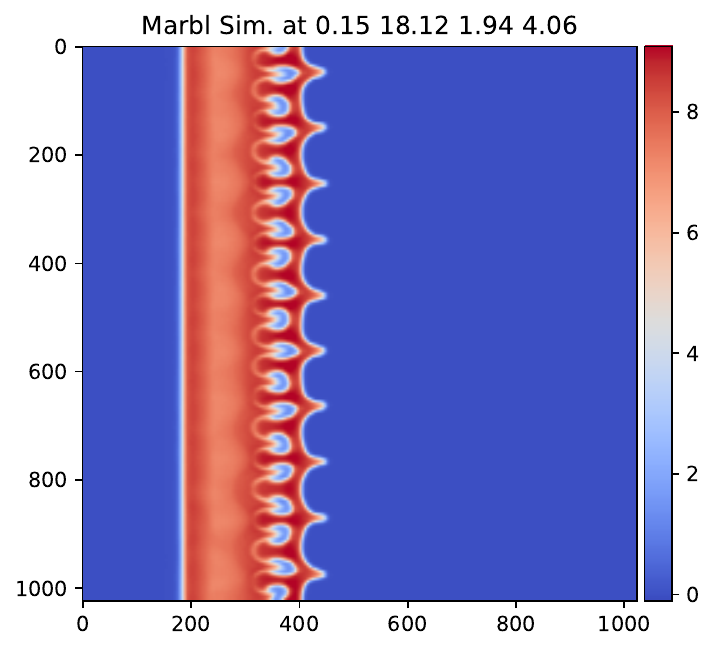} & \includegraphics[width=0.175\textwidth]{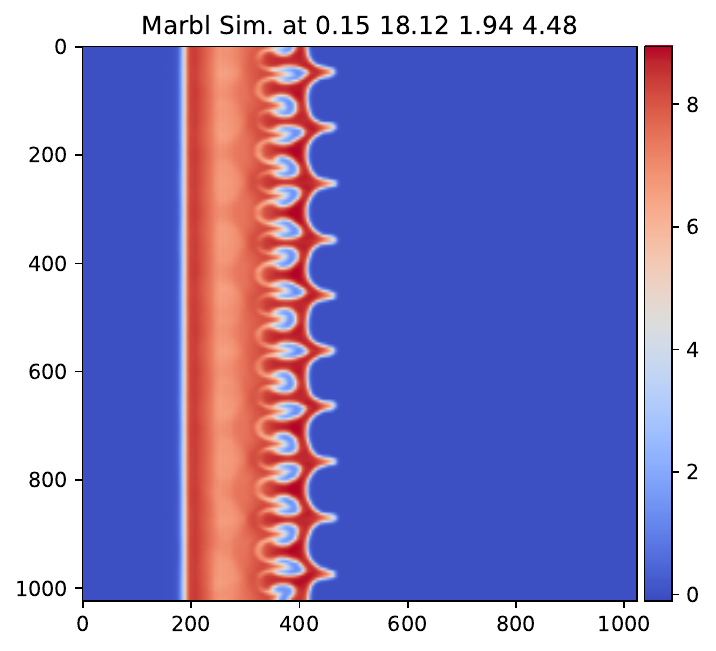} & \includegraphics[width=0.175\textwidth]{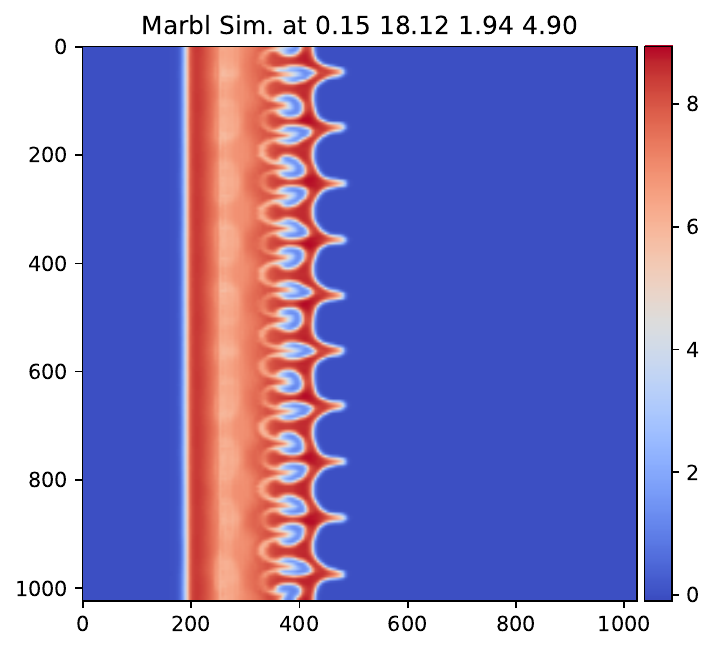}   \\
  \hline
  Ml & \includegraphics[width=0.175\textwidth]{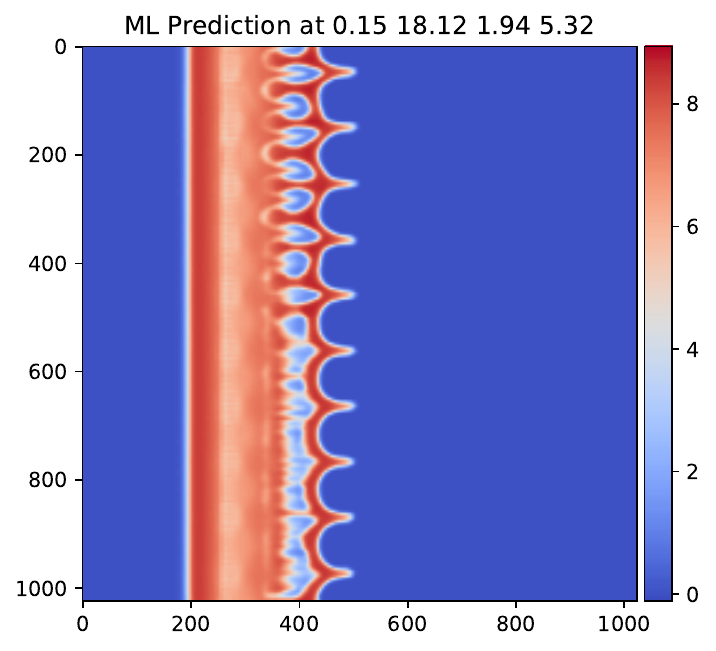} & \includegraphics[width=0.175\textwidth]{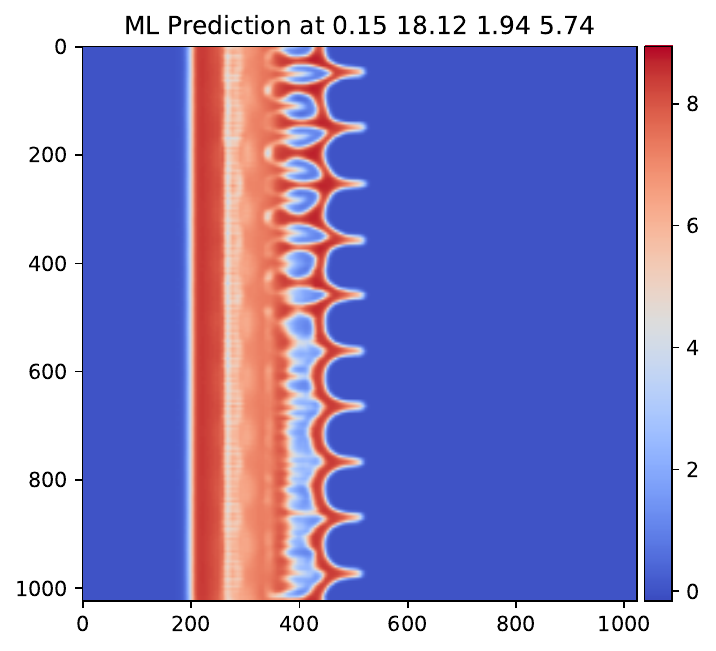} & \includegraphics[width=0.175\textwidth]{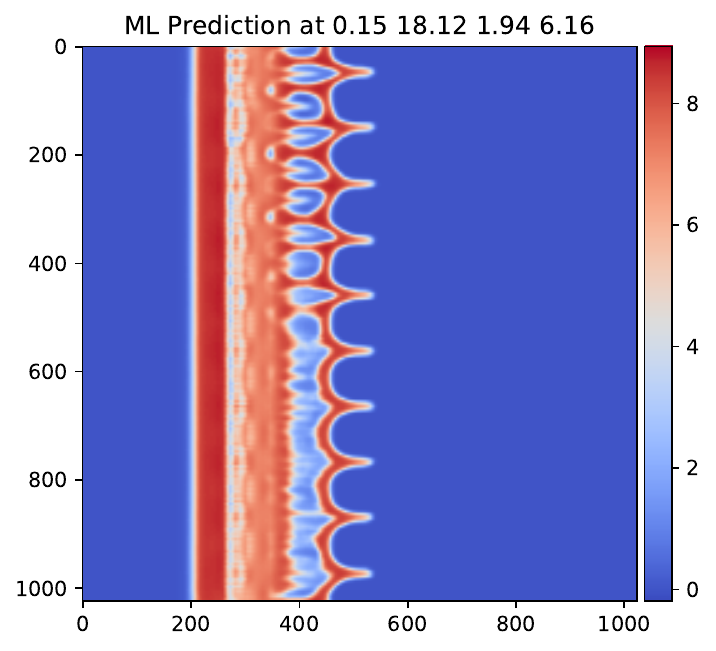} & \includegraphics[width=0.175\textwidth]{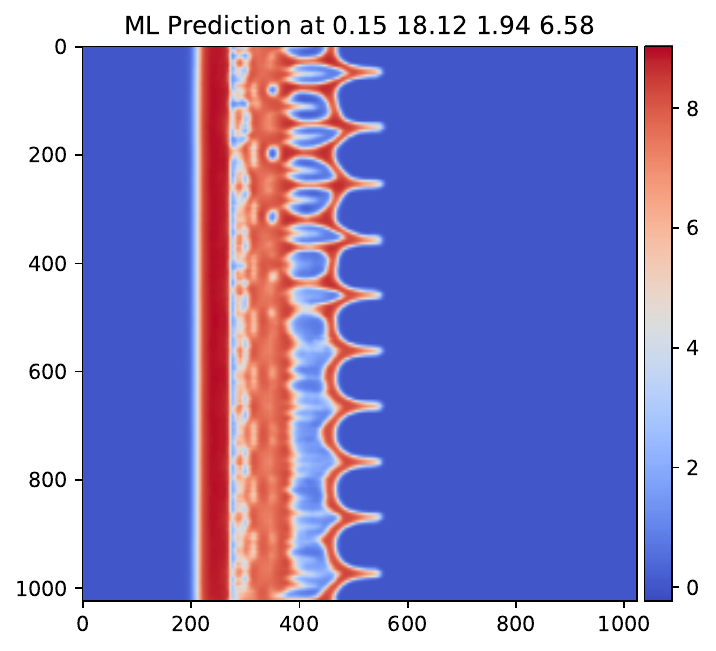} \\
  Sim & \includegraphics[width=0.175\textwidth]{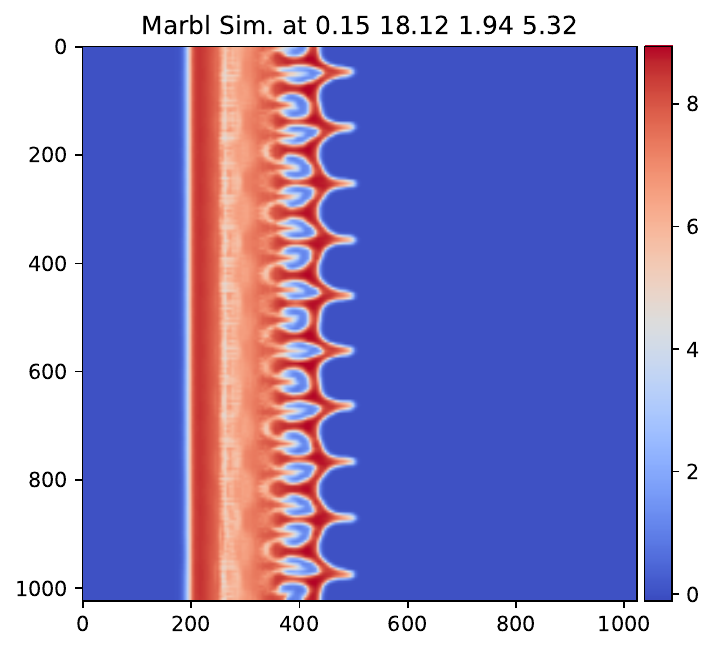} & \includegraphics[width=0.175\textwidth]{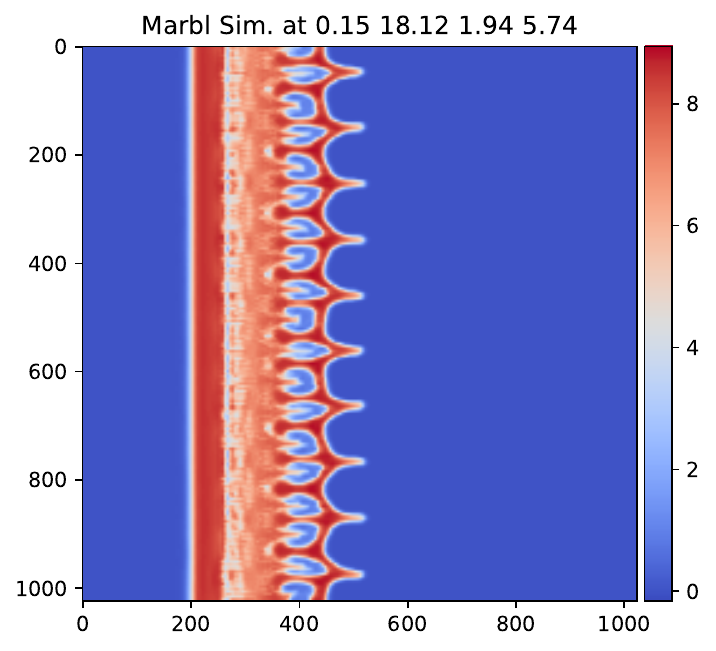} & \includegraphics[width=0.175\textwidth]{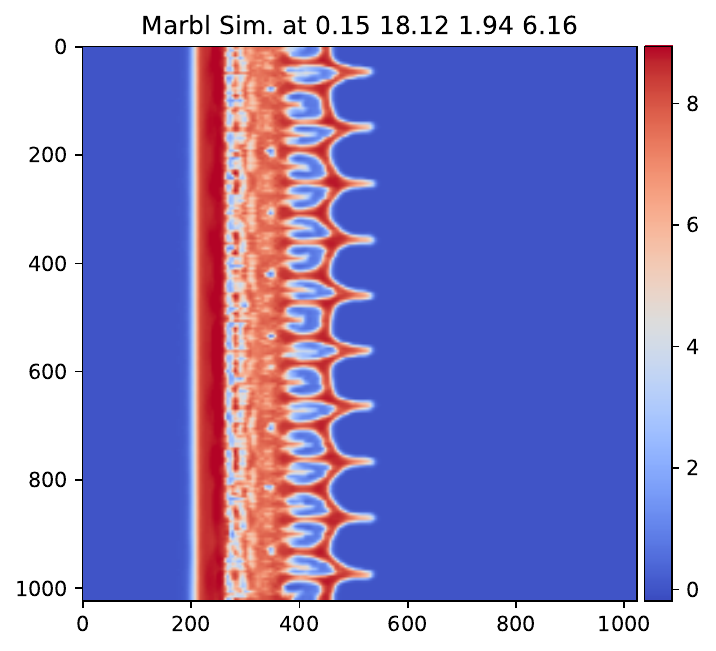} & \includegraphics[width=0.175\textwidth]{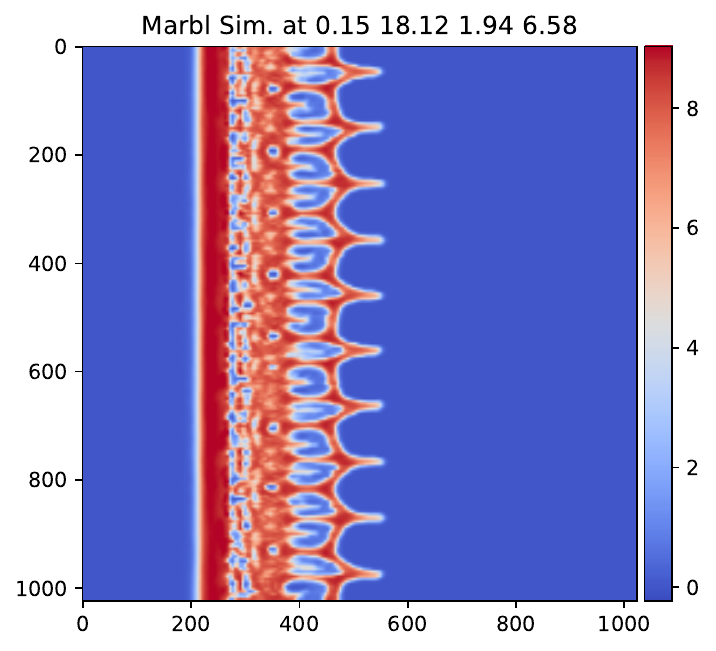}   \\
  \hline
  \end{tabular}
  \caption{
      Figure of density predictions and  truth for the double sine wave impact. Rows show machine learning model (ML) next to simulation (Sim) results.
      }\label{fig:ds_density}
  \end{center}
\end{figure}

\begin{figure}[!htb]
  \begin{center}
  \begin{tabular}{*{2}{c}}
    Ml prediction & Simulation \\
  \includegraphics[width=0.33\textwidth]{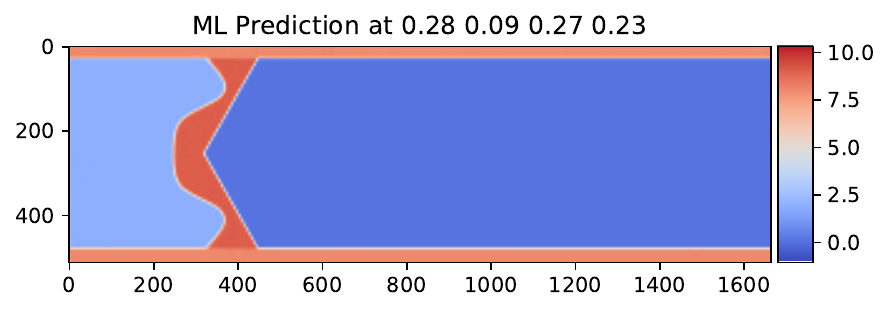} & \includegraphics[width=0.33\textwidth]{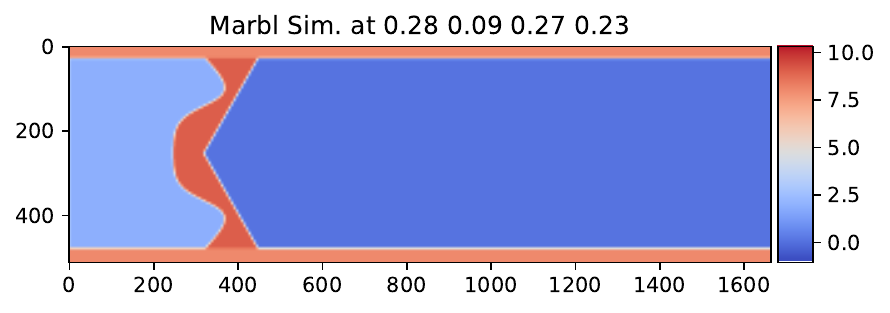} \\
  \includegraphics[width=0.33\textwidth]{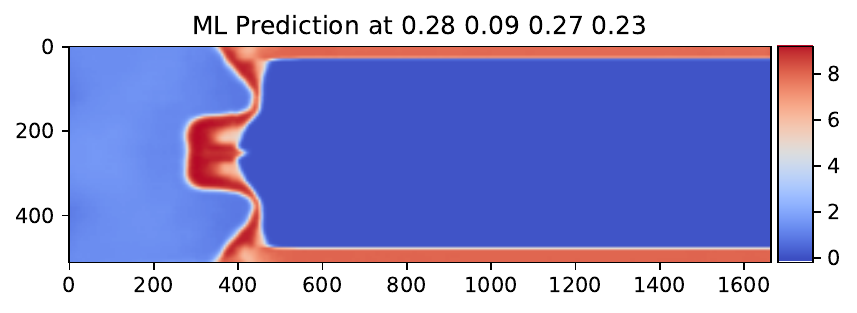} & \includegraphics[width=0.33\textwidth]{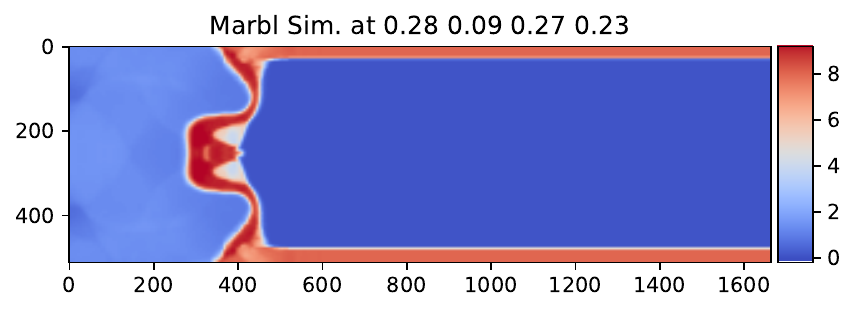} \\
  \includegraphics[width=0.33\textwidth]{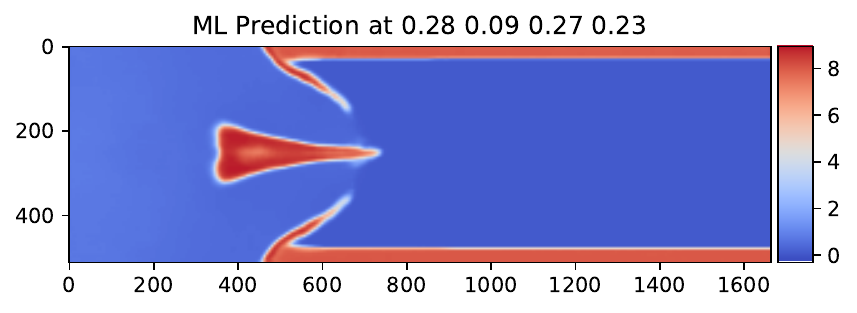} & \includegraphics[width=0.33\textwidth]{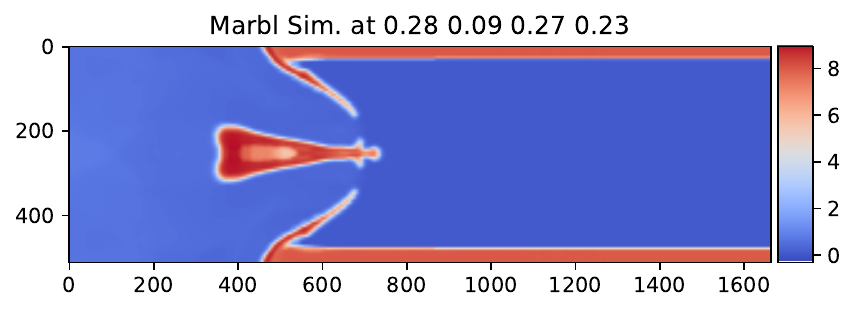} \\
  \includegraphics[width=0.33\textwidth]{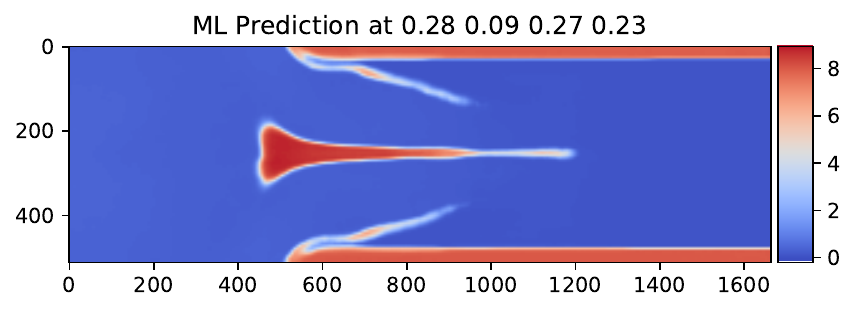} & \includegraphics[width=0.33\textwidth]{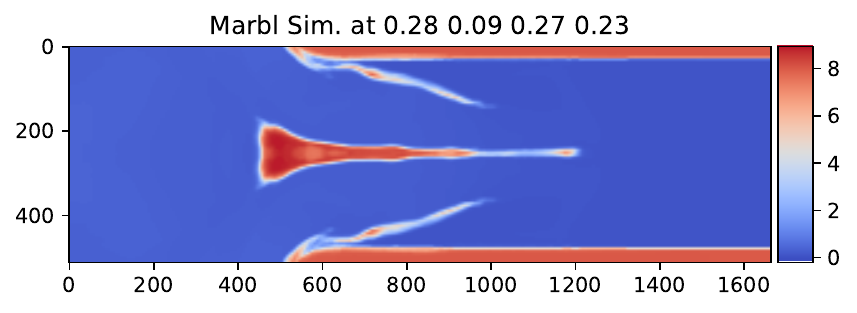} \\
  \includegraphics[width=0.33\textwidth]{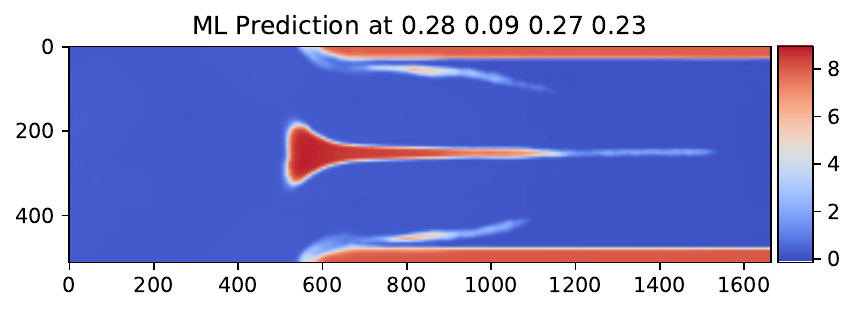} & \includegraphics[width=0.33\textwidth]{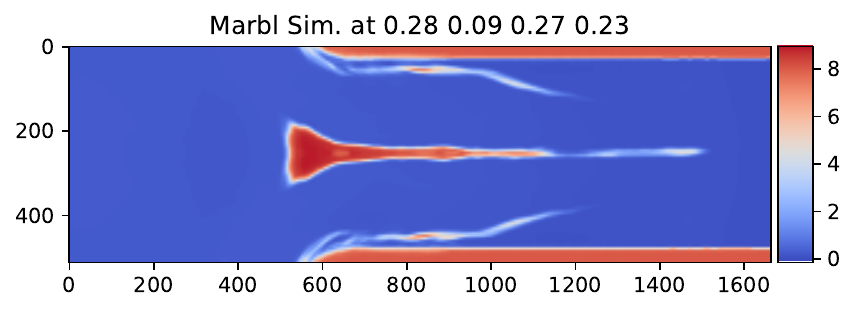} \\
  \includegraphics[width=0.33\textwidth]{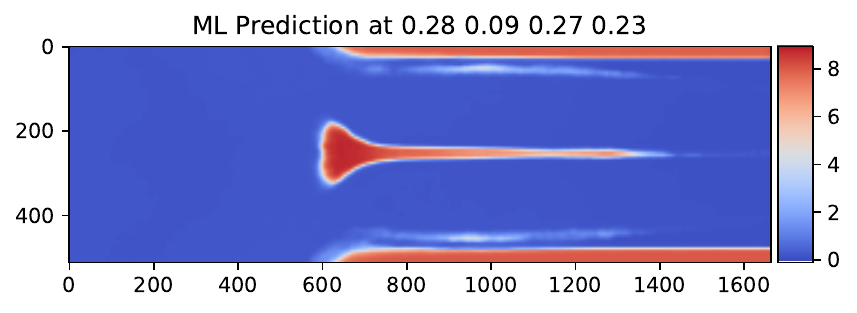} & \includegraphics[width=0.33\textwidth]{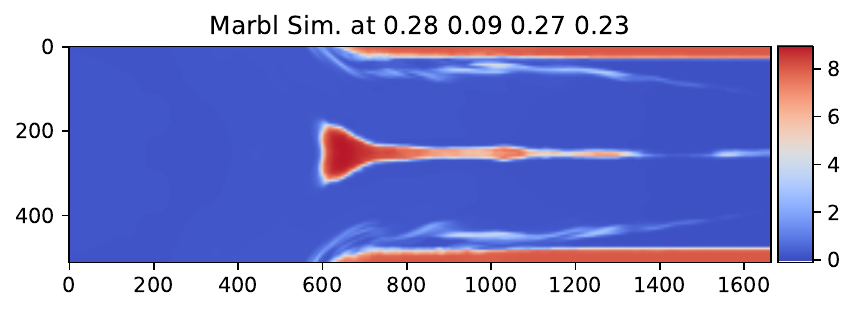} \\
  \includegraphics[width=0.33\textwidth]{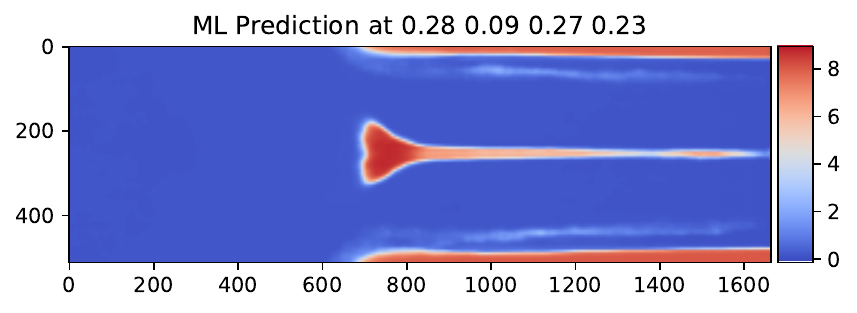} & \includegraphics[width=0.33\textwidth]{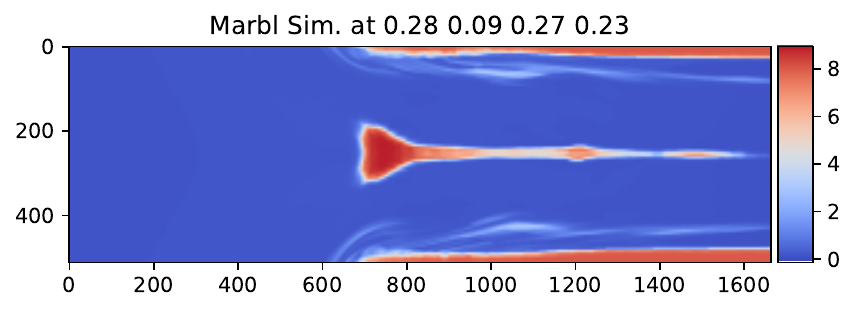} \\
  \hline
  \end{tabular}
  \caption{
      Figure of density predictions and truth for the linear shaped charge.
      }\label{fig:lsc_density}
  \end{center}
\end{figure}

\begin{figure}[!htb]
  \begin{center}
  \begin{tabular}{*{6}{c}}
  Ml & \includegraphics[width=0.175\textwidth]{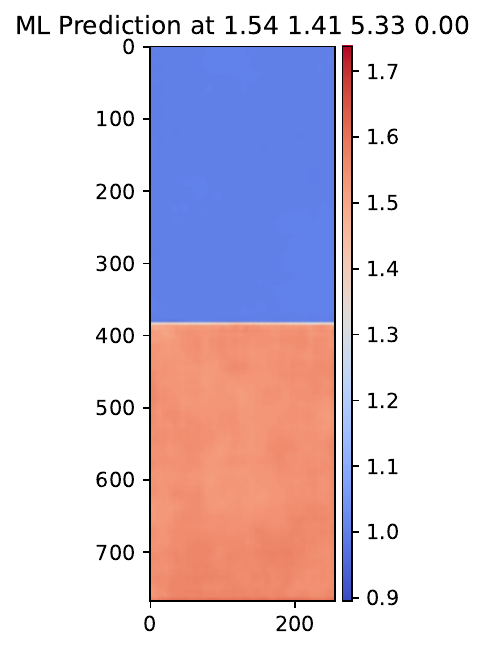} & \includegraphics[width=0.175\textwidth]{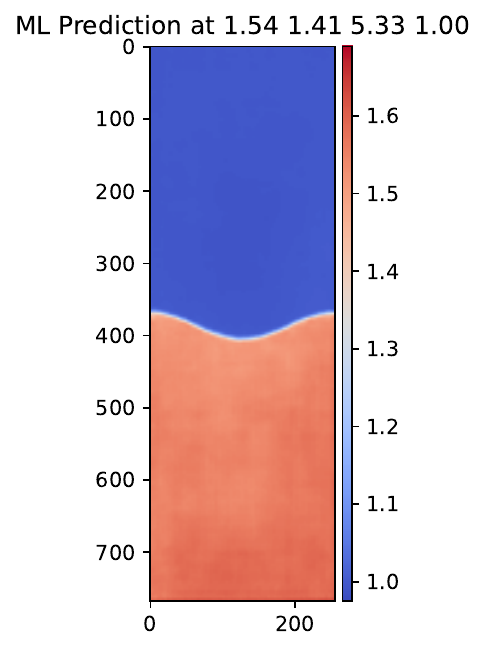}  & \includegraphics[width=0.175\textwidth]{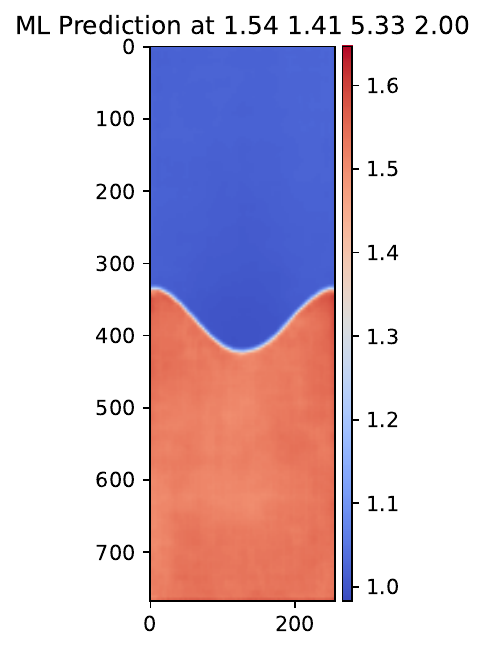}  & \includegraphics[width=0.175\textwidth]{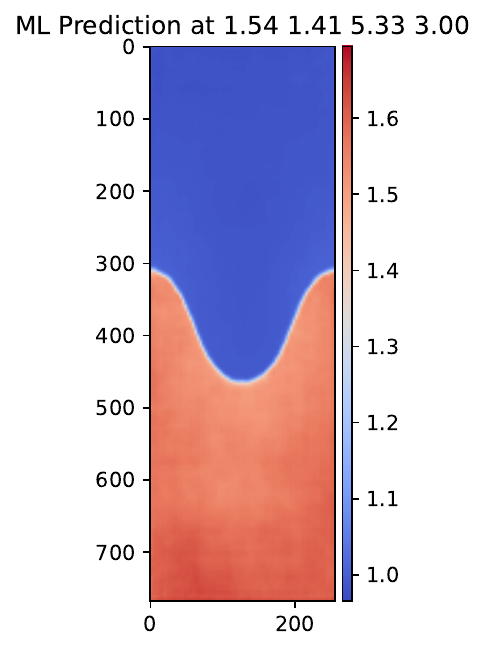}  & \includegraphics[width=0.175\textwidth]{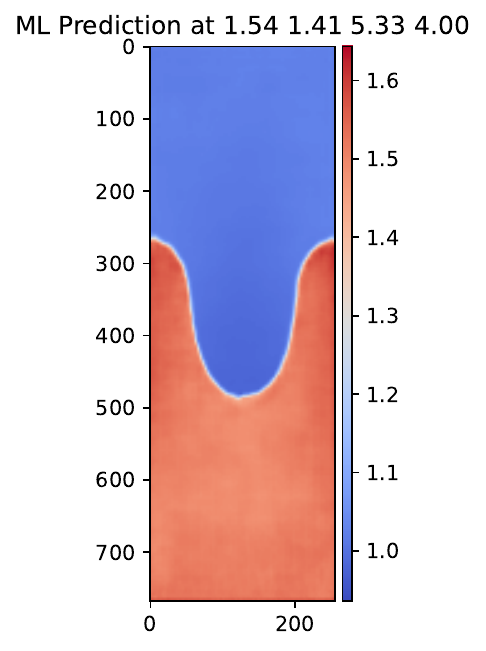} \\
  Sim & \includegraphics[width=0.175\textwidth]{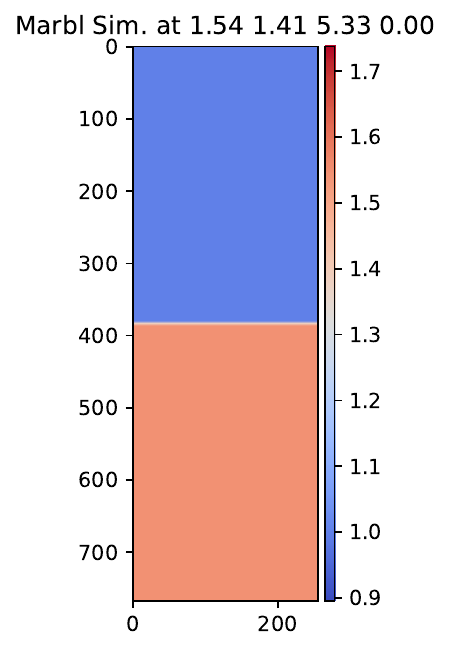} & \includegraphics[width=0.175\textwidth]{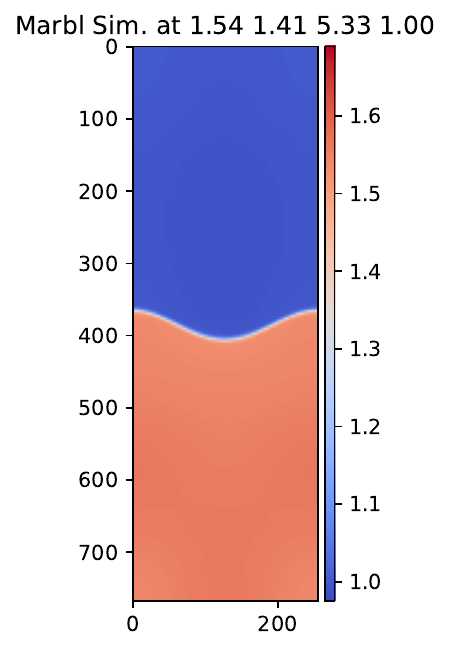}  & \includegraphics[width=0.175\textwidth]{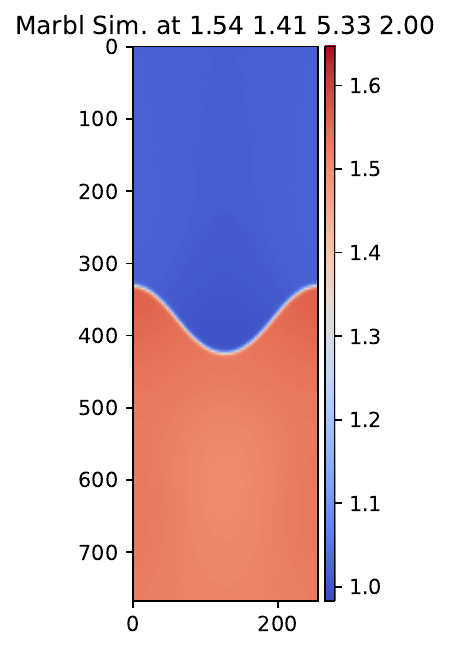}  & \includegraphics[width=0.175\textwidth]{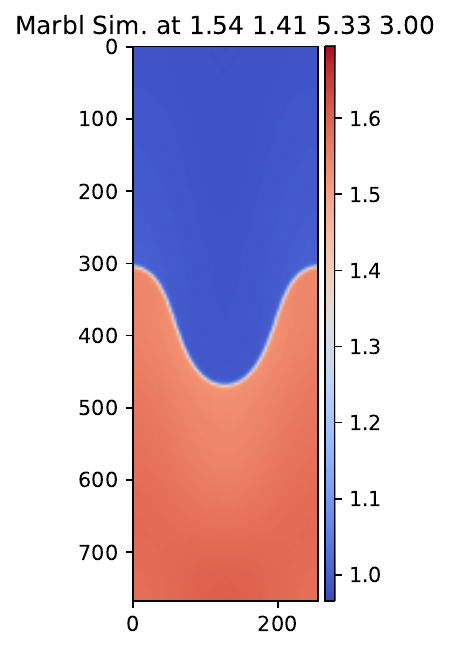}  & \includegraphics[width=0.175\textwidth]{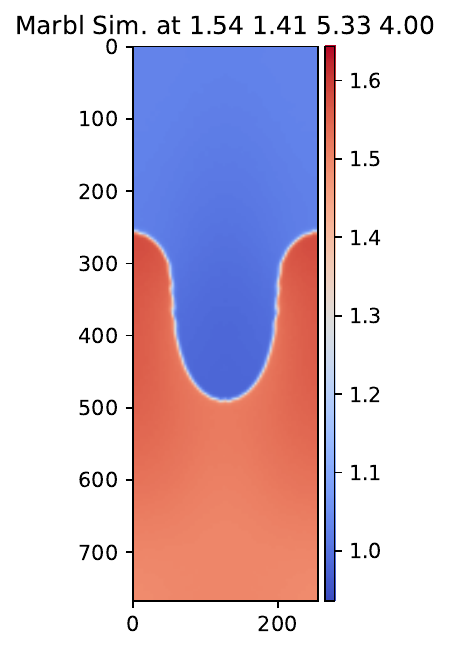} \\
  \hline
  Ml & \includegraphics[width=0.175\textwidth]{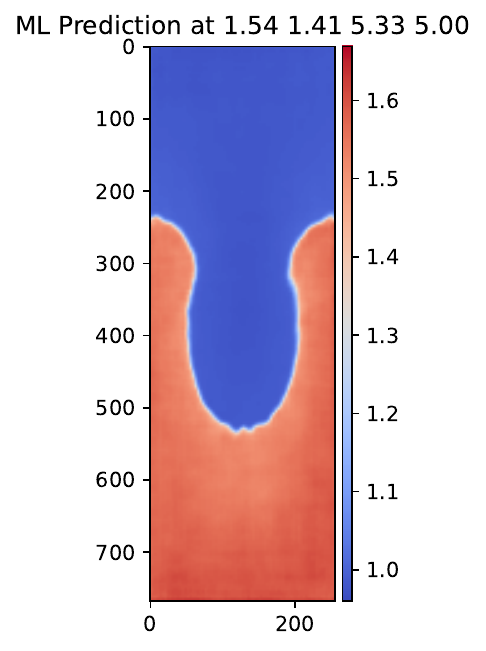} & \includegraphics[width=0.175\textwidth]{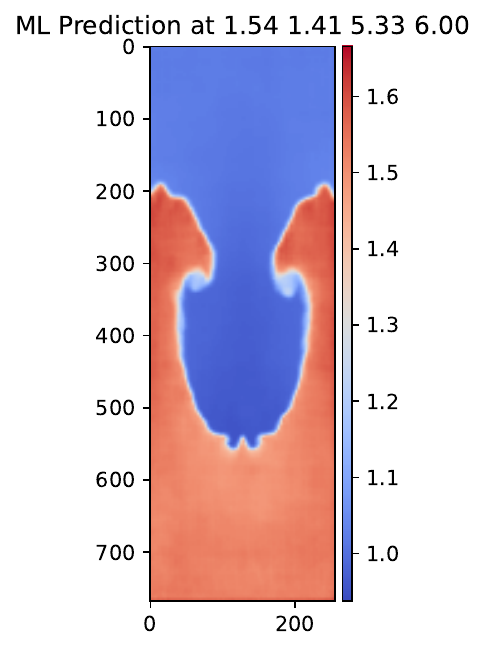}  & \includegraphics[width=0.175\textwidth]{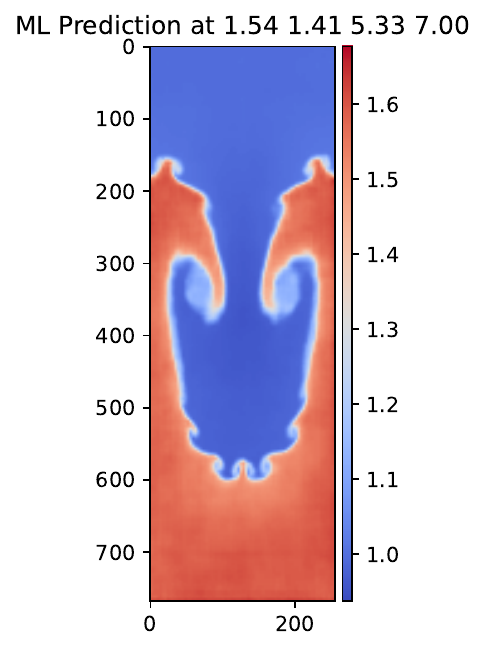}  & \includegraphics[width=0.175\textwidth]{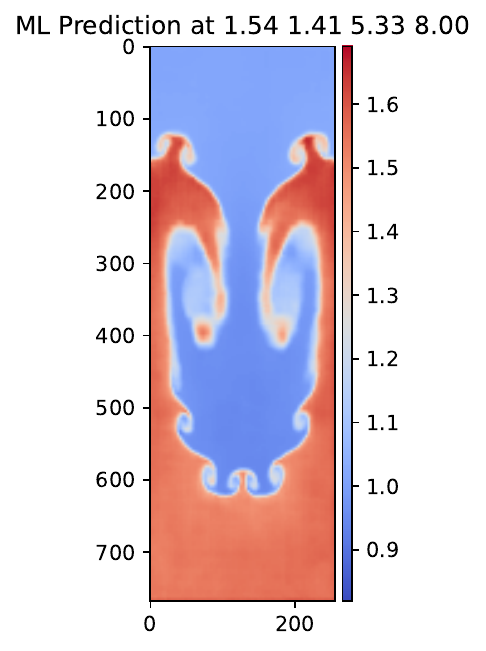}  & \includegraphics[width=0.175\textwidth]{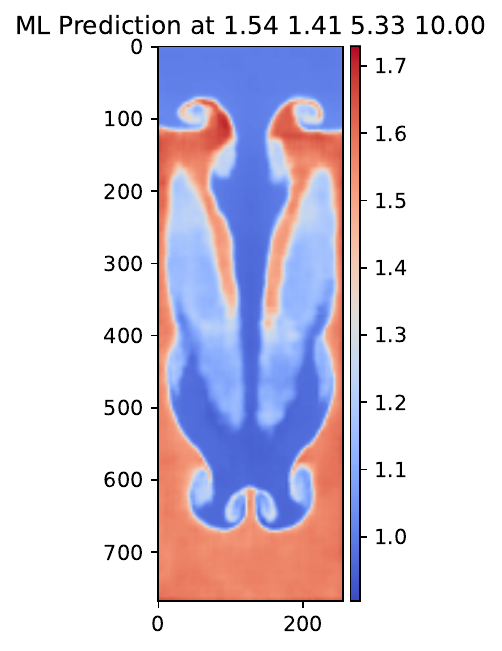} \\
  Sim & \includegraphics[width=0.175\textwidth]{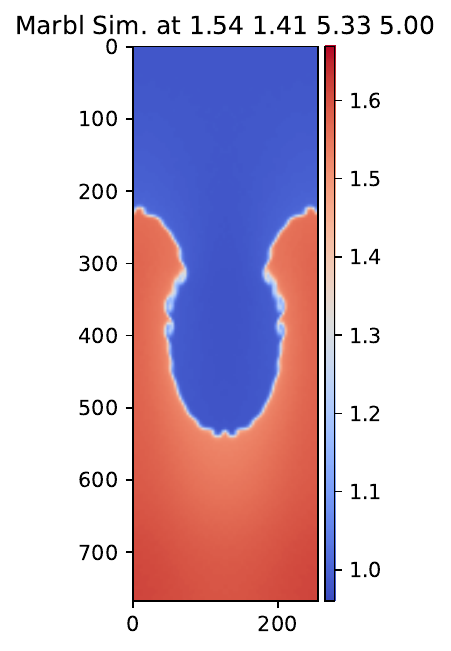} & \includegraphics[width=0.175\textwidth]{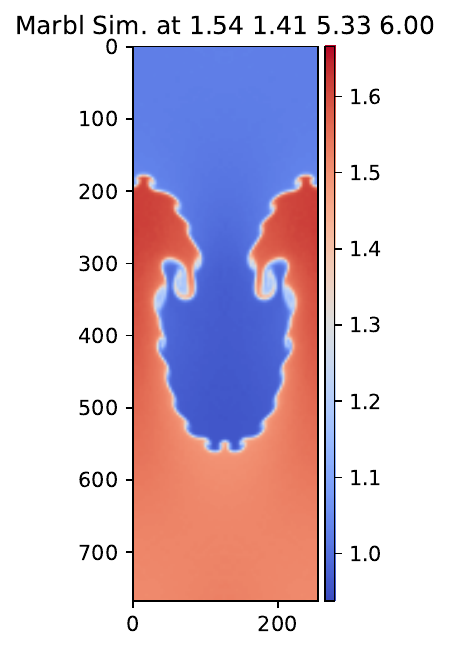}  & \includegraphics[width=0.175\textwidth]{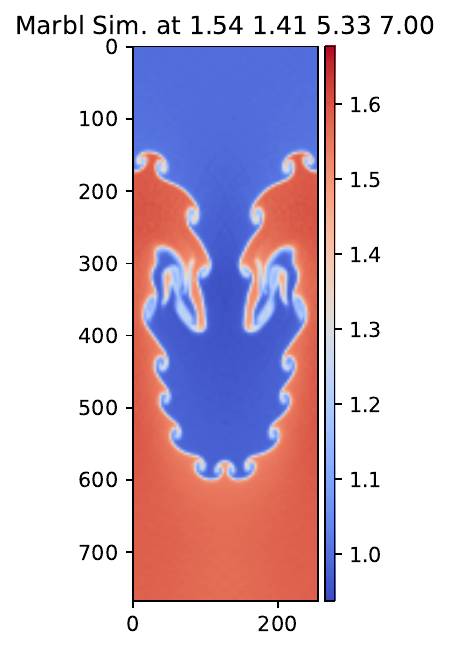}  & \includegraphics[width=0.175\textwidth]{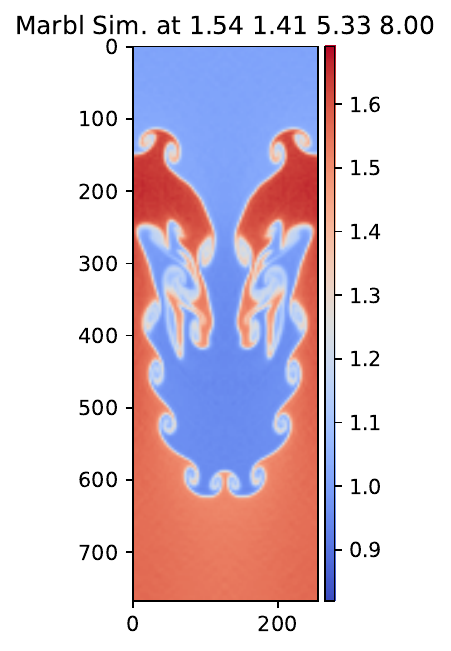}  & \includegraphics[width=0.175\textwidth]{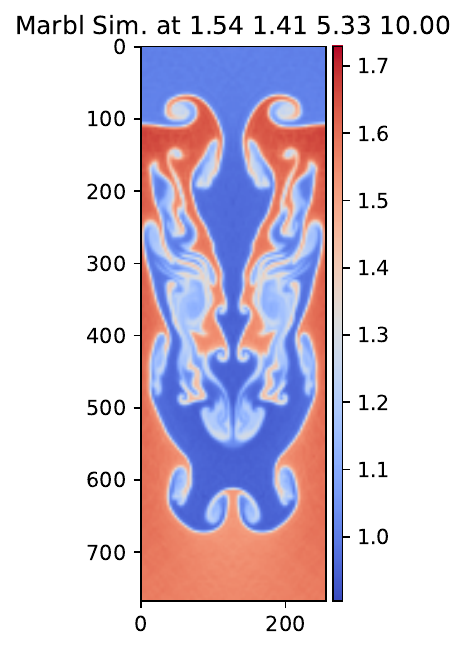} \\
  \hline
  \end{tabular}
  \caption{
      Figure of density predictions and truth for the Rayleigh-Taylor simulation. Rows show machine learning model (ML) next to simulation (Sim) results.
      }\label{fig:rt_density}
  \end{center}
\end{figure}

For all models, a simulation not in the training set is compared to the ML model's density predictions in Figures~\ref{fig:pchip_density}-\ref{fig:rt_density}. Density is just one of the multi-field outputs from these models. The density field provides a good illustration on how the materials are mixing in the hydrodynamic simulations. All of the ML predictions appear to be an excellent representation of the actual field. It is nearly impossible to notice the errors in the ML predictions of the PCHIP impact. The double sine wave impact has much finer Richtmyer-Meshkov instabilities than the PCHIP impact, and the ML model blurs a high-low density interface of the finer instabilities (e.g. at the copper-lexan-copper interface). The ML predictions of the linear shaped charge look excellent, with only some minor fine details missing on the jet as it forms.

The Rayleigh-Taylor results (shown in Figure~\ref{fig:rt_density}) are perhaps the most interesting. For all times the ML prediction is tracking the overall interface between the high-low density fluids. However, as the simulation time progresses the overall interface becomes significantly more complicated with many very fine perturbations along the mixing interface. The ML model appears incapable of modeling these finer features and appears to blur the many very fine high-low density features with a smooth "middle" density prediction. This spatial averaging behavior of the ML predictions is similar to the expected behavior of a RANS solver \cite{PhysRevE.106.025101}. 

Additional results showing predictions and errors for each field are shown in Appendix~\ref{app:modelres}. Predictions are compared to the simulation results for a simulation not in the training set. All model fields are shown at different times within the simulation domain. The mean absolute error for each field is roughly two orders of magnitude better than the ML model initiated with random weights.
Additional results showing predictions and errors for each field are shown in Appendix~\ref{app:modelres}. Predictions are compared to the simulation results for a simulation not in the training set. All model fields are shown at different times within the simulation domain. The mean absolute error for each field is roughly two orders of magnitude better than the ML model initiated with random weights.

\section{Real time visualization}\label{sec:visualization}

A desired application that the ML models enable is real-time visualization and exploration. The ML model can be queried at any place within the parameterized space. These full-field solutions out of the ML model can be thought of as an interpolation in the parameterized space. While it would be possible to quickly view results from an ensemble dataset in a similar fashion, the ML model offers the ability to quickly explore anywhere in the high dimensional space. Additionally, inference from the ML model can be performed on a portable laptop computer, while both the dataset and simulations are solidly in the realm of HPC. The full-field inference for the PCHIP impact model takes 0.2~s using an Apple M2 Max on battery power.

\begin{figure}[!htb]
  \begin{center}
  \begin{tabular}{*{3}{c}}
  \includegraphics[width=0.32\textwidth]{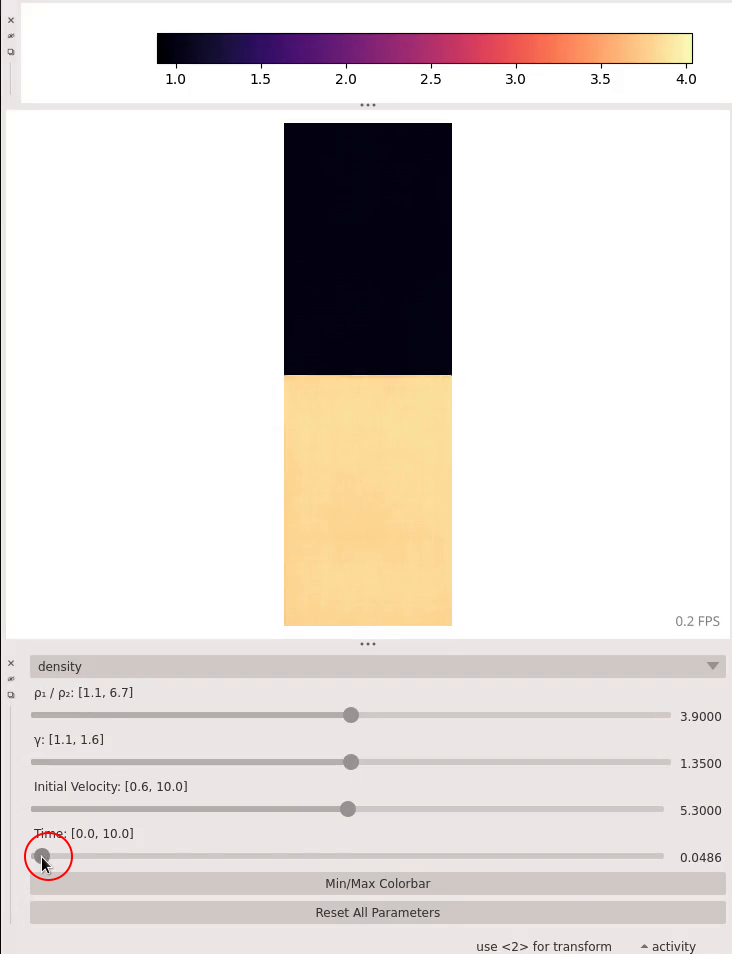} & \includegraphics[width=0.32\textwidth]{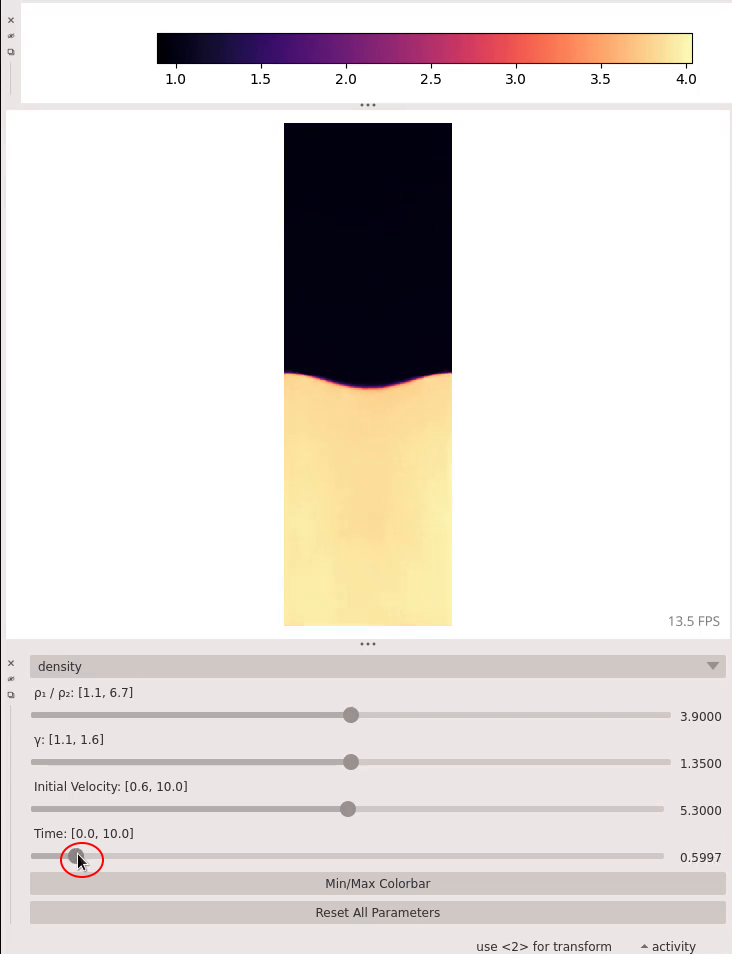} & \includegraphics[width=0.32\textwidth]{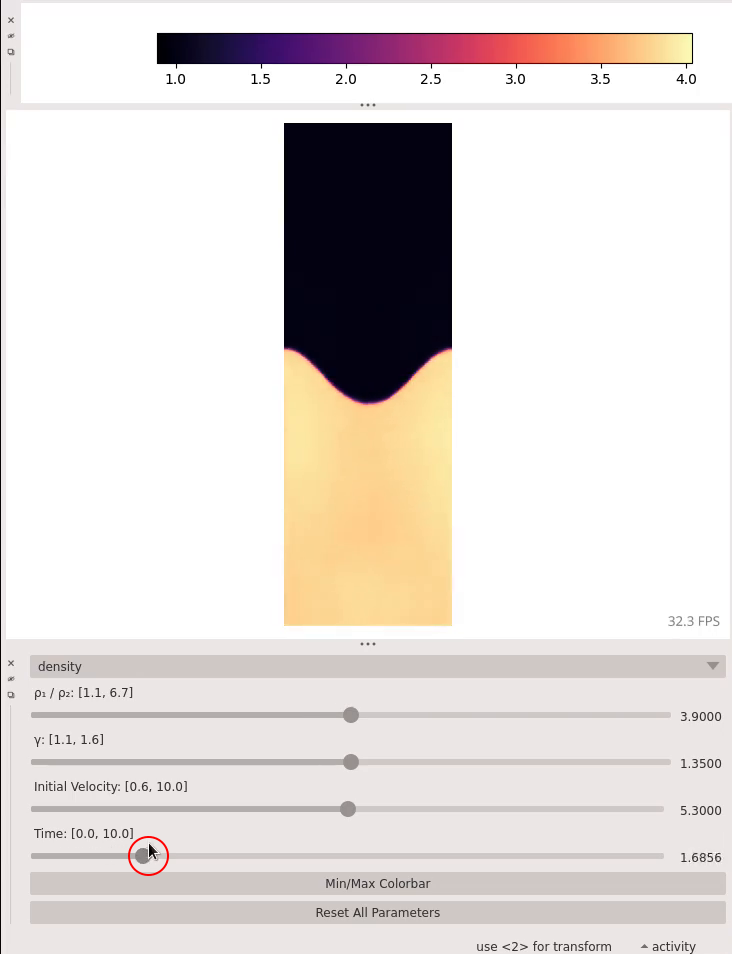} \\
  \includegraphics[width=0.32\textwidth]{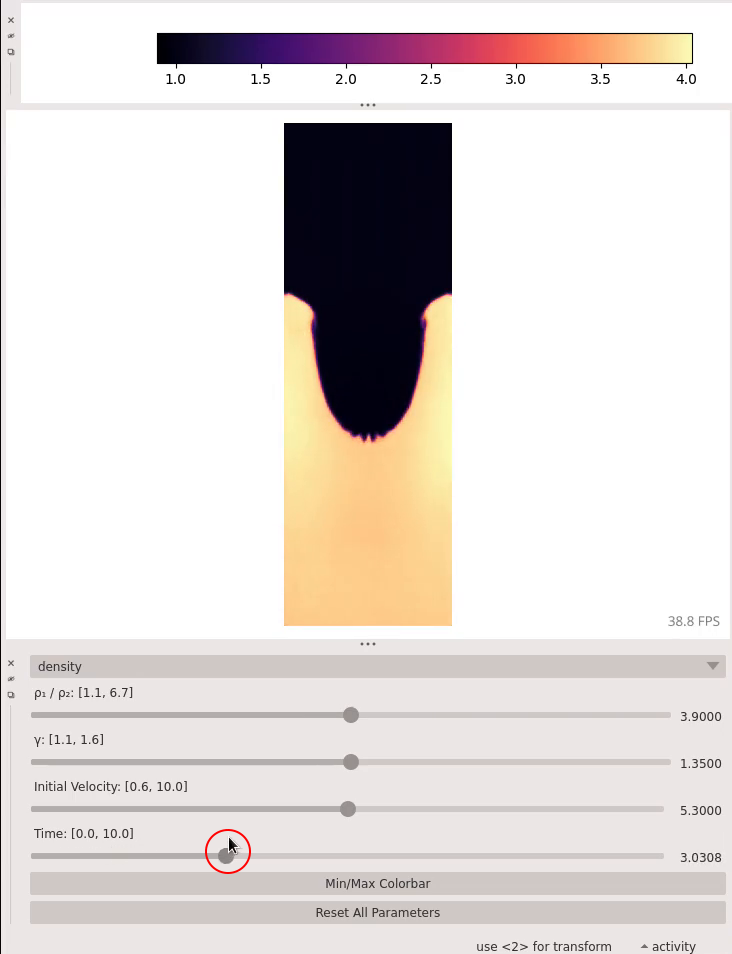} & \includegraphics[width=0.32\textwidth]{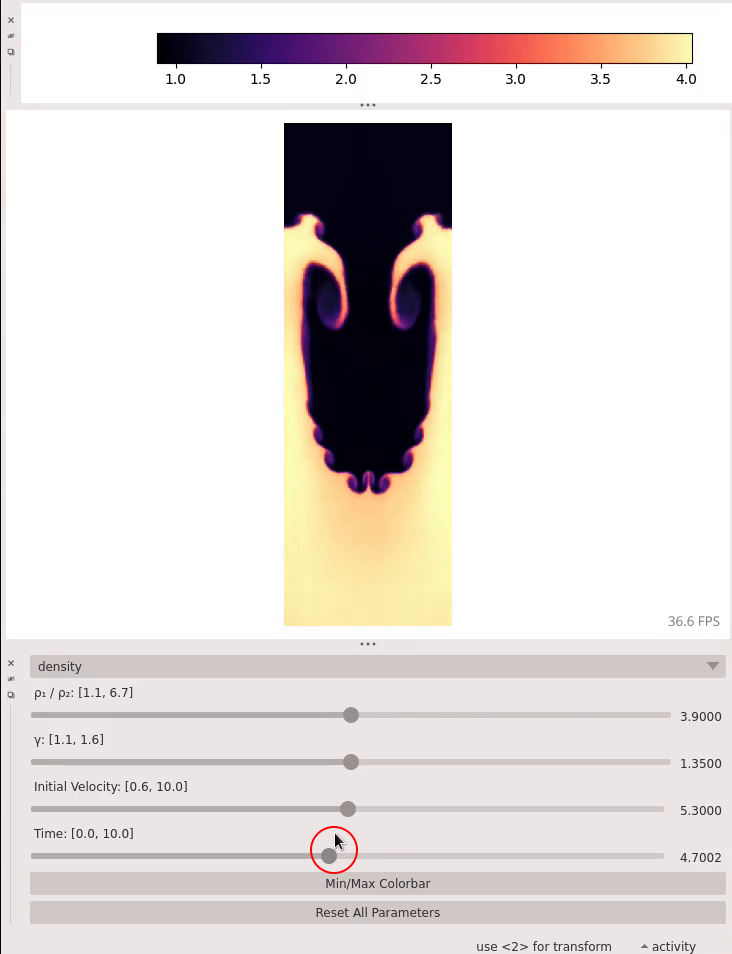} &
  \includegraphics[width=0.32\textwidth]{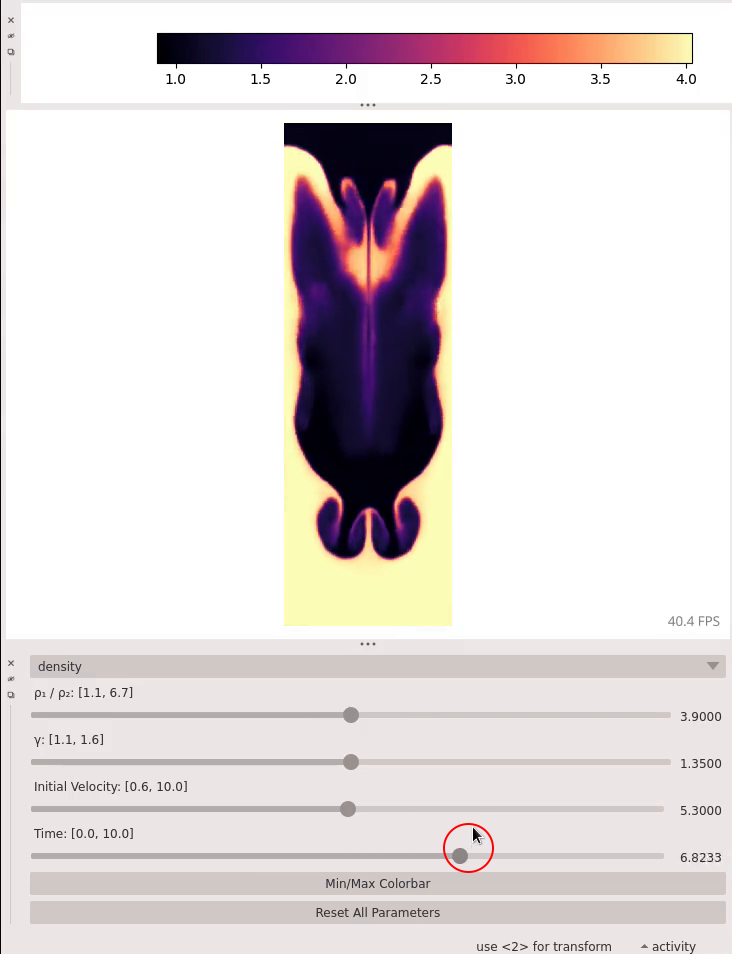} \\
  \includegraphics[width=0.32\textwidth]{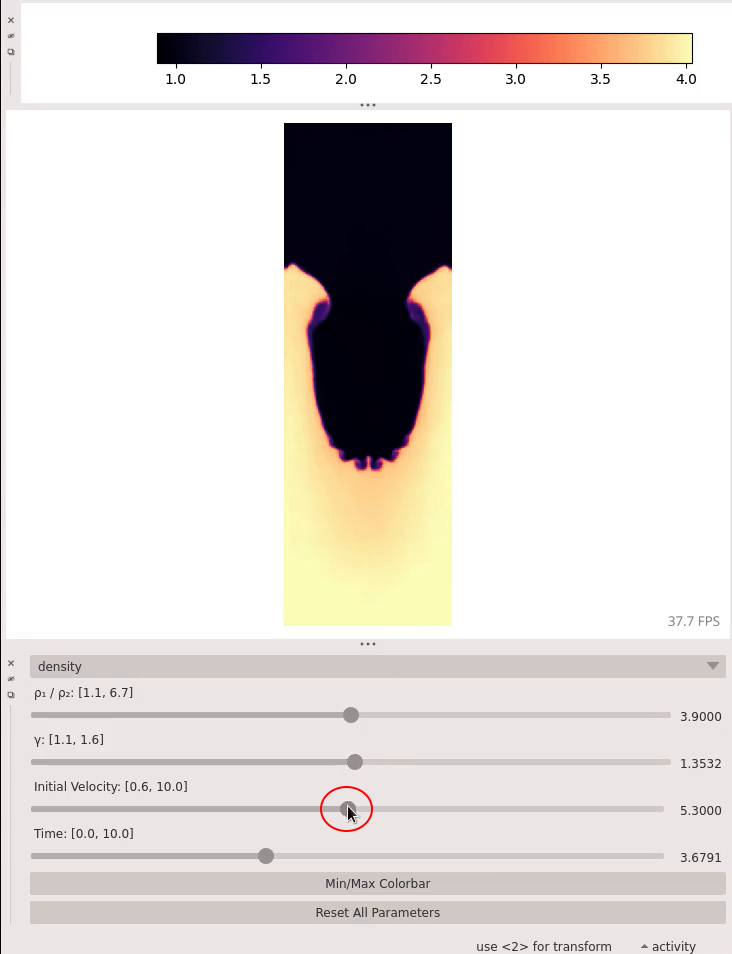} & \includegraphics[width=0.32\textwidth]{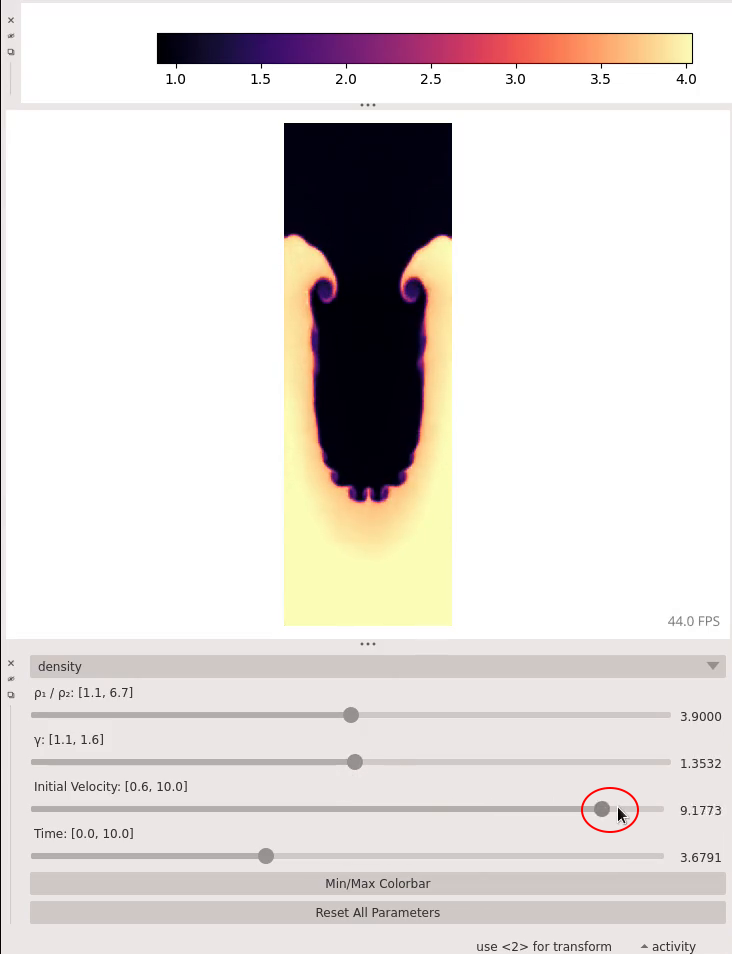} &
  \includegraphics[width=0.32\textwidth]{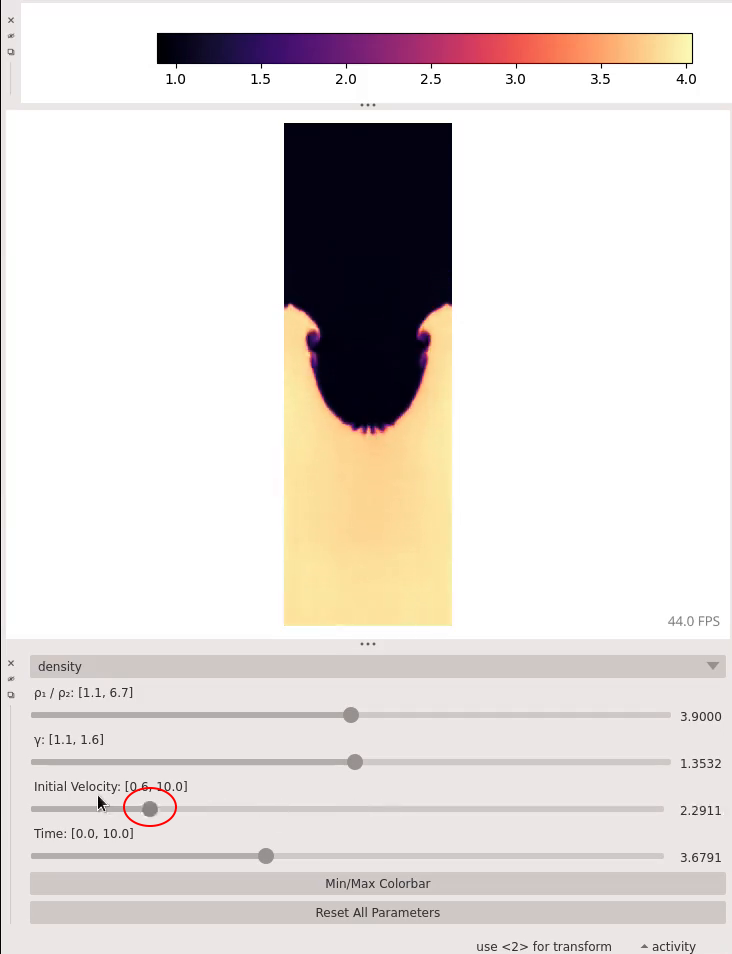} \\
  \end{tabular}
  \caption{
      Real-time visualization of the Rayleigh-Taylor model showing the full-field density solution. The first two rows show a user sliding through time for a particular parameterization that the model was not trained on. The last row shows the user sliding through the input velocity at a fixed time. These are still frames from a video of a user interacting with the tool.
      }\label{fig:rt_slider}
  \end{center}
\end{figure}

A real-time visualization tool was created using napari\cite{contributors2019napari}. Static images taken from the tool are shown in Figure~\ref{fig:rt_slider}. The user is presented with 4 inputs to the Rayleigh-Taylor ML model (density ratio, heat capacity ratio $\gamma$, initial velocity, and time) as a slider interface. The user can either click or slide through the input parameters and view the resulting full-field solutions in real-time. There is a dropdown menu to switch between which output field is displayed (e.g. density, velocity, pressure). Inference was done using an AMD MI250X, which takes an average time of 0.002~ms using float16 precision. Having fast ML inference is key to visualize the full-field solutions in real-time.

Videos of the slider interface while doing live ML inference are included in the supplemental material for all four ML models. The interface allows for the interpolation of the physical fields at any point within the ensemble domain. The predictions from the tool look excellent. With the linear shaped charge model, users can interactively explore how the liner shape influences the long-term jet development, to build intuition on how these parameters work together to enhance or mitigate the jet. With the double sine wave and PCHIP models, users can gain intuition on how the initial shape of the copper target influences the development of the RMI. This interactive ability allows users to explore and interpolate an ensemble of hydrodynamic simulations to further their understanding of complex physical instabilities.

The real-time visualization tool also makes it easy to explore how well ML models behave when extrapolating outside of the ensemble domain. While it would be incredibly desirable for a fast-running hydrodynamics tool to extrapolate, unfortunately these ML models possess no ability to extrapolate the hydrodynamic instabilities. As all input parameters are extrapolated, the ML model's predictions go to zero everywhere. Even with subtle extrapolations like 10\%, users will see predictions quickly begin to look unrealistic as interfaces begin to blur with cloud-like pockets of material disappearing. The material interfaces do not move when extrapolating in simulation time, but rather begin to disappear. In summary as the model begins to extrapolate then the predictions quickly look both nonphysical and unbelievable.


\section{Conclusion}

Performing ensembles of simulations is one way to understand the complex sensitivity of physical instabilities to initial conditions. We present results from four ensemble datasets involving the Richtmyer-Meshkov Instability (RMI) and Rayleigh-Taylor instabilities. A machine leaning (ML) model framework was proposed to learn the full-field solutions as functions of initial physical and geometric conditions. The ML model weights were a thousand times smaller than the datasets (as a form of lossy compression). Additionally inference from the ML model can be millions of times faster than a full hydrodynamic solution, while also not requiring HPC resources demanded by the ensemble. The ML model enables real-time visualization of the instabilities by sliding through both initial conditions and time. This enables users to explore and gain intuition about complicated hydrodynamic relationships. Such tools also enable the dissemination of hydrodynamic simulations and results to additional audiences (e.g. experimentalists) with reduced barriers of entry (e.g. high performance computing). Lastly, the output of the ML model is fully differentiable with respect to the input, which can be useful for applications in design optimization or uncertainty quantification.

\FloatBarrier

\section*{Acknowledgments}
This work was performed under the auspices of the U.S. Department of Energy by Lawrence Livermore National Laboratory under Contract DE-AC52-07NA27344 and was supported by the LLNL-LDRD Program under Project No. 21-SI-006.

We would like to acknowledge Kareem Hegazy for pointing out that placing the multiple physics fields into the image channel would enable the model to learn the correlations between physical fields, which thus resulted in more accurate ML models.

\section*{Data Availability Statement}
The Rayleigh-Taylor and PCHIP datasets and ML model weights will be available upon request.

\appendix

\section{Model architectures}\label{app:modelarch}

\begin{table}
  \center
  \caption{\label{tab:modelpchip}PCHIP impact ML model architecture.}
  \begin{tabular}{lcc}
    Layer name & Output shape & \#~of learnable parameters \\ 
    \hline
    Input parameters & [6, 1, 1] & - \\
    ConvTranspose2d $\to$ BatchNorm2d $\to$ ReLU & [1024, 4, 4] & 100,352 \\
    ConvTranspose2d $\to$ BatchNorm2d $\to$ ReLU & [1024, 8, 8] & 16,779,264 \\
    ConvTranspose2d $\to$ BatchNorm2d $\to$ ReLU & [1024, 16, 16] & 16,779,264 \\
    ConvTranspose2d $\to$ BatchNorm2d $\to$ ReLU & [1024, 32, 32] & 16,779,264 \\
    ConvTranspose2d $\to$ BatchNorm2d $\to$ ReLU & [1024, 64, 64] & 16,779,264 \\
    ConvTranspose2d $\to$ BatchNorm2d $\to$ ReLU & [512, 128, 128] & 8,389,632 \\
    ConvTranspose2d $\to$ BatchNorm2d $\to$ ReLU & [256, 256, 256] & 2,097,664 \\
    ConvTranspose2d $\to$ BatchNorm2d $\to$ ReLU & [128, 512, 512] & 524,544 \\
    ConvTranspose2d & [6, 1024, 1024] & 12,288 \\
    Linear map & [6, 1024, 1024] & 12 \\
    \hline
    Total number of learnable parameters & & 78,241,548 \\
  \end{tabular}
\end{table}

\begin{table}
  \center
  \caption{\label{tab:modelDS}Double sine wave ML model architecture.}
  \begin{tabular}{lcc}
    Layer name & Output shape & \#~of learnable parameters \\ 
    \hline
    Input parameters & [4, 1, 1] & - \\
    ConvTranspose2d $\to$ BatchNorm2d $\to$ ReLU & [1024, 4, 4] & 67,584 \\
    ConvTranspose2d $\to$ BatchNorm2d $\to$ ReLU & [1024, 8, 8] & 16,779,264 \\
    ConvTranspose2d $\to$ BatchNorm2d $\to$ ReLU & [1024, 16, 16] & 16,779,264 \\
    ConvTranspose2d $\to$ BatchNorm2d $\to$ ReLU & [1024, 32, 32] & 16,779,264 \\
    ConvTranspose2d $\to$ BatchNorm2d $\to$ ReLU & [1024, 64, 64] & 16,779,264 \\
    ConvTranspose2d $\to$ BatchNorm2d $\to$ ReLU & [512, 128, 128] & 8,389,632 \\
    ConvTranspose2d $\to$ BatchNorm2d $\to$ ReLU & [256, 256, 256] & 2,097,664 \\
    ConvTranspose2d $\to$ BatchNorm2d $\to$ ReLU & [128, 512, 512] & 524,544 \\
    ConvTranspose2d & [3, 1024, 1024] & 6,144 \\
    Linear map & [3, 1024, 1024] & 6 \\
    \hline
    Total number of learnable parameters & & 78,202,630 \\
  \end{tabular}
\end{table}

\begin{table}
  \center
  \caption{\label{tab:modellin}Linear shaped charge ML model architecture. }
  \begin{tabular}{lcc}
    Layer name & Output shape & \#~of learnable parameters \\ 
    \hline
    Input parameters & [6, 1, 1] & - \\
    ConvTranspose2d $\to$ BatchNorm2d $\to$ ReLU & [1024, 4, 13] & 321,536 \\
    ConvTranspose2d $\to$ BatchNorm2d $\to$ ReLU & [1024, 8, 26] & 16,779,264 \\
    ConvTranspose2d $\to$ BatchNorm2d $\to$ ReLU & [1024, 16, 52] & 16,779,264 \\
    ConvTranspose2d $\to$ BatchNorm2d $\to$ ReLU & [1024, 32, 104] & 16,779,264 \\
    ConvTranspose2d $\to$ BatchNorm2d $\to$ ReLU & [512, 64, 208] & 8,389,632\\
    ConvTranspose2d $\to$ BatchNorm2d $\to$ ReLU & [256, 128, 416] & 2,097,664  \\
    ConvTranspose2d $\to$ BatchNorm2d $\to$ ReLU & [128, 256, 832] & 524,544 \\
    ConvTranspose2d & [128, 512, 1664] & 14,343 \\
    Linear map & [6, 1024, 1024] & 14 \\
    \hline
    Total number of learnable parameters & & 61,685,518 \\
  \end{tabular}
\end{table}

\begin{table}
  \center
  \caption{\label{tab:modelrt}Rayleigh Taylor ML model architecture. }
  \begin{tabular}{lcc}
    Layer name & Output shape & \#~of learnable parameters \\ 
    \hline
    Input parameters & [4, 1, 1] & - \\
    ConvTranspose2d $\to$ BatchNorm2d $\to$ ReLU & [1024, 3, 1] & 14,336 \\
    ConvTranspose2d $\to$ BatchNorm2d $\to$ ReLU & [1024, 6, 2] & 16,779,264 \\
    ConvTranspose2d $\to$ BatchNorm2d $\to$ ReLU & [1024, 12, 4] & 16,779,264 \\
    ConvTranspose2d $\to$ BatchNorm2d $\to$ ReLU & [1024, 24, 8] & 16,779,264 \\
    ConvTranspose2d $\to$ BatchNorm2d $\to$ ReLU & [1024, 48, 16] & 16,779,264 \\
    ConvTranspose2d $\to$ BatchNorm2d $\to$ ReLU & [512, 96, 32] & 8,389,632 \\
    ConvTranspose2d $\to$ BatchNorm2d $\to$ ReLU & [256, 192, 64] & 2,097,664 \\
    ConvTranspose2d $\to$ BatchNorm2d $\to$ ReLU & [128, 384, 128] & 524,544 \\
    ConvTranspose2d & [6, 768, 256] & 12,288 \\
    Linear map & [6, 768, 256] & 12 \\
    \hline
    Total number of learnable parameters & & 78,155,532 \\
  \end{tabular}
\end{table}

\FloatBarrier

\section{Model results}\label{app:modelres}

\subsection{PCHIP impact results}

\begin{figure}[!htb]
  \centering
    \includegraphics[width=1.0\textwidth]{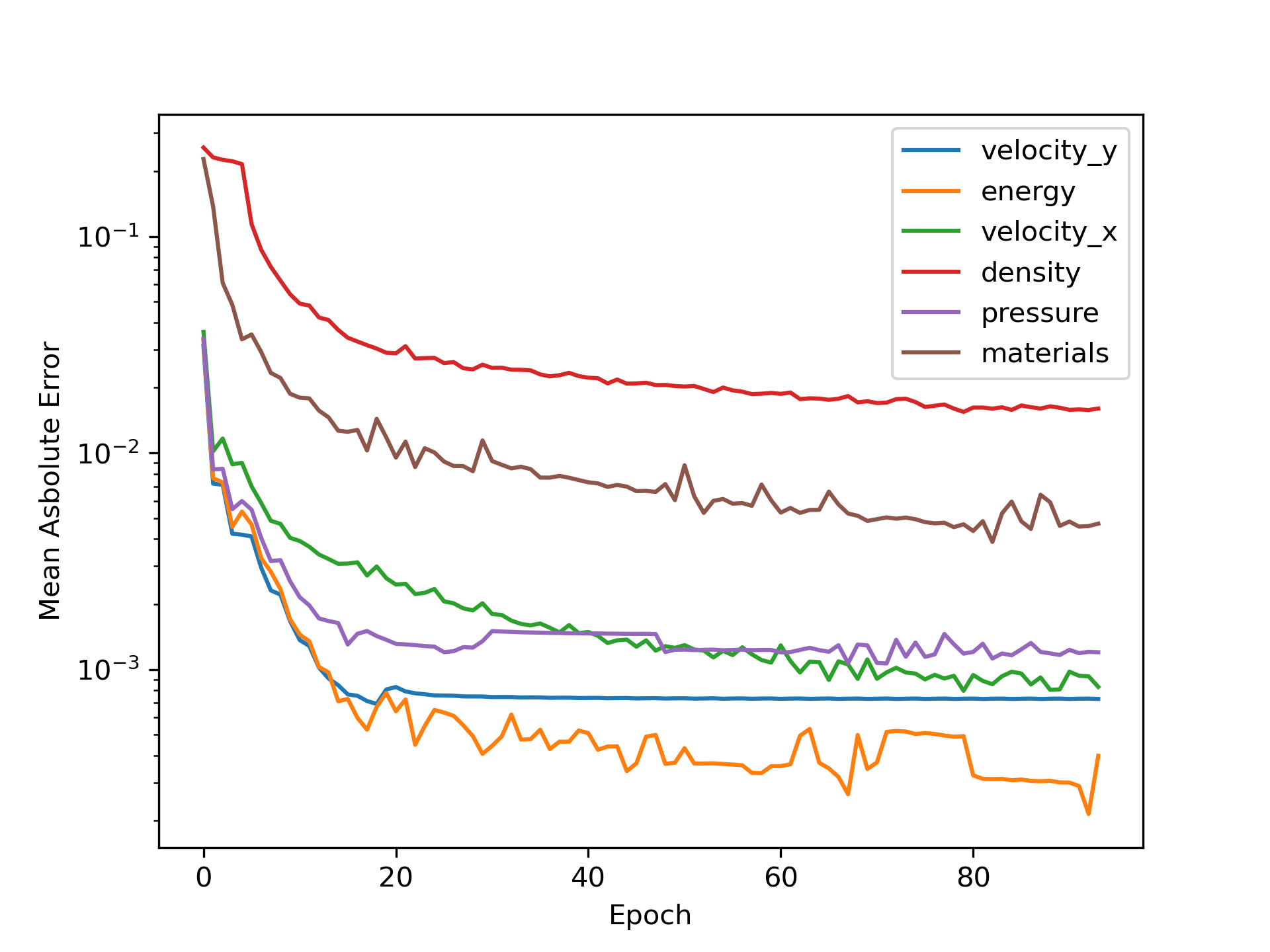}
  \caption{
      The epoch vs mean absolute error for each field while training the PCHIP impact model.
      }
  \label{fig:pchiptraining}
\end{figure}

\begin{figure}[!htb]
  \begin{center}
  \begin{tabular}{*{7}{c}}
  Field & Ml prediction & Simulation & Abs. Error \\
  \hline
  density & \includegraphics[width=0.24\textwidth]{figs/pchip/density_yhat_00.pdf} & \includegraphics[width=0.24\textwidth]{figs/pchip/density_y_00.pdf} & \includegraphics[width=0.24\textwidth]{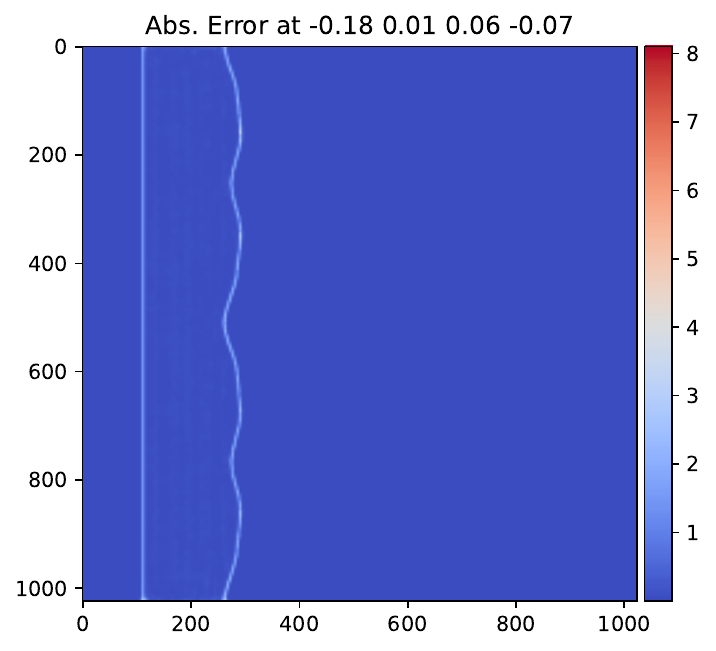} \\
  \hline
  materials & \includegraphics[width=0.24\textwidth]{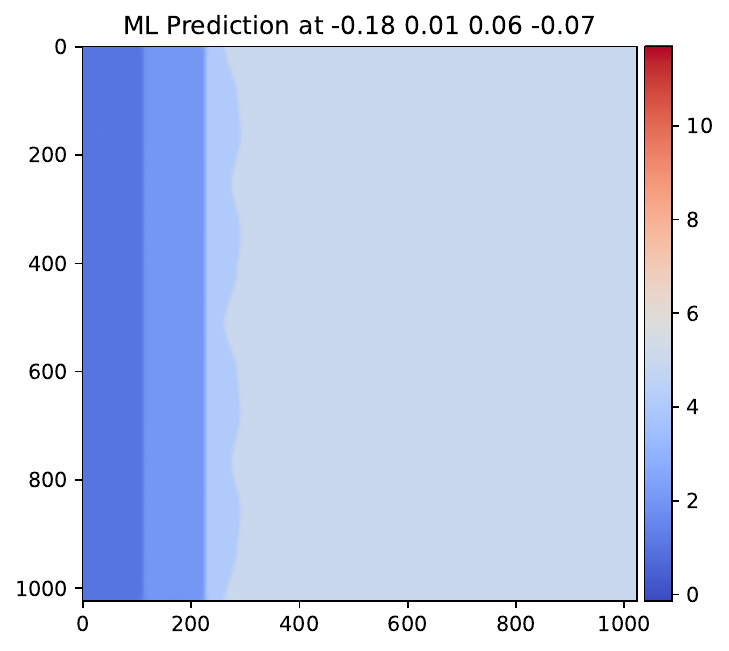} & \includegraphics[width=0.24\textwidth]{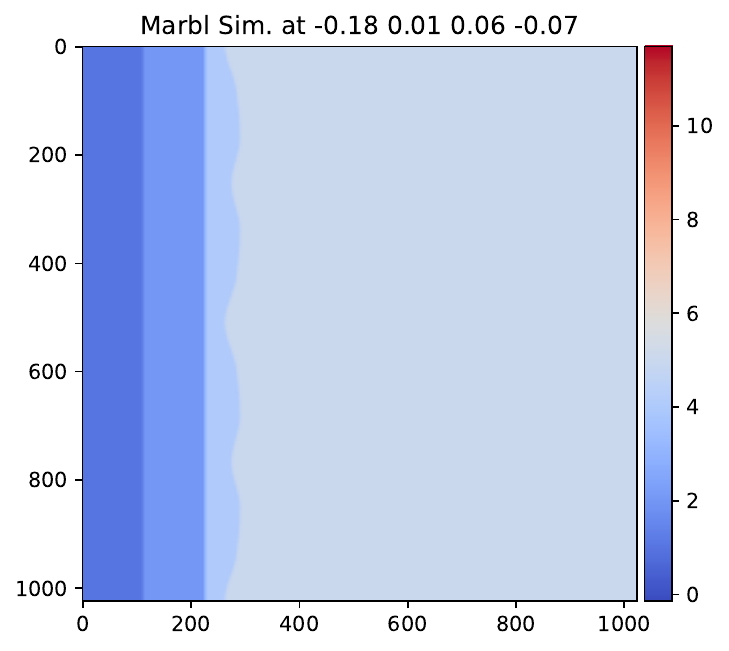} & \includegraphics[width=0.24\textwidth]{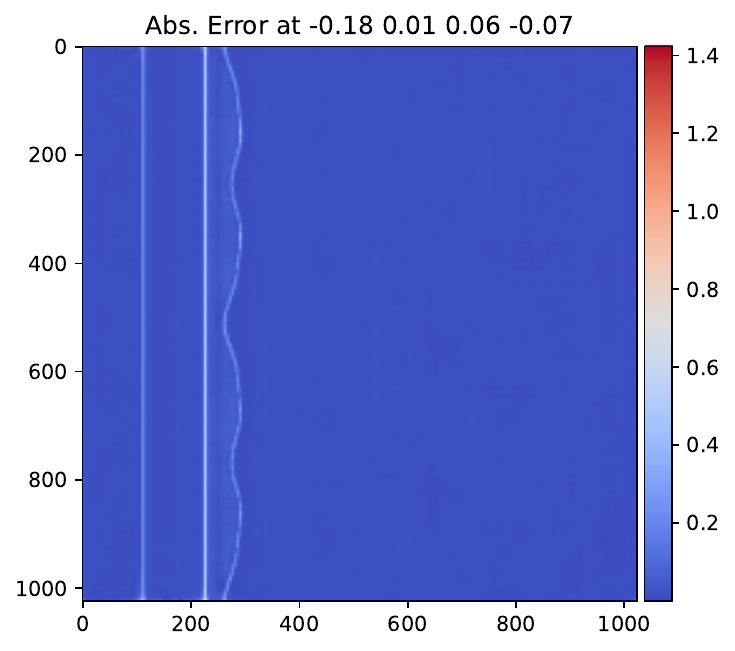} \\
  \hline
  energy & \includegraphics[width=0.24\textwidth]{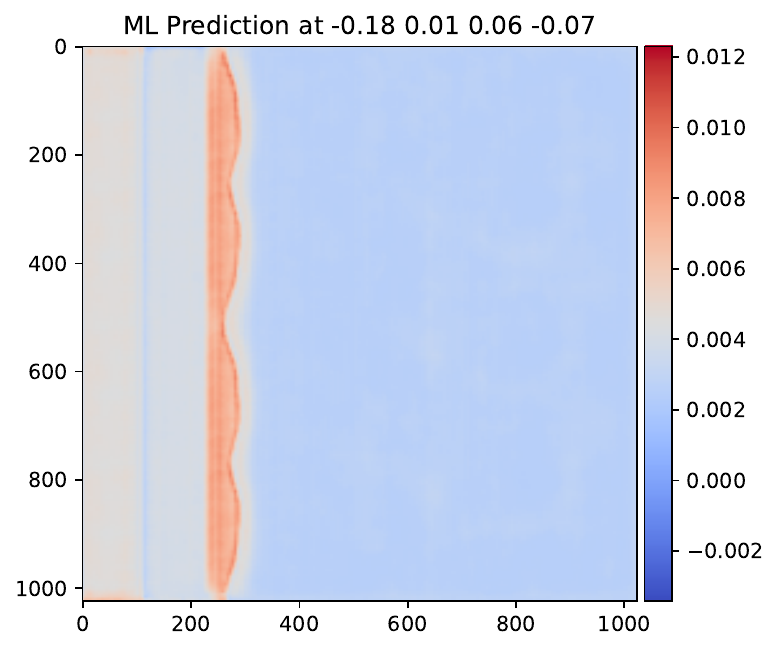} & \includegraphics[width=0.24\textwidth]{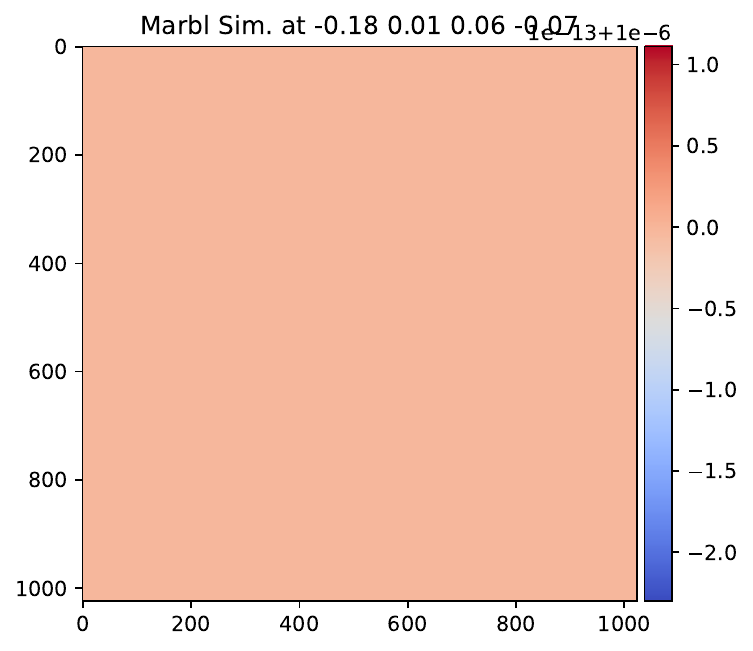} & \includegraphics[width=0.24\textwidth]{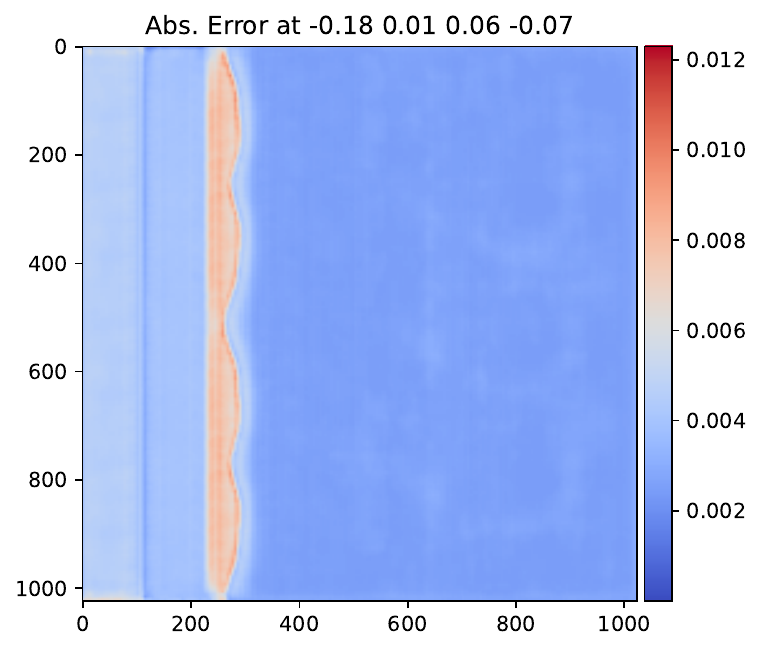} \\
  \hline
  pressure & \includegraphics[width=0.24\textwidth]{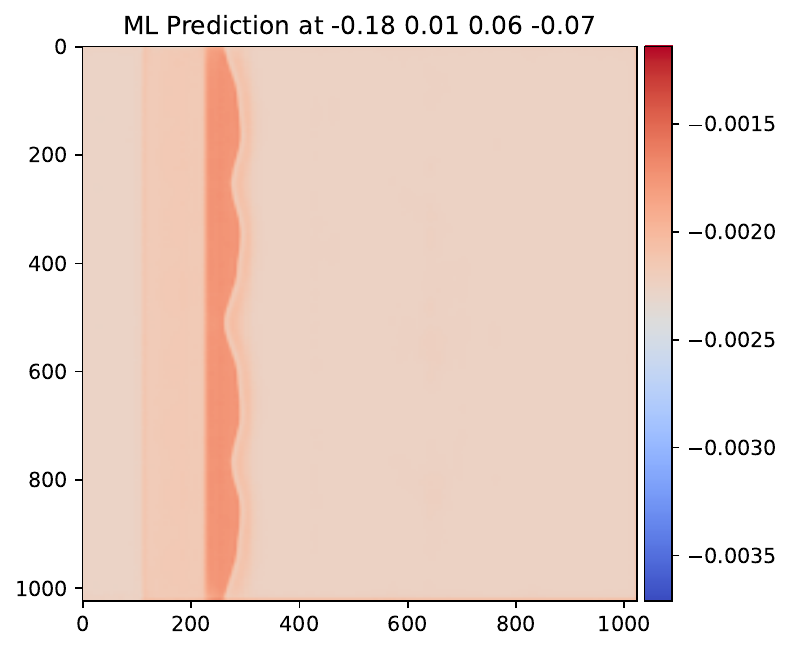} & \includegraphics[width=0.24\textwidth]{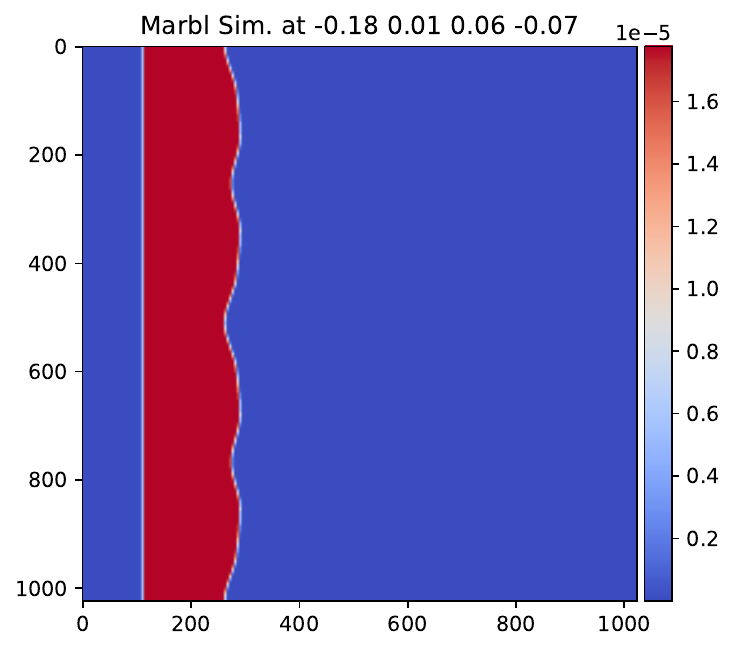} & \includegraphics[width=0.24\textwidth]{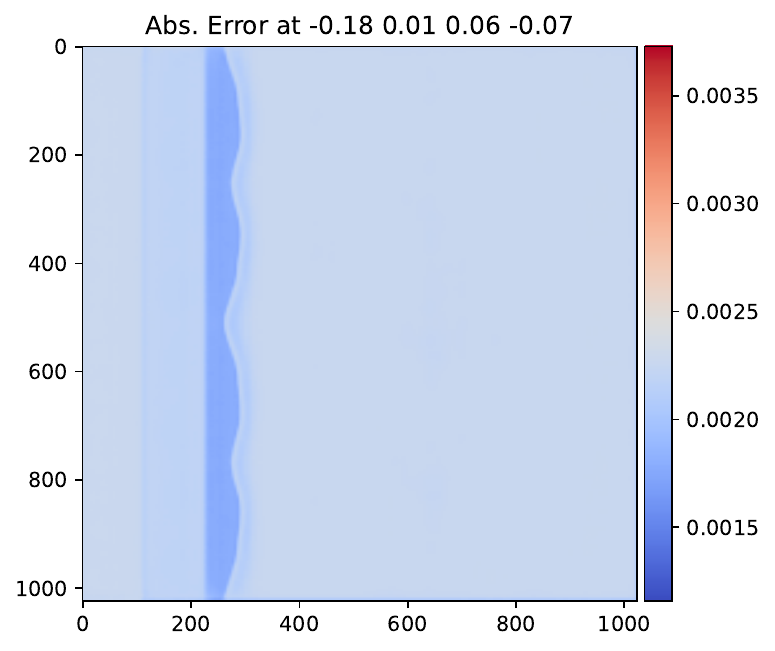} \\
  \hline
  velocity $x$ & \includegraphics[width=0.24\textwidth]{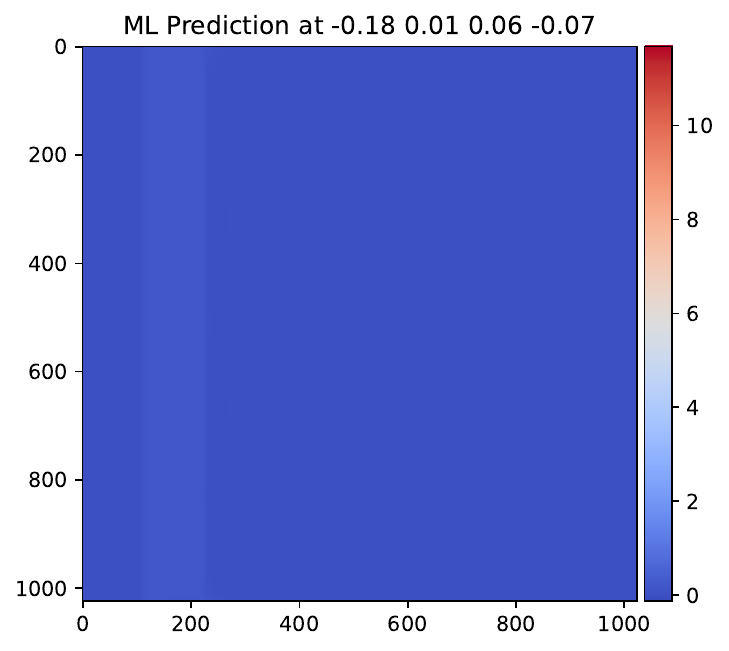} & \includegraphics[width=0.24\textwidth]{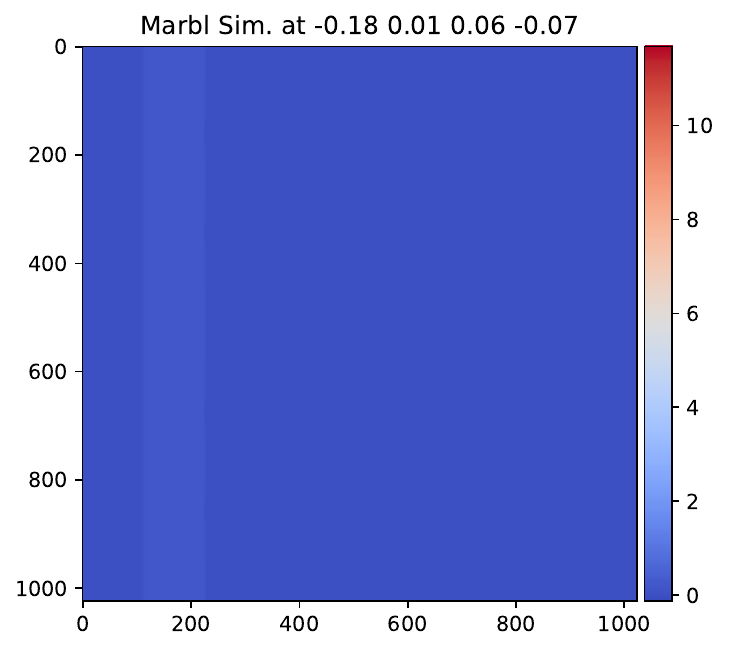} & \includegraphics[width=0.24\textwidth]{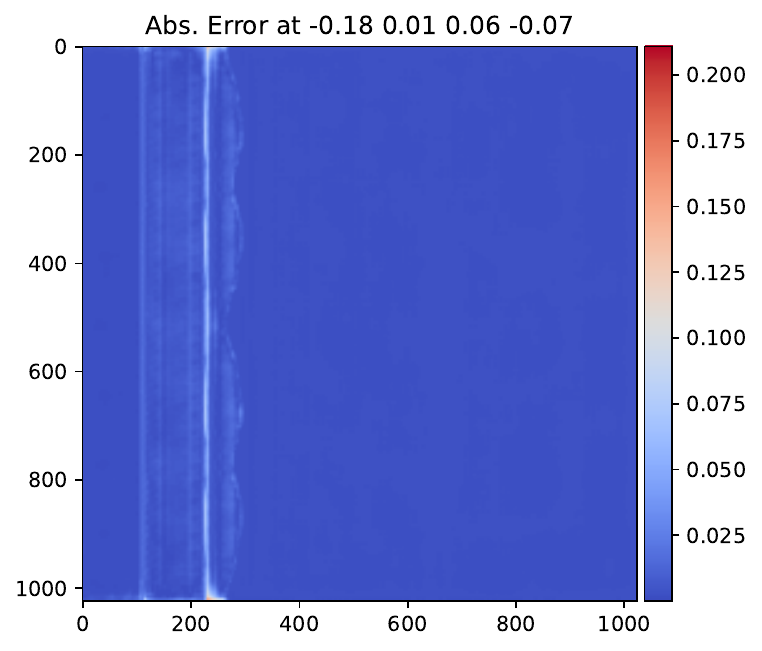} \\
  \hline
  velocity $y$ & \includegraphics[width=0.24\textwidth]{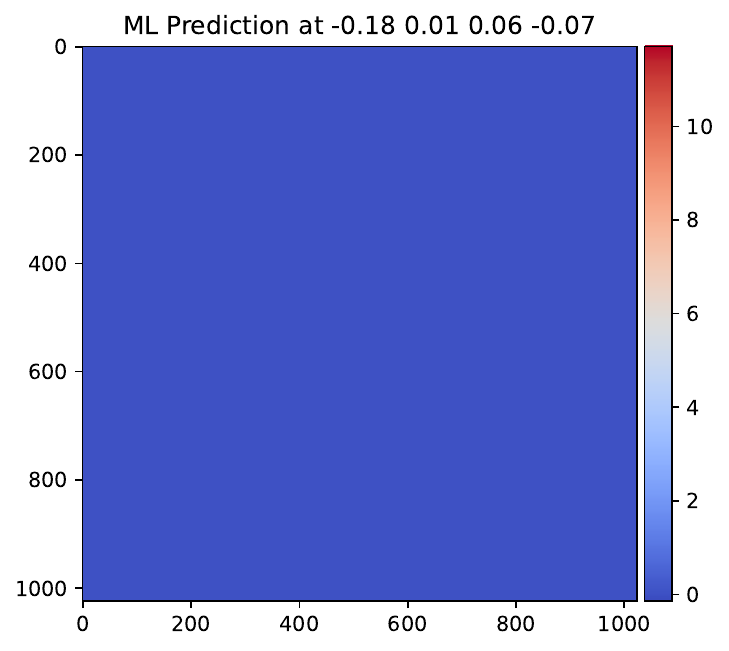} & \includegraphics[width=0.24\textwidth]{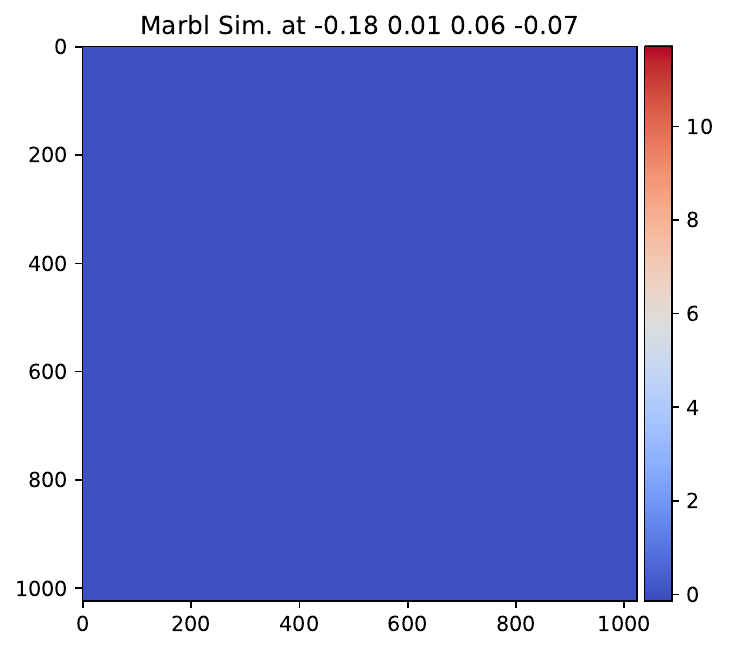} & \includegraphics[width=0.24\textwidth]{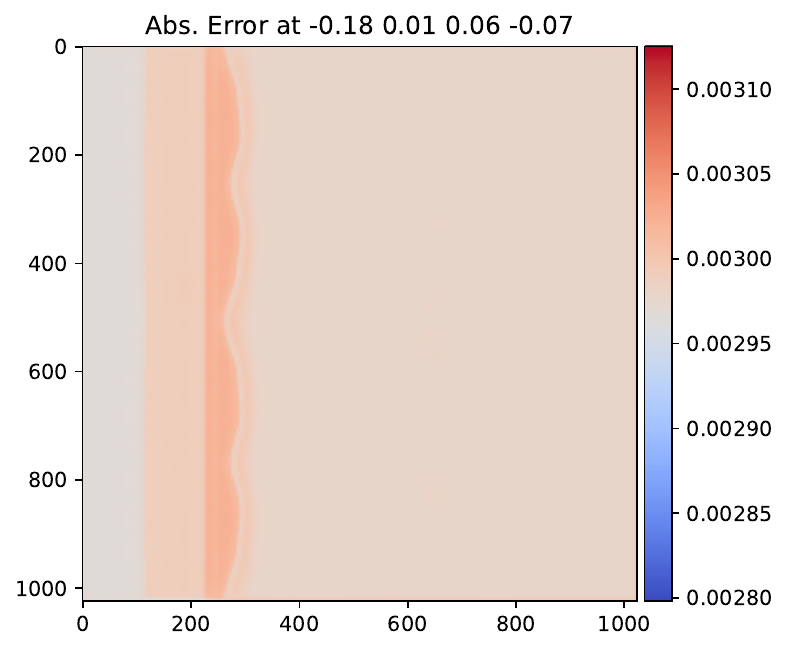} \\
  \end{tabular}
  \caption{
      Figure of predictions, truth, and absolute error for linear shaped charge predictions at $t$=0.
      }\label{fig:pchip_predictions_0}
  \end{center}
\end{figure}

\begin{figure}[!htb]
  \begin{center}
  \begin{tabular}{*{7}{c}}
  Field & Ml prediction & Simulation & Abs. Error \\
  \hline
  density & \includegraphics[width=0.24\textwidth]{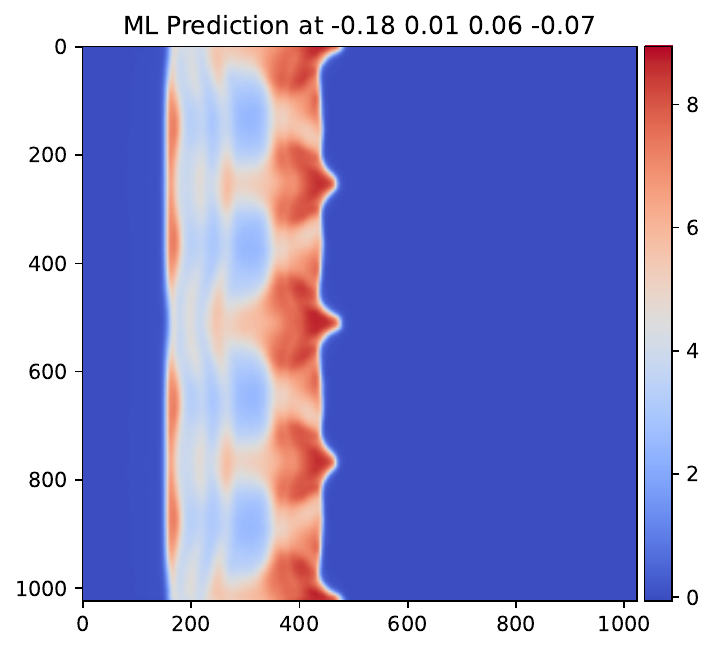} & \includegraphics[width=0.24\textwidth]{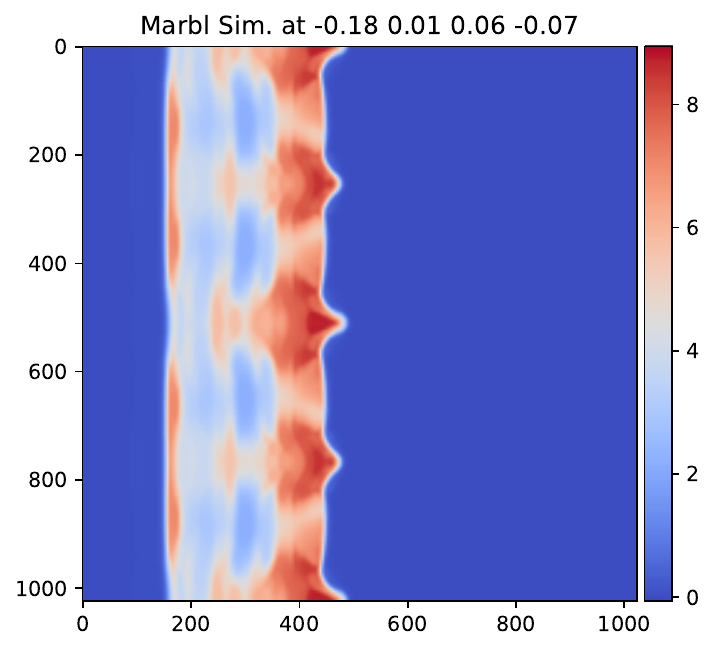} & \includegraphics[width=0.24\textwidth]{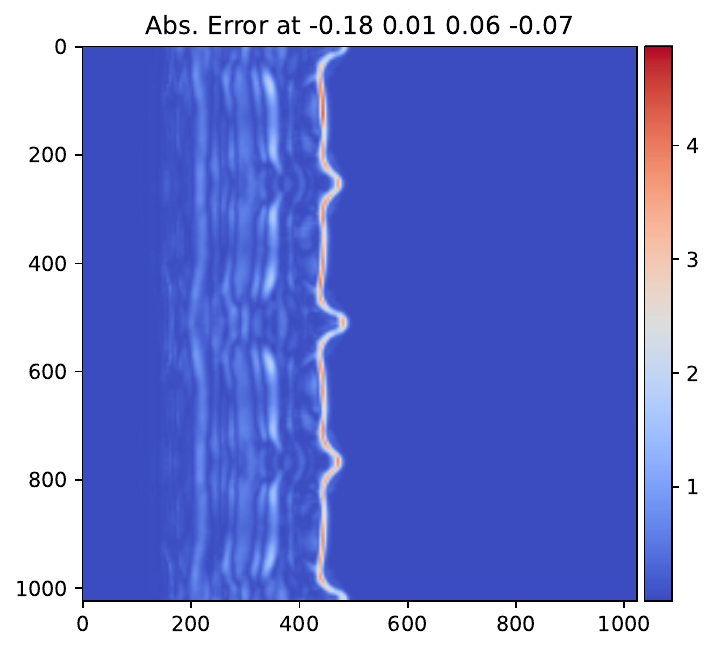} \\
  \hline
  materials & \includegraphics[width=0.24\textwidth]{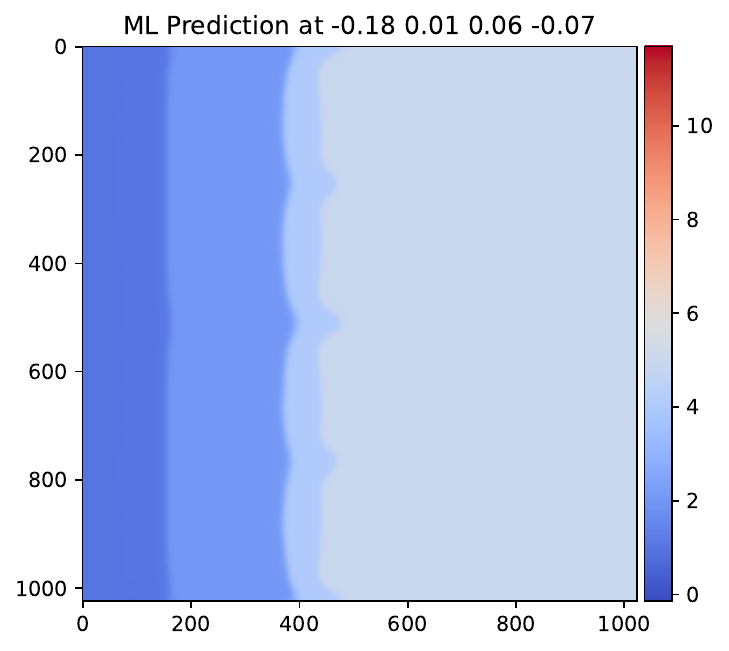} & \includegraphics[width=0.24\textwidth]{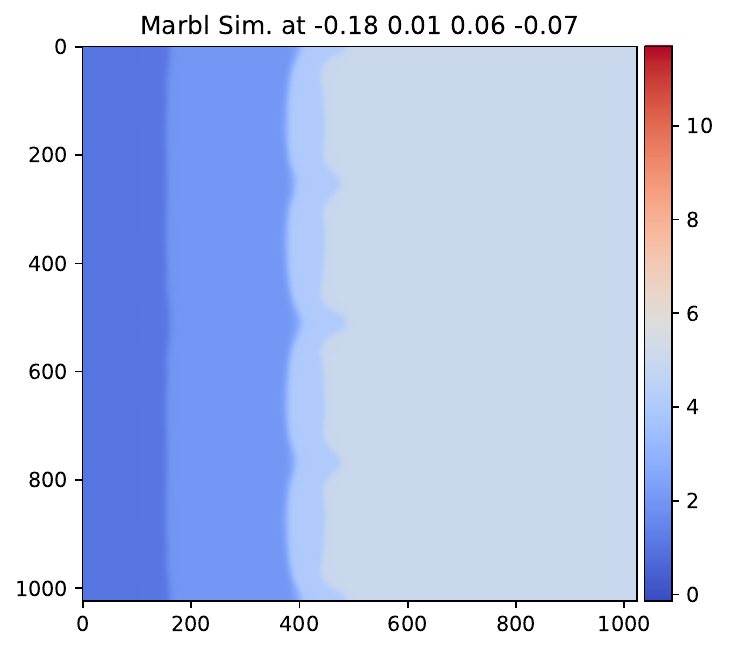} & \includegraphics[width=0.24\textwidth]{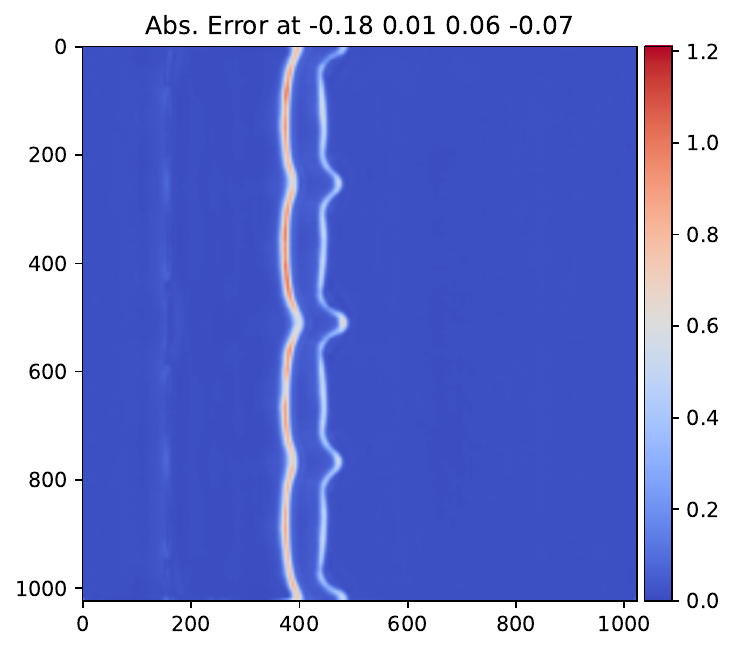} \\
  \hline
  energy & \includegraphics[width=0.24\textwidth]{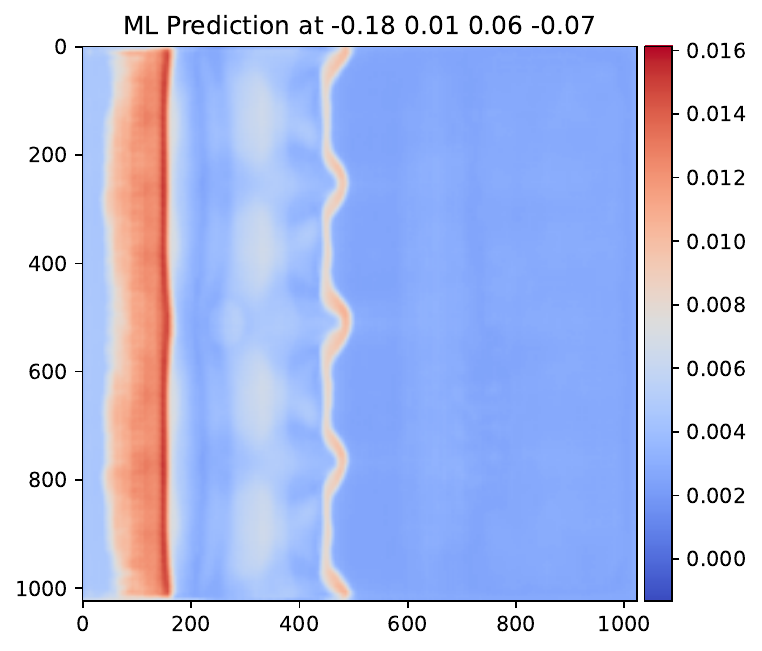} & \includegraphics[width=0.24\textwidth]{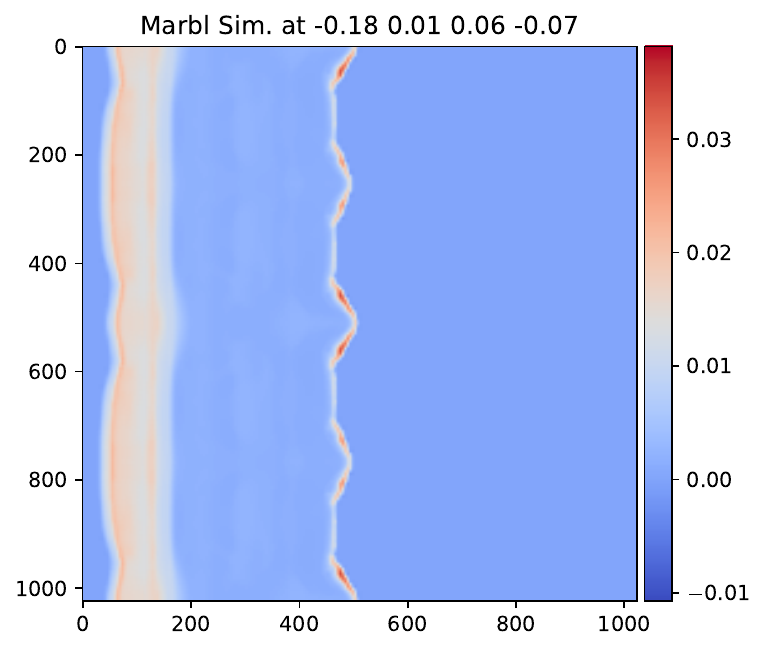} & \includegraphics[width=0.24\textwidth]{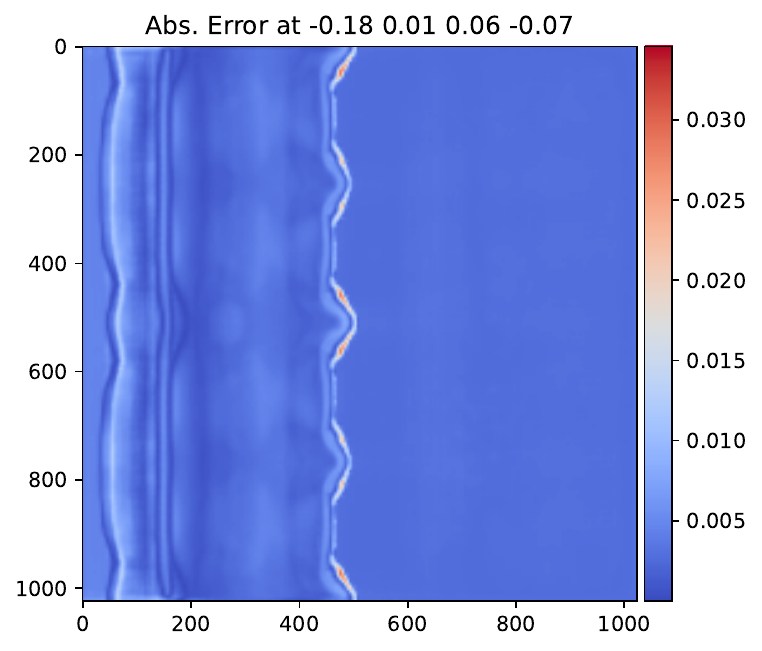} \\
  \hline
  pressure & \includegraphics[width=0.24\textwidth]{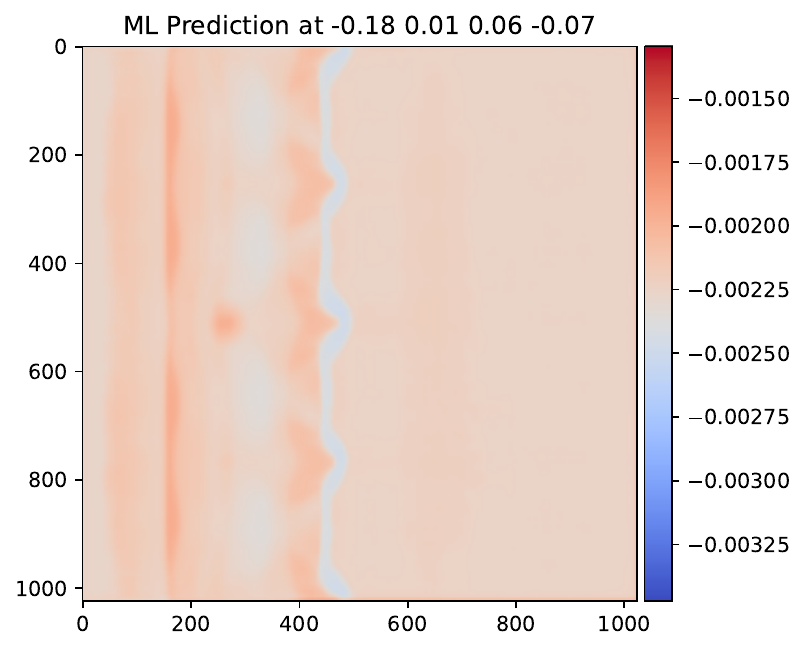} & \includegraphics[width=0.24\textwidth]{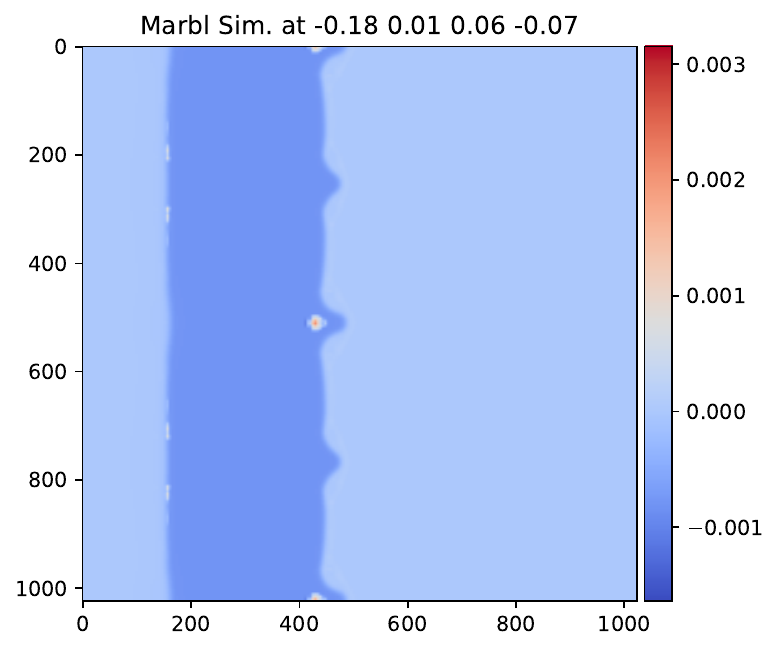} & \includegraphics[width=0.24\textwidth]{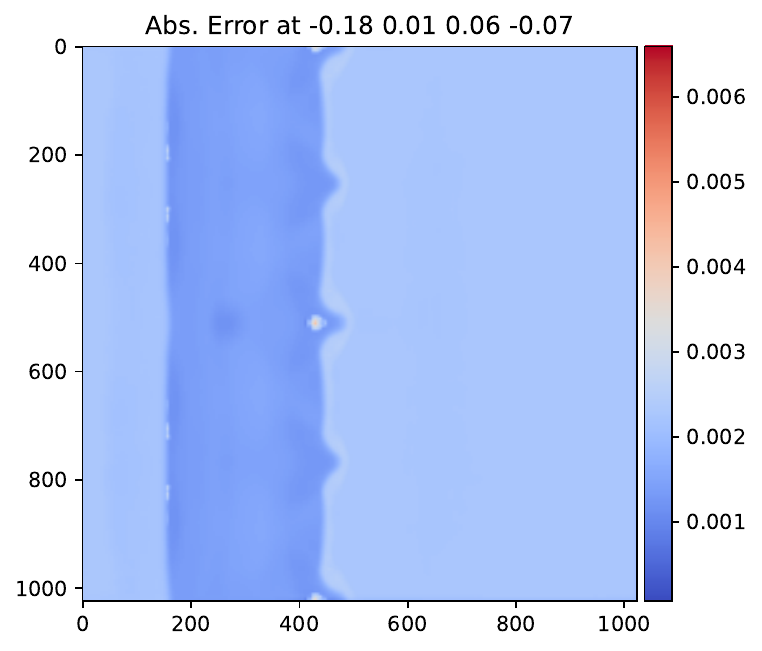} \\
  \hline
  velocity $x$ & \includegraphics[width=0.24\textwidth]{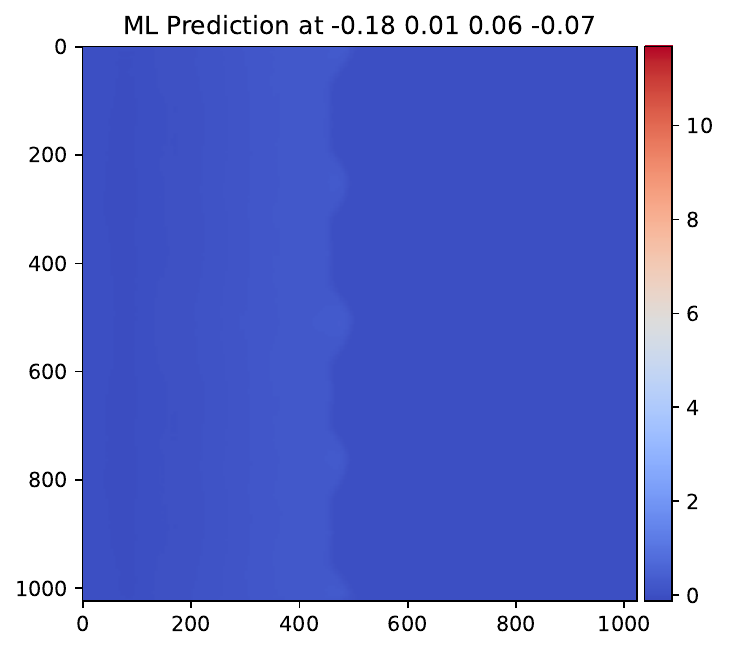} & \includegraphics[width=0.24\textwidth]{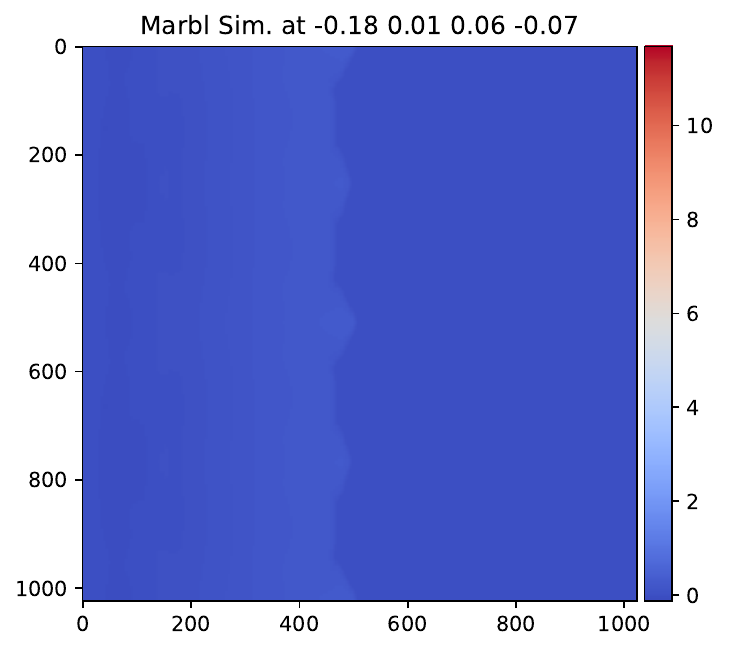} & \includegraphics[width=0.24\textwidth]{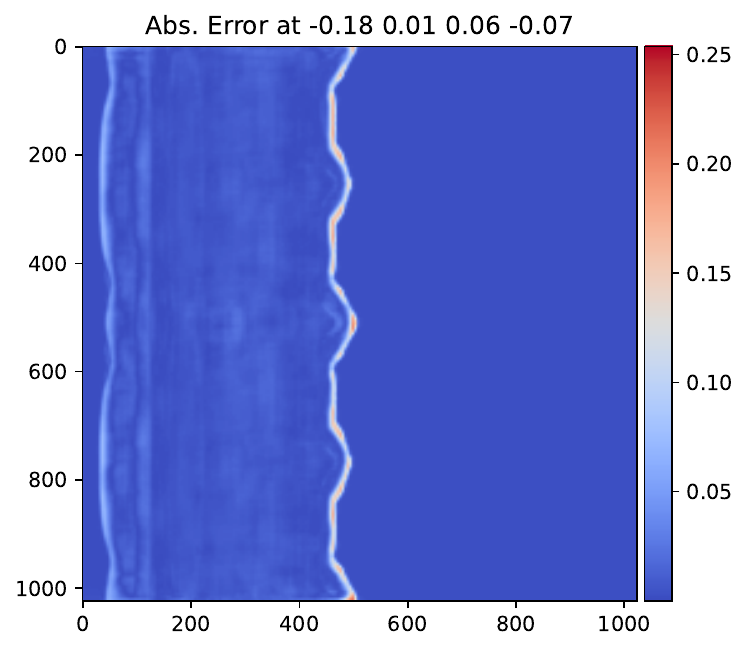} \\
  \hline
  velocity $y$ & \includegraphics[width=0.24\textwidth]{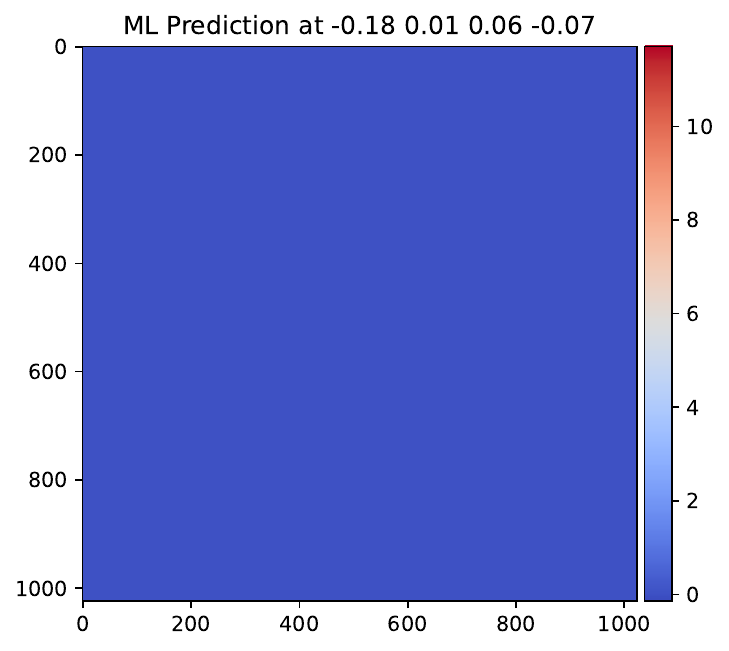} & \includegraphics[width=0.24\textwidth]{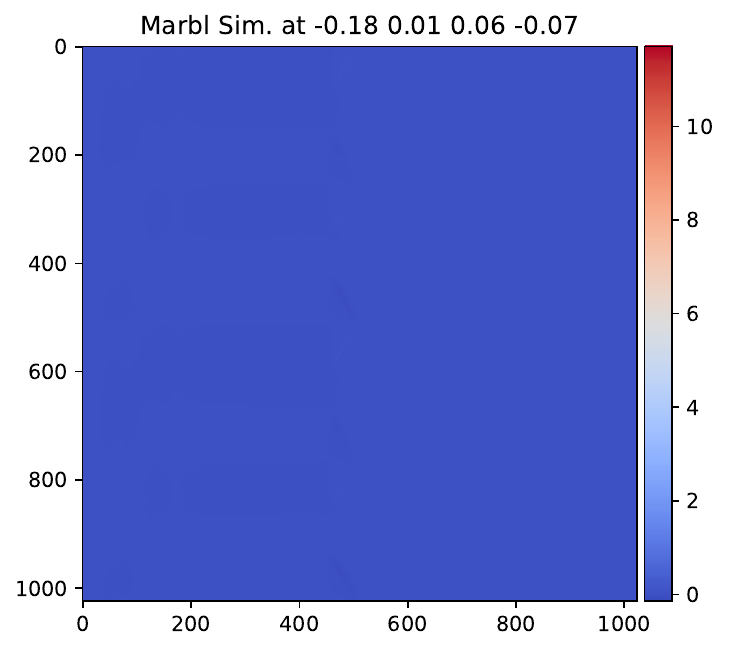} & \includegraphics[width=0.24\textwidth]{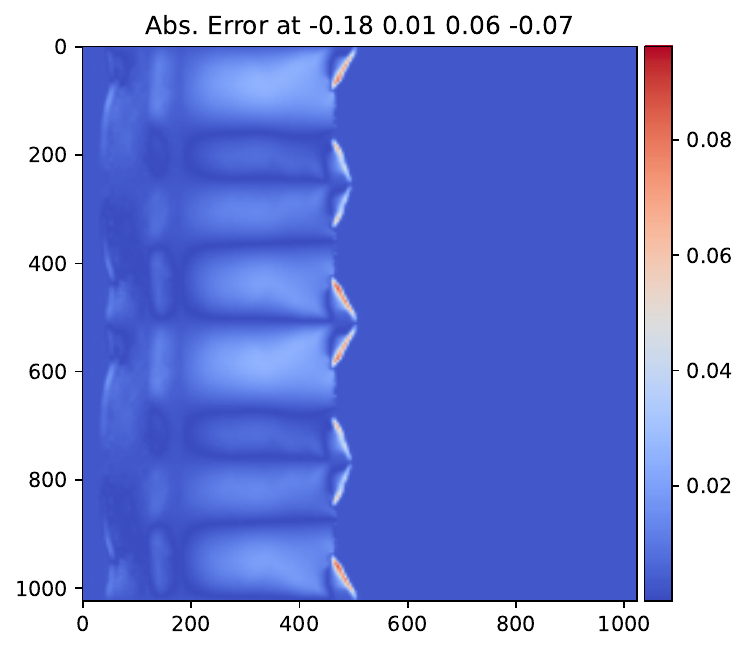} \\
  \end{tabular}
  \caption{
      Figure of predictions, truth, and absolute error for linear shaped charge predictions at $t$=7.
      }\label{fig:pchip_predictions_1}
  \end{center}
\end{figure}

\begin{figure}[!htb]
  \begin{center}
  \begin{tabular}{*{7}{c}}
  Field & Ml prediction & Simulation & Abs. Error \\
  \hline
  density & \includegraphics[width=0.24\textwidth]{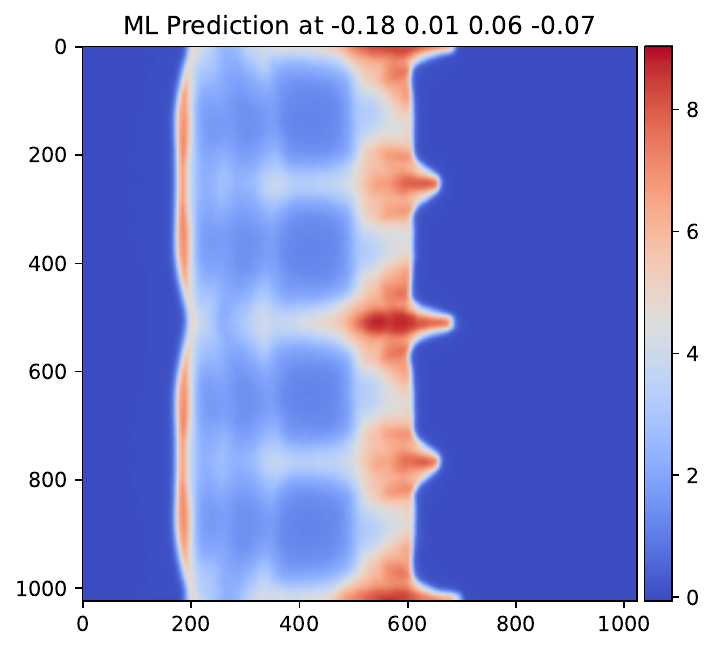} & \includegraphics[width=0.24\textwidth]{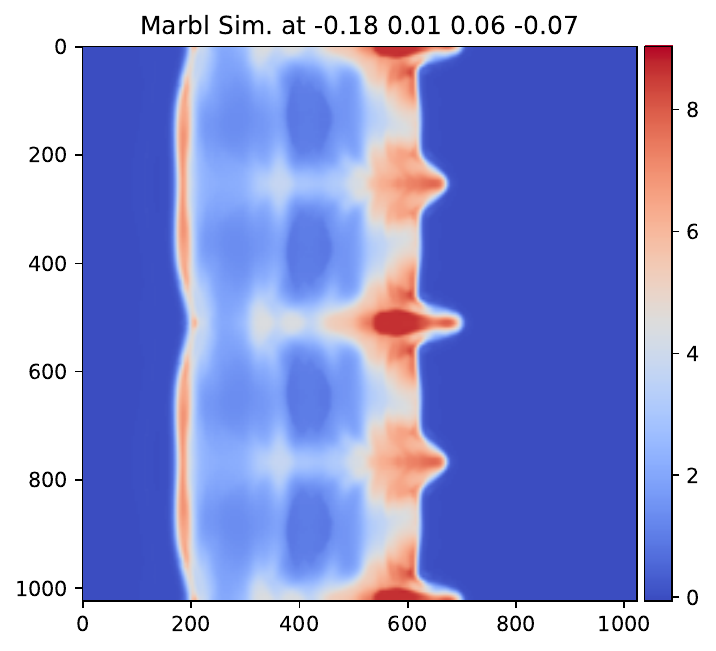} & \includegraphics[width=0.24\textwidth]{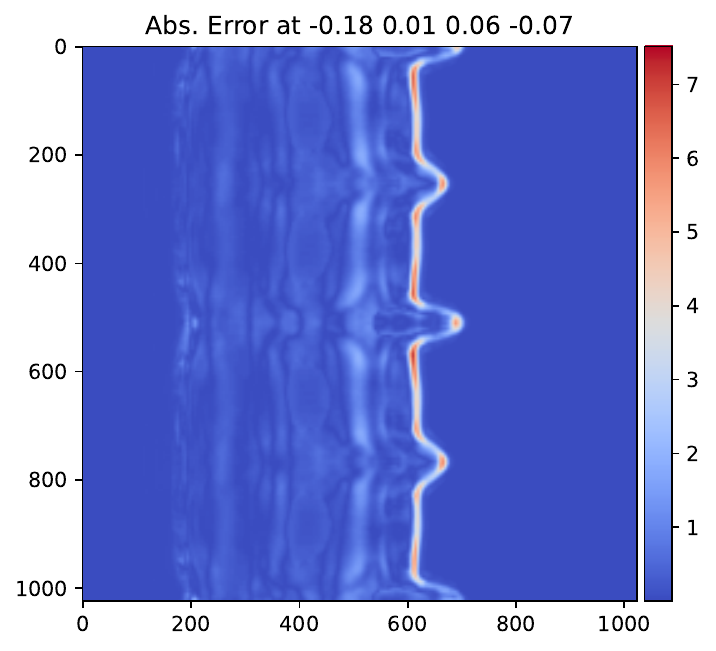} \\
  \hline
  materials & \includegraphics[width=0.24\textwidth]{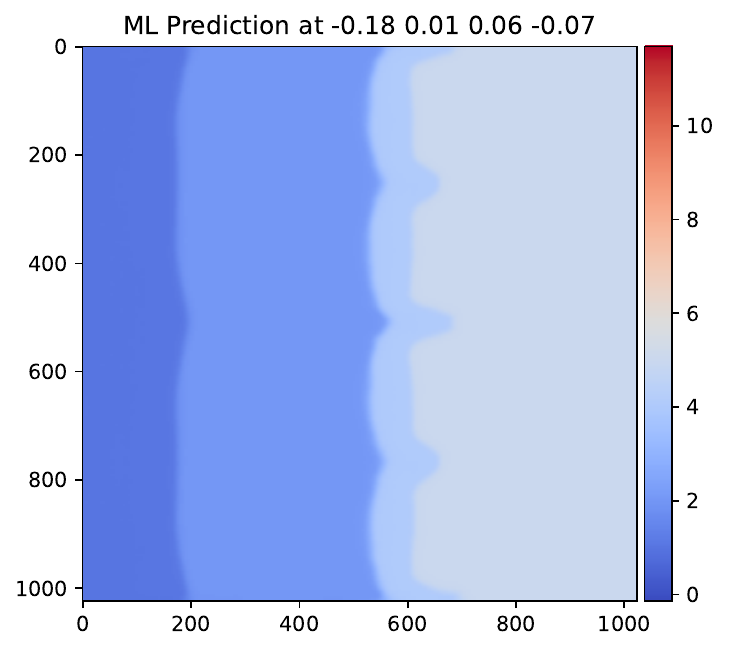} & \includegraphics[width=0.24\textwidth]{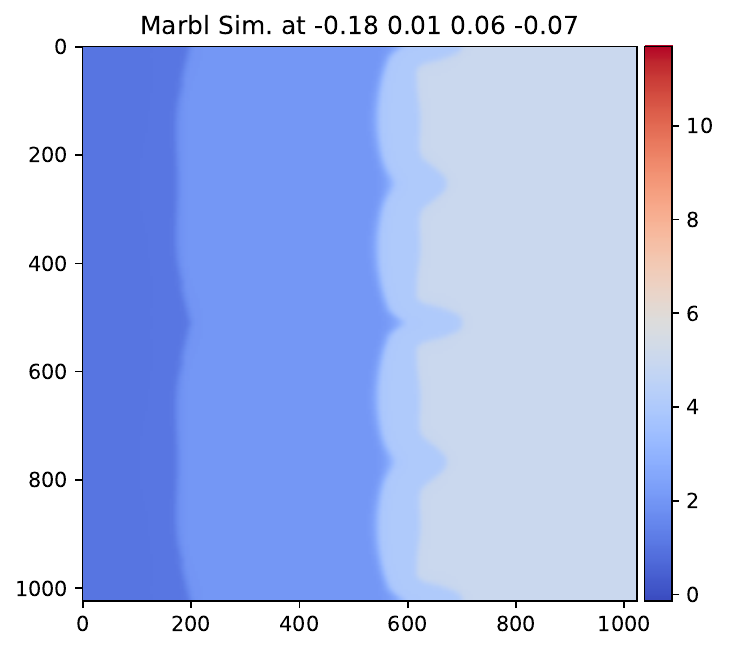} & \includegraphics[width=0.24\textwidth]{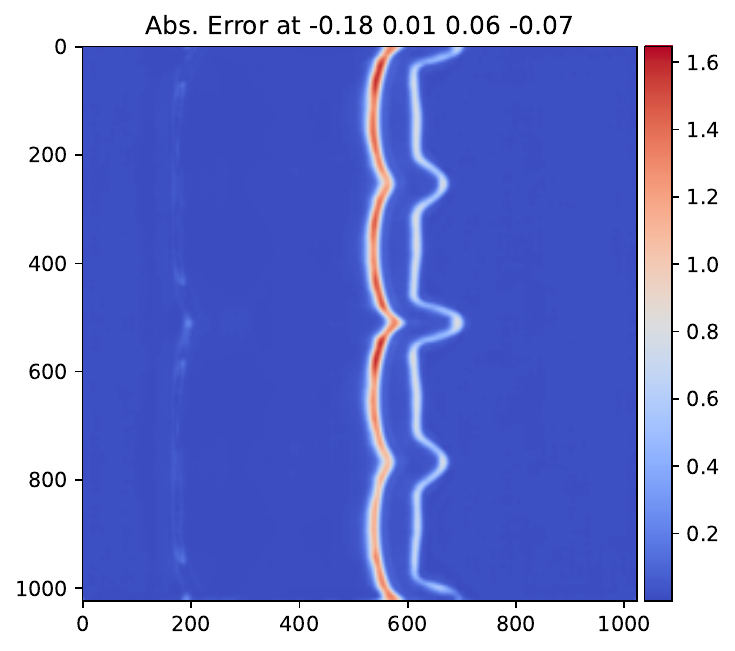} \\
  \hline
  energy & \includegraphics[width=0.24\textwidth]{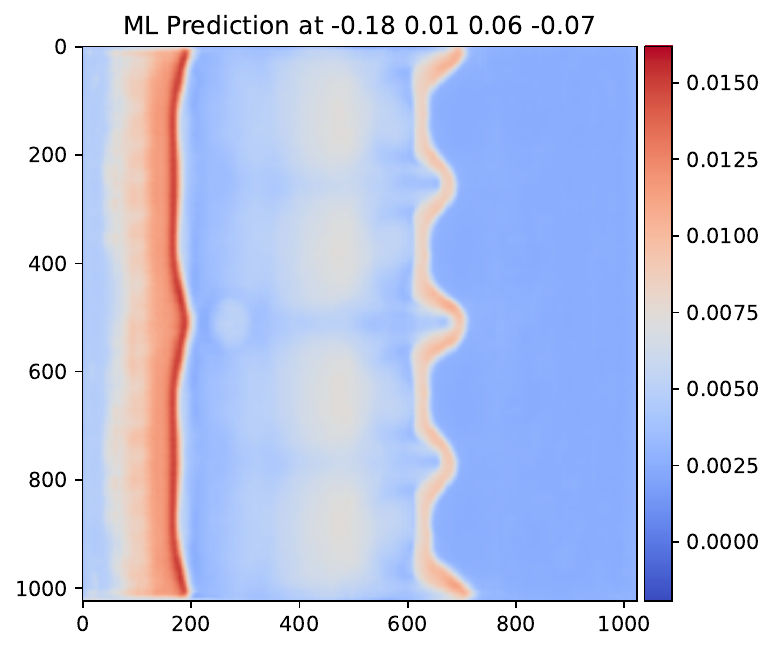} & \includegraphics[width=0.24\textwidth]{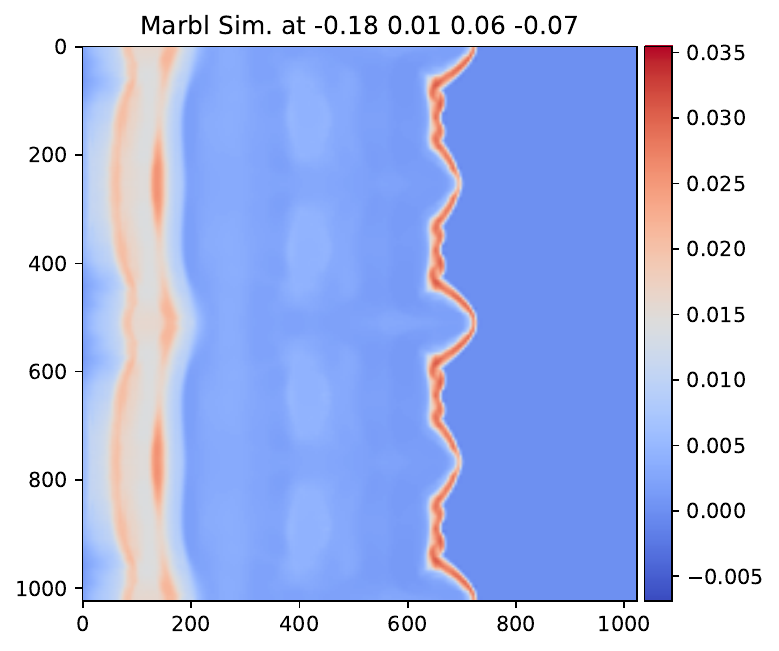} & \includegraphics[width=0.24\textwidth]{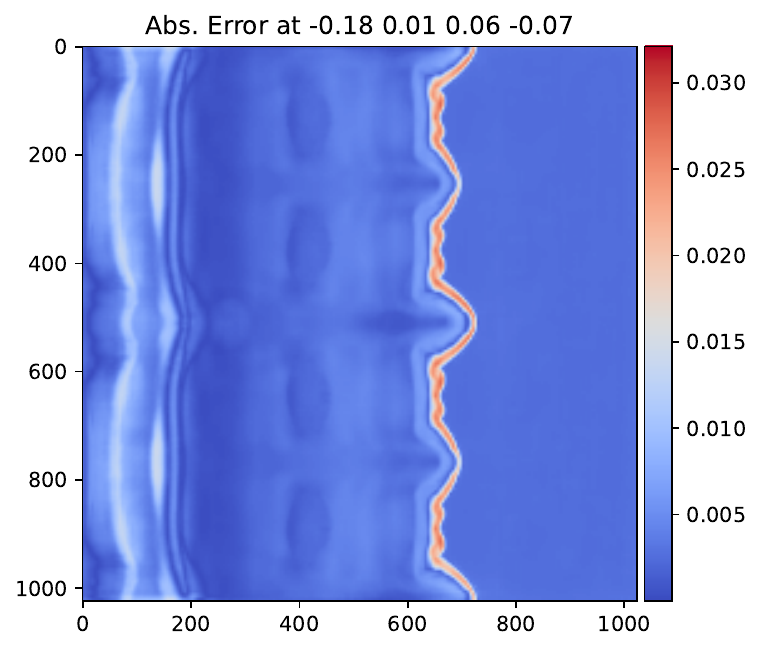} \\
  \hline
  pressure & \includegraphics[width=0.24\textwidth]{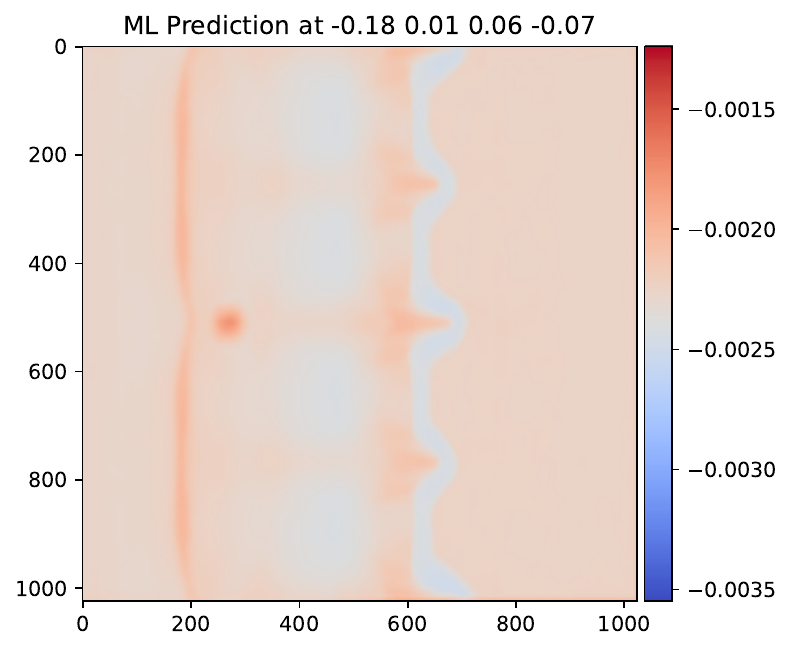} & \includegraphics[width=0.24\textwidth]{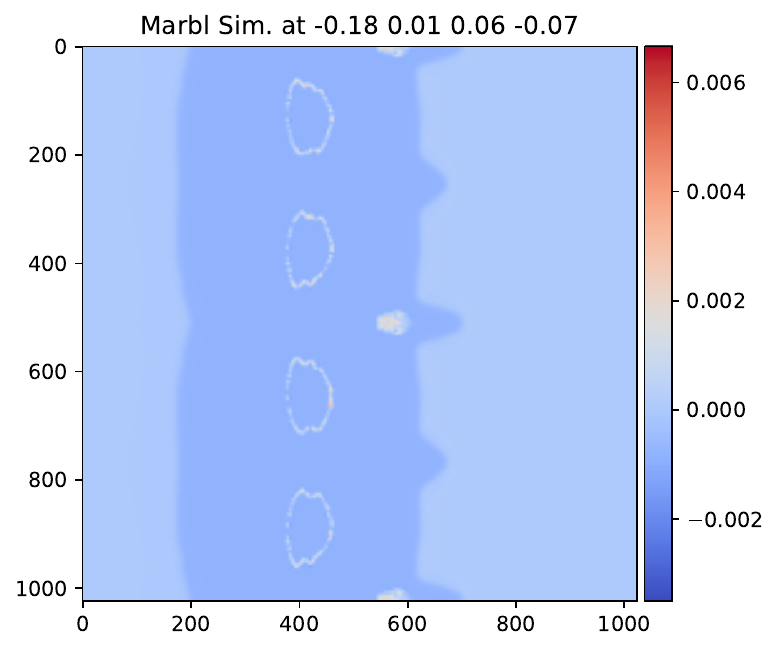} & \includegraphics[width=0.24\textwidth]{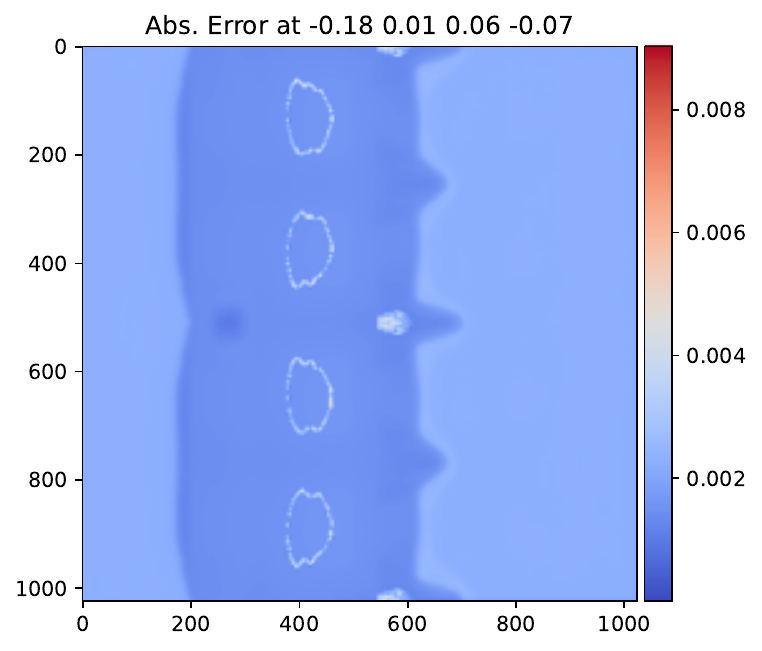} \\
  \hline
  velocity $x$ & \includegraphics[width=0.24\textwidth]{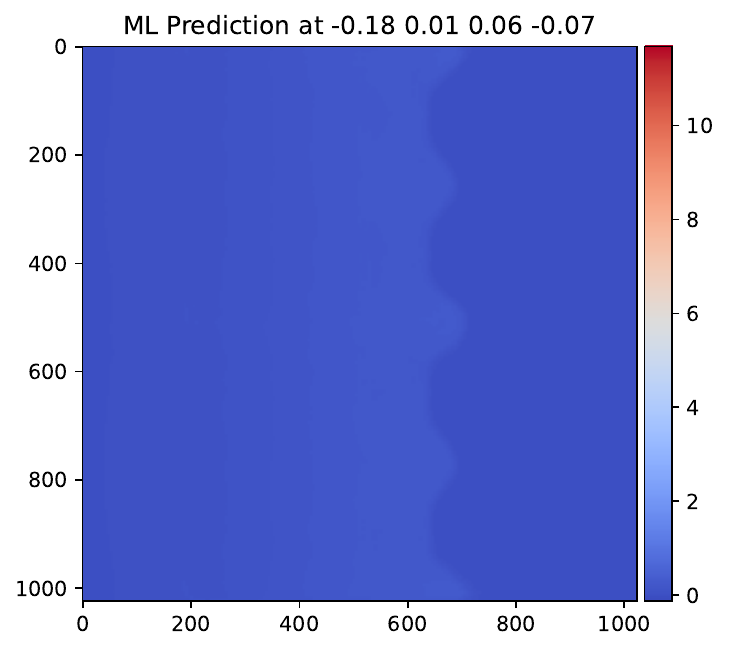} & \includegraphics[width=0.24\textwidth]{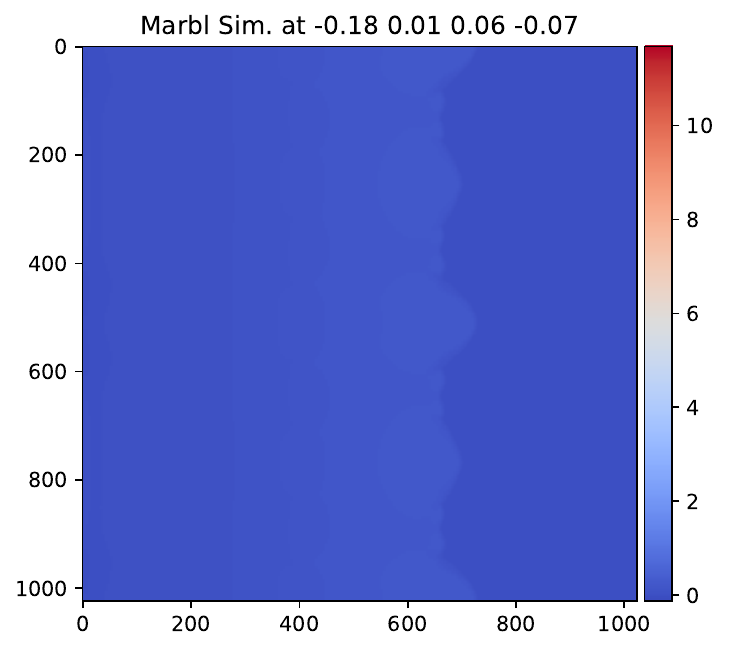} & \includegraphics[width=0.24\textwidth]{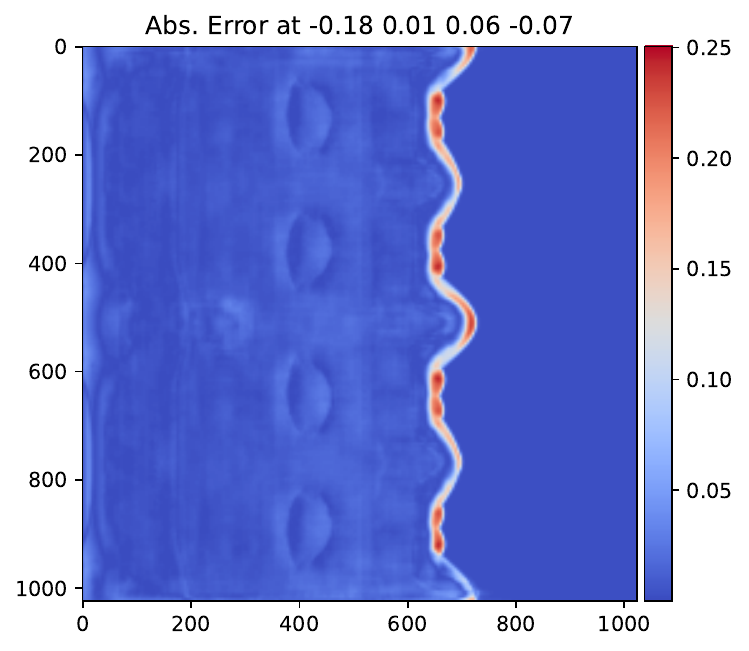} \\
  \hline
  velocity $y$ & \includegraphics[width=0.24\textwidth]{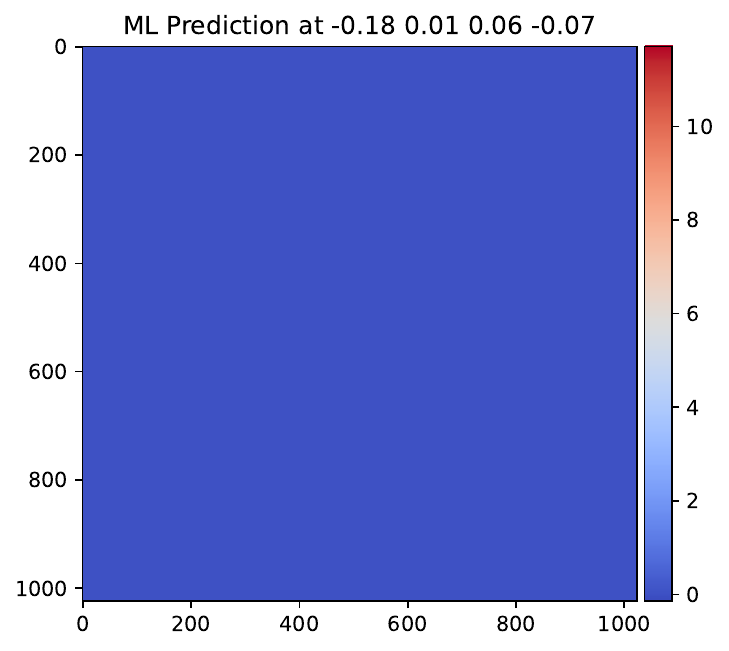} & \includegraphics[width=0.24\textwidth]{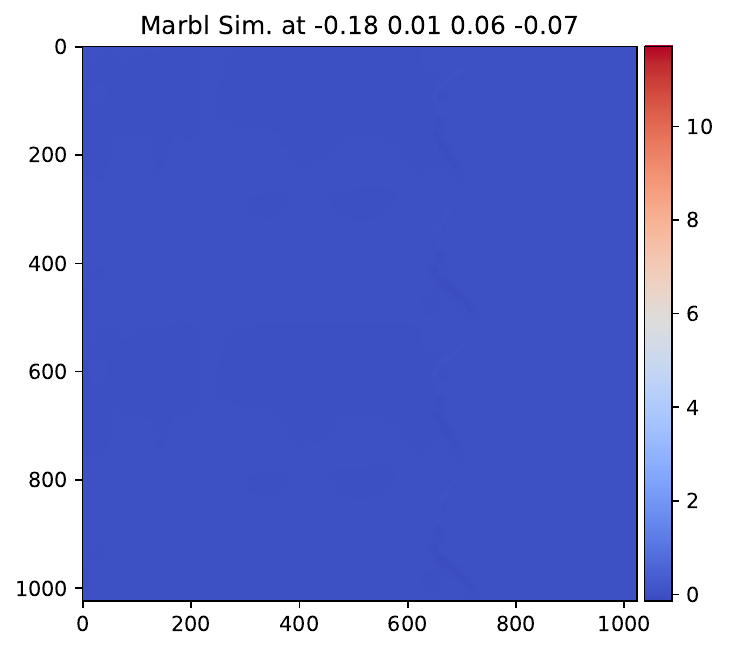} & \includegraphics[width=0.24\textwidth]{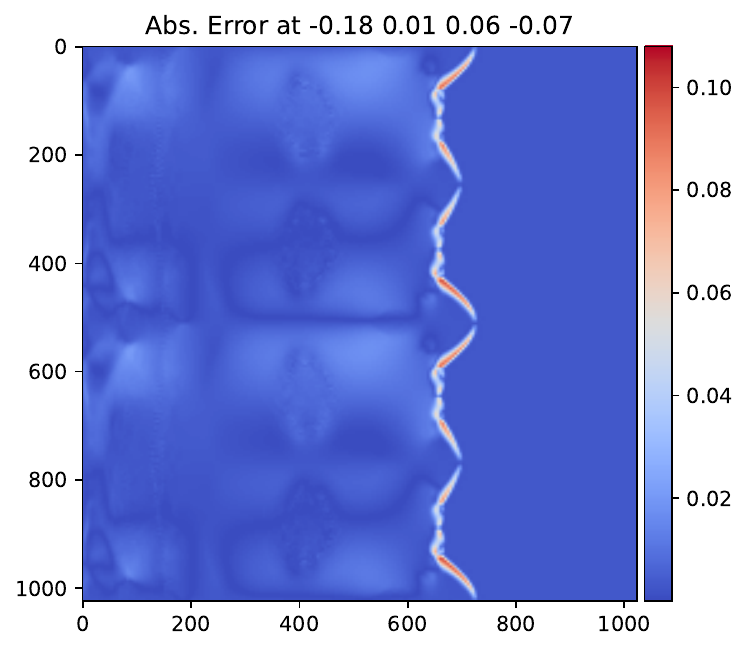} \\
  \end{tabular}
  \caption{
      Figure of predictions, truth, and absolute error for linear shaped charge predictions at $t$=15.
      }\label{fig:pchip_predictions_2}
  \end{center}
\end{figure}

\FloatBarrier

\subsection{Double sine wave results}

\begin{figure}[!htb]
  \centering
    \includegraphics[width=1.0\textwidth]{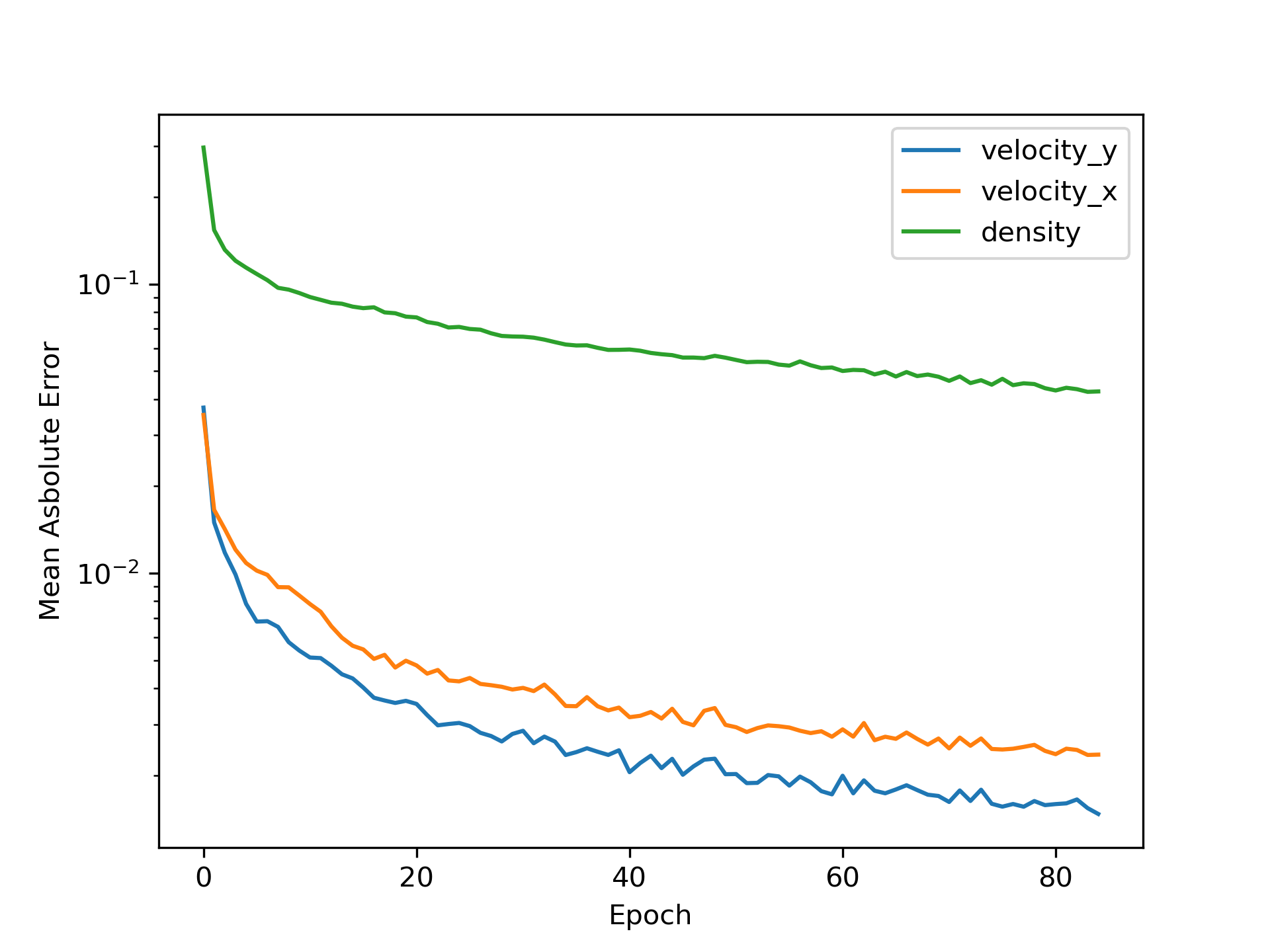}
  \caption{
      The epoch vs mean absolute error for each field while training the double sine wave model.
      }
  \label{fig:dstraining}
\end{figure}

\begin{figure}[!htb]
  \begin{center}
  \begin{tabular}{*{7}{c}}
  Field & Ml prediction & Simulation & Abs. Error \\
  \hline
  density & \includegraphics[width=0.29\textwidth]{figs/ds/density_yhat_00.pdf} & \includegraphics[width=0.29\textwidth]{figs/ds/density_y_00.pdf} & \includegraphics[width=0.29\textwidth]{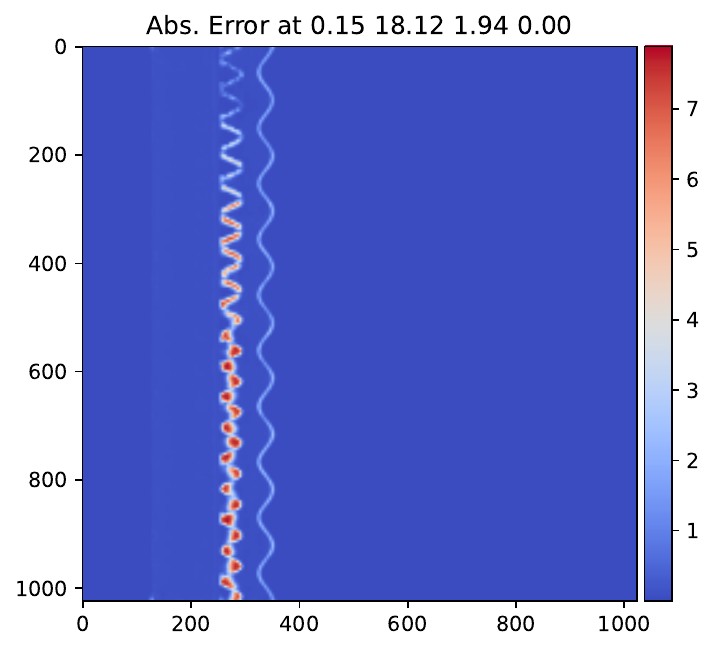} \\
  \hline
  velocity $x$ & \includegraphics[width=0.29\textwidth]{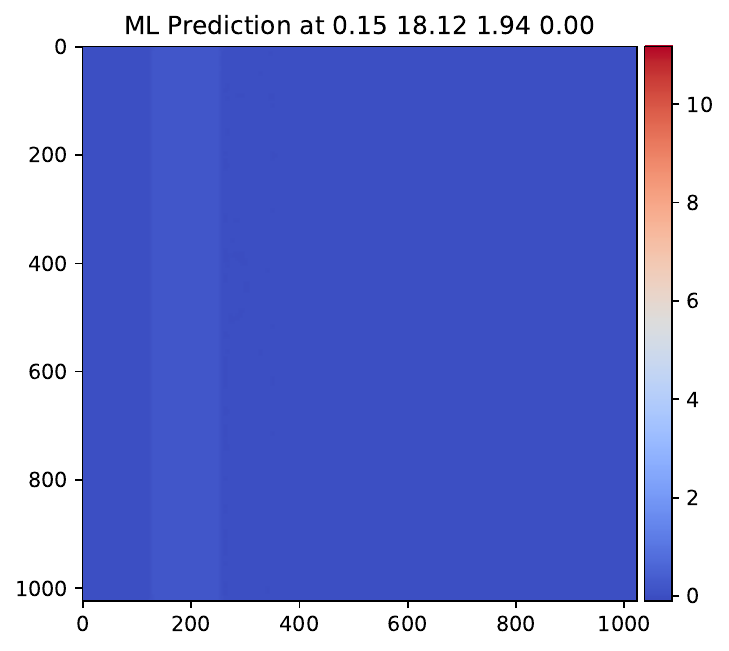} & \includegraphics[width=0.29\textwidth]{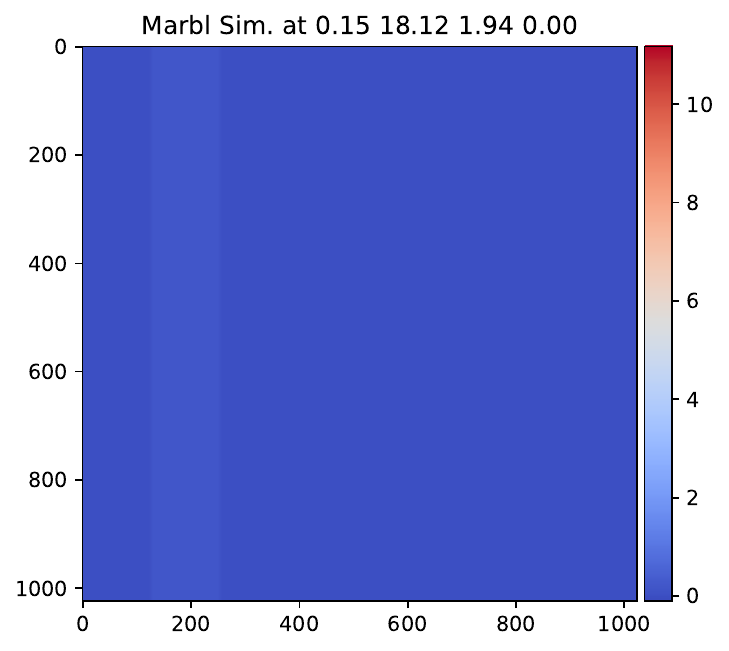} & \includegraphics[width=0.29\textwidth]{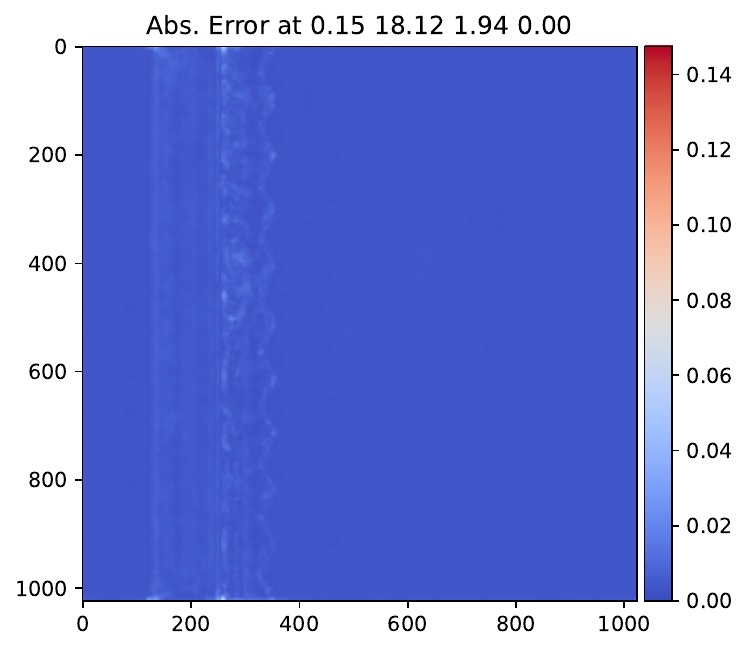} \\
  \hline
  velocity $y$ & \includegraphics[width=0.29\textwidth]{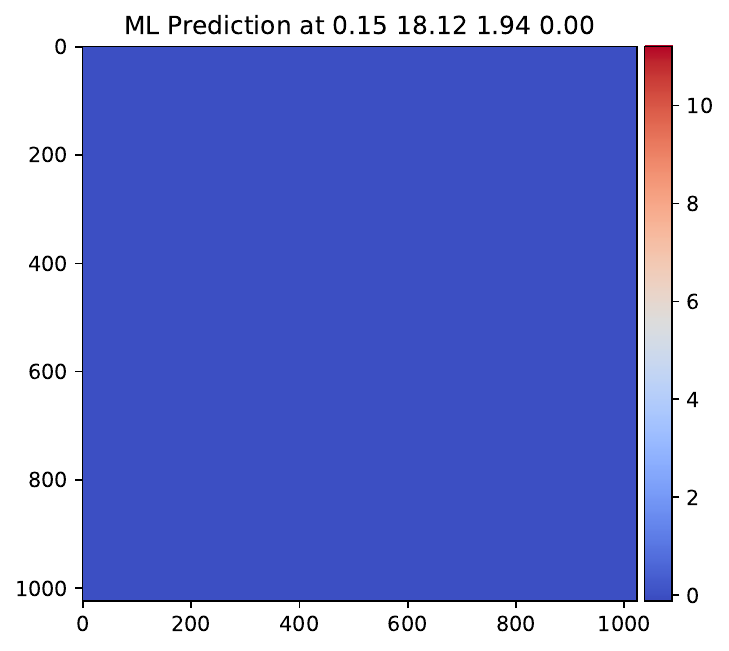} & \includegraphics[width=0.29\textwidth]{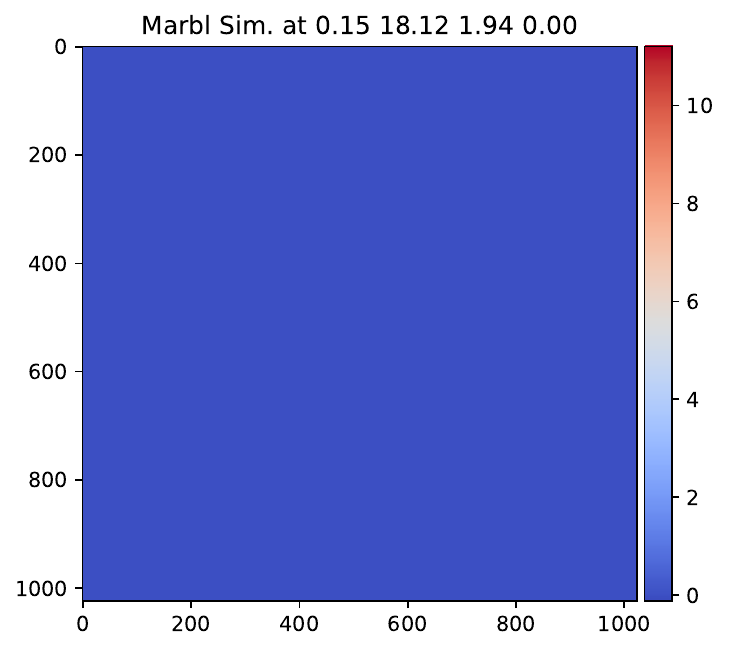} & \includegraphics[width=0.29\textwidth]{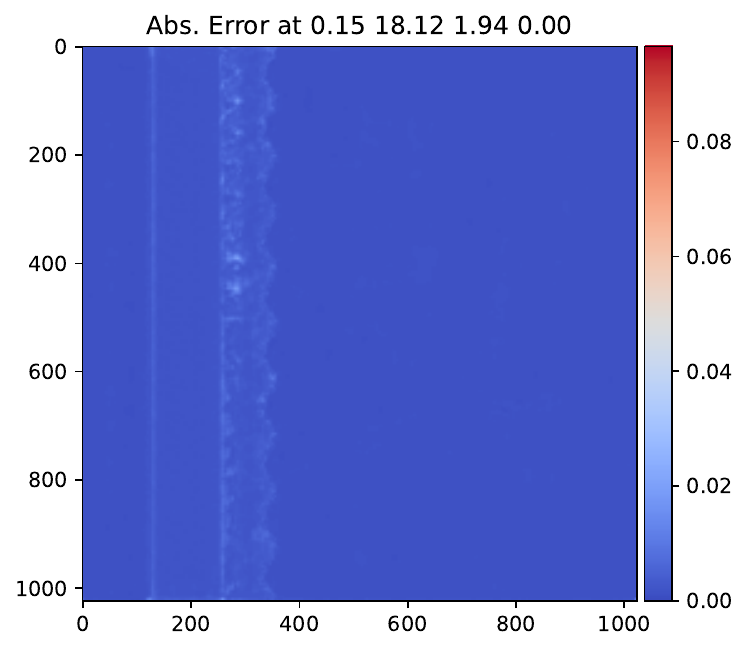} \\
  \end{tabular}
  \caption{
      Figure of predictions, truth, and absolute error for double sine wave predictions at $t$=0.
      }\label{fig:ds_predictions_0}
  \end{center}
\end{figure}

\begin{figure}[!htb]
  \begin{center}
  \begin{tabular}{*{7}{c}}
  Field & Ml prediction & Simulation & Abs. Error \\
  \hline
  density & \includegraphics[width=0.29\textwidth]{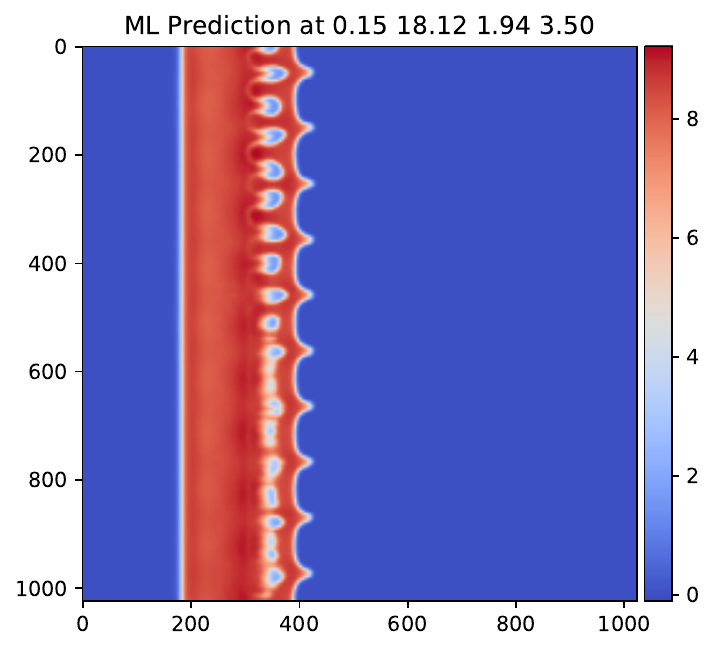} & \includegraphics[width=0.29\textwidth]{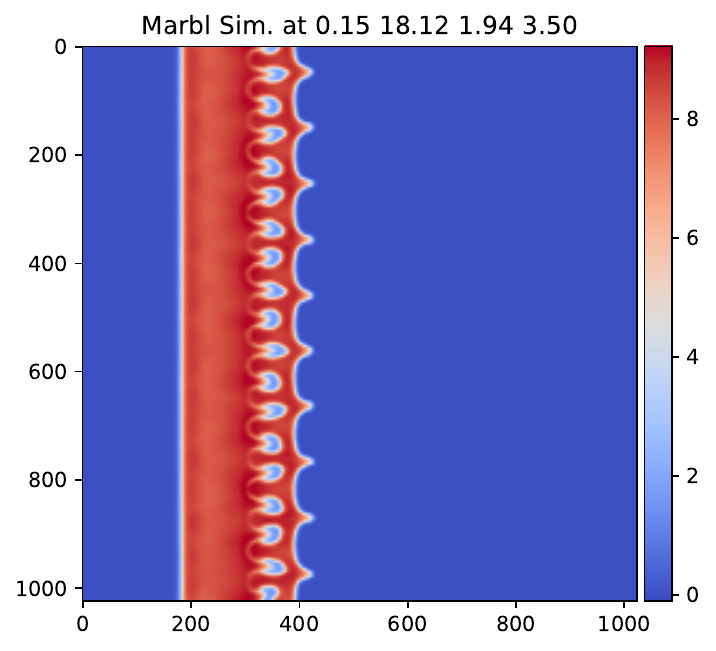} & \includegraphics[width=0.29\textwidth]{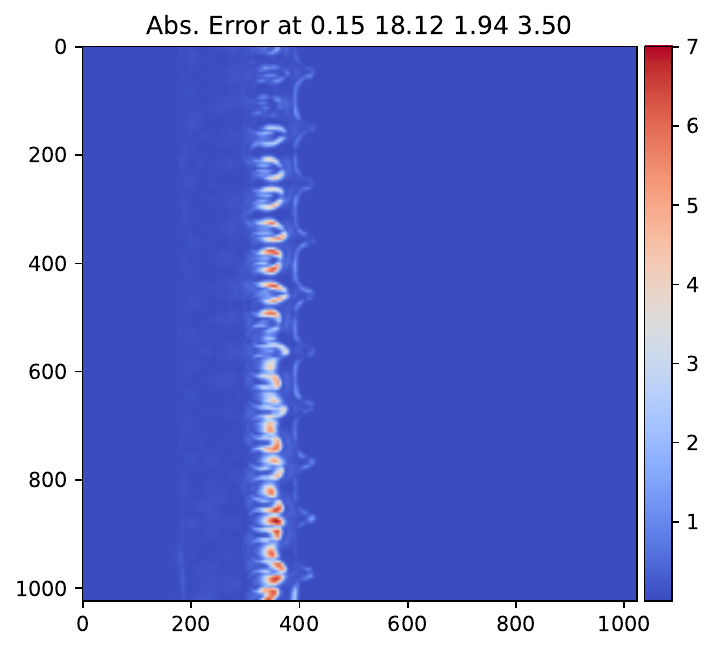} \\
  \hline
  velocity $x$ & \includegraphics[width=0.29\textwidth]{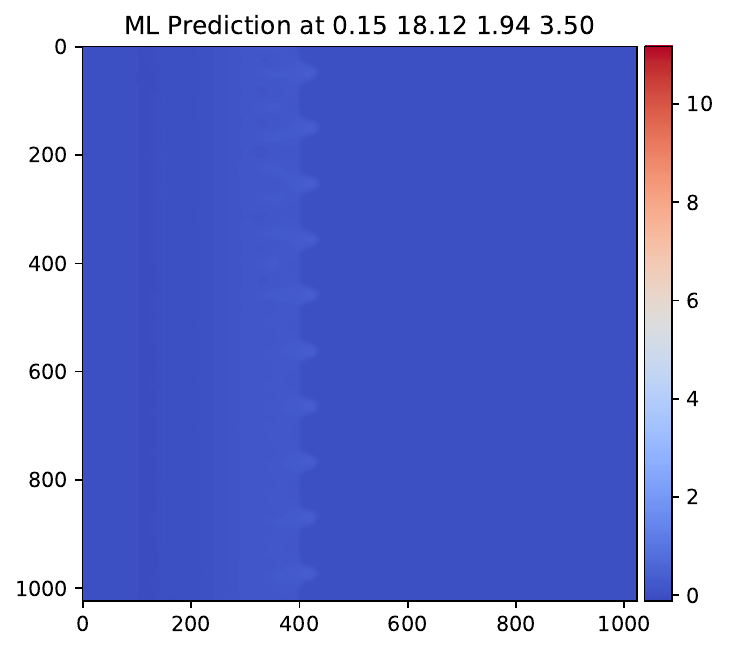} & \includegraphics[width=0.29\textwidth]{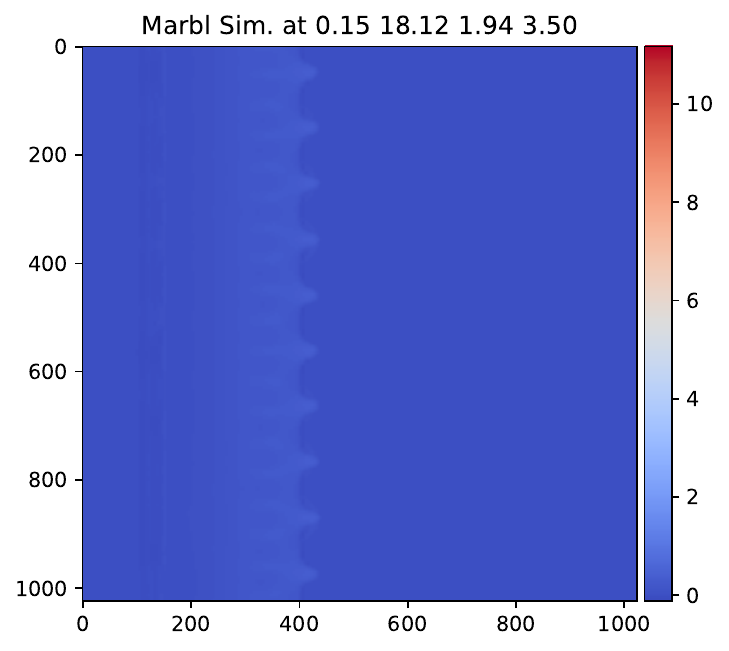} & \includegraphics[width=0.29\textwidth]{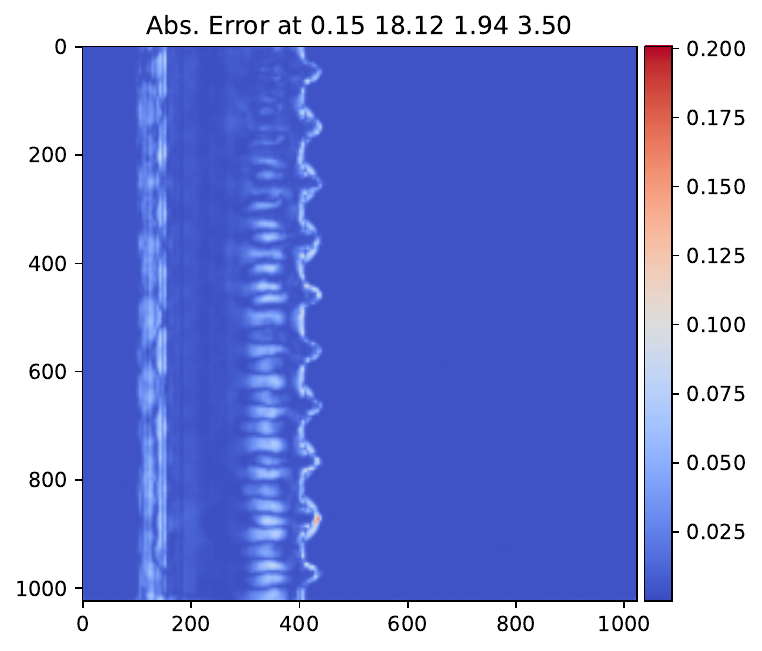} \\
  \hline
  velocity $y$ & \includegraphics[width=0.29\textwidth]{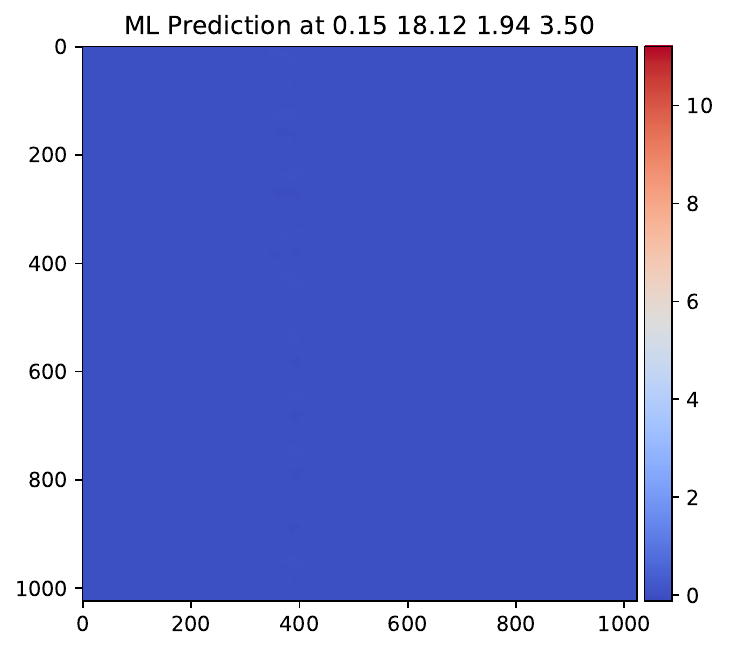} & \includegraphics[width=0.29\textwidth]{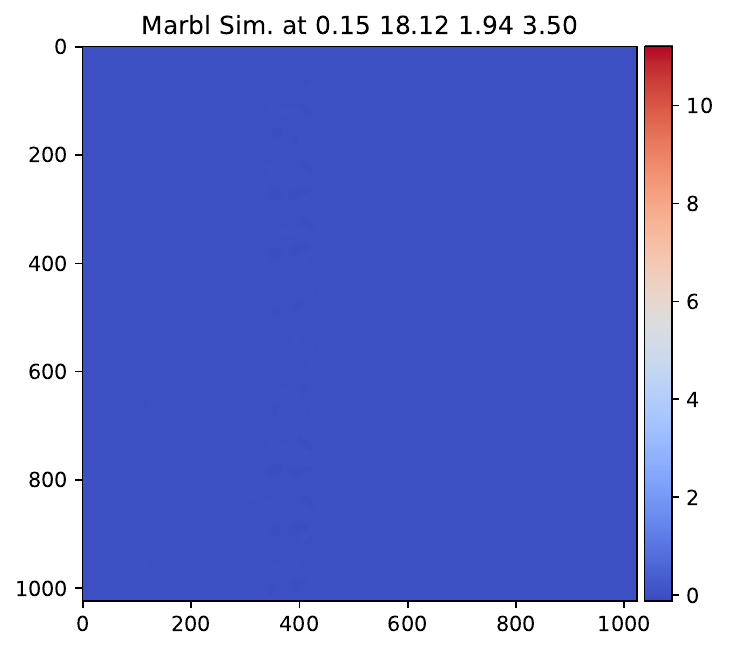} & \includegraphics[width=0.29\textwidth]{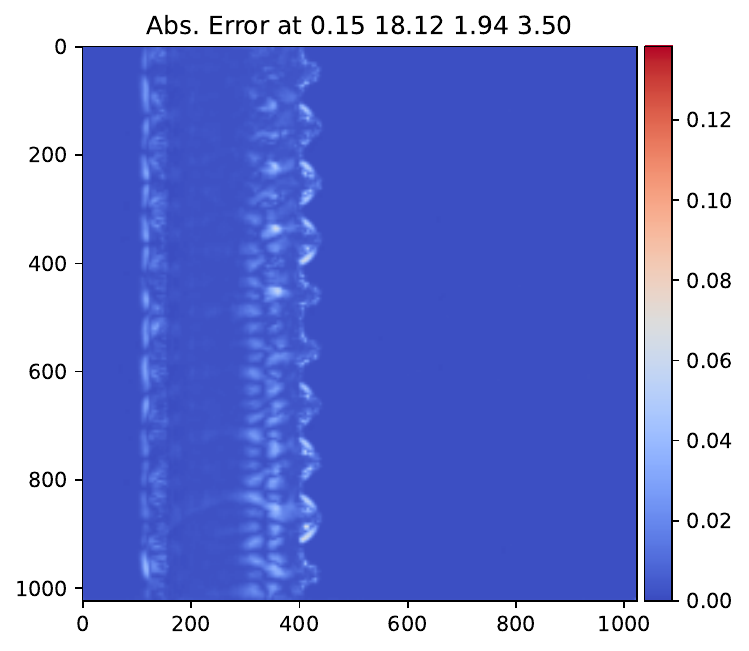} \\
  \end{tabular}
  \caption{
      Figure of predictions, truth, and absolute error for double sine wave predictions at $t$=3.5.
      }\label{fig:ds_predictions_1}
  \end{center}
\end{figure}

\begin{figure}[!htb]
  \begin{center}
  \begin{tabular}{*{7}{c}}
  Field & Ml prediction & Simulation & Abs. Error \\
  \hline
  density & \includegraphics[width=0.29\textwidth]{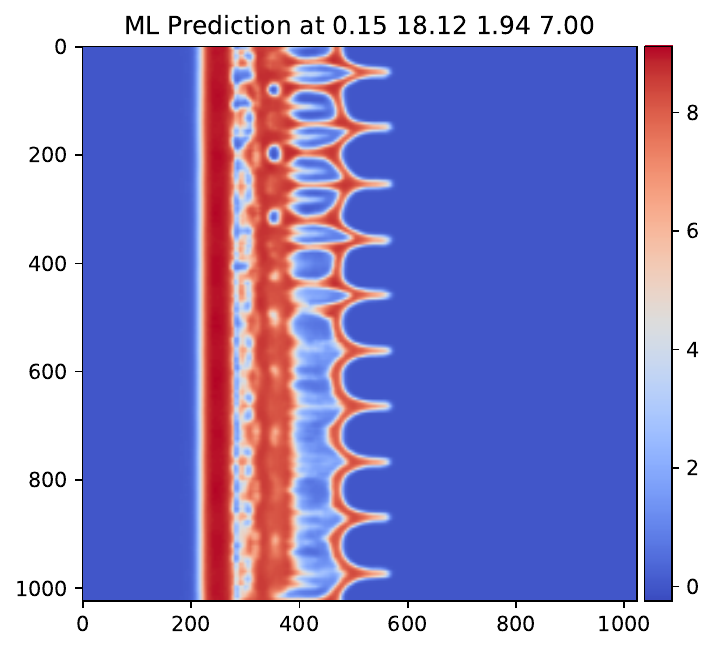} & \includegraphics[width=0.29\textwidth]{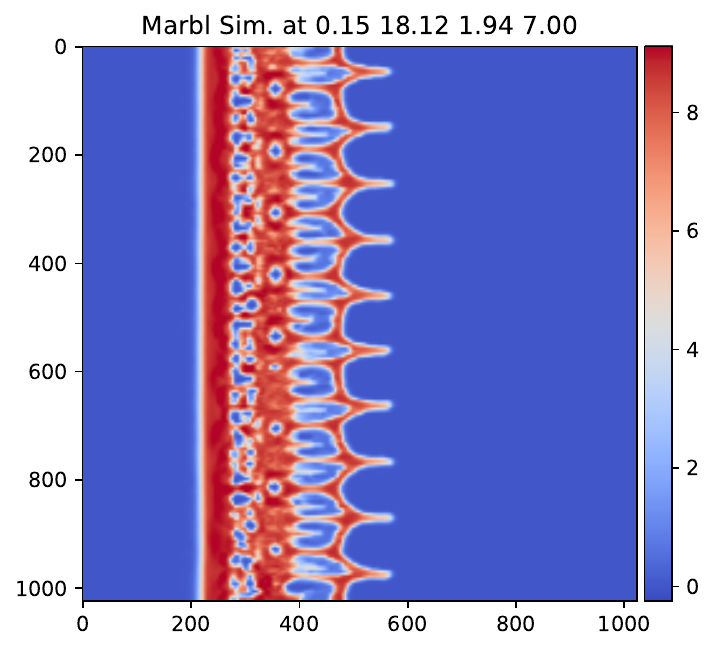} & \includegraphics[width=0.29\textwidth]{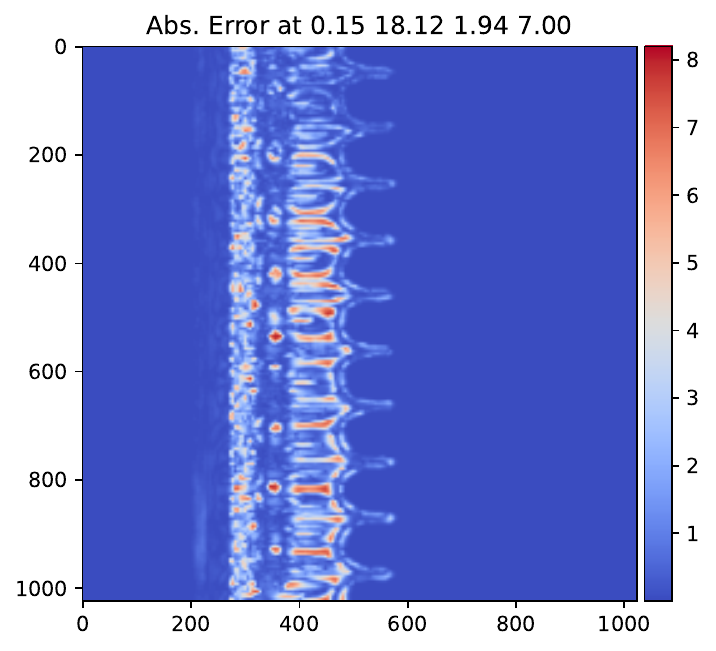} \\
  \hline
  velocity $x$ & \includegraphics[width=0.29\textwidth]{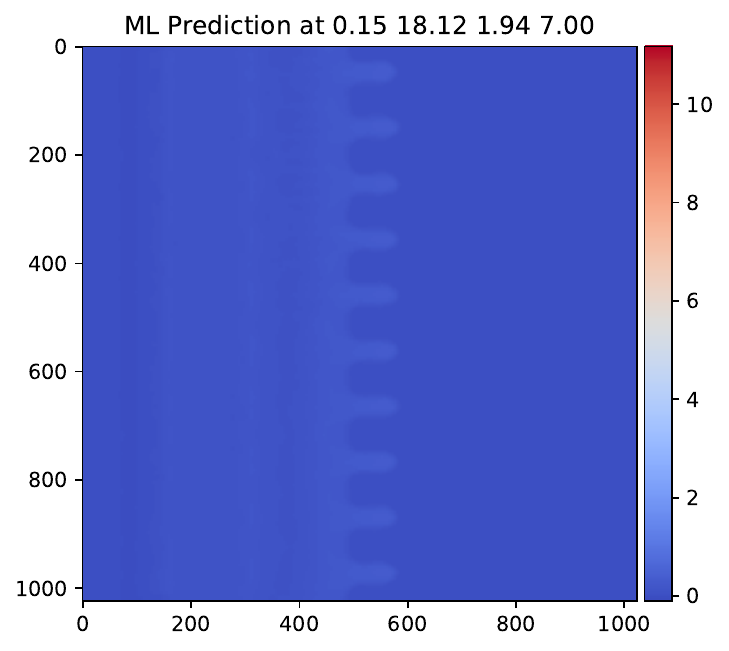} & \includegraphics[width=0.29\textwidth]{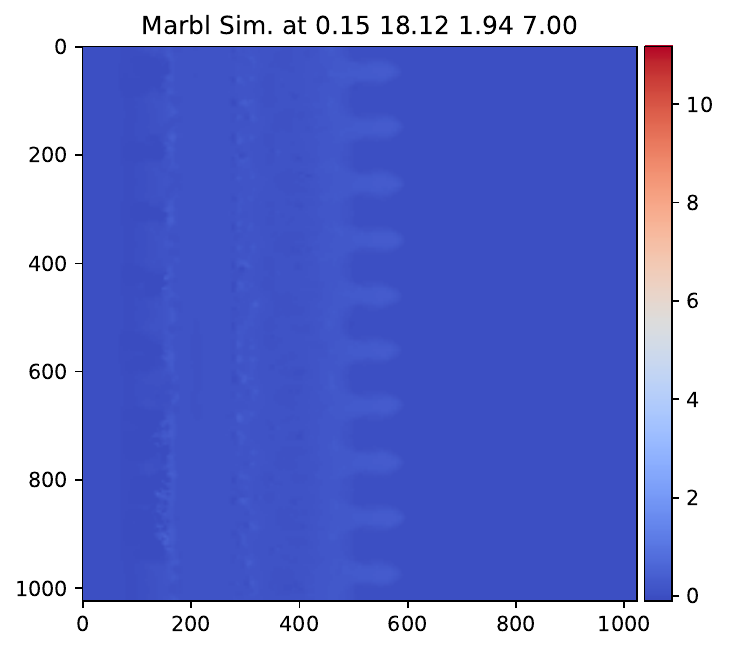} & \includegraphics[width=0.29\textwidth]{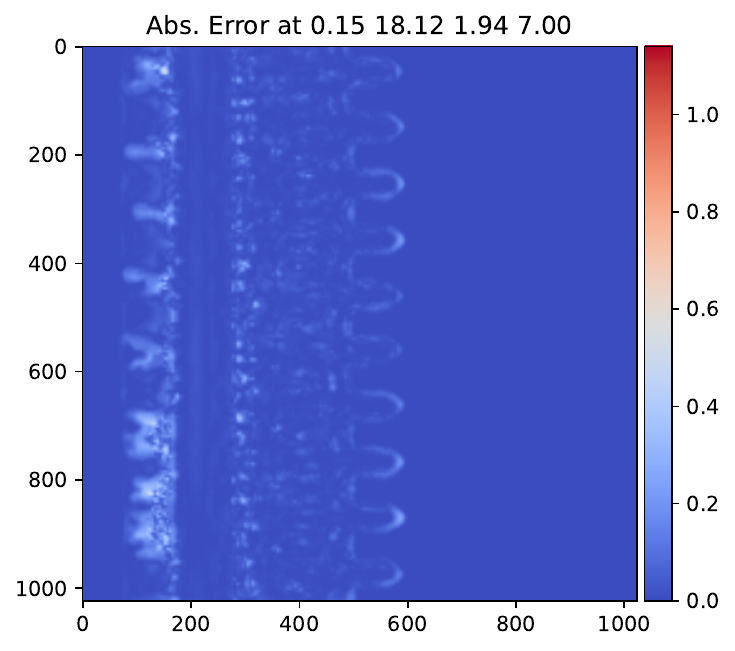} \\
  \hline
  velocity $y$ & \includegraphics[width=0.29\textwidth]{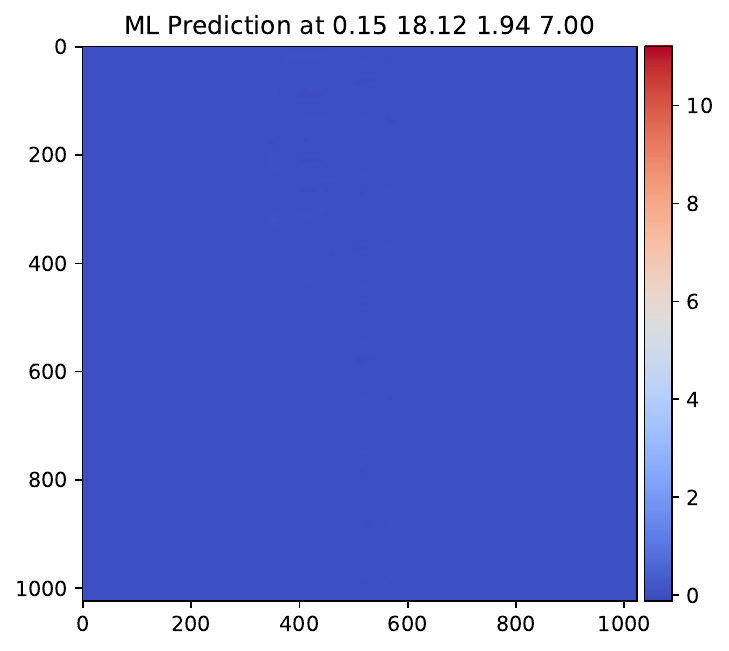} & \includegraphics[width=0.29\textwidth]{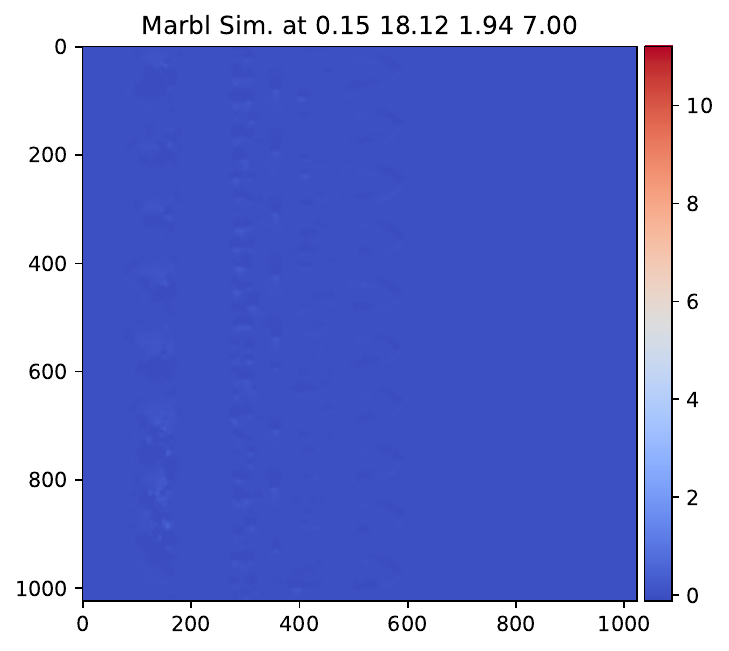} & \includegraphics[width=0.29\textwidth]{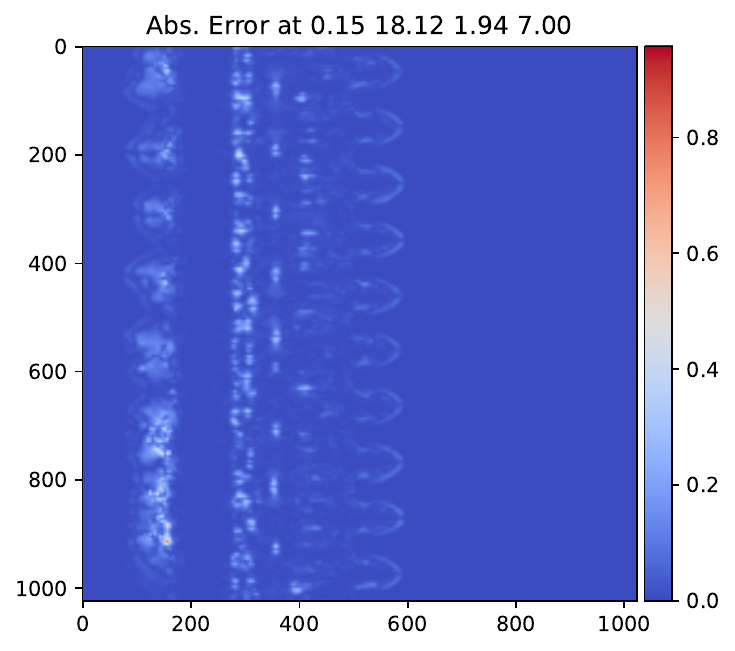} \\
  \end{tabular}
  \caption{
      Figure of predictions, truth, and absolute error for double sine wave predictions at $t$=7.
      }\label{fig:ds_predictions_2}
  \end{center}
\end{figure}

\FloatBarrier

\subsection{Linear shaped charge results}

\begin{figure}[!htb]
  \centering
    \includegraphics[width=1.0\textwidth]{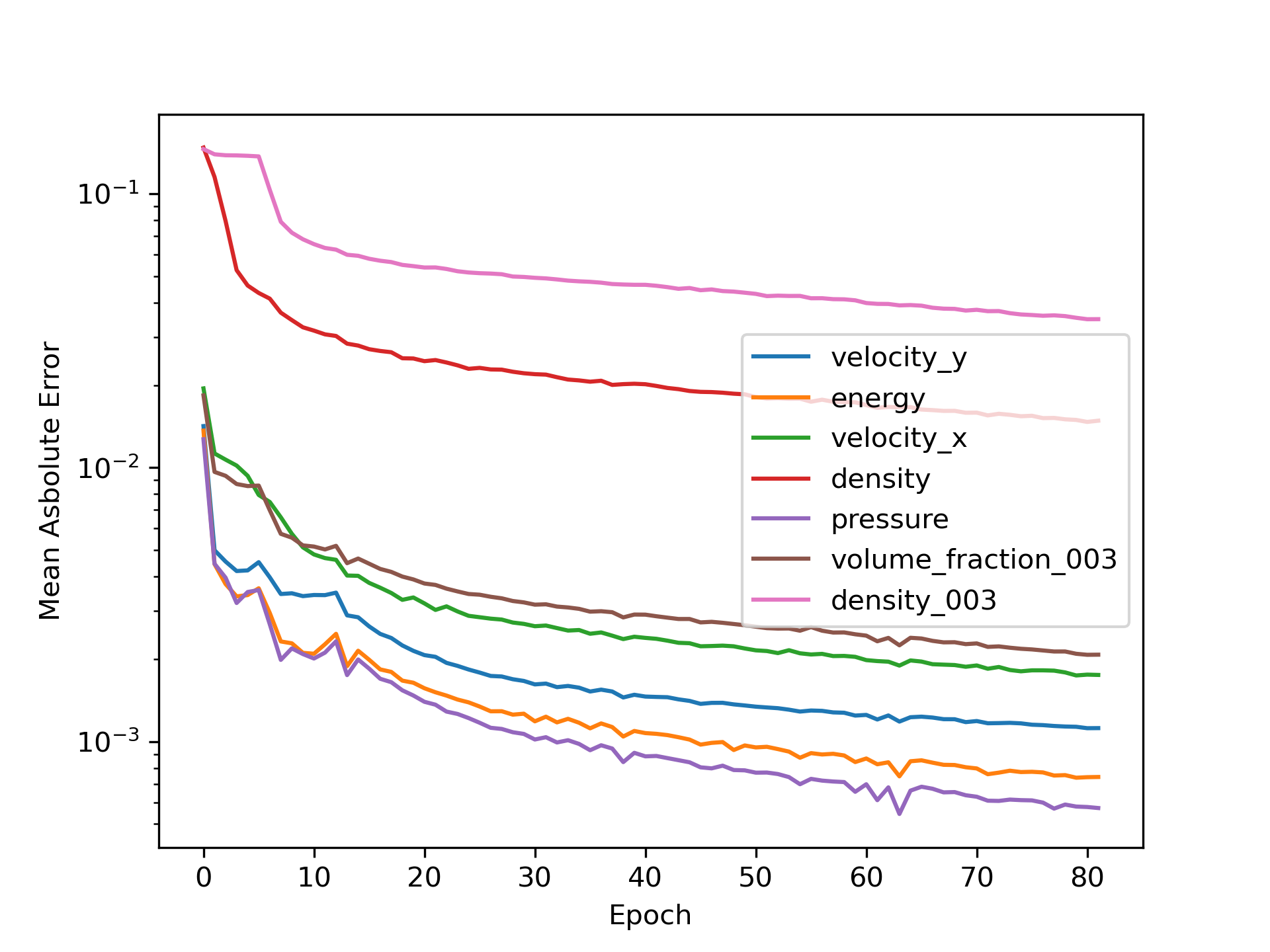}
  \caption{
      The epoch vs mean absolute error for each field while training the linear shaped charge model.
      }
  \label{fig:lsctraining}
\end{figure}

\begin{figure}[!htb]
  \begin{center}
  \begin{tabular}{*{7}{c}}
  Field & Ml prediction & Simulation & Abs. Error \\
  \hline
  density copper (Cu) & \includegraphics[width=0.27\textwidth]{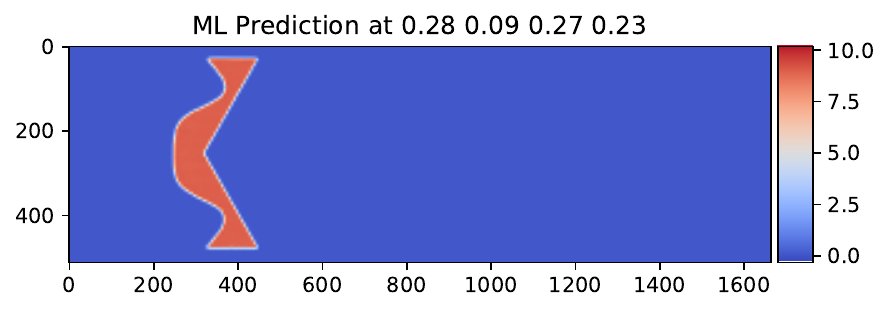} & \includegraphics[width=0.27\textwidth]{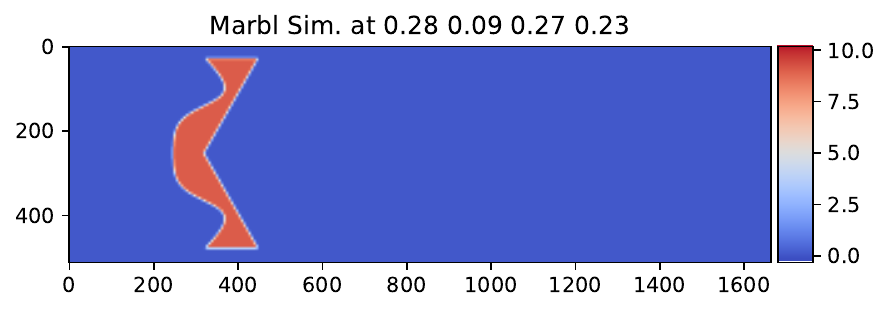} & \includegraphics[width=0.27\textwidth]{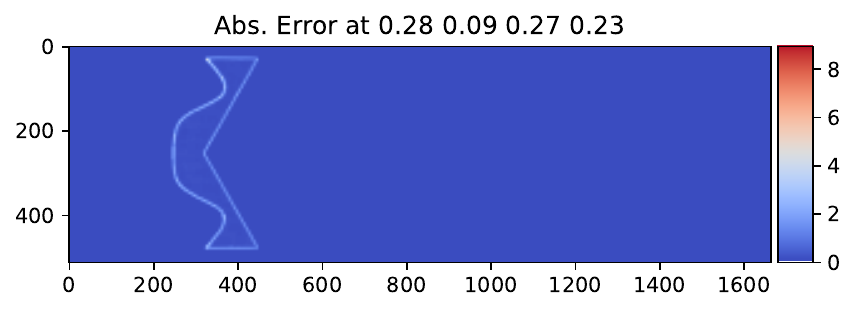} \\
  \hline
  density & \includegraphics[width=0.27\textwidth]{figs/lsc/density_yhat_00.pdf} & \includegraphics[width=0.27\textwidth]{figs/lsc/density_y_00.pdf} & \includegraphics[width=0.27\textwidth]{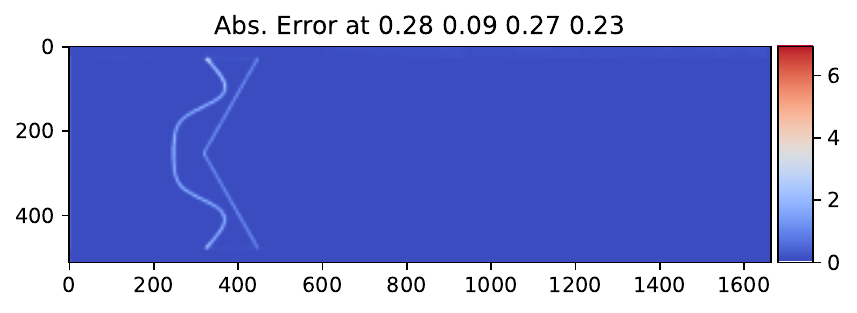} \\
  \hline
  energy & \includegraphics[width=0.27\textwidth]{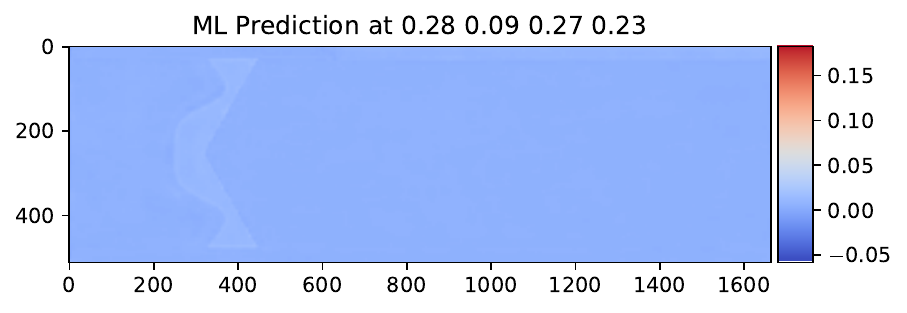} & \includegraphics[width=0.27\textwidth]{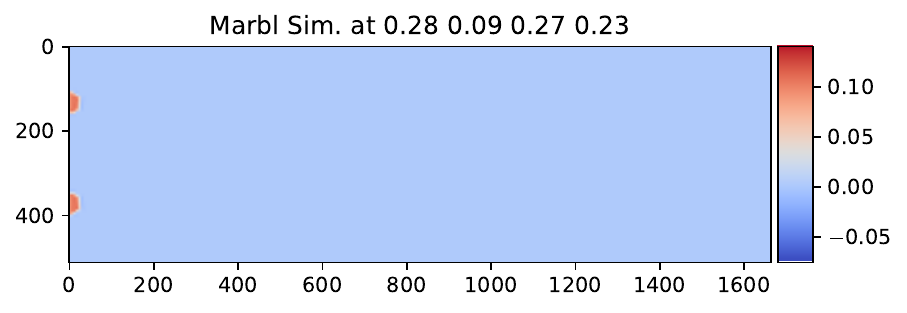} & \includegraphics[width=0.27\textwidth]{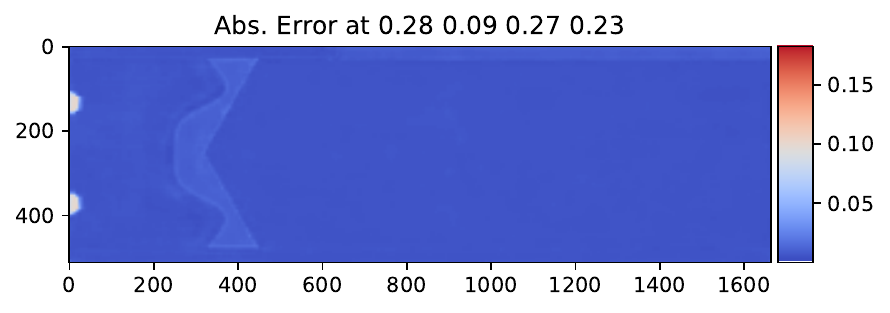} \\
  \hline
  pressure & \includegraphics[width=0.27\textwidth]{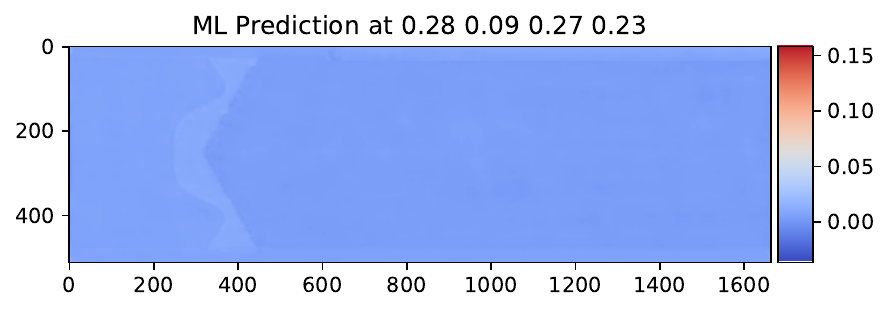} & \includegraphics[width=0.27\textwidth]{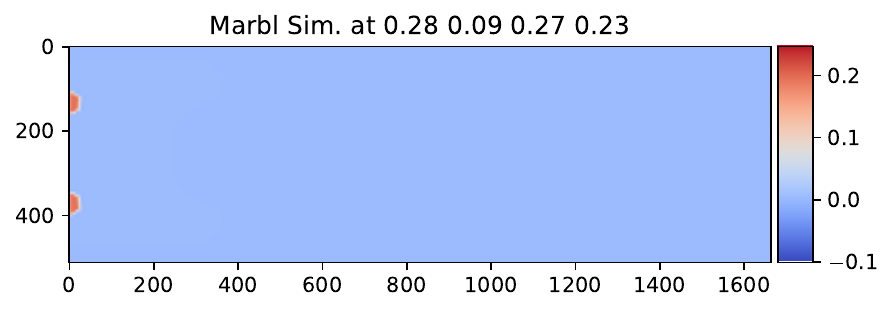} & \includegraphics[width=0.27\textwidth]{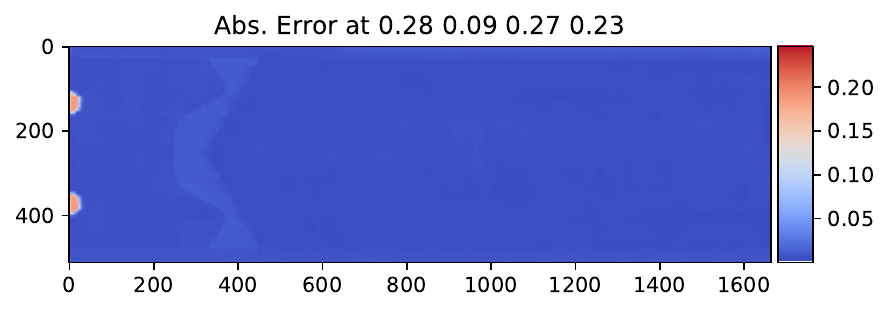} \\
  \hline
  velocity $x$ & \includegraphics[width=0.27\textwidth]{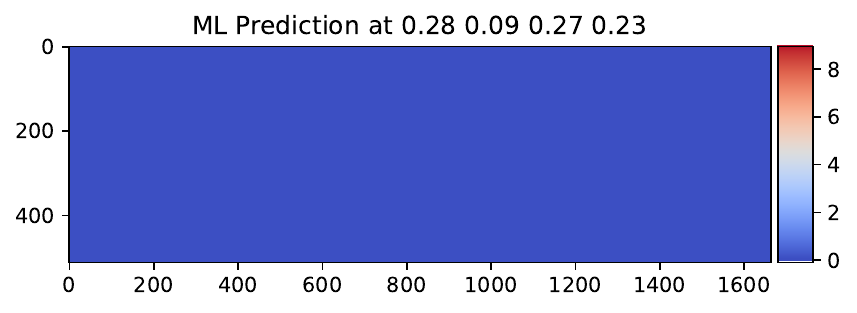} & \includegraphics[width=0.27\textwidth]{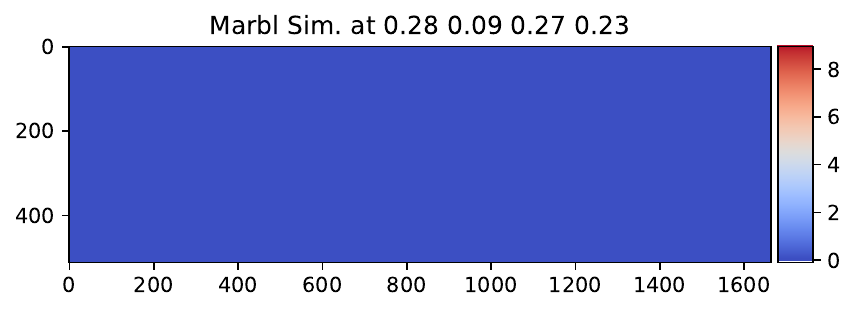} & \includegraphics[width=0.27\textwidth]{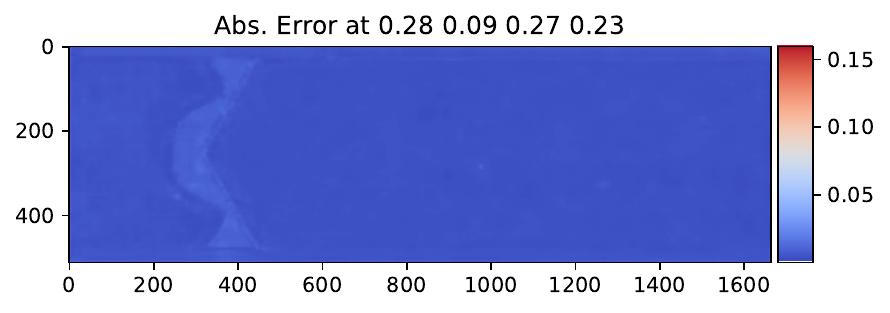} \\
  \hline
  velocity $y$ & \includegraphics[width=0.27\textwidth]{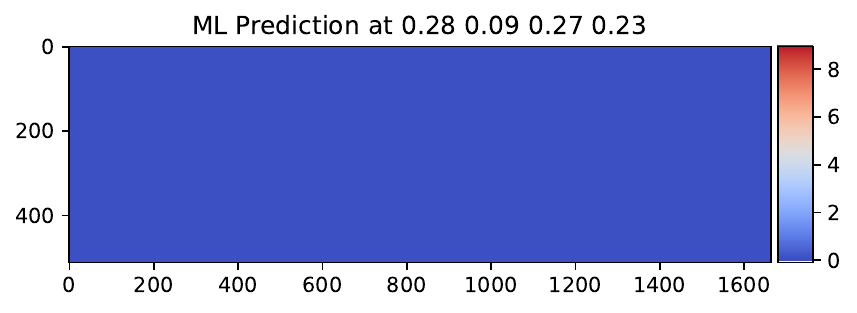} & \includegraphics[width=0.27\textwidth]{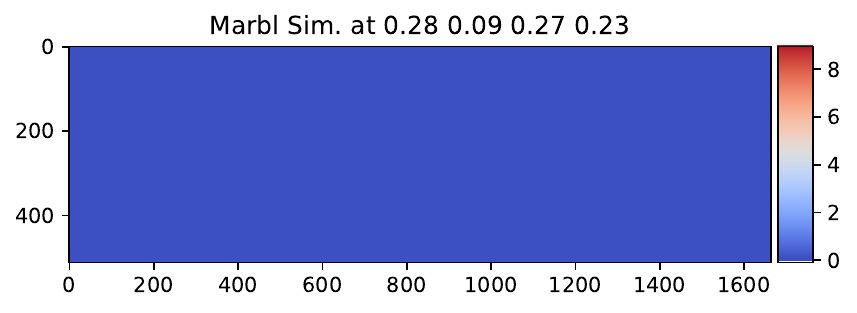} & \includegraphics[width=0.27\textwidth]{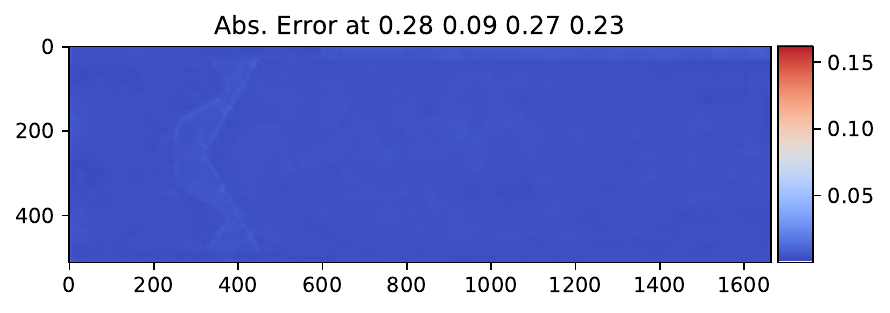} \\
  \hline
  volume fraction Cu & \includegraphics[width=0.27\textwidth]{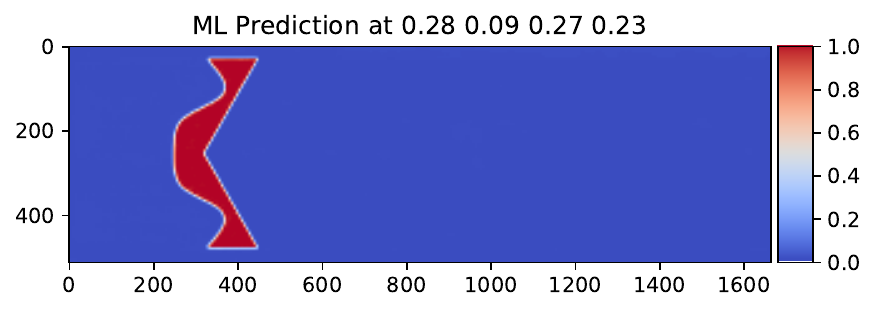} & \includegraphics[width=0.27\textwidth]{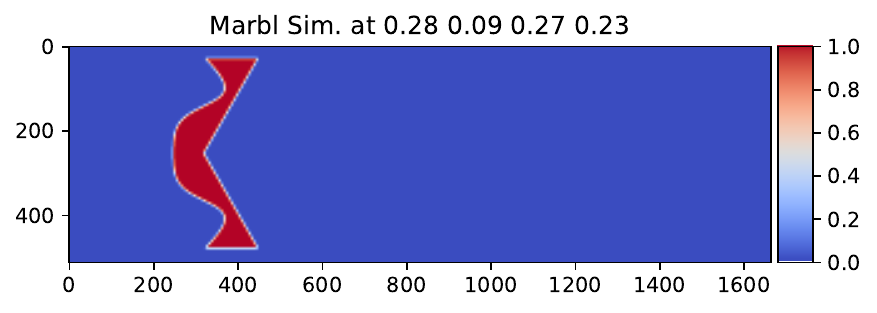} & \includegraphics[width=0.27\textwidth]{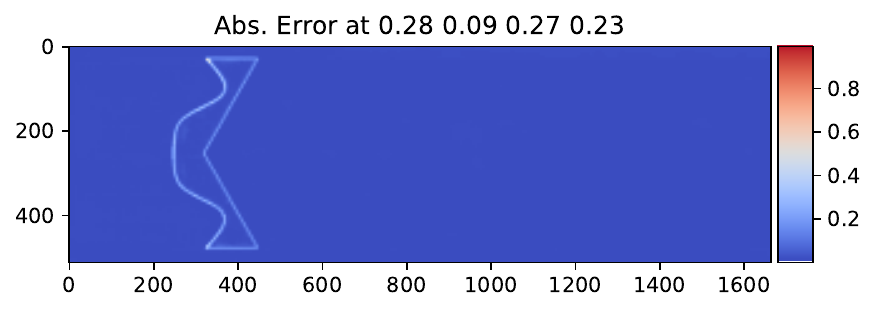} \\
  \end{tabular}
  \caption{
      Figure of predictions, truth, and absolute error for linear shaped charge predictions at $t$=0.
      }\label{fig:lsc_predictions_0}
  \end{center}
\end{figure}

\begin{figure}[!htb]
  \begin{center}
  \begin{tabular}{*{7}{c}}
  Field & Ml prediction & Simulation & Abs. Error \\
  \hline
  density copper (Cu) & \includegraphics[width=0.27\textwidth]{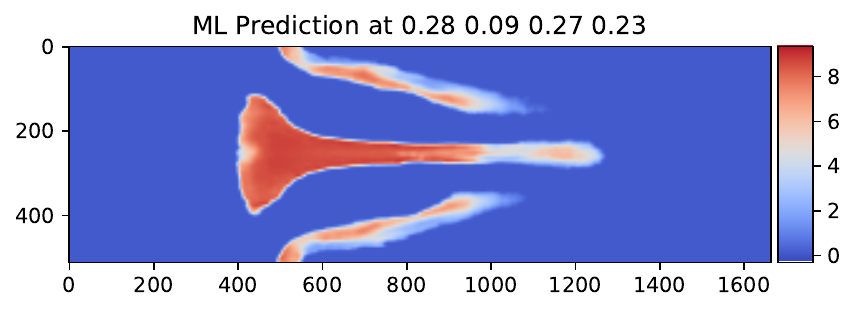} & \includegraphics[width=0.27\textwidth]{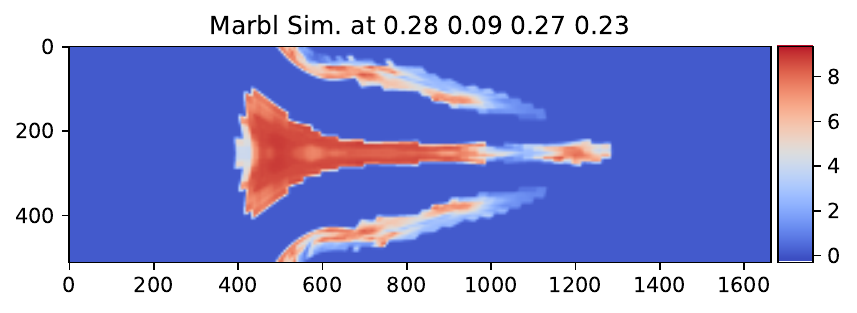} & \includegraphics[width=0.27\textwidth]{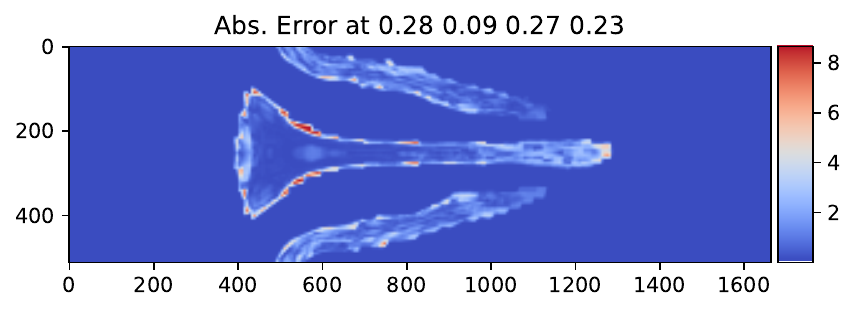} \\
  \hline
  density & \includegraphics[width=0.27\textwidth]{figs/lsc/density_yhat_23.pdf} & \includegraphics[width=0.27\textwidth]{figs/lsc/density_y_23.pdf} & \includegraphics[width=0.27\textwidth]{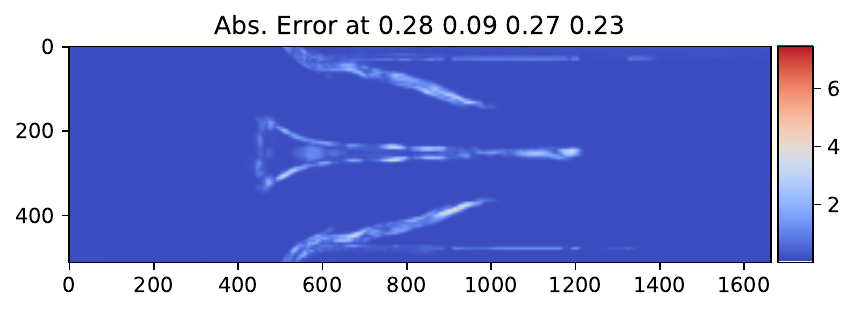} \\
  \hline
  energy & \includegraphics[width=0.27\textwidth]{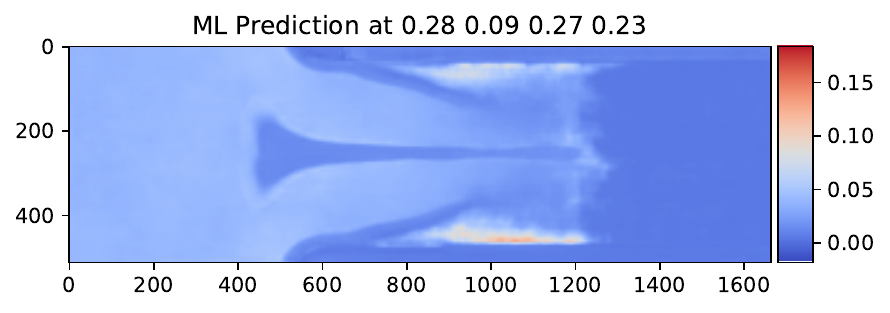} & \includegraphics[width=0.27\textwidth]{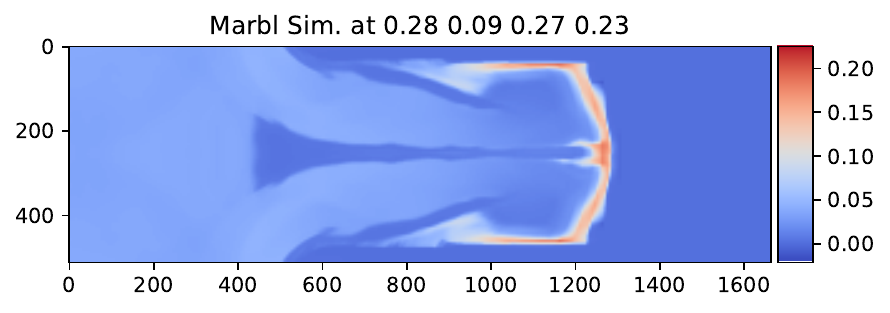} & \includegraphics[width=0.27\textwidth]{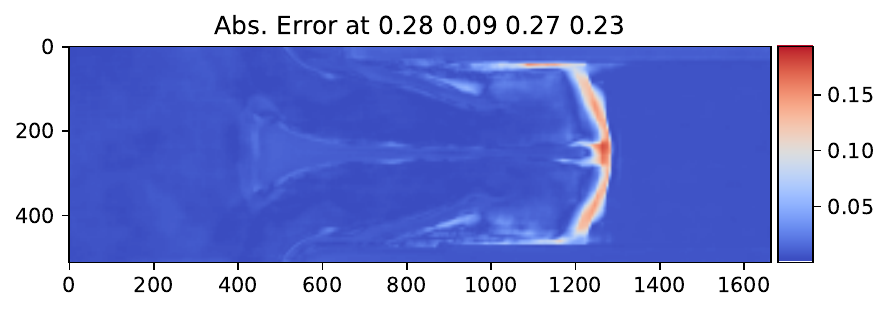} \\
  \hline
  pressure & \includegraphics[width=0.27\textwidth]{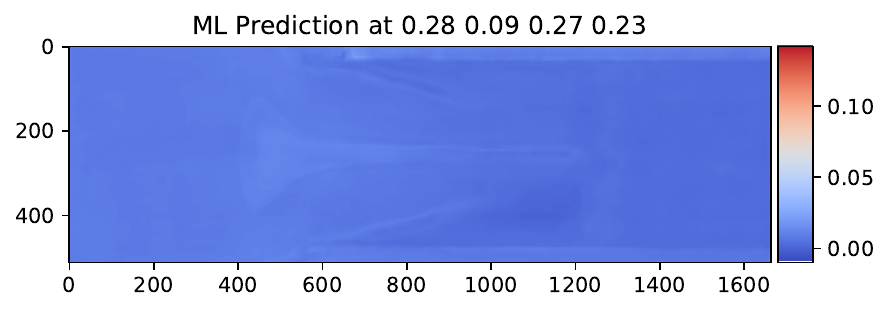} & \includegraphics[width=0.27\textwidth]{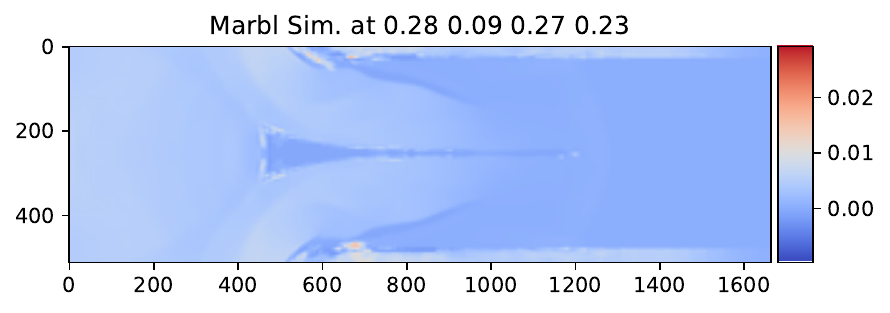} & \includegraphics[width=0.27\textwidth]{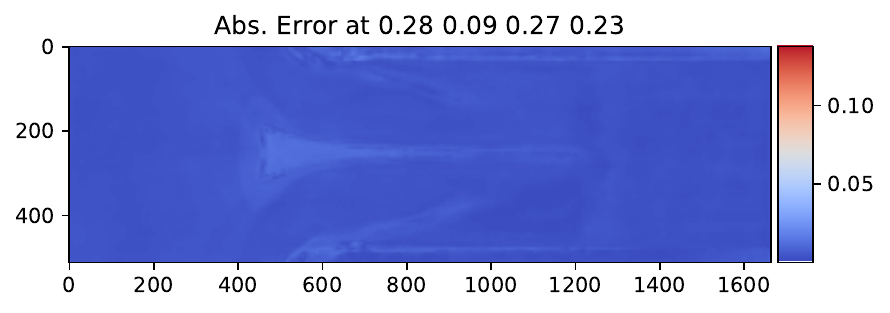} \\
  \hline
  velocity $x$ & \includegraphics[width=0.27\textwidth]{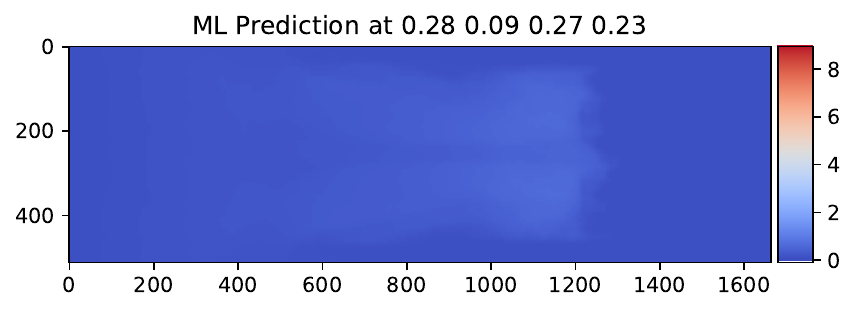} & \includegraphics[width=0.27\textwidth]{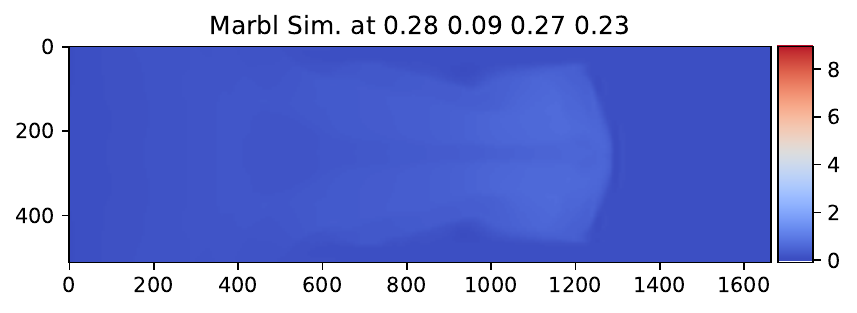} & \includegraphics[width=0.27\textwidth]{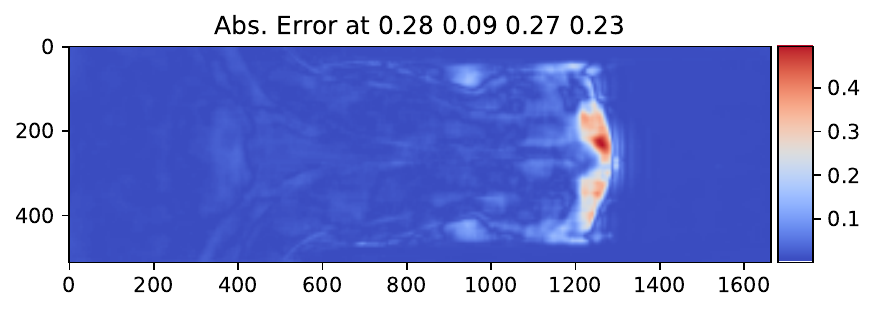} \\
  \hline
  velocity $y$ & \includegraphics[width=0.27\textwidth]{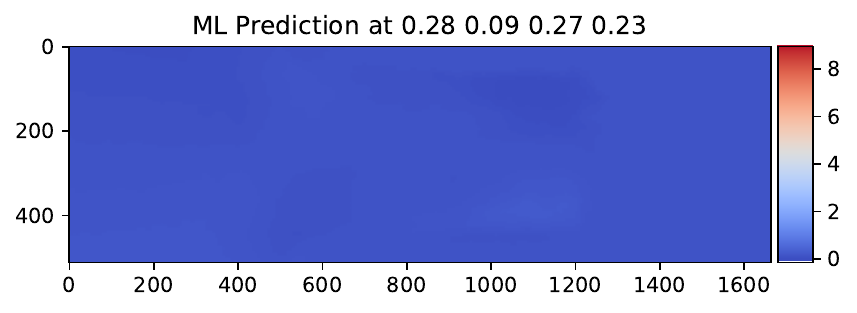} & \includegraphics[width=0.27\textwidth]{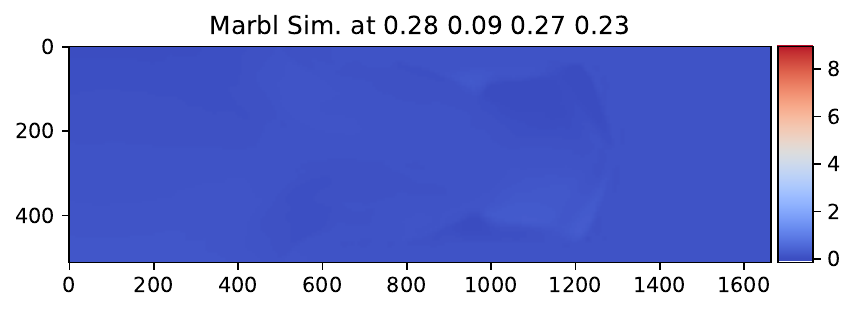} & \includegraphics[width=0.27\textwidth]{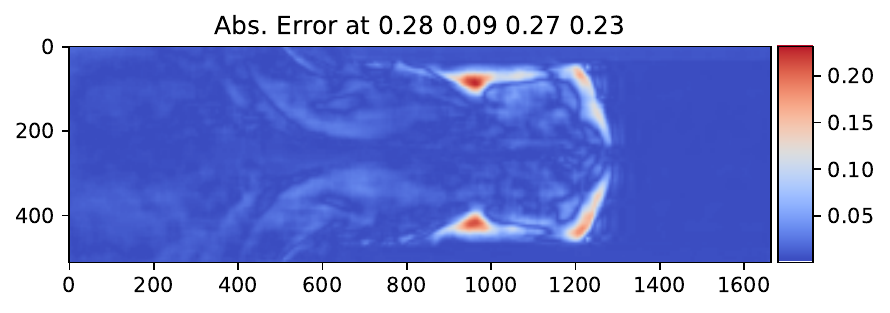} \\
  \hline
  volume fraction Cu & \includegraphics[width=0.27\textwidth]{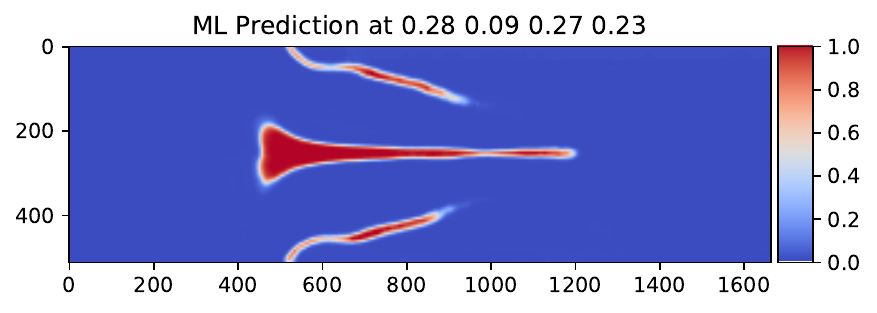} & \includegraphics[width=0.27\textwidth]{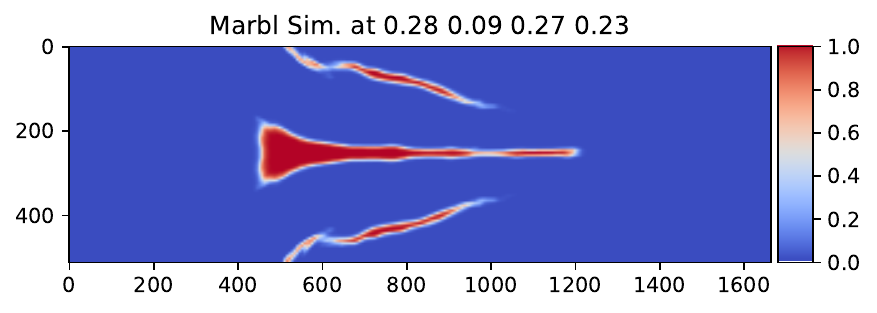} & \includegraphics[width=0.27\textwidth]{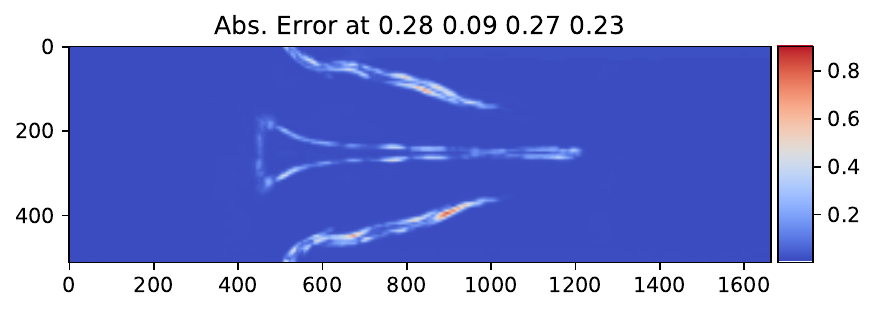} \\
  \end{tabular}
  \caption{
      Figure of predictions, truth, and absolute error for linear shaped charge predictions at $t$=0.
      }\label{fig:lsc_predictions_1}
  \end{center}
\end{figure}

\begin{figure}[!htb]
  \begin{center}
  \begin{tabular}{*{7}{c}}
  Field & Ml prediction & Simulation & Abs. Error \\
  \hline
  density copper (Cu) & \includegraphics[width=0.27\textwidth]{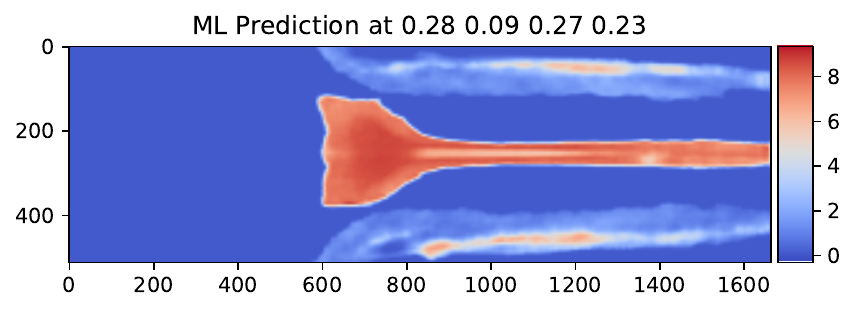} & \includegraphics[width=0.27\textwidth]{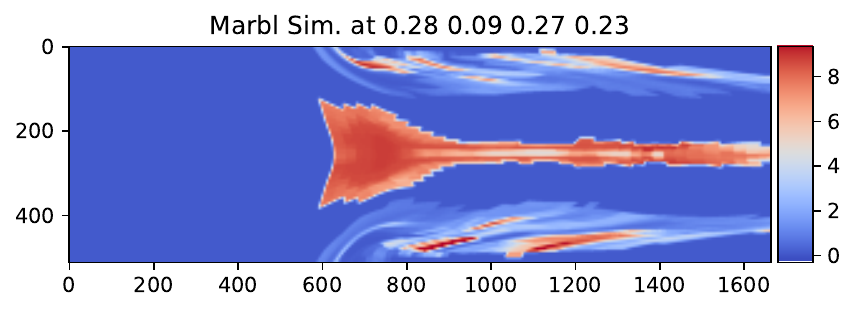} & \includegraphics[width=0.27\textwidth]{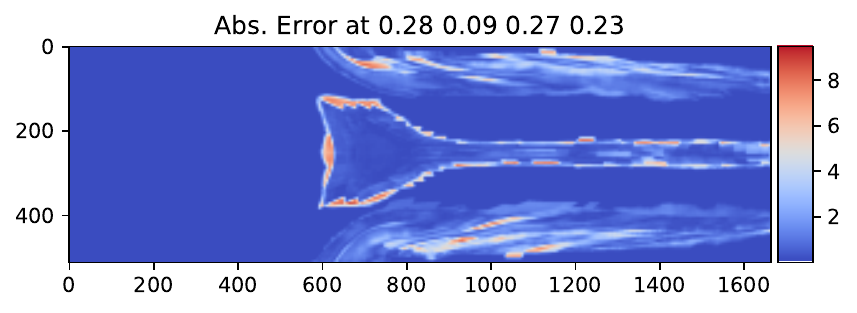} \\
  \hline
  density & \includegraphics[width=0.27\textwidth]{figs/lsc/density_yhat_46.pdf} & \includegraphics[width=0.27\textwidth]{figs/lsc/density_y_46.pdf} & \includegraphics[width=0.27\textwidth]{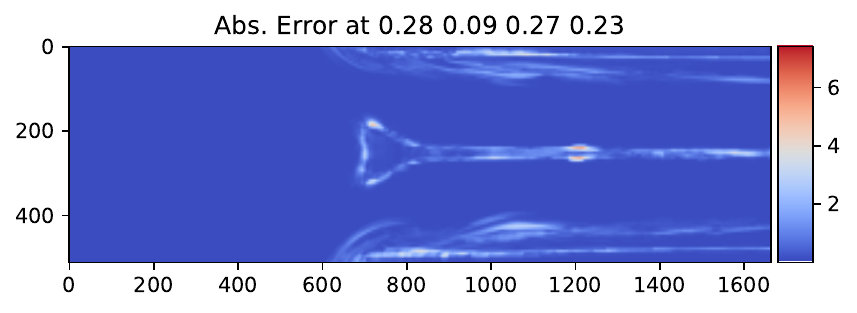} \\
  \hline
  energy & \includegraphics[width=0.27\textwidth]{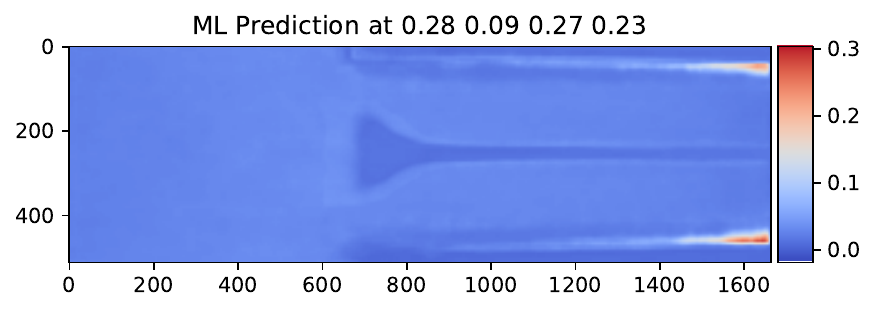} & \includegraphics[width=0.27\textwidth]{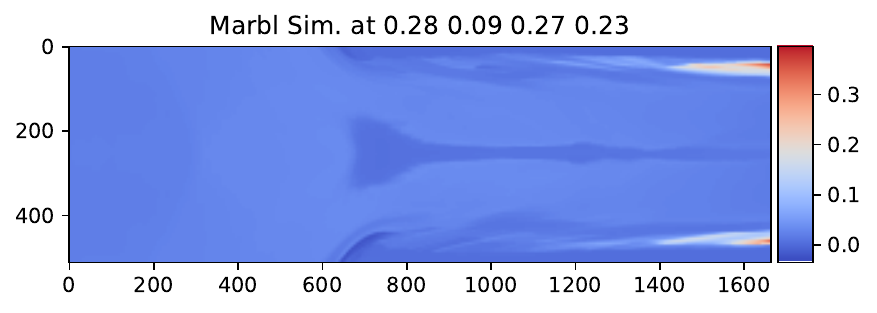} & \includegraphics[width=0.27\textwidth]{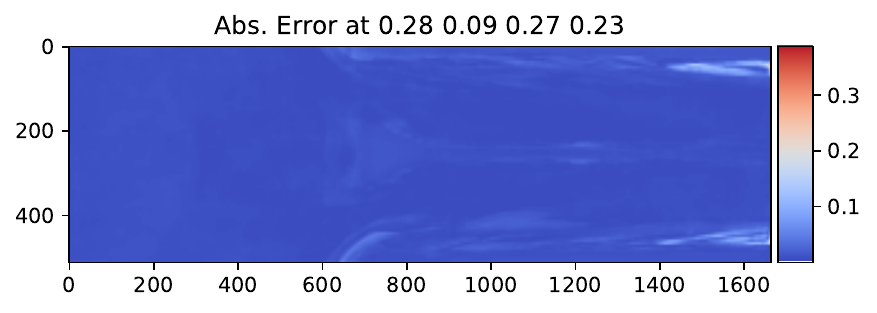} \\
  \hline
  pressure & \includegraphics[width=0.27\textwidth]{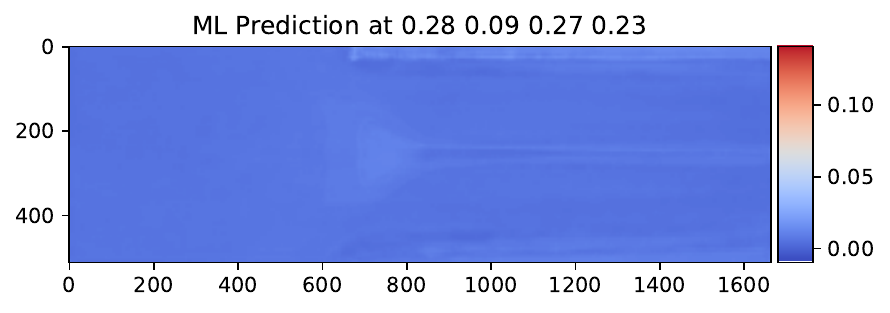} & \includegraphics[width=0.27\textwidth]{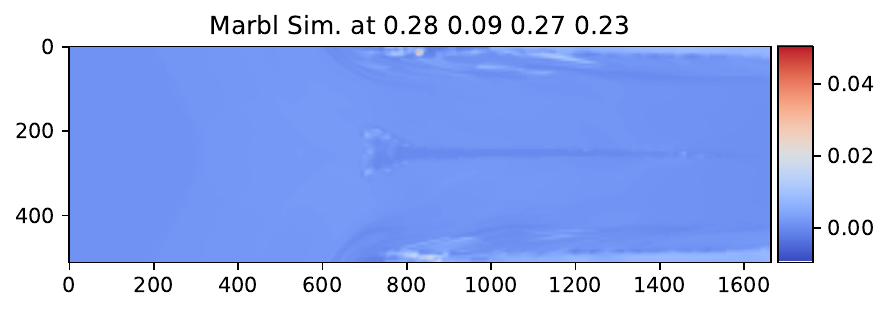} & \includegraphics[width=0.27\textwidth]{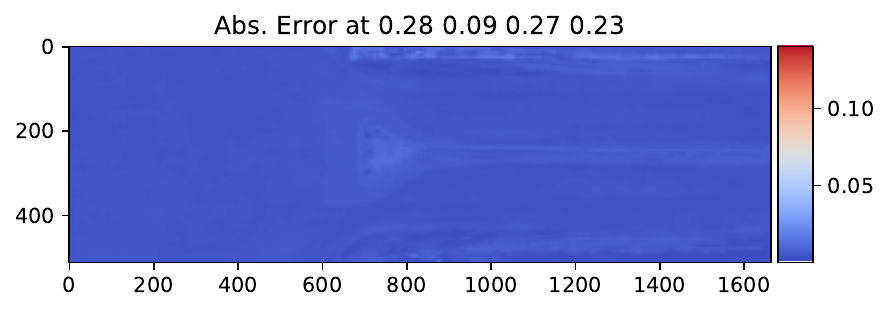} \\
  \hline
  velocity $x$ & \includegraphics[width=0.27\textwidth]{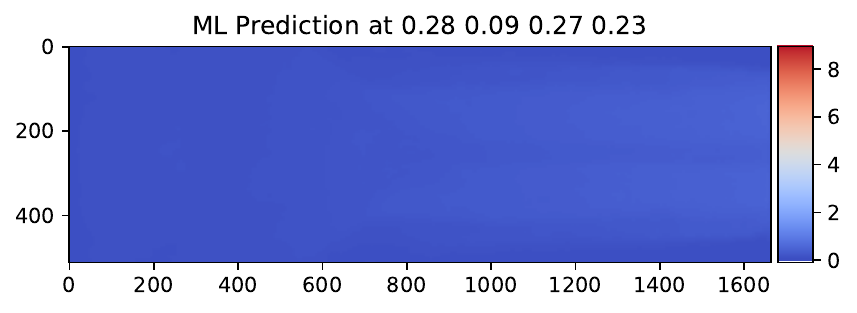} & \includegraphics[width=0.27\textwidth]{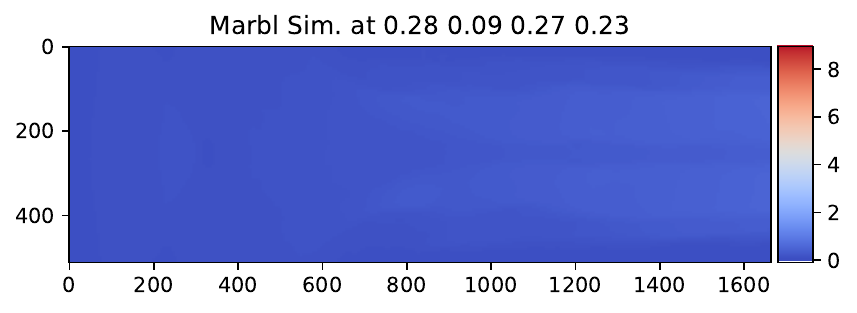} & \includegraphics[width=0.27\textwidth]{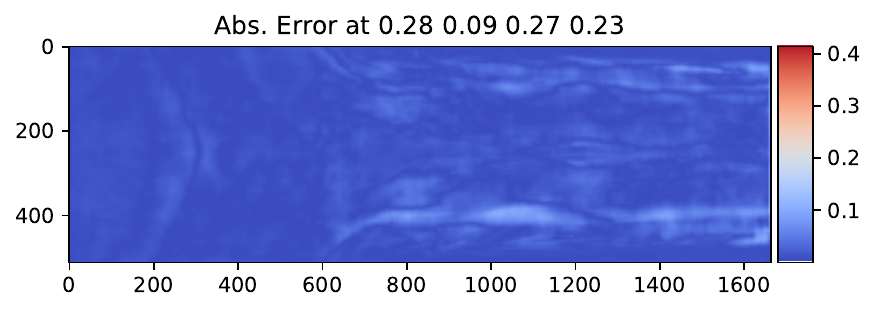} \\
  \hline
  velocity $y$ & \includegraphics[width=0.27\textwidth]{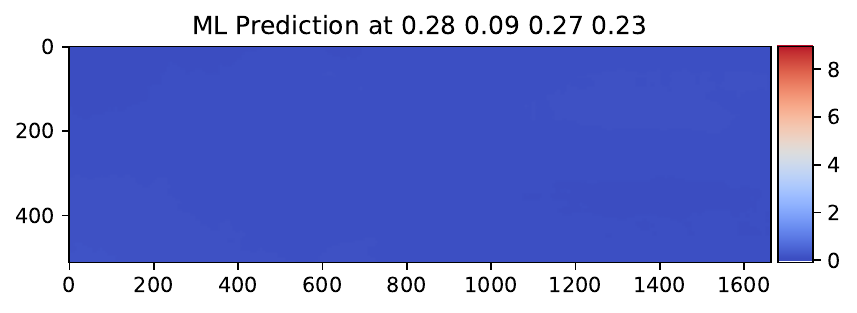} & \includegraphics[width=0.27\textwidth]{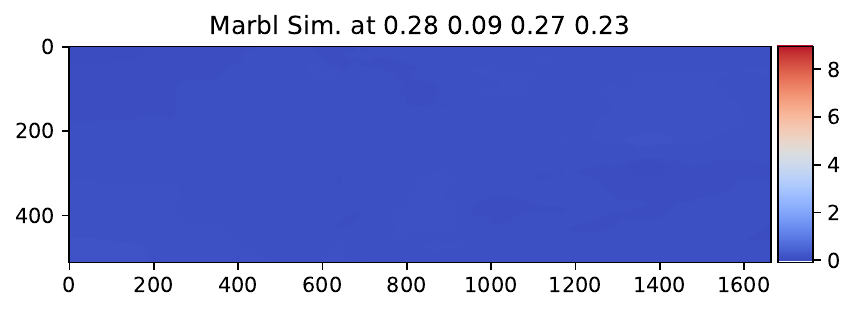} & \includegraphics[width=0.27\textwidth]{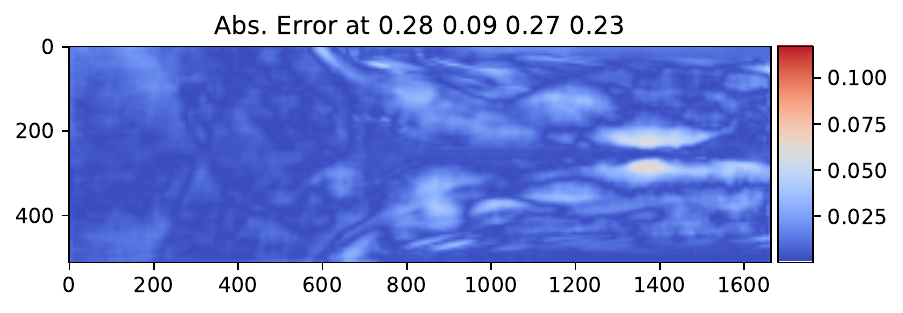} \\
  \hline
  volume fraction Cu & \includegraphics[width=0.27\textwidth]{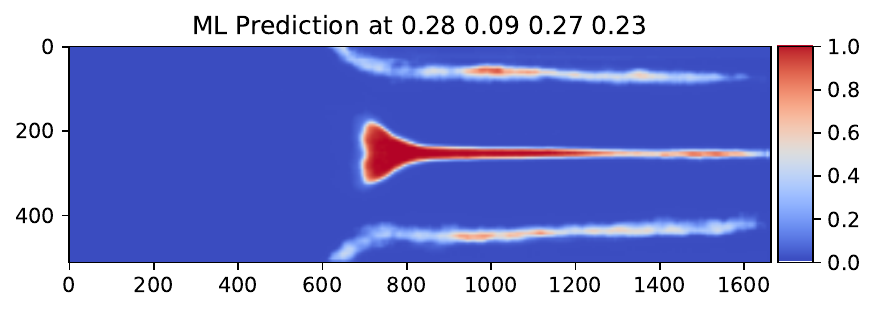} & \includegraphics[width=0.27\textwidth]{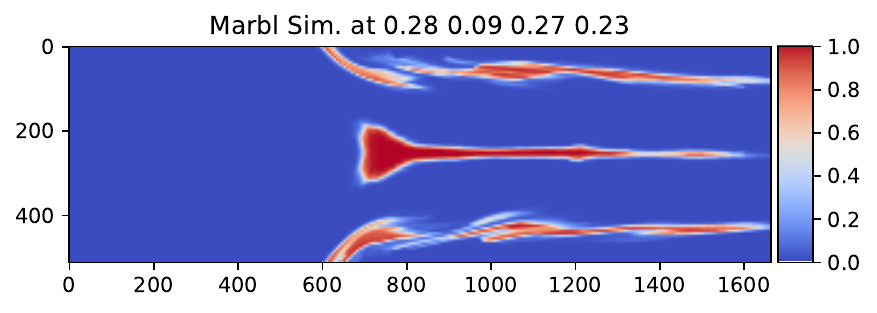} & \includegraphics[width=0.27\textwidth]{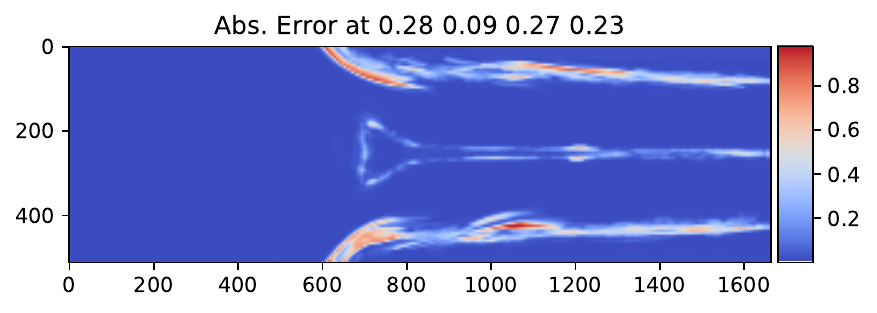} \\
  \end{tabular}
  \caption{
      Figure of predictions, truth, and absolute error for linear shaped charge predictions at $t$=0.
      }\label{fig:lsc_predictions_2}
  \end{center}
\end{figure}

\FloatBarrier

\subsection{Rayleigh-Taylor results}

\begin{figure}[!htb]
  \centering
    \includegraphics[width=1.0\textwidth]{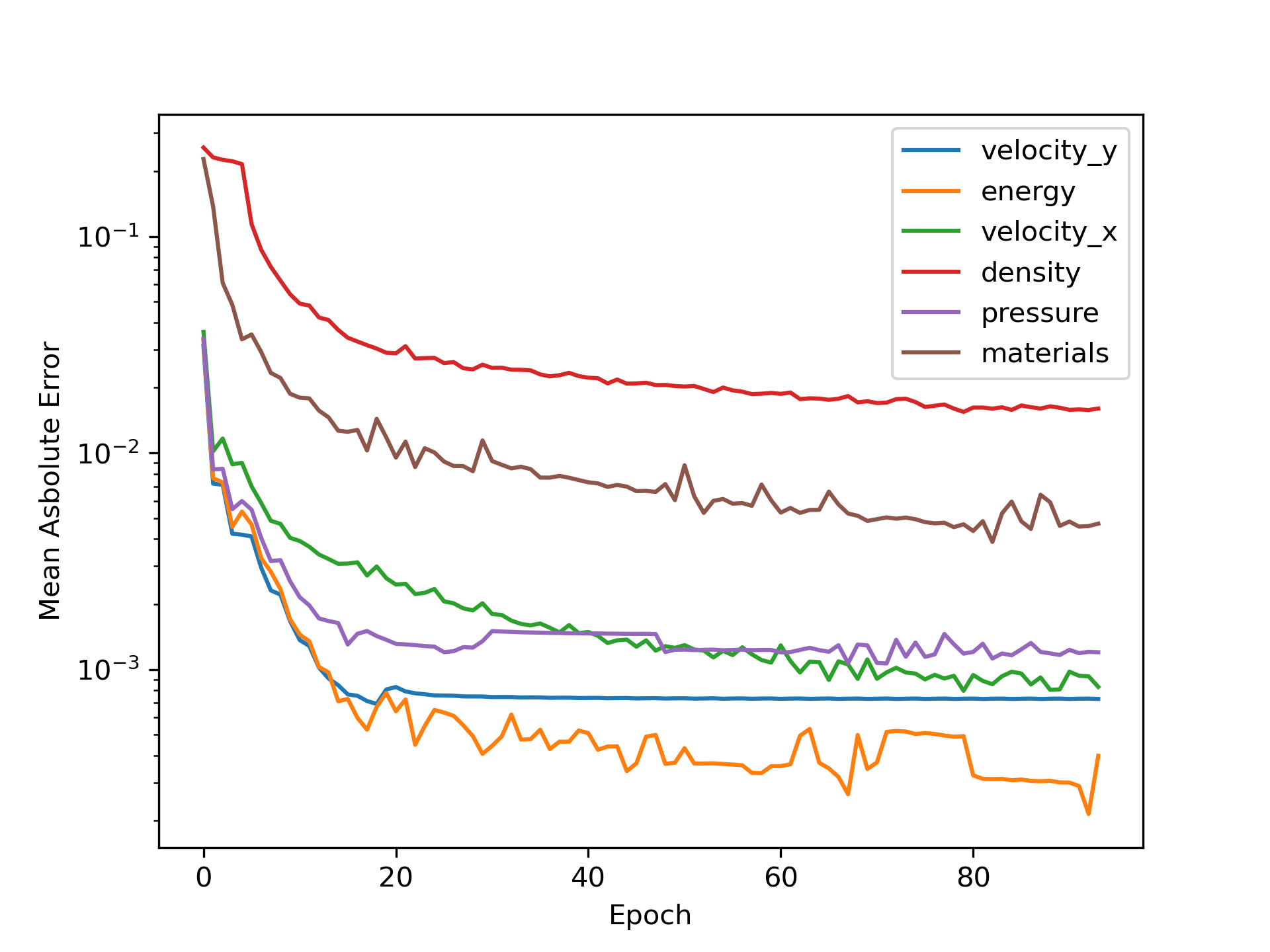}
  \caption{
      The epoch vs mean absolute error for each field while training the Rayleigh-Taylor model.
      }
  \label{fig:rttraining}
\end{figure}

The mean absolute error for each field separately in training the Rayleigh Taylor simulation is shown in Figure~\ref{fig:rttraining}. The final fields errors (from highest to lowest) are density, energy, materials, velocity~$y$, velocity~$x$, and pressure. The error in density was roughly an order of magnitude larger than pressure and velocity~$x$.

\begin{figure}[!htb]
  \begin{center}
  \begin{tabular}{*{7}{|c|}}
     & density & velocity $x$ & velocity $y$ & pressure & energy & materials \\
  ML prediction & \includegraphics[width=0.140\textwidth]{figs/rt/density_yhat_00.pdf} & \includegraphics[width=0.140\textwidth]{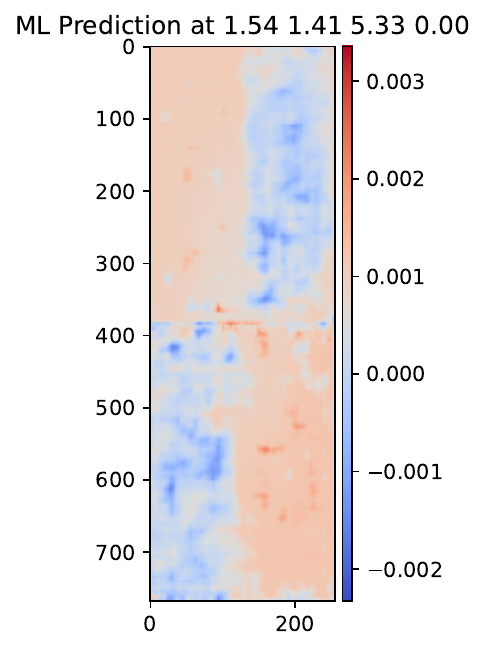} & \includegraphics[width=0.140\textwidth]{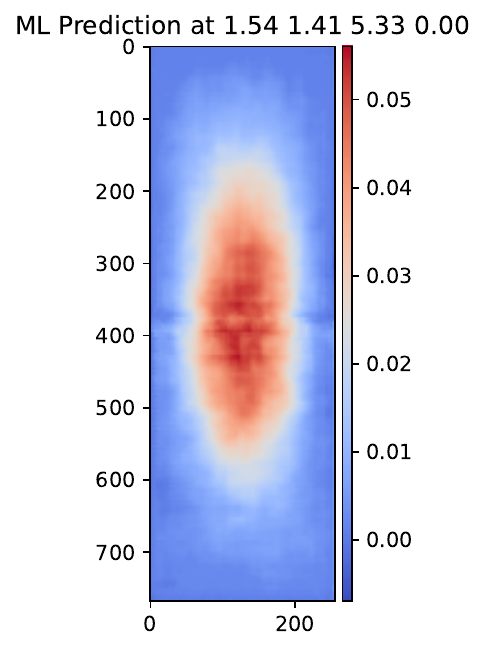} & \includegraphics[width=0.140\textwidth]{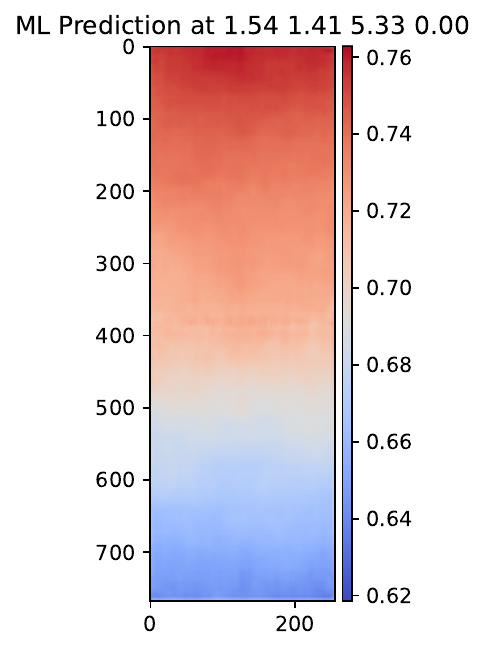} & \includegraphics[width=0.140\textwidth]{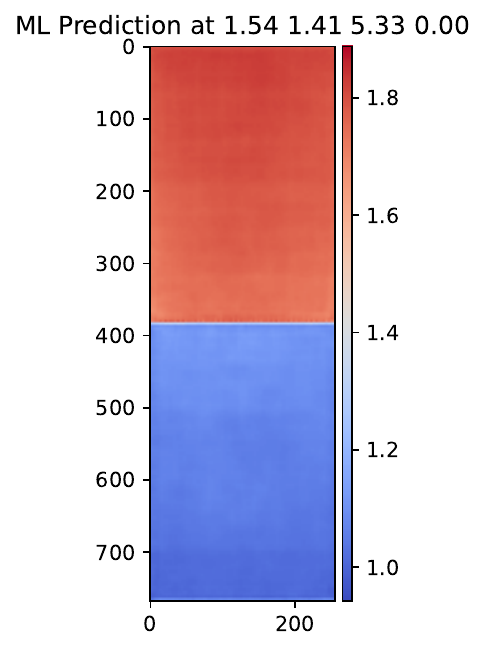} & \includegraphics[width=0.140\textwidth]{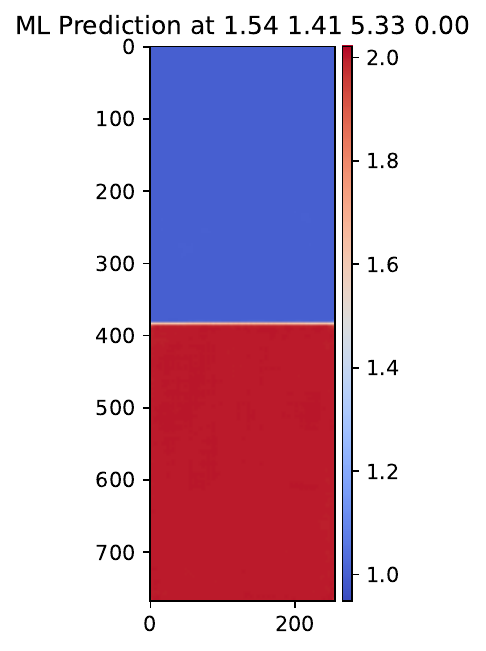} \\
  Simulation & \includegraphics[width=0.140\textwidth]{figs/rt/density_y_00.pdf} & \includegraphics[width=0.140\textwidth]{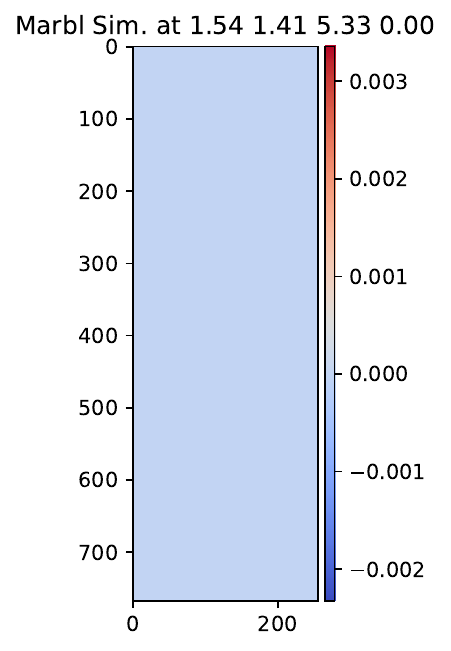} & \includegraphics[width=0.140\textwidth]{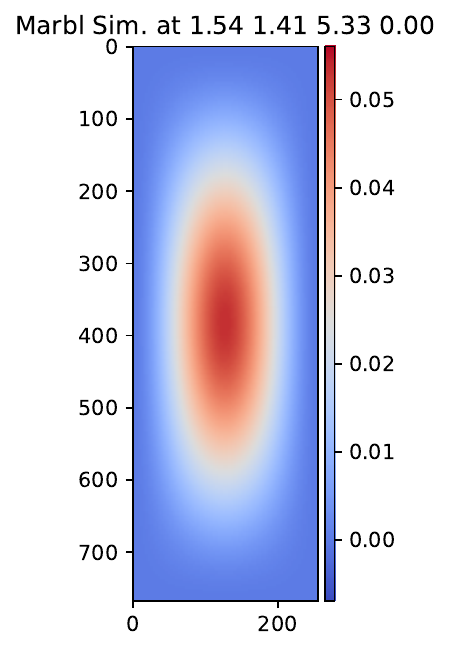} & \includegraphics[width=0.140\textwidth]{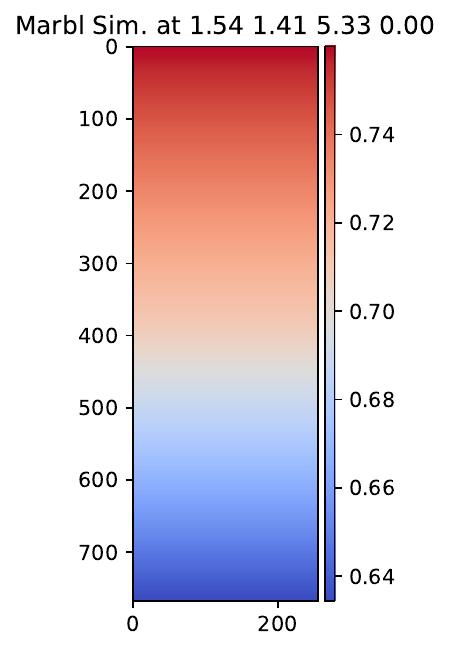} & \includegraphics[width=0.140\textwidth]{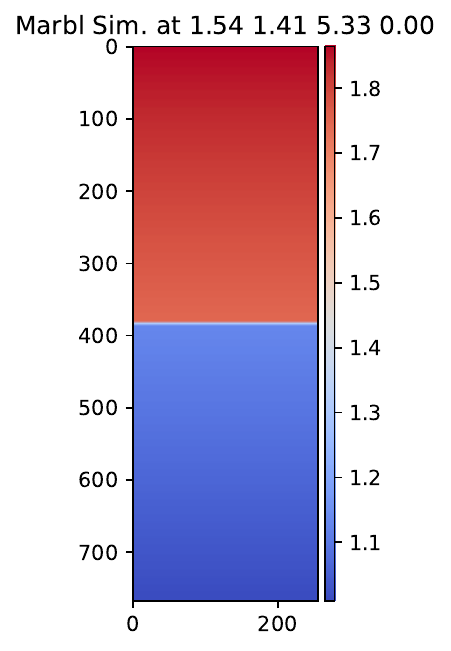} & \includegraphics[width=0.140\textwidth]{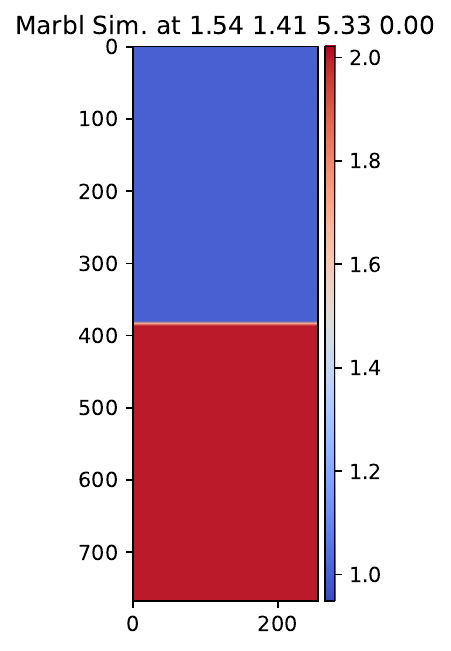} \\
  Abs. Error & \includegraphics[width=0.140\textwidth]{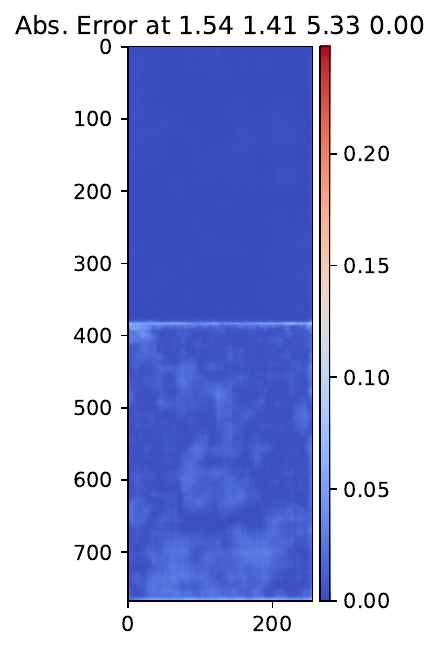} & \includegraphics[width=0.140\textwidth]{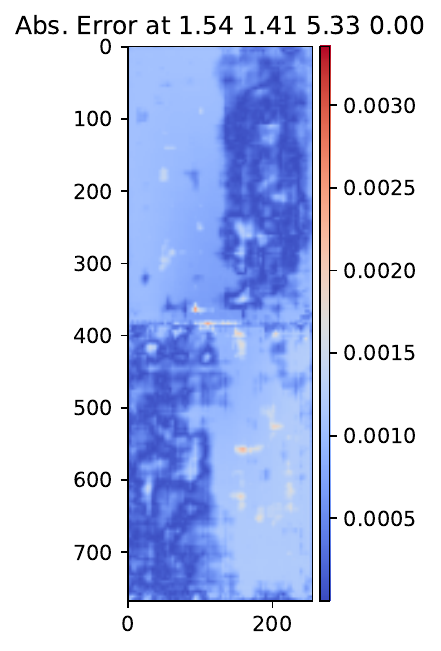} & \includegraphics[width=0.140\textwidth]{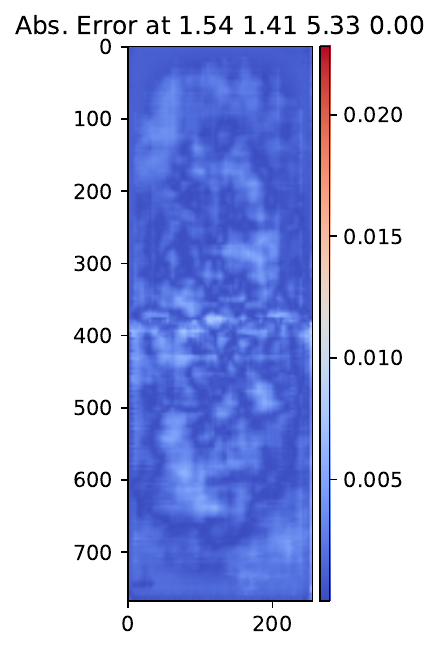} & \includegraphics[width=0.140\textwidth]{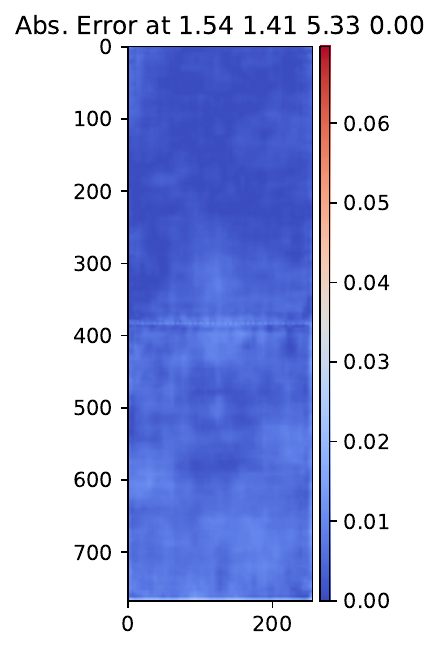} & \includegraphics[width=0.140\textwidth]{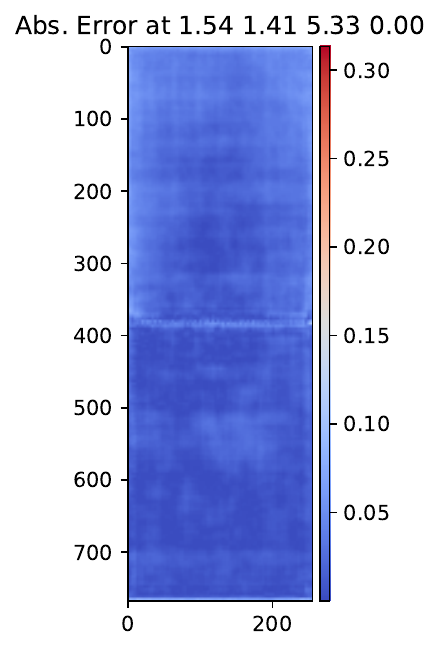} & \includegraphics[width=0.140\textwidth]{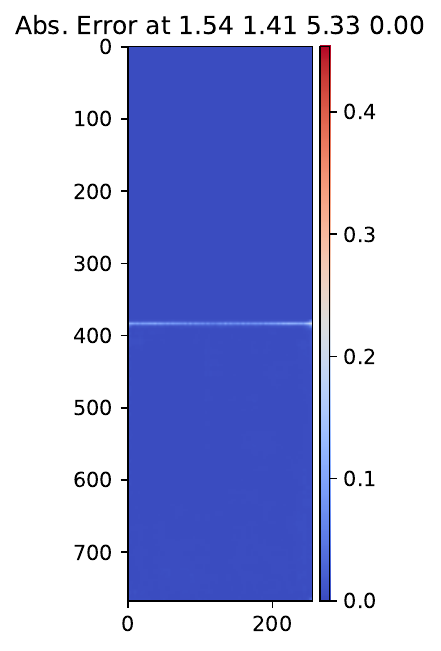} \\
  \end{tabular}
  \caption{
      Figure of predictions, truth, and absolute error for Rayleigh-Taylor predictions at $t$=0.
      }\label{fig:rt_predictions_0}
  \end{center}
\end{figure}

\begin{figure}[!htb]
  \begin{center}
  \begin{tabular}{*{7}{|c|}}
     & density & velocity $x$ & velocity $y$ & pressure & energy & materials \\
  ML prediction & \includegraphics[width=0.140\textwidth]{figs/rt/density_yhat_25.pdf} & \includegraphics[width=0.140\textwidth]{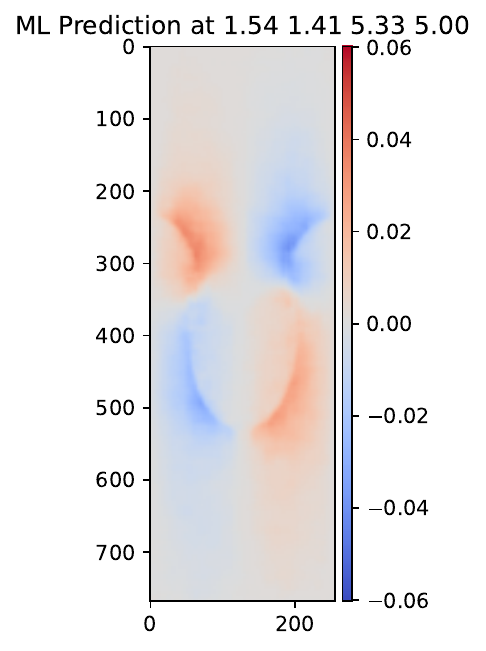} & \includegraphics[width=0.140\textwidth]{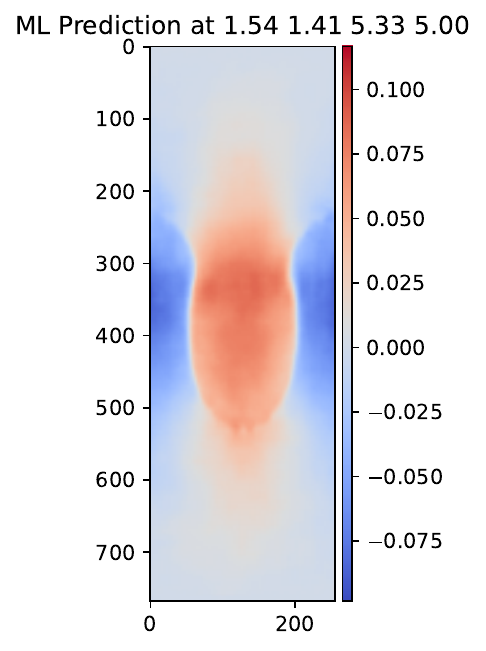} & \includegraphics[width=0.140\textwidth]{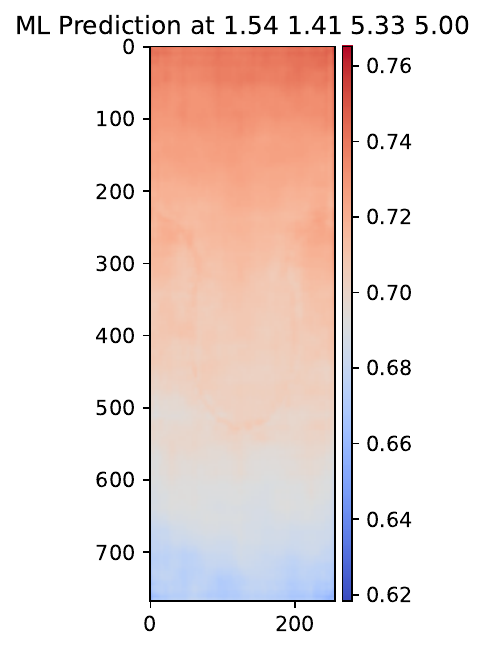} & \includegraphics[width=0.140\textwidth]{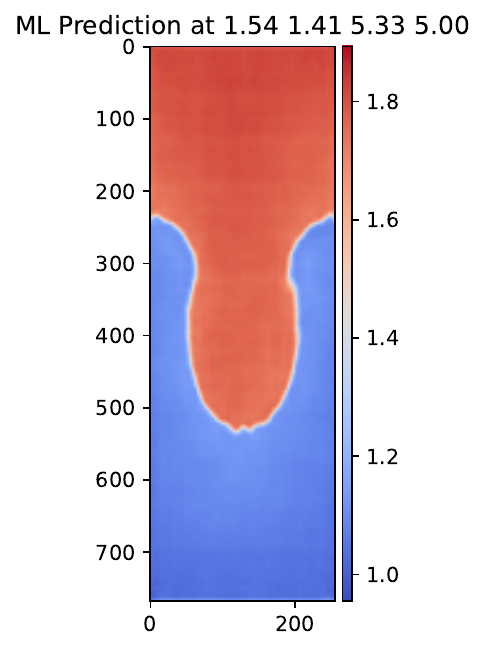} & \includegraphics[width=0.140\textwidth]{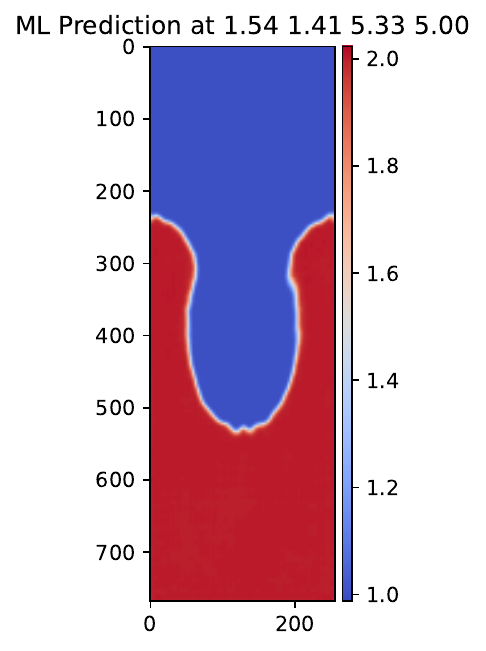} \\
  Simulation & \includegraphics[width=0.140\textwidth]{figs/rt/density_y_25.pdf} & \includegraphics[width=0.140\textwidth]{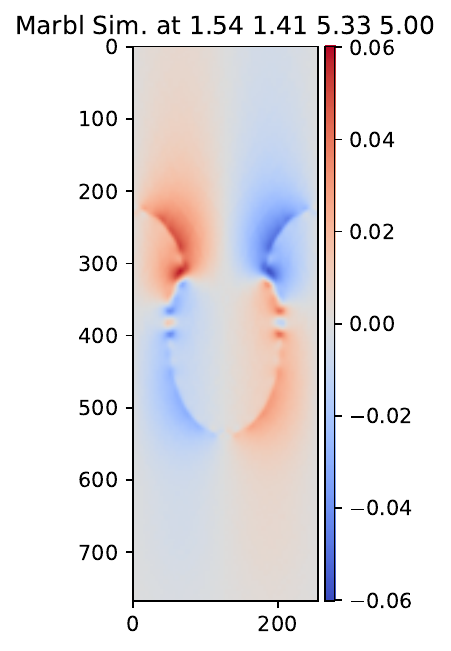} & \includegraphics[width=0.140\textwidth]{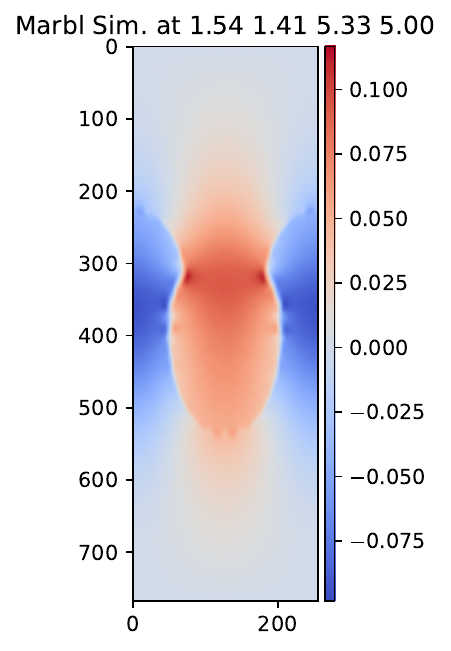} & \includegraphics[width=0.140\textwidth]{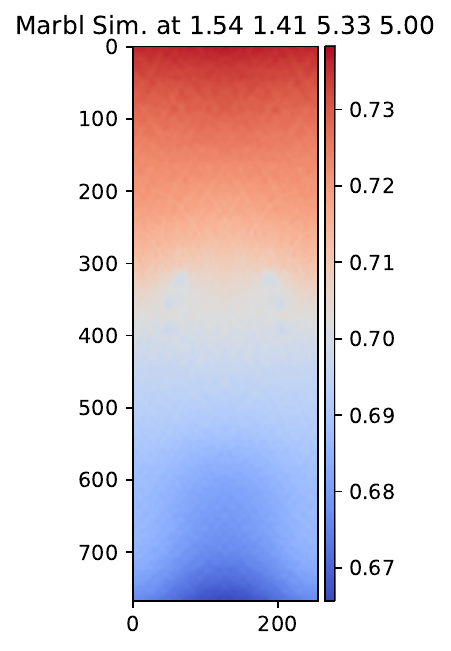} & \includegraphics[width=0.140\textwidth]{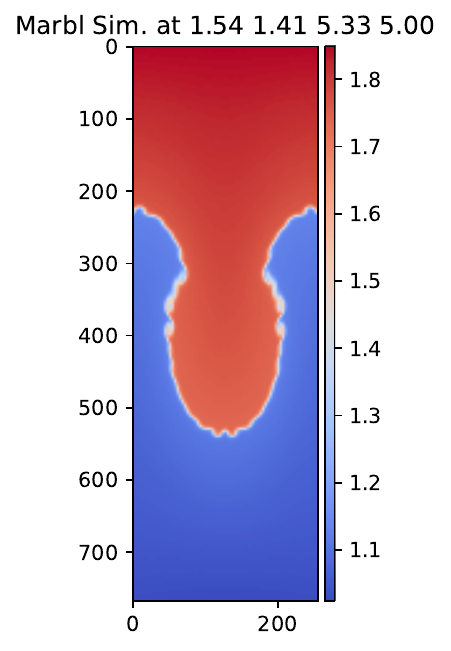} & \includegraphics[width=0.140\textwidth]{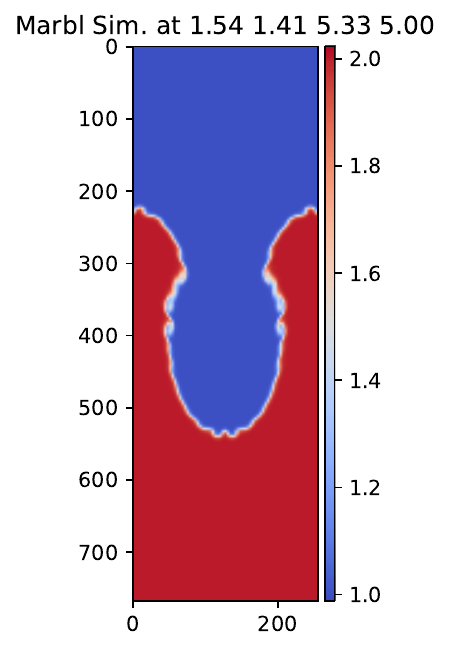} \\
  Abs. Error & \includegraphics[width=0.140\textwidth]{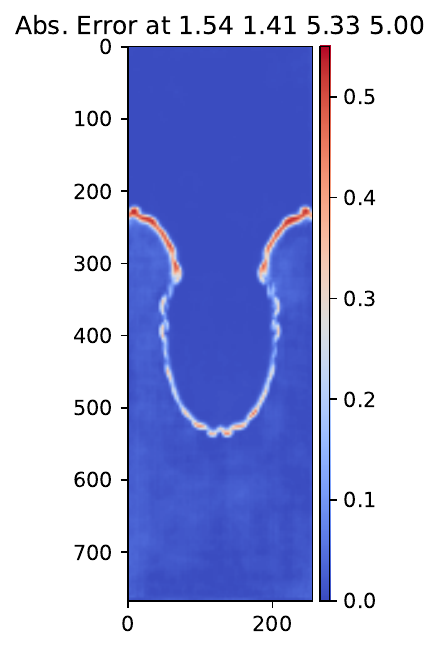} & \includegraphics[width=0.140\textwidth]{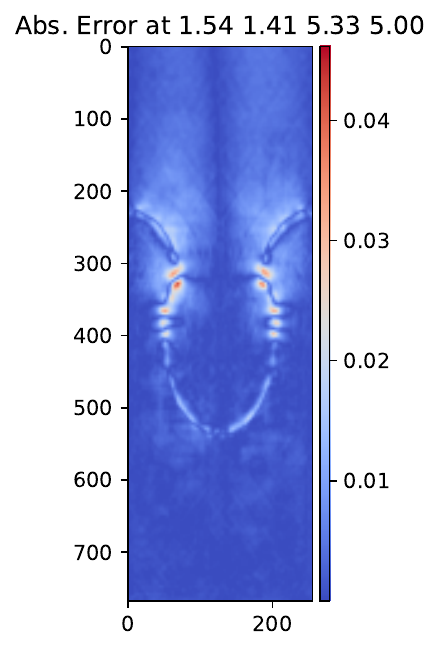} & \includegraphics[width=0.140\textwidth]{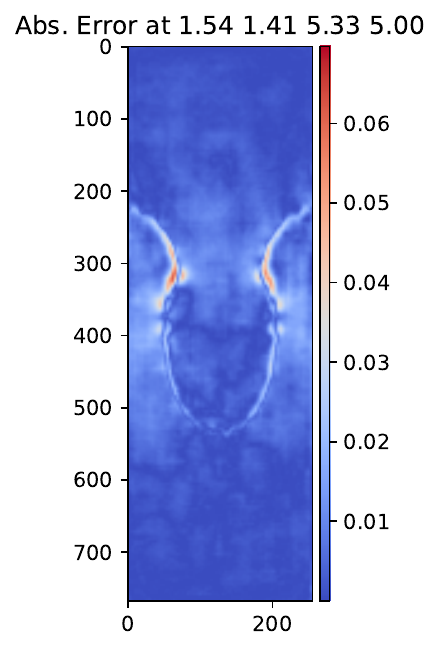} & \includegraphics[width=0.140\textwidth]{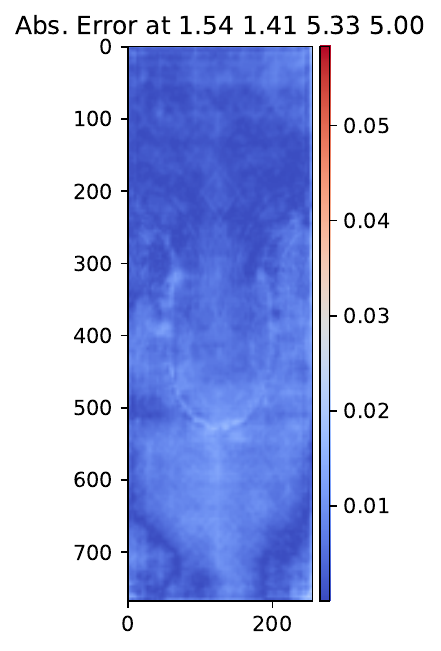} & \includegraphics[width=0.140\textwidth]{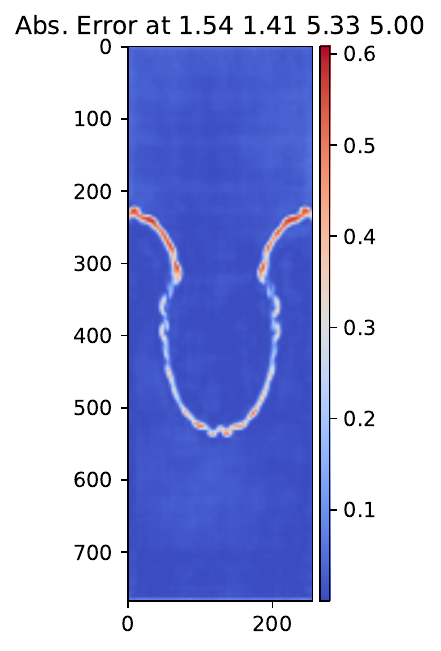} & \includegraphics[width=0.140\textwidth]{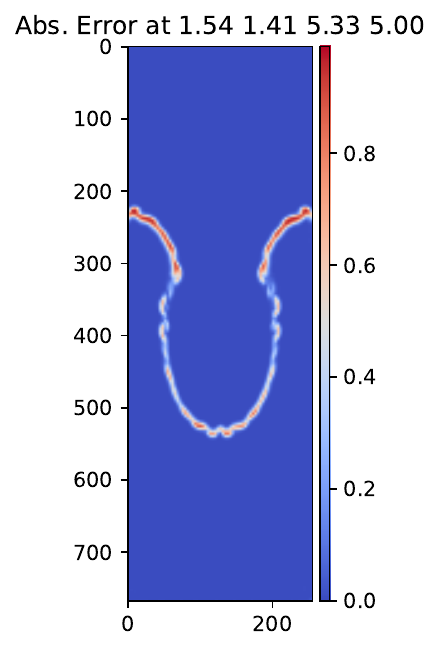} \\
  \end{tabular}
  \caption{
      Figure of predictions, truth, and absolute error for Rayleigh-Taylor predictions at $t$=5.
      }\label{fig:rt_predictions_5}
  \end{center}
\end{figure}

\begin{figure}[!htb]
  \begin{center}
  \begin{tabular}{*{7}{|c|}}
     & density & velocity $x$ & velocity $y$ & pressure & energy & materials \\
  ML prediction & \includegraphics[width=0.140\textwidth]{figs/rt/density_yhat_50.pdf} & \includegraphics[width=0.140\textwidth]{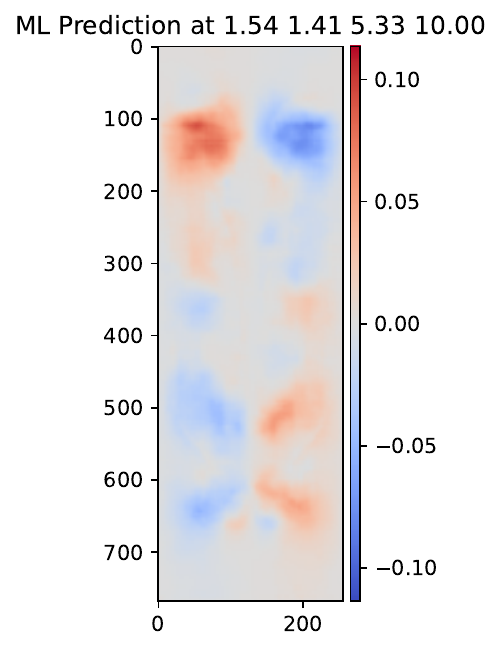} & \includegraphics[width=0.140\textwidth]{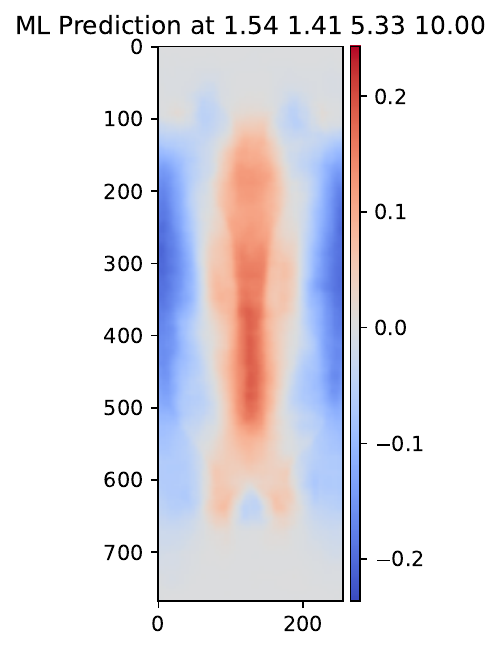} & \includegraphics[width=0.140\textwidth]{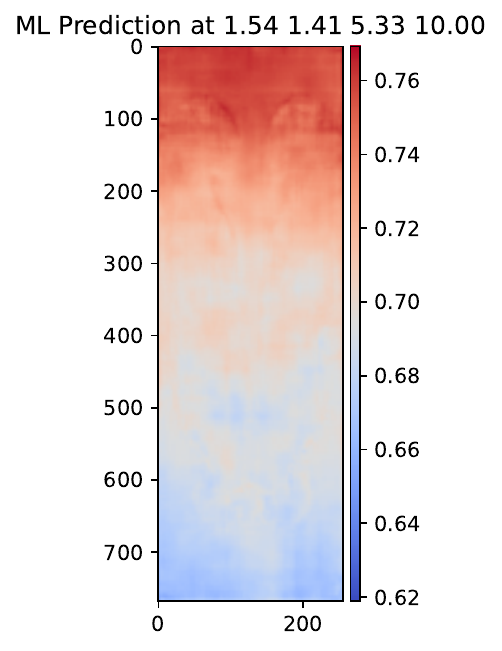} & \includegraphics[width=0.140\textwidth]{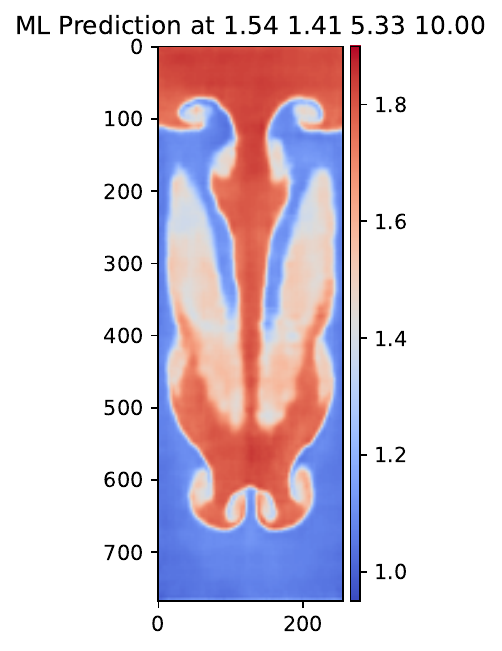} & \includegraphics[width=0.140\textwidth]{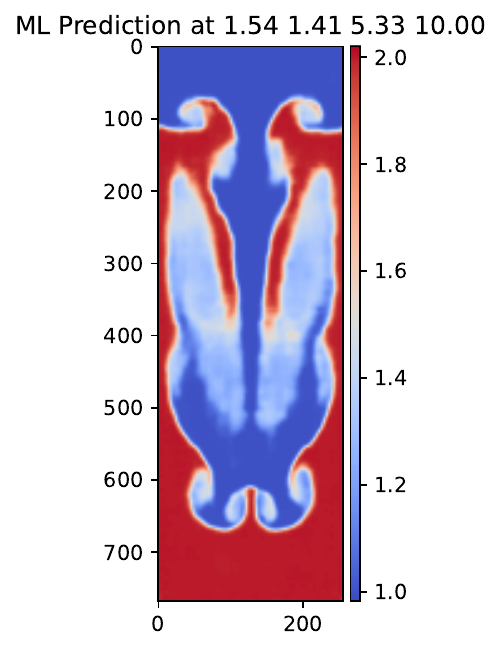} \\
  Simulation & \includegraphics[width=0.140\textwidth]{figs/rt/density_y_50.pdf} & \includegraphics[width=0.140\textwidth]{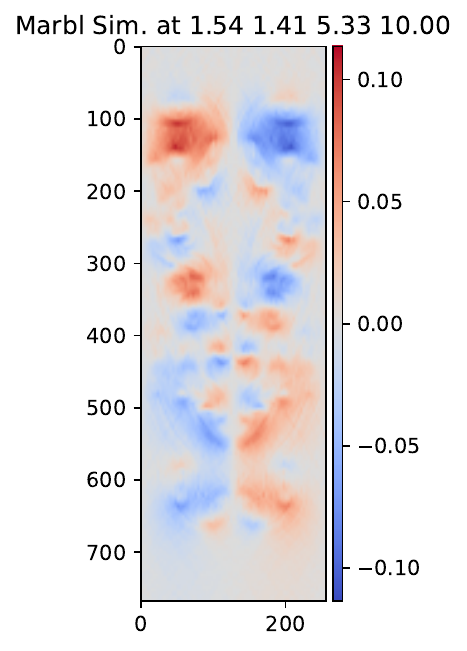} & \includegraphics[width=0.140\textwidth]{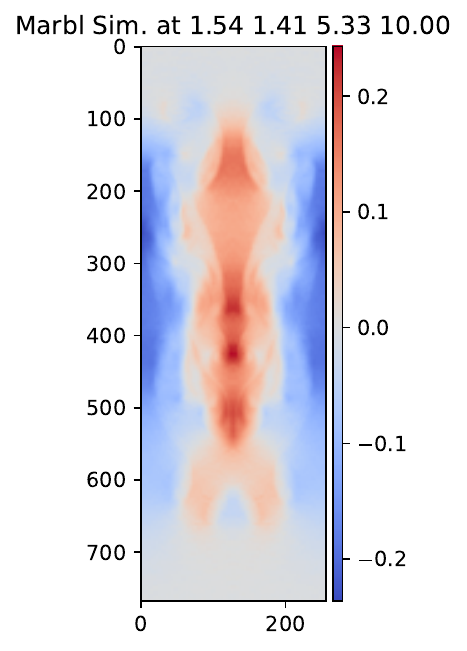} & \includegraphics[width=0.140\textwidth]{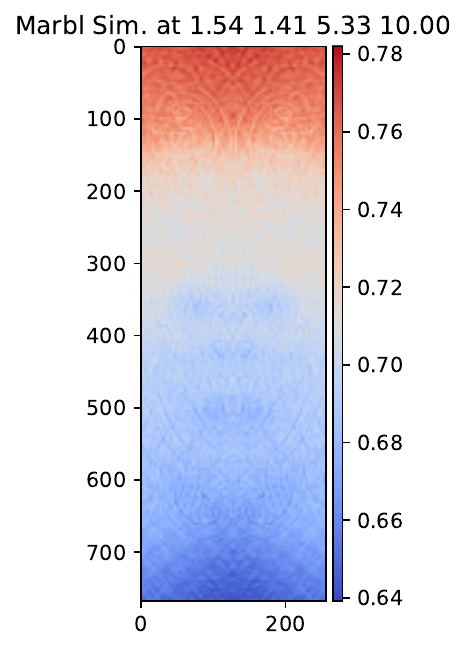} & \includegraphics[width=0.140\textwidth]{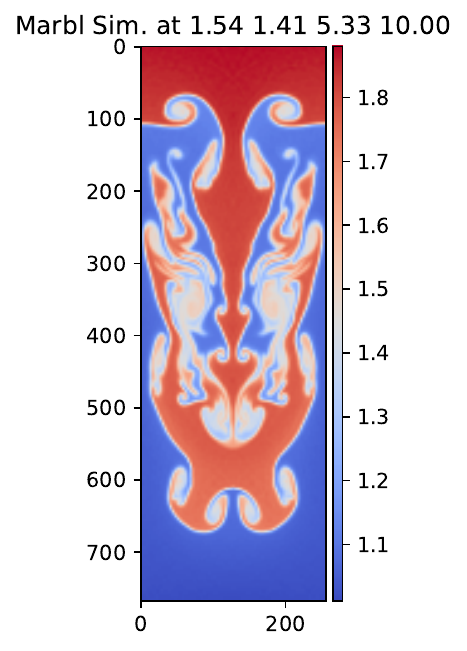} & \includegraphics[width=0.140\textwidth]{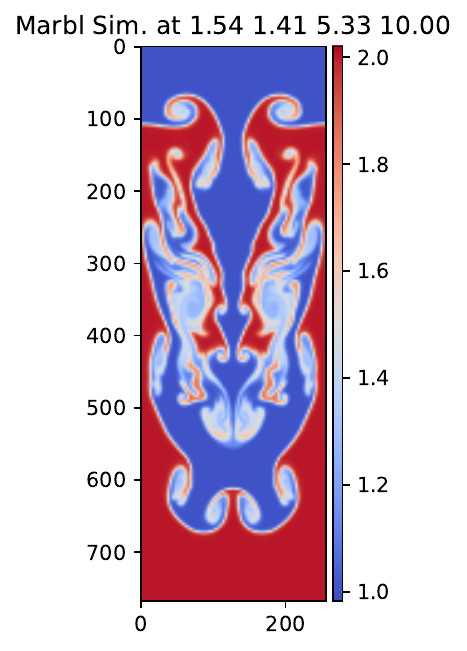} \\
  Abs. Error & \includegraphics[width=0.140\textwidth]{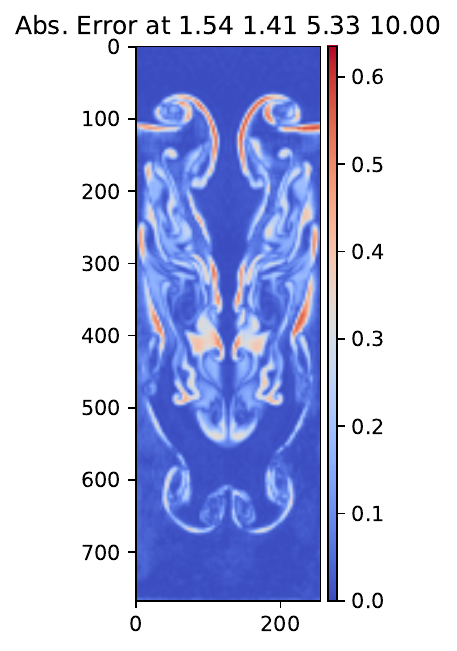} & \includegraphics[width=0.140\textwidth]{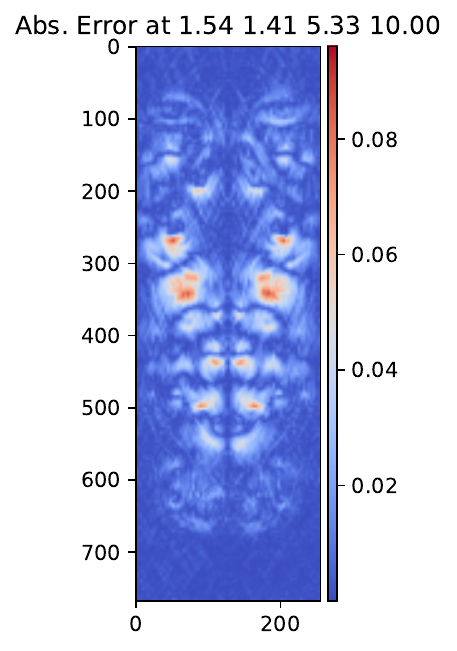} & \includegraphics[width=0.140\textwidth]{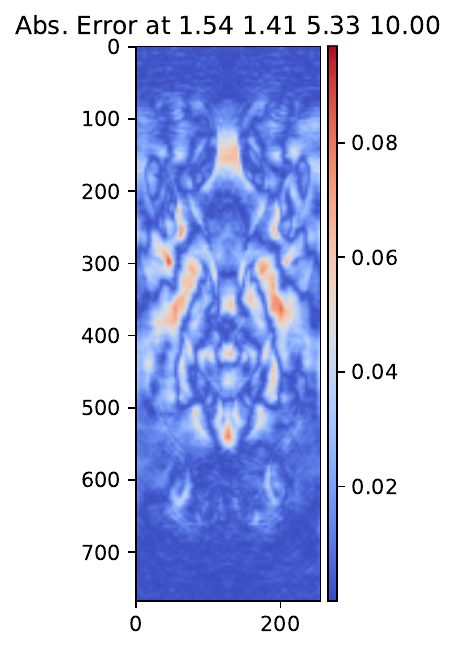} & \includegraphics[width=0.140\textwidth]{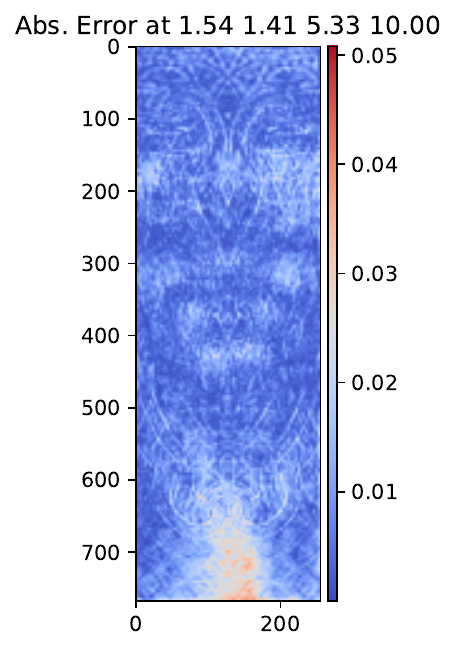} & \includegraphics[width=0.140\textwidth]{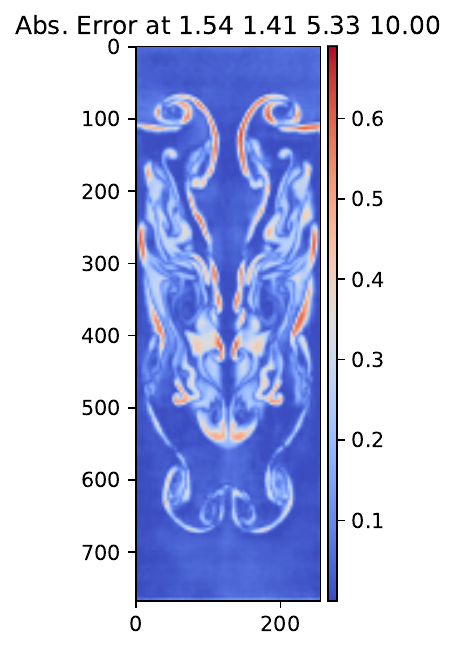} & \includegraphics[width=0.140\textwidth]{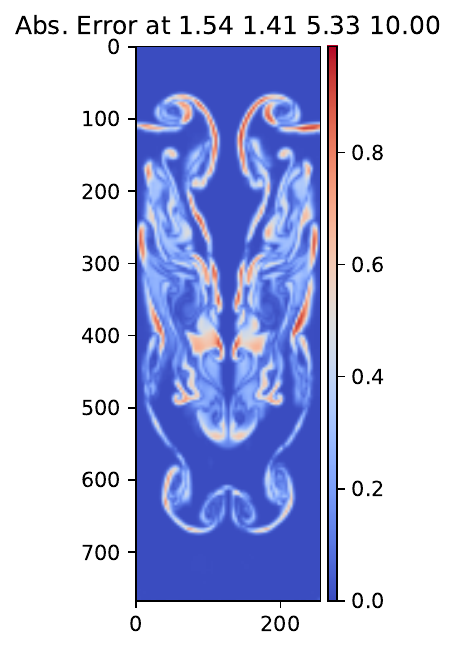} \\
  \end{tabular}
  \caption{
      Figure of predictions, truth, and absolute error for Rayleigh-Taylor predictions at $t$=10.
      }\label{fig:rt_predictions_10}
  \end{center}
\end{figure}

Samples of ML predictions for multiple fields are plotted in Figures~\ref{fig:rt_predictions_0}-\ref{fig:rt_predictions_10} with the simulation results and the absolute error (between simulation and prediction). This was a simulation within the training set, that is representative of the general trends within the Rayleigh-Taylor results. Essentially at early to middle times (Figure~\ref{fig:rt_predictions_0} and Figure~\ref{fig:rt_predictions_5}) we see the ML predictions are excellent at tracking the Rayleigh-Taylor instability. The errors are largest at the interface, which is generally tracked well. However, at late times as shown in Figure~\ref{fig:rt_predictions_10}, a large number of eddies are formed along the interface of the two gases where the ML predictions are less accurate at maintaining a crisp interface. For these high detail and complex predictions, it appears as if the ML model is locally averaging or smoothing the complicated interface.

\FloatBarrier

\bibliographystyle{unsrtnat}
\bibliography{references}

\end{document}